\documentclass[prd, aps, twocolumn, preprintnumbers, showpacs, nofootinbib, superscriptaddress, notitlepage]{revtex4-1}
\usepackage{mathrsfs}
\usepackage{amsfonts}
\usepackage{amsmath}
\usepackage{slashed}
\usepackage{array}
\usepackage{verbatim}
\usepackage{epsfig}
\usepackage{graphicx}
\usepackage{color}
\usepackage{hyperref}
\usepackage[dvipsnames]{xcolor}
\usepackage{subfigure}
\usepackage{graphicx}   % figures
\usepackage{color}      % color is used in text
\usepackage{slashed}    % Feynman slash
\usepackage{verbatim}
\usepackage[normalem]{ulem}
\usepackage{rotating}   % to rotate tables
\usepackage{multirow}   % multicolumn and multirow
\usepackage{inputenc}
\usepackage{bm}
\usepackage{booktabs}  % 用于 \toprule, \midrule, \bottomrule
\usepackage{slashed}
\usepackage{multirow}
\usepackage{subfigure}
\usepackage{dsfont}

\allowdisplaybreaks
\usepackage{graphicx}
\usepackage{floatrow}
\usepackage[section]{placeins}
\usepackage{cleveref}
\usepackage{tikz}
\usepackage{natbib} % 用于引用管理

\begin{document}

\title{Lattice QCD Determination of the Collins–Soper Kernel in the Continuum and Physical Mass Limits}

\author{Jin-Xin Tan}
\affiliation{State Key Laboratory of Dark Matter Physics, Key Laboratory for Particle Astrophysics and Cosmology (MOE), Shanghai Key Laboratory for Particle Physics and Cosmology, School of Physics and Astronomy, Shanghai Jiao Tong University, Shanghai 200240, China}
\affiliation{Tsung-Dao Lee Institute, Shanghai Jiao Tong University, Shanghai 201210, China}

\author{Zhi-Chao Gong}
\affiliation{State Key Laboratory of Dark Matter Physics, Key Laboratory for Particle Astrophysics and Cosmology (MOE), Shanghai Key Laboratory for Particle Physics and Cosmology, School of Physics and Astronomy, Shanghai Jiao Tong University, Shanghai 200240, China}
\affiliation{Tsung-Dao Lee Institute, Shanghai Jiao Tong University, Shanghai 201210, China}

\author{Jun Hua}
\email{Corresponding author: junhua@scnu.edu.cn}
\affiliation{State Key Laboratory of Nuclear Physics and Technology, Institute of Quantum Matter, South China Normal University, Guangzhou 510006, China}
\affiliation{Guangdong Basic Research Center of Excellence for Structure and Fundamental Interactions of Matter, Guangdong Provincial Key Laboratory of Nuclear Science, Guangzhou 510006, China}

\author{Xiangdong Ji}
\affiliation{Department of Physics, University of Maryland, College 
Park, MD 20742, USA}

\author{Xiangyu Jiang}
\affiliation{CAS Key Laboratory of Theoretical Physics, Institute of Theoretical Physics, Chinese Academy of Sciences, Beijing 100190, China}

\author{Hang Liu}
\affiliation{Department of Physics, Shanghai Normal University, Shanghai 200234, China}

\author{Andreas Sch\"afer}
\affiliation{Institut f\"ur Theoretische Physik, Universit\"at Regensburg, D-93040 Regensburg, Germany}
\affiliation{Department of Physics, National Taiwan University, Taipei, Taiwan 106, China}

\author{Yushan Su}
\affiliation{Department of Physics, University of Maryland, College 
Park, MD 20742, USA}

\author{Han-Zhang Wang}
\affiliation{State Key Laboratory of Dark Matter Physics, Key Laboratory for Particle Astrophysics and Cosmology (MOE), Shanghai Key Laboratory for Particle Physics and Cosmology, School of Physics and Astronomy, Shanghai Jiao Tong University, Shanghai 200240, China}
\affiliation{Zhiyuan College, Shanghai Jiao Tong University, Shanghai 200240, China}

\author{Wei Wang}
\email{Corresponding author: wei.wang@sjtu.edu.cn}
\affiliation{State Key Laboratory of Dark Matter Physics, Key Laboratory for Particle Astrophysics and Cosmology (MOE), Shanghai Key Laboratory for Particle Physics and Cosmology, School of Physics and Astronomy, Shanghai Jiao Tong University, Shanghai 200240, China}
\affiliation{Southern Center for Nuclear-Science Theory (SCNT), Institute of Modern Physics, Chinese Academy of Sciences, Huizhou 516000, Guangdong Province, China}

\author{Yi-Bo Yang}
\affiliation{CAS Key Laboratory of Theoretical Physics, Institute of Theoretical Physics, Chinese Academy of Sciences, Beijing 100190, China}
\affiliation{School of Fundamental Physics and Mathematical Sciences, Hangzhou Institute for Advanced Study, UCAS, Hangzhou 310024, China}
\affiliation{International Centre for Theoretical Physics Asia-Pacific, Beijing/Hangzhou, China}
\affiliation{School of Physical Sciences, University of Chinese Academy of Sciences, 
Beijing 100049, China}

\author{Jun Zeng}
\affiliation{College of Physics and Electronic Engineering, Hainan Normal University, Haikou 571158, Hainan, China}

\author{Jian-Hui Zhang}
\affiliation{School of Science and Engineering, The Chinese University of Hong Kong, Shenzhen 518172, China}

\author{Jia-Lu Zhang}
\affiliation{Tsung-Dao Lee Institute, Shanghai Jiao Tong University, Shanghai 201210, China}
\affiliation{State Key Laboratory of Dark Matter Physics, Key Laboratory for Particle Astrophysics and Cosmology (MOE), Shanghai Key Laboratory for Particle Physics and Cosmology, School of Physics and Astronomy, Shanghai Jiao Tong University, Shanghai 200240, China}

\author{Qi-An Zhang}
\email{Corresponding author: zhangqa@buaa.edu.cn}
\affiliation{School of Physics, Beihang University, Beijing 102206, China}

\begin{abstract}
The Collins-Soper (CS) kernel governs the rapidity evolution of transverse-momentum-dependent (TMD) parton distributions, a cornerstone for QCD factorization and linking nucleon structure data across scales. Its nonperturbative behavior at large transverse separations ($b_{\perp}$) remains weakly constrained due to phenomenological model dependencies. We present a first-principles determination of the CS kernel at the continuum limit and physical pion mass from lattice QCD in the large-momentum effective theory framework. Using (2+1)-flavor  configurations (lattice spacings $a \in[0.052, 0.105]$ fm, and pion mass $m_{\pi} \approx ( 136, 230, 300, 320)$ MeV), we simulating the nonlocal equal-time correlation function and extract the  quasi-TMD wave functions. Taking into account systematic  improvements including hypercubic  smearing, nonperturbative renormalization, and a $b_{\perp}$-unexpanded matching kernel, we obtain the CS kernel at the continuum, chiral, and infinite-momentum limits. Our results are determined up to $b_{\perp} \sim 1$ fm, with controllable  uncertainties, and agree with  perturbative QCD at small $b_{\perp}$ and global TMD phenomenological extractions. We conduct a global analysis integrated with phenomenological fits and demonstrate the impact of our results on such fits. This work yields the most precise nonperturbative constraint on the CS kernel’s long-distance behavior from Lattice QCD, which not only bridges Lattice QCD, perturbation theory, and nucleon structure experiments for TMD studies, but also boosts the utility of our constraint for future global TMD analyses.
\end{abstract}

\maketitle

\section{Introduction}

Understanding the three-dimensional (3D) structure of the nucleon is a central objective of modern hadronic physics. While the collinear parton distribution functions (PDFs) provide crucial information on the longitudinal momentum fractions carried by quarks and gluons, they offer only a one-dimensional projection of the nucleon’s internal structure. To achieve a more complete spatial and dynamical understanding, one must go beyond the collinear approximation and explore the transverse-momentum-dependent (TMD) regime. The TMD parton distribution functions (TMDPDFs) and TMD wave functions (TMDWFs) encode not only the longitudinal motion but also the intrinsic transverse dynamics of partons. In particular, TMDPDFs constitute essential theoretical ingredients for describing processes such as semi-inclusive deep-inelastic scattering (SIDIS) and Drell–Yan production, and TMDWFs play a key role in the study of exclusive production processes and hadron structures~\cite{Boussarie:2023izj}.

A central theoretical challenge in TMD physics lies in their rapidity evolution, which supplements the usual ultraviolet (UV) renormalization by connecting different rapidity scales. This evolution is governed by the Collins–Soper (CS) kernel~\cite{Collins:1981uk, Collins:1981va, Collins:2011zzd}, which encapsulates the interplay between soft and collinear dynamics. The CS kernel is perturbatively calculable at small transverse separations and has been determined at three loops and four loops in QCD~\cite{Li:2016ctv, Moch:2017uml, Vladimirov:2016dll, Moult:2022xzt, Duhr:2022yyp}, but becomes intrinsically nonperturbative at large  $b_\perp$ . Its precise determination is therefore crucial for achieving accurate TMD factorization and for linking data from different experiments.

Extensive experimental efforts over the past decades, including SIDIS at HERMES, COMPASS, and Jefferson Lab, as well as Drell–Yan and  W/Z  production at RHIC and the LHC, have provided rich information on TMD-sensitive observables~\cite{HERMES:2009lmz, JeffersonLabHallA:2011ayy, COMPASS:2012dmt, HERMES:2012uyd, Anselmino:2013lza, STAR:2015vmv, COMPASS:2016led, Anselmino:2016uie, HERMES:2020ifk, Liu:2021boj, Cruz-Martinez:2023sdv, ATLAS:2023lsr, Isaacson:2023iui, COMPASS:2023vqt, Niemiec:2024eie}. Global phenomenological analyses~\cite{Scimemi:2019cmh, Bacchetta:2022awv, Moos:2023yfa, Bacchetta:2024qre, Kang:2024dja, Martinez:2024mou, Moos:2025sal, Bacchetta:2025ara,Camarda:2025lbt} have extracted the CS kernel from these data, revealing consistent results in the perturbative region but leaving sizable uncertainties in the nonperturbative domain due to model dependencies and parametrization assumptions.

To overcome these limitations, a first-principles determination of the Collins–Soper  kernel and the soft functions from lattice QCD has become an urgent task. However, direct lattice calculations of light-cone TMDs are prohibited by the presence of (two) lightlike Wilson lines, which cannot be realized in Euclidean space. The large-momentum effective theory (LaMET)~\cite{Ji:2013dva, Ji:2014gla, Ji:2020ect, Ji:2022ezo} provides a practical solution by establishing a systematic factorization between Euclidean and light-cone observables. In this framework, one considers Euclidean equal-time correlators of hadrons boosted to large longitudinal momentum, known as quasi-distributions. These quantities can be directly computed on the lattice and are related to their light-cone counterparts through a perturbative matching at large $P^z$.
Based on this framework, one can further utilize the dependence of quasi-distributions on the momentum $P^z$ to extract the CS kernel nonperturbatively~\cite{Ebert:2018gzl, Ji:2019sxk, Ji:2020ect}. Specifically, the variation of quasi-distributions with respect to $P^z$ encodes the same rapidity evolution that governs the TMDs in continuum QCD. By comparing results at different boosts, the Collins–Soper kernel can be determined from first principles. This approach thus provides a nonperturbative determination of the CS kernel and establishes a unified bridge connecting lattice observables with the phenomenological framework of TMD evolution.

Recent lattice studies~\cite{Ebert:2018gzl, Shanahan:2020zxr, LatticeParton:2020uhz, Schlemmer:2021aij, Shanahan:2021tst, Li:2021wvl, LatticePartonLPC:2022eev, LatticePartonLPC:2023pdv, Shu:2023cot, Avkhadiev:2023poz, Liu:2024sqj, Avkhadiev:2024mgd, Alexandrou:2025xci} have convincingly demonstrated the feasibility of this strategy. By implementing improved operator constructions, refined renormalization schemes, and simulations at higher hadron momenta, these works have shown that the Collins–Soper kernel can be reliably extracted from lattice data with steadily increasing precision. The extracted CS kernels exhibit remarkable consistency with perturbative QCD predictions and phenomenological extractions in the overlapping kinematic regions. These developments highlight the growing predictive power of the lattice approach and its promising role in providing model-independent inputs for global analyses of TMD evolution.

However, existing calculations of the CS kernel were limited in precision, particularly at large transverse separations, where the signal rapidly deteriorates, making it difficult to access the long-range correlations relevant to nucleon-size scales. 
A recent study~\cite{Avkhadiev:2024mgd} performed the first continuum-extrapolated calculation at the physical pion mass using multiple lattice spacings in the range $a\in[0.09,0.15]~\mathrm{fm}$. Their continuum extrapolation, however, relied on an explicit parametrization of the $b_\perp$ dependence of the CS kernel, because discretization effects and power corrections could not be disentangled simultaneously at each $b_\perp$ within the available data constraints.
In this work, we employ ensembles with multiple lattice spacings spanning $a\in[0.052,0.105]~\mathrm{fm}$ and multiple pion masses, including the physical one, together with substantially improved statistical precision. This enables controlled continuum and chiral extrapolations and allows us to determine the CS kernel from first principles up to transverse separations of $b_{\perp}\simeq 1~\mathrm{fm}$.  Our results therefore provide, to date, the most precise determination of the long-distance CS kernel and the first continuum-extrapolated extraction achieved without introducing the explicit parametrization of its $b_\perp$ dependence, thereby offering a robust nonperturbative input for future global TMD analyses.

The rest of this paper is organized as follows. In Sec.~\ref{sec:Theoretical Framework}, we present the theoretical formalism, including the definition of quasi-TMDWFs, their factorization properties, and the connection to the CS kernel. In Sec.~\ref{sec:Lattice Simulations and Results}, we provide a detailed description of the lattice calculation of quasi-TMDWFs and the extraction of the CS kernel. We give our framework of the
calculation from correlation functions to the bare quasi-TMDWFs and their subsequent renormalization. In addition, by comparing the renormalized quasi-TMDWFs, we demonstrate the significant improvement in the signal-to-noise ratio achieved through HYP smearing. Using results obtained from ensembles with different lattice parameters, we extract the CS kernel within the LaMET framework and perform extrapolations to the infinite-momentum, continuum, and chiral limits to obtain the final physical results. Finally, in Sec.~\ref{sec:Summary and outlook}, we summarize our main results and discuss their implications for phenomenology, as well as prospects for future theoretical and experimental studies. Some details in the calculation and supplementary results   are collected in the Appendix.

\section{Theoretical Framework}
\label{sec:Theoretical Framework}

In this section, we present the theoretical framework underlying our calculation. 
We begin by defining the CS kernel and discussing its fundamental role in the rapidity evolution of transverse-momentum-dependent observables. 
We then introduce the quasi-TMDWFs within the LaMET framework and illustrate their factorization properties, highlighting how they enable the extraction of the CS kernel from lattice QCD.

\subsection{Collins-Soper Kernel and Rapidity Evolution}
CS kernel is introduced to characterize the rapidity evolution of TMD distributions.
Rapidity divergences arise from gluon radiation that is collinear to the lightlike gauge links, leading to an unbounded growth in the large-rapidity region and necessitating additional parameters and mechanisms to control this dependence. 
Within the TMD framework, the CS kernel $K(b_\perp, \mu)$ functions as the rapidity anomalous dimension, governing the evolution of TMDs across different rapidity scales, and is defined as~\cite{Collins:1981uk, Collins:1981va, Collins:2011zzd, Ji:2004wu}
\begin{align}
    K(b_\perp, \mu) = 2 \zeta \frac{d}{d \zeta} \ln f^{\mathrm{TMD}}(x, b_\perp, \mu, \zeta).
\end{align}
where $f^{\mathrm{TMD}}(x, b_\perp, \mu, \zeta)$ denotes physical TMD distribution with longitudinal momentum fraction $x$ and transverse displacement $b_{\perp}$, at rapidity scale $\zeta$, and renormalization scale $\mu$.

The CS kernel governs the universal rapidity evolution of transverse-momentum-dependent  distributions, connecting the soft and collinear sectors in QCD. At small transverse separations $b_\perp$, it can be reliably computed perturbatively and is currently known at three and four loops~\cite{Li:2016ctv, Moch:2017uml, Vladimirov:2016dll, Moult:2022xzt, Duhr:2022yyp}. In contrast, at large $b_\perp$, the CS kernel encodes essential nonperturbative physics that cannot be accessed perturbatively. A precise determination of $K(b_\perp, \mu)$ is therefore crucial, both for relating TMD observables measured at different energy scales and for achieving high-precision resummation in QCD phenomenology.

The scale dependence of the CS kernel can be described by the renormalization group equation (RGE)~\cite{Collins:1981uk, Collins:1981uw, Collins:1984kg, Collins:2011zzd, Becher:2010tm}:
\begin{align}
    \frac{d}{d\operatorname{ln}\mu} K(b_\perp, \mu) = -2\, \Gamma_{\rm cusp}(\alpha_s), 
\end{align}
where $\Gamma_{\rm cusp}$ denotes the cusp anomalous dimension. 
This equation reflects that the ultraviolet behavior of the CS kernel is governed by a universal quantity, $\Gamma_{\rm cusp}$, which also appears in the evolution of Wilson lines and other QCD observables. 
Physically, the CS kernel characterizes how TMD distributions evolve with {changes in} the rapidity scale, providing a universal evolution mechanism that connects TMDs defined at different rapidity regimes and ensuring the consistency of QCD factorization across energy scales.

\subsection{Factorization in LaMET}

Within the LaMET framework, the infrared (IR) behavior of quasi-distributions can be systematically factorized and matched to their corresponding light-front (LF) quantities. {In particular, the unsubtracted TMDs, $f(x, b_\perp, \mu, \zeta)$, together with the intrinsic soft function $S_I(b_\perp, \mu)$, } can be factorized in terms of the quasi-TMDs computed at finite hadron momentum on the lattice~\cite{Ji:2019ewn, Ji:2021znw, LatticePartonCollaborationLPC:2022myp, LatticeParton:2024mxp, LatticeParton:2025eui}:
\begin{align}
    &\frac{f(x, b_\perp, \mu, \zeta)}{\sqrt{S_I(b_\perp, \mu)}}=\exp\left[- \frac{1}{2} K(b_\perp, \mu) \ln\left( \frac{- \zeta_z + i\epsilon}{\zeta} \right) \right] \notag\\
    &\qquad\times H(\zeta_z, \bar{\zeta}_z, b_\perp, \mu)\tilde{f}(x, b_\perp, \mu, \zeta_z) + \mathcal{O}\left( \frac{\Lambda_{\text{QCD}}^2}{\zeta_z}, \frac{1}{b_\perp^2 \zeta_z} \right), \label{factorizaton of quasi-TMDWFs}
\end{align}
where $\zeta = (2xP^+)^2$ and  $\zeta_z = (2xP^z)^2$ denotes the rapidity scales of the TMDs and quasi-TMDs, respectively. Their evolution is connected through the exponential factor involving the CS kernel $K(b_\perp, \mu)$. $H(\zeta_z, \bar{\zeta}_z, b_\perp, \mu)$ denotes the perturbative matching kernel, while the terms $\mathcal{O}(\Lambda_{\text{QCD}}^2/\zeta_z, 1/(b_\perp^2\zeta_z))$ represent higher-twist corrections arising from finite-momentum and nonperturbative effects, which vanish in the large-$P^z$ limit.

Eq.(\ref{factorizaton of quasi-TMDWFs}) thus provides the theoretical foundation for extracting the CS kernel from lattice QCD. By computing the quasi-TMDs at multiple hadron momenta and investigating their dependence on $\zeta_z$, one can eliminate the $\zeta_z$-independent factors, $f(x, b_\perp, \mu, \zeta)$ and $S_I(b_\perp, \mu)$, from the ratio of quasi-TMDs at different $\zeta_z$. The CS kernel can then be extracted as
\begin{align}
K(b_\perp, &\mu, x, P_1^z, P_2^z)=\frac{1}{\ln(P_1^z/P_2^z)}\notag \\
&\times\left[\ln\frac{H(\zeta_{z, 2}, \bar{\zeta}_{z, 2}, b_\perp, \mu)\tilde f(x, b_\perp, \mu, \zeta_{z, 1})}{H(\zeta_{z, 1}, \bar{\zeta}_{z, 1}, b_\perp, \mu)\tilde f(x, b_\perp, \mu, \zeta_{z, 2})}\right] \notag \\
&+ \mathcal{O}\!\left( \frac{\Lambda_{\text{QCD}}^2}{\zeta_z}, \, \frac{1}{b_\perp^2 \zeta_z} \right)
\label{cs kernel of ratio}
\end{align}
with condition $P_1^z\neq P_2^z\gg1/b_\perp$. At large $P^z$, power corrections are  suppressed by powers of  $1/(P^z)^2$, as will be discussed in Sec.~\ref{subsec:lattice CS kernel}.

It should be noted that the rapidity evolution of TMDs is universally valid for both TMDPDFs and TMDWFs. Consequently, Eq.(\ref{cs kernel of ratio}) can be applied to extract the CS kernel either from the quasi-beam function or from the unsubtracted quasi-TMDWF. Previous studies have shown that the quasi-TMDWF, which extracted from two-point correlation functions \cite{LatticeParton:2020uhz, Li:2021wvl, LatticePartonLPC:2022eev, LatticePartonLPC:2023pdv, Avkhadiev:2023poz, Avkhadiev:2024mgd, Alexandrou:2025xci}, exhibits a significantly better signal-to-noise ratio than the quasi-beam function obtained from three-point functions \cite{Shanahan:2020zxr, Schlemmer:2021aij, Shanahan:2021tst, Shu:2023cot}. Therefore, in this work, we choose to determine the CS kernel using the pion quasi-TMDWF.

\begin{figure}[http]
\centering
\includegraphics[width=0.95\textwidth]{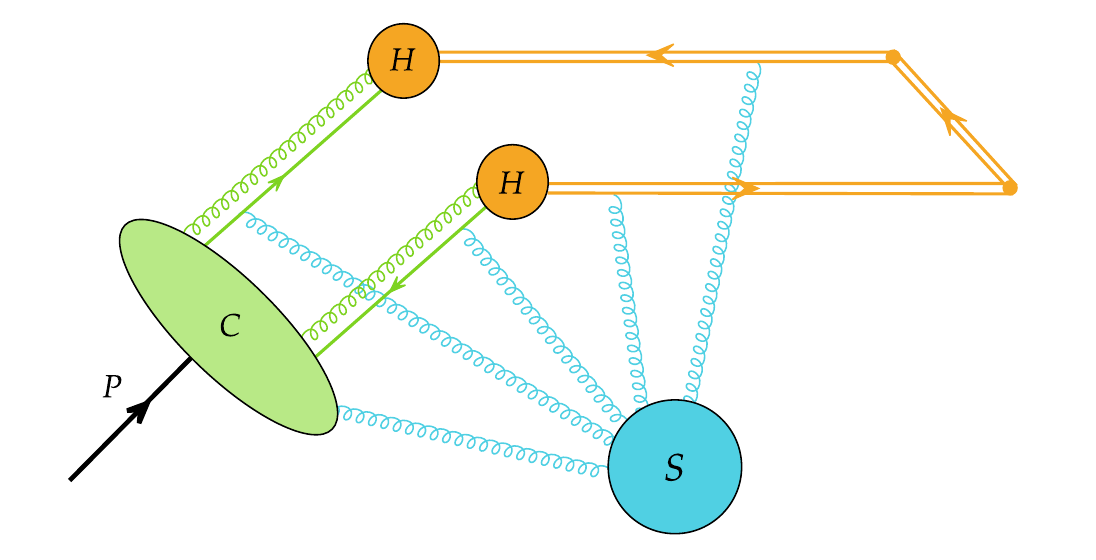}
\caption{The leading-power reduced graph for pseudoscalar meson quasi-TMDWFs.} 
\label{fig:factorization of TMD}
\end{figure}

Furthermore, the factorization structure of the quasi-TMDWFs can be visualized through the leading-power reduced diagram shown in Fig.~\ref{fig:factorization of TMD}. In this picture, the quasi-TMDWF is decomposed into distinct collinear ($C$), soft ($S$), and hard ($H$) sub-diagrams, each corresponding to a specific momentum region in QCD.
In contrast to processes such as SIDIS or Drell–Yan, the hard-gluon exchange between the quark and antiquark inside a meson is power-suppressed at large momentum.
Consequently, the two hard sub-diagrams become disconnected, and the quasi-TMDWF amplitude factorizes multiplicatively into the hard, collinear, and soft components~\cite{Ji:2019ewn, Deng:2022gzi}.
This feature naturally leads to the multiplicative structure of the LaMET factorization formula given in Eq.(\ref{factorizaton of quasi-TMDWFs}).

\subsection{Definition of Quasi-TMD Wave Functions}

The quasi-TMDWF for a meson boosted to a large momentum $P^z$ along the $z$-direction is defined as
\begin{align}
\tilde{f}\left(x, b_\perp, \mu, \zeta_z\right)&=\lim_{L\to\infty}\int\frac{dz}{2\pi}e^{i \left(x-\frac{1}{2}\right) zP^z} \notag \\
&\times \frac{\tilde{\Phi}^{0}\left(z, b_\perp, P^z, L\right)}{Z_O(1/a, \mu)\sqrt{Z_E(2L+z, b_\perp)}}, 
\label{definition of quasi-tmdwf in mom-space}
\end{align}
where the bare matrix element $\tilde{\Phi}^{0}\left(z, b_\perp, P^z, L\right)$ is constructed from a gauge-invariant nonlocal operator sandwiched between the vacuum and a highly boosted meson state~\cite{LatticePartonLPC:2021gpi, LatticePartonLPC:2022eev, LatticePartonLPC:2023pdv}
\begin{align}
    \tilde{\Phi}^{0}\left(z, b_\perp, P^z, L\right) &= \langle 0 | \overline{q}(z\hat{n}_{z}/2 + b_{\perp}\hat{n}_{\perp}) \Gamma \notag\\
    &\times U_{\sqsupset}(L, z, b_\perp)  q(-z\hat{n}_{z}/2) | P^{z} \rangle.
\label{matrix elements}    
\end{align}
where $U_{\sqsupset}$ denotes the staple-shaped Wilson link connecting the quark and antiquark fields, consisting of both longitudinal ($z$) and transverse ($b_\perp$) spatial separations on the Euclidean lattice
\begin{align}
        &U_{\sqsupset}(L, z, b_\perp) \equiv U_{z}^\dagger\left[\frac{z}{2}\hat{n}_z+b_\perp \hat{n}_\perp;\left(L+\frac{z}{2}\right)\hat{n}_z+b_\perp \hat{n}_\perp\right] \notag \\ 
    &\qquad \times U_{\perp}\left[\left(L+\frac{z}{2}\right)\hat{n}_z+b_\perp \hat{n}_\perp;\left(L+\frac{z}{2}\right)\hat{n}_z\right]  \notag \\ 
    &\qquad \times U_{z}\left[\left(L+\frac{z}{2}\right)\hat{n}_z;-\frac{z}{2}\hat{n}_z\right], 
    \label{eq:Wilson_link_define}
\end{align}
with the Euclidean gauge link along the $\mu$-direction is defined as
\begin{align}
    U_{\mu}(\vec{\xi}+\lambda\hat{n}_{\mu};\vec{\xi})\equiv \mathcal{P}\exp\left[ -ig\int_0^1 dt \lambda \hat{n} \cdot A(\vec{\xi} + t \lambda \hat{n}_{\mu}) \right].
\end{align}

For a highly boosted pseudoscalar meson state, we employ an interpolating operator with the Dirac structure $\Gamma = \gamma^t\gamma_5$ to construct the quasi-TMDWFs. At sufficiently large hadron momenta, the leading-twist component of the boosted correlator is dominated by the $\gamma^+\gamma_5$ structure, which has a strong overlap with $\gamma^t\gamma_5$ in the large-momentum limit \cite{Zhang:2025hyo}. In our simulations, the pion momenta reach up to $P^z \sim 10m_\pi$ on each ensemble, corresponding to a sizable Lorentz boost, ensuring that the $\gamma^t\gamma_5$ projection accurately captures the leading-twist contribution.

The linear divergence in the bare matrix element $\tilde{\Phi}^{0}$ originates from the self-energy correction of the Wilson link and can be removed by dividing by a Wilson link with identical geometry. To preserve gauge symmetry, we introduce a rectangular Wilson loop that is exactly formed by gluing together two identical staple-shaped Wilson links into a rectangle of longitudinal extent $2L+z$ and transverse width $b_\perp$. We define
\begin{align}
    &Z_E(2L+z, b_\perp) = {\frac{1}{N_c}} \left\langle  
    U_z\left[ (L+\frac{z}{2})\hat{n}_z ; (-L-\frac{z}{2})\hat{n}_z \right] \right. \notag \\
    & \quad \times U_{\perp}^{\dagger} \left[ (-L-\frac{z}{2})\hat{n}_z + b_\perp \hat{n}_\perp; (-L-\frac{z}{2})\hat{n}_z \right] \notag \\
    & \quad \times U_z^{\dagger}\left[(L+\frac{z}{2})\hat{n}_z +b_\perp \hat{n}_\perp; (-L-\frac{z}{2})\hat{n}_z +b_\perp \hat{n}_\perp\right] \notag \\
    & \quad \times  \left. U_{\perp}\left[ (L+\frac{z}{2})\hat{n}_z +b_\perp \hat{n}_\perp; (L+\frac{z}{2})\hat{n}_z \right] \right\rangle.
\label{eq:wilson_loop_define}    
\end{align}
By construction, the square root $\sqrt{Z_E}$ furnishes precisely the staple-shaped Wilson-line renormalization factor and thus cancels the linear divergence of $\tilde{\Phi}^{0}$.

Besides, the logarithmic divergence arises from vertex corrections involving the Wilson line and the light-quark field. We introduce the corresponding renormalization factor $Z_O$, which also serves to connect the lattice regularization used for quasi-TMDWFs to the perturbative $\overline{\text{MS}}$ scheme. The parametrization of $Z_O$ follows Refs.~\cite{Ji:1991pr, LatticePartonLPC:2021gpi}:
\begin{align}
    \operatorname{ln}Z_O(1/a, \, \mu)  =& \frac{\gamma_0}{\beta_0}\operatorname{ln} \biggl[ \operatorname{ln}[1/(a\Lambda_{\mathrm{QCD}} )]   \biggr] \notag\\
     &+ \frac{c_1}{\operatorname{ln}[1/(a\Lambda_{\mathrm{QCD}} )]} + d'(\mu), 
\end{align}
where $a$ denotes the lattice spacing. The first two terms capture the characteristic logarithmic behavior, with $\gamma_0$ the anomalous dimension of the operator and $\beta_0$ the QCD $\beta$-function coefficient, and the constant $c_1$ is determined from lattice data. All explicit $\mu$-dependence is absorbed into the finite term $d'(\mu)$. Although $Z_O$ cancels in the extraction of the CS kernel, we still determine it in order to obtain renormalized quasi-TMDWFs and to enable meaningful comparisons under different kinematic and lattice conditions.

\section{Lattice Simulation and Result}
\label{sec:Lattice Simulations and Results}

\subsection{Lattice Setup}

%%%%%%%%%%%%%%%%%%%%%%%%%%%%%%%%%%%%%%%%
\begin{table*}[ht]
    \centering
    \renewcommand{\arraystretch}{1.5}
      \setlength{\tabcolsep}{3mm}
    \begin{tabular}{c c c c c c c}
        \hline\hline
        Ensemble & $a$ (fm) & $n_s^3 \times n_t$  & $m_{\pi}$ (MeV)  &$P^z$ (GeV) & $N_{\mathrm{cfg}}\times N_{\mathrm{meas}}$ \\
         \hline
        C32P29  & 0.10530 & $32^3 \times 64$  & 292.4 &1.47, ~1.84, ~2.21  & $984\times 4$ \\

        C32P23 & 0.10530 & $32^3\times64$ &228.0 &1.47, ~1.84, ~2.21 &$448\times10$\\

        C48P14 & 0.10530 & $48^3\times96$ &135.5 &0.98, ~1.22, ~1.47 &$204\times24$\\
       
        F32P30  & 0.07746 & $32^3  \times 96$  & 303.2  &2.00, ~2.50, ~3.00 & $1153\times 6$\\
        
        H48P32  & 0.05187 & $48^3 \times 144$  & 317.2  &1.99, ~2.49, ~2.99 & $550\times 8$\\
        \hline
    \end{tabular}
    \caption{Details of the parameters used for the calculations on each ensembles. The number of gauge configurations is denoted as $N_{\mathrm{cfg}}$, and for each configuration, $N_{\mathrm{meas}}$ measurements are performed.}  
    \label{tab:conf inf}
\end{table*} 
%%%%%%%%%%%%%%%%%%%%%%%%%%%%%%%%%%%%%%%%

The numerical simulations in this work are performed using the gauge configurations generated by the China Lattice QCD (CLQCD) collaboration, employing $N_f=2 + 1$ flavor stout smeared clover fermions and the Symanzik gauge action \cite{CLQCD:2023sdb}. One step of stout link smearing is applied to the gauge field used in the clover action to enhance the stability of the pion mass for a given bare quark mass. 
We utilize multiple ensembles with lattice spacings $a=\{0.05187, ~0.07746, ~0.1053\}\rm\, fm$ and pion masses $m_\pi \simeq \{136, ~230, ~300, ~320\}\rm MeV$. 
To explore the large-momentum limit and investigate the momentum dependence, we employ three different momenta, $P^z\in\{4, 5, 6\}\times{2\pi}/{(n_s a)} \rm \, GeV$ for each ensemble. The detailed simulation parameters are listed in Table \ref{tab:conf inf}.

To calculate the nonlocal matrix elements in Eq.~(\ref{matrix elements}), we generate the wall-source propagator
\begin{align}
    S_{w}\left( \vec{x}, t, t^{\prime};\vec{p}\right) = \sum_{\vec{y}} S\left( t, \vec{x};t^{\prime}, \vec{y}\right) e^{i\vec{p}\cdot(\vec{y}-\vec{x})}, 
\end{align}
on the Coulomb gauge-fixed configurations. This propagator starts from the source at $t'$, sums over all spatial positions $\vec{y}$, and propagates to the sink at $(\vec{x}, t)$, carrying momentum $\vec{p}$.

%%%%%%%%%%%%%%%%%%%%%%%%
\begin{figure}[htbp]
\centering
\includegraphics[width=0.95\textwidth]{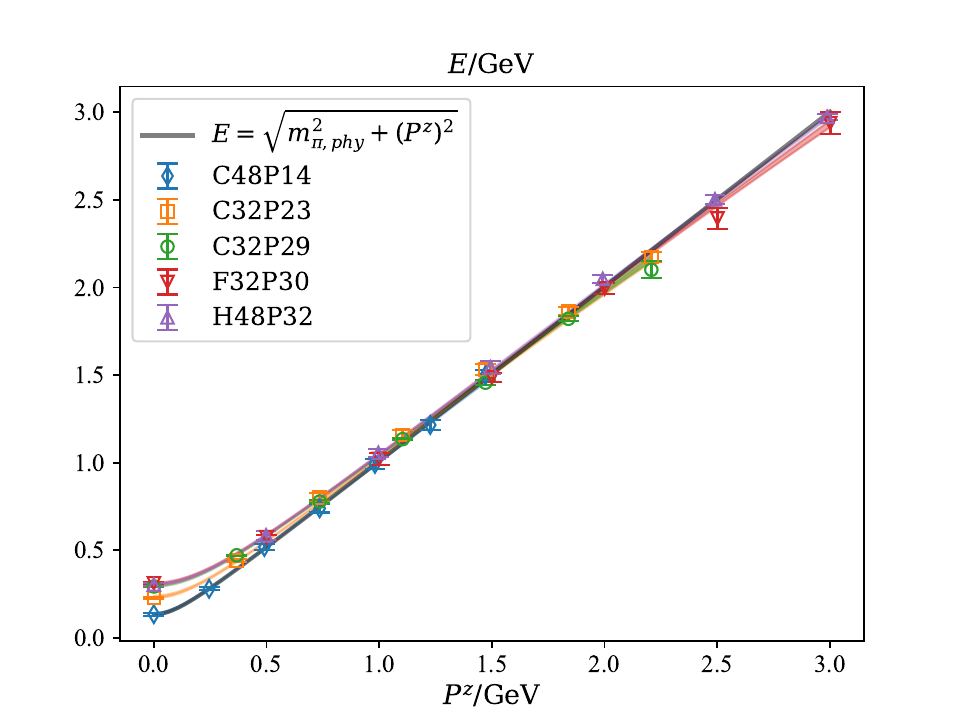}
\caption{The dispersion relation of the pion state. The data up to 3~GeV can be described with the formula in Eq.(\ref{eq:dispersion relation}) with $m_\pi=136.0(7.2)$, $b_1=1.0034 (83)$ and $b_2=-0.042 (14)$.} 
\label{fig:dispersion}
\end{figure}

%%%%%%%%%%%%%%%%%%%%%%%%

By extracting the ground-state energies $E(P^z)$ at different momenta, we determine the pion dispersion relation to ensure that lattice discretization errors are under control. The dispersion relation is parametrized as
\begin{align}
    E(P^z) = \sqrt{m_\pi^2 + b_1 (P^z)^2 +  b_2  (P^z)^4 a^2}, 
    \label{eq:dispersion relation}
\end{align}
where $m_\pi$ denotes the pion mass, $b_1$ accounts for deviations from continuum relativistic kinematics, and the term $(P^z)^4a^2$ parametrizes discretization effects. Fig.~\ref{fig:dispersion} shows the pion dispersion relation across the five ensembles.
The joint fit gives results of $m_{\pi}=136.0(7.2)$MeV, consistent with the physical pion mass. The value $b_1=1.0034 (83)$ is in agreement with the square of the speed of light ($c^2=1$ at the natural units) within the error range, and $b_2=-0.042 (14)$ indicates small discretization effects. The fit quality is characterized by $\chi^2=0.94$.

To improve the statistical quality of the Wilson loop and nonlocal correlation functions, we apply the hypercubic (HYP) smearing technique~\cite{Hasenfratz:2001hp}. This method modifies the gauge links by replacing each original link with a new link constructed from its surrounding staple-shaped neighbors. Through iterative averaging of the gauge fields within local hypercubic neighborhoods, HYP smearing effectively suppresses short-distance ultraviolet fluctuations while preserving the long-distance physics and maintaining gauge invariance. As a result, the technique significantly enhances the signal-to-noise ratio, and stabilizes operator renormalization and matching. It is worth noting that, as discussed later, the renormalized matrix elements are found to be insensitive to the number of HYP smearing steps applied.

\subsection{Quasi-TMD Wave Functions}

\subsubsection{Correlation Functions}

%%%%%%%%%%%%%%%%%%%%%%%%%%%%%%%%%%%%%%%%
\begin{figure*}[http]
\centering
\includegraphics[width=0.48\textwidth]{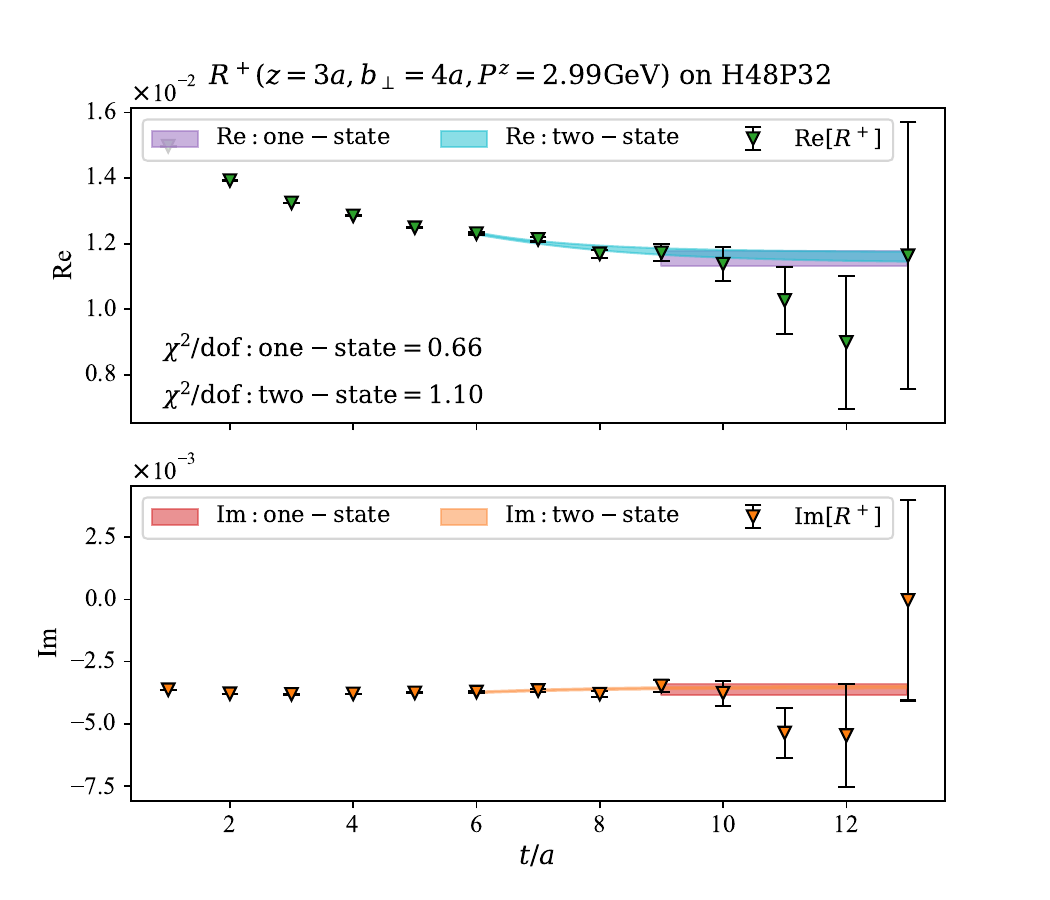}
\includegraphics[width=0.48\textwidth]{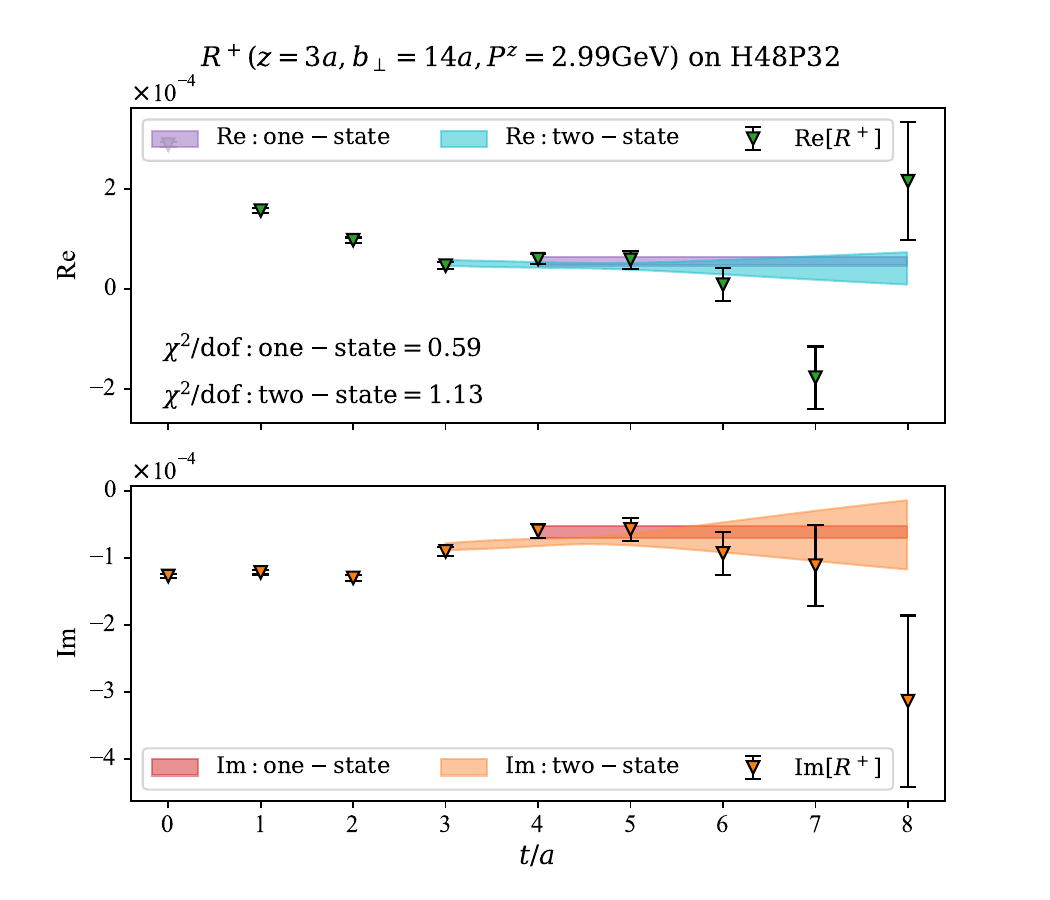}
\includegraphics[width=0.48\textwidth]{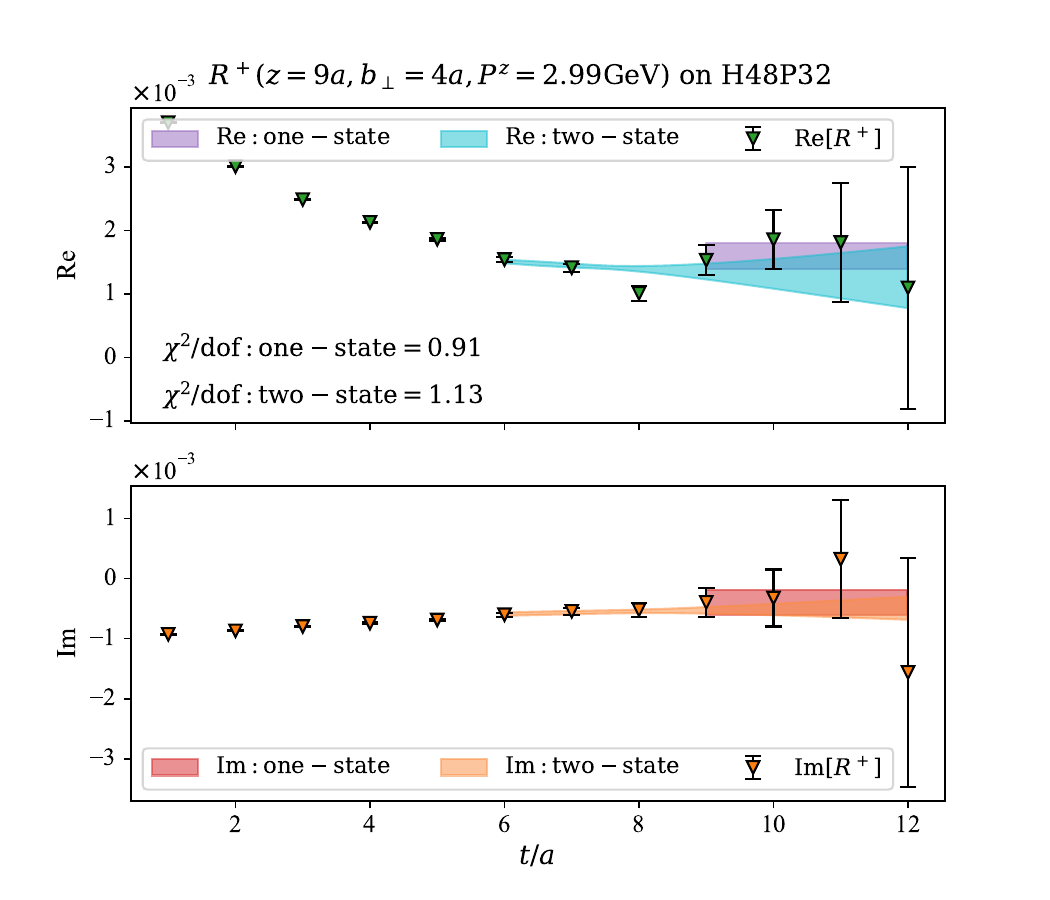}
\includegraphics[width=0.48\textwidth]{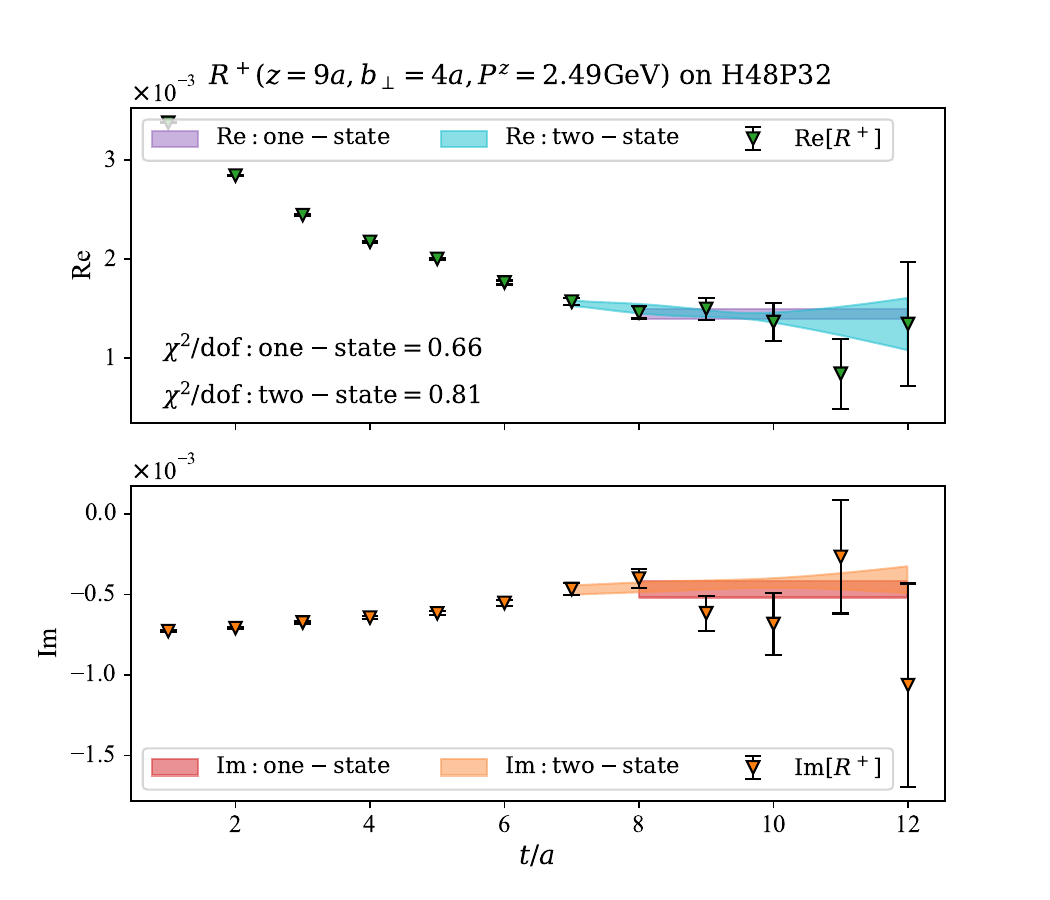}
\caption{Comparison of the lattice data $R^+(z, b_\perp, P^z, t; L=12a)$ along with the fit results of bare matrix element $\tilde{\Phi}^{0}$ obtained using different fitting strategies. The figures present results from the H48P32 ensemble with 1-step HYP smearing, and compare the combinations for $(z, b_\perp, P^z)=(3a, 4a, 2.99\mathrm{GeV})$, $(3a, 14a, 2.99\mathrm{GeV})$, $(9a, 4a, 2.99\mathrm{GeV})$ and $(9a, 4a, 2.49\mathrm{GeV})$ as examples to demonstrate the consistency of the results obtained one-state and two-state fits.
} 
\label{fig:one/two state fit}
\end{figure*}
%%%%%%%%%%%%%%%%%%%%%%%%%%%%%%%%%%%%%%%%

We construct the two-point correlation function corresponding to the bare matrix elements of the quasi-TMDWF as
\begin{align}
   & C_2(z, b_\perp, P, p^z, L, t, t') =\frac{1}{n^3_s} \sum _ { \vec{x} }  e^{i \vec{P} \cdot \vec{x}    }  \operatorname{Tr} \langle S_w(t';\vec{x}_1, t;-p^z)  \notag \\  
    &\qquad\qquad \times  U_{\sqsupset}(L, z, b_\perp)\Gamma_1    S_w(\vec{x}_2, t;t';p^z)  \Gamma_2 \rangle , 
    \label{nonlocal C2}
\end{align}
where $\vec{x}_1 = \Vec{x} + L \hat{n}_z + b \hat{n}_{\perp} $ and $\vec{x}_2 = \Vec{x} + (L+z) \hat{n}_z$. The quark and antiquark are assigned momenta $p^z$ and $-p^z$ along the $z$-direction, respectively, and $P$ denotes the hadron momentum, which satisfies $P = (0, 0, 0, 2p^z)$.

Since the Dirac structure of pseudoscalar meson interpolators at the source does not influence the matrix elements, we choose $\Gamma_2 = (\gamma^t\gamma_5+ \gamma^z\gamma_5)/2$~\cite{Zhang:2025hyo}. At the sink, as mentioned earlier, we select $\Gamma_1 = \gamma^t\gamma_5$, which will projects onto the leading-twist light-cone operator $\gamma^+\gamma_5$ in the large-momentum limit. Employing the reduction formula, the parametrized form of the two-point correlation function can be expressed as
\begin{align}
    C_2(z, b_\perp, P^z, L, t)&=\frac { A_w(P^z) A_p } { 2 E  } \tilde{\Phi}^{0}\left(z, b_\perp, P^z, L\right) \notag\\
    & \times e^{-Et}[1+c_0(z, b_\perp, P^z, L)e^{-\Delta Et}], 
\end{align}
where $A_w(P^z)$ and $A_p$ denote the local matrix element of the meson interpolator from wall source and point sink, respectively, and these factors can be canceled by the local two-point correlation function from the ratio:
\begin{align}
 &R^{\pm}(z, b_\perp, P_z, t;L)\equiv\frac{C_2(z, b_\perp, P^z, L, t)}{C_2(0, 0, P^z, L, t)}\notag\\
 &\quad=\tilde{\Phi}^{\pm0}(z, b_\perp, P^z, L)[1+c'(z, b_\perp, P^z, L)e^{-\Delta Et}], 
 \label{eq:ratio 2pt}
\end{align}
in which the superscript $\pm$ in $R^{\pm}$ and $\tilde{\Phi}^{\pm0}$ denote the $z$-direction of the staple-shape Wilson link, and $c'$ term denotes the contribution from excited states.

\begin{figure*}[ht]
\centering
\includegraphics[width=0.48\textwidth]{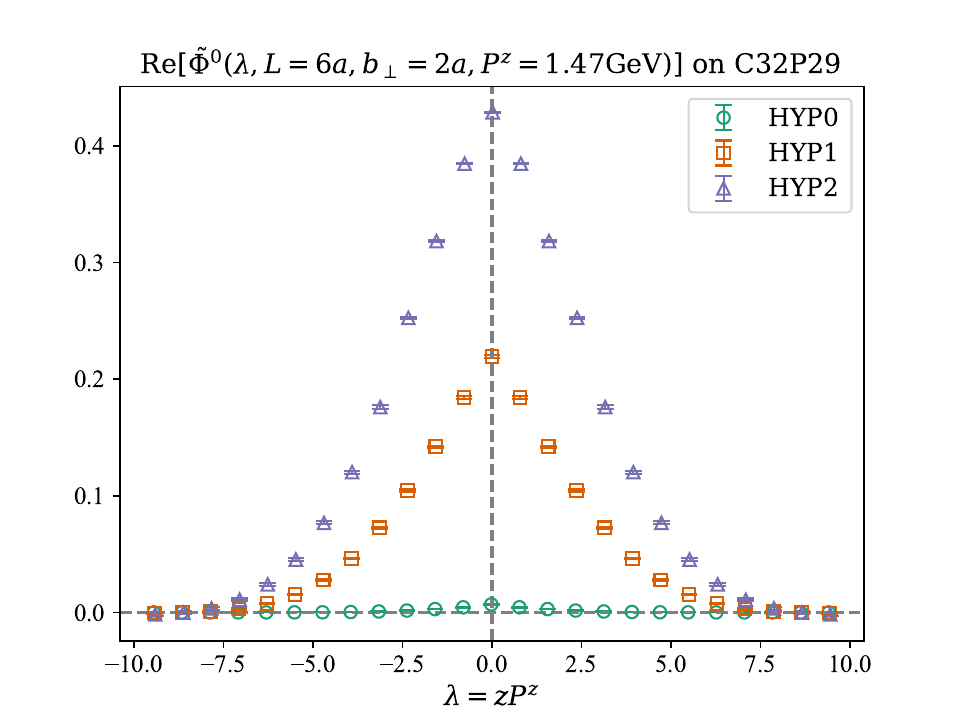}
\includegraphics[width=0.48\textwidth]{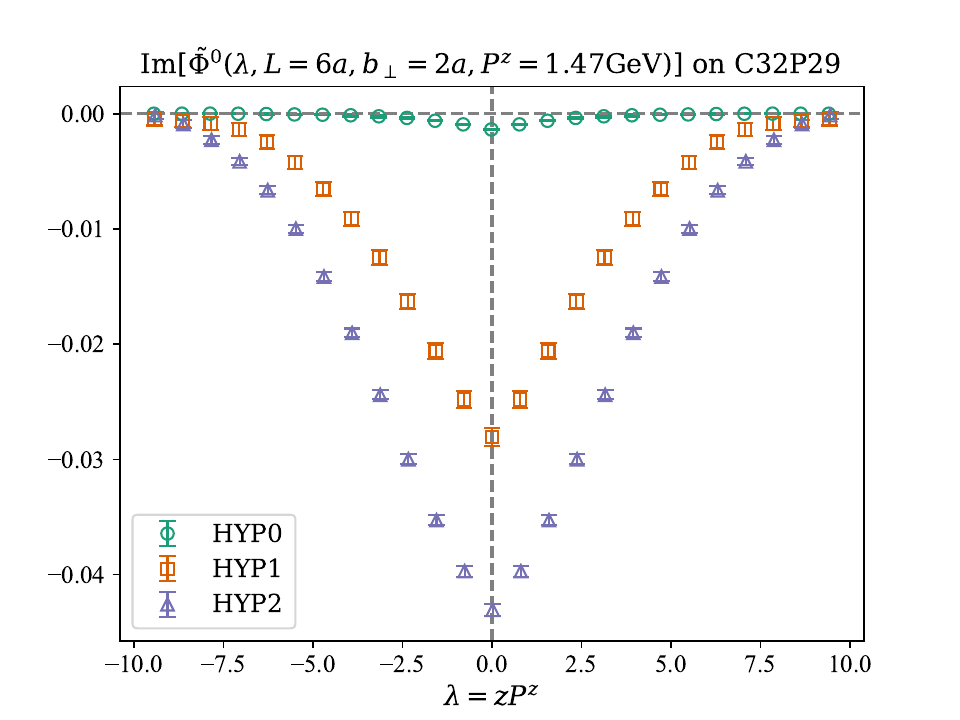}
\caption{Comparison of the real part (left panel) and imaginary part (right panel) of the bare matrix element from different HYP smearing steps. In this figure, we use the case with $(L, b_\perp, P^z)=(6a, 1a, 1.47\mathrm{GeV})$ from the C32P29 ensemble as an example, with similar results for other cases.}
\label{fig:bare_TMDWFs}
\end{figure*}

In this work, we calculate the nonlocal two-point correlation functions on all five configurations with the $P^z$ settings listed in Tab.~\ref{tab:conf inf}. As shown in Eq.(\ref{eq:ratio 2pt}), we extract the ground-state matrix elements from the ratio of the nonlocal and local two-point correlation functions. 
In practice, we compare the results of the one-state fit and the two-state fit, as shown in Fig.~\ref{fig:one/two state fit}. Typically, the one-state fit is applied when the ratio has reached a well-defined plateau region, while the two-state fit is used when residual excited-state contamination remains visible within the plateau. From the cases shown in Fig.~\ref{fig:one/two state fit}, the results from the one-state and two-state fits are in good agreement. To conservatively estimate the statistical errors of the fit results, we adopt the two-state fit results for the subsequent analysis.

In the definition of physical TMDs in Minkowski space, the longitudianl direction of the staple-shaped Wilson link is nontrivial. The past-pointing and future-pointing link directions correspond to different physical processes, and they are not easily related through simple symmetries. However, in quasi-TMDs, the staple-shaped Wilson link is independent of time, the $O(3)$ spatial symmetry and isospin symmetry ensure the symmetry of the results when the staple-shaped Wilson link is oriented along different longitudinal directions ($+z$ or $-z$). 
This symmetry was observed in Refs.~\cite{LatticePartonLPC:2022eev, LatticePartonLPC:2023pdv}, where it was used to improve their analysis. In the Appendix~\ref{ax:The symmetry of quasi-TMDWFs} of this paper, we provide a proof of this symmetry: the coordinate-space matrix element defined in Eq.(\ref{matrix elements}) is symmetric under $\pm z$, and correspondingly, the momentum-space distribution in Eq.(\ref{definition of quasi-tmdwf in mom-space}) is symmetric about $x=1/2$. Based on this symmetry, in the following discussion, we will no longer distinguish between quantities along the $\pm z$ directions (e.g. $R^{\pm}$, $\tilde{\Phi}^{\pm0}$) or their averages ($R$, $\tilde{\Phi}^{0}$). In the final results, we use the averaged values to reduce statistical fluctuations.

In lattice calculations of quasi-TMDs, the most significant statistical error arises from the long Wilson link. In Eq. (\ref{nonlocal C2}), the statistical fluctuations originating from the gauge field sampling accumulate will eventually dominate the statistical error of the nonlocal correlation functions as the length of the Wilson line increases. This is one of the main challenges in lattice calculations of TMDs. In this study, we employ the HYP smearing procedure to improve the statistical accuracy of the gauge field. In practice, we apply different steps of HYP smearing to $U_{\sqsupset}(L, z, b_\perp)$ (we test 1 step (HYP1) and 2 steps (HYP2), and compare these with the case without any HYP smearing (HYP0)) to examine the impact of HYP smearing on statistical fluctuations and the resulting measurements. 
Fig.~\ref{fig:bare_TMDWFs} shows a comparison of the bare matrix element results for different steps of HYP smearing. At the bare matrix element level, this comparison is not significant. Since HYP smearing modifies the short-distance ultraviolet fluctuations of the gauge field \cite{Hasenfratz:2001hp, Hasenfratz:2001tw}, one can observe noticeable differences in the bare matrix elements at different smearing steps, which arise from changes in the ultraviolet behavior. This difference will be corrected through nonperturbative renormalization, and the comparison of the renormalized results will be presented in the following section.

\subsubsection{Nonperturbative Renormalization and Large-$L$ Saturation}

\begin{figure*}[http]
\centering
\includegraphics[width=0.32\textwidth]{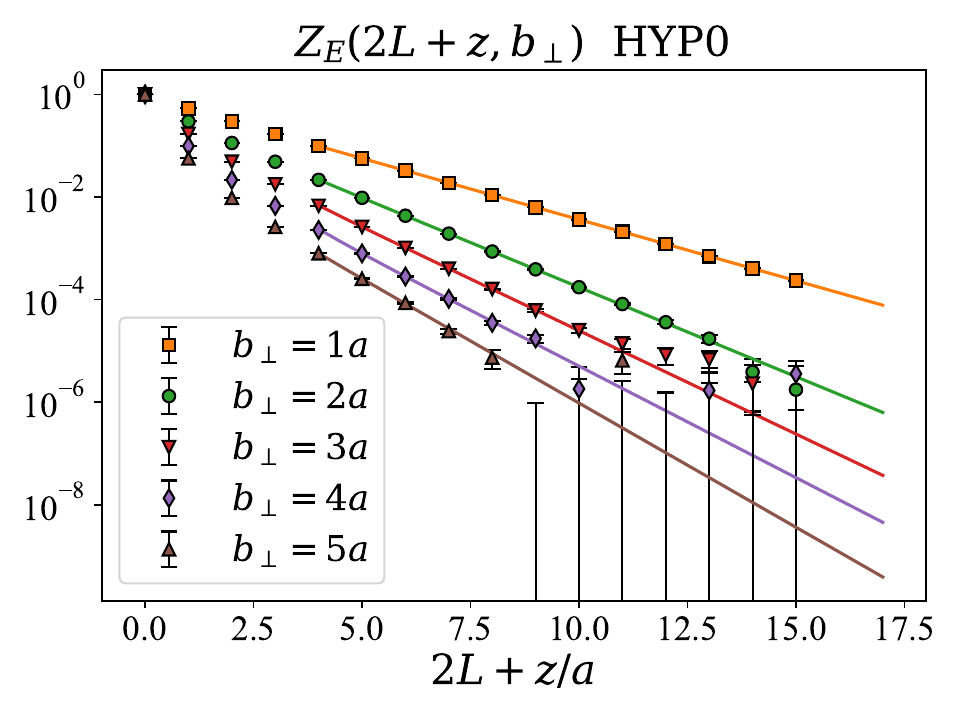}
\includegraphics[width=0.32\textwidth]{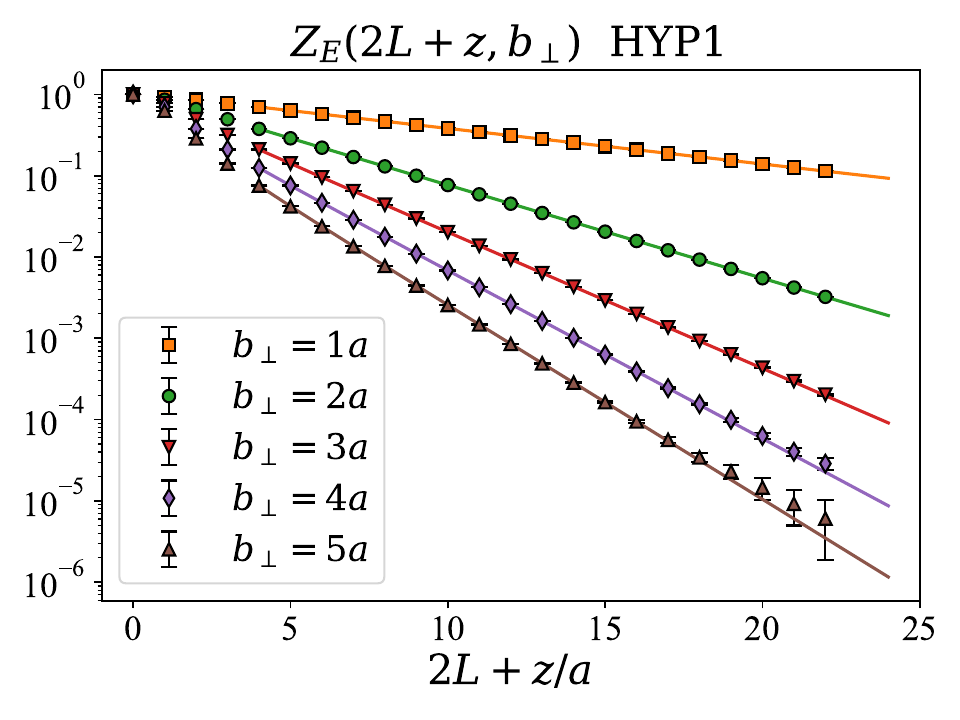}
\includegraphics[width=0.32\textwidth]{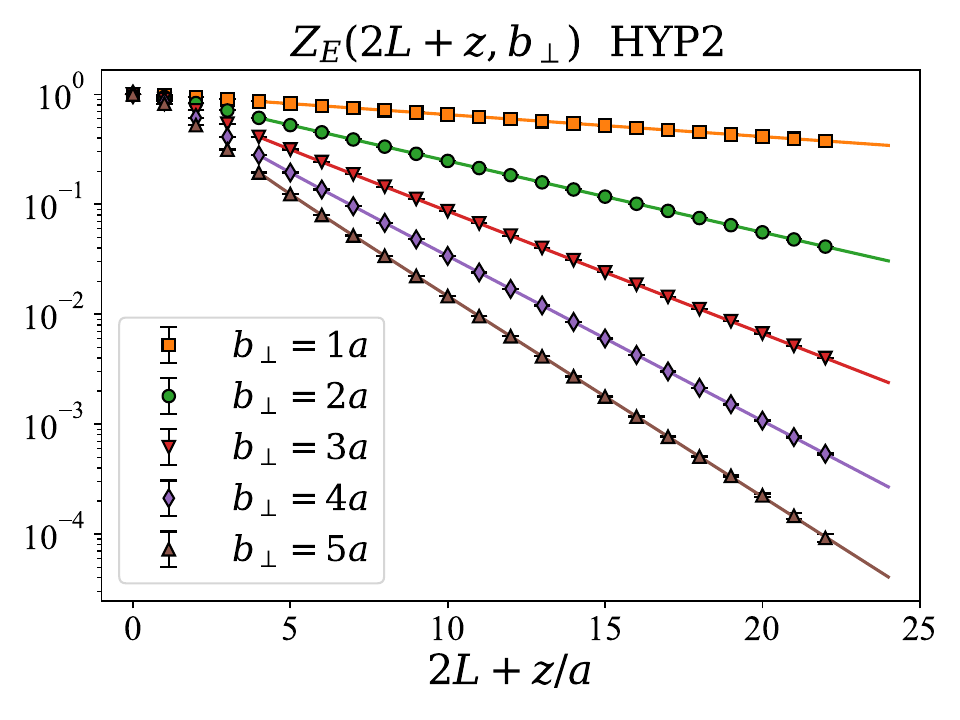}
\caption{Results of the Wilson loop for different shapes $(2L+z)\times b_\perp$ on the C32P29 ensemble, with three panels showing a comparison of the results from different step of HYP smearing. The solid lines represent the extrapolated results based on Eq.(\ref{Wloop extrapolation}). It can be seen that the extrapolated results agree well with the behavior of the data at large spatial separations.}
\label{fig:Wilson loop}
\end{figure*}

%%%%%%%%%%%%%%%%%%%%%%%%%%%%%%%%%%%%%%%%%%%%%
\begin{figure*}[http]
\centering
\includegraphics[width=0.95\textwidth]{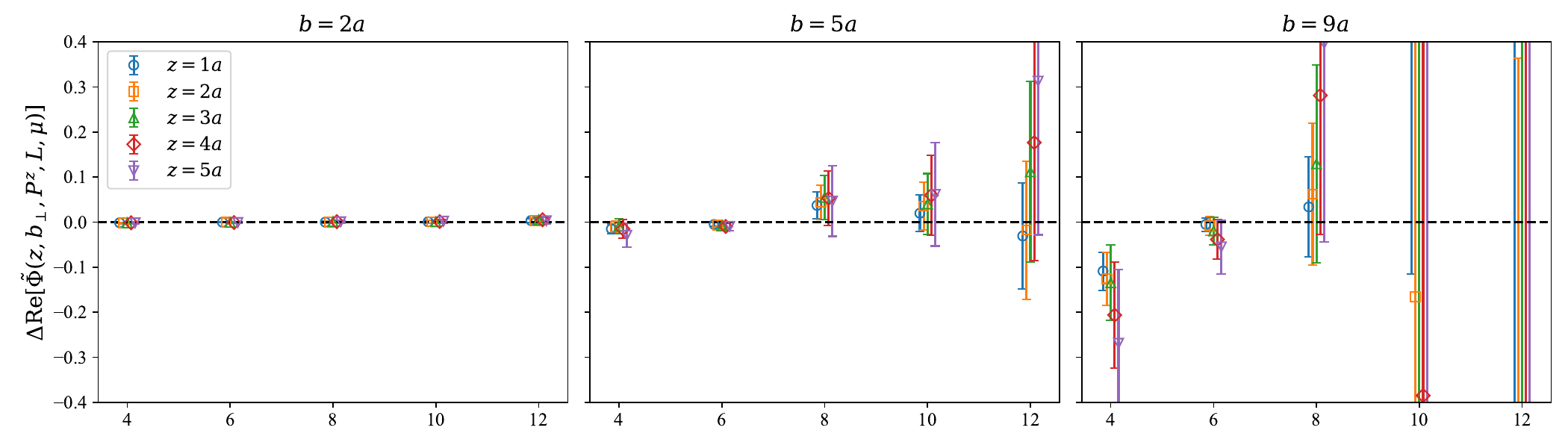}
\includegraphics[width=0.95\textwidth]{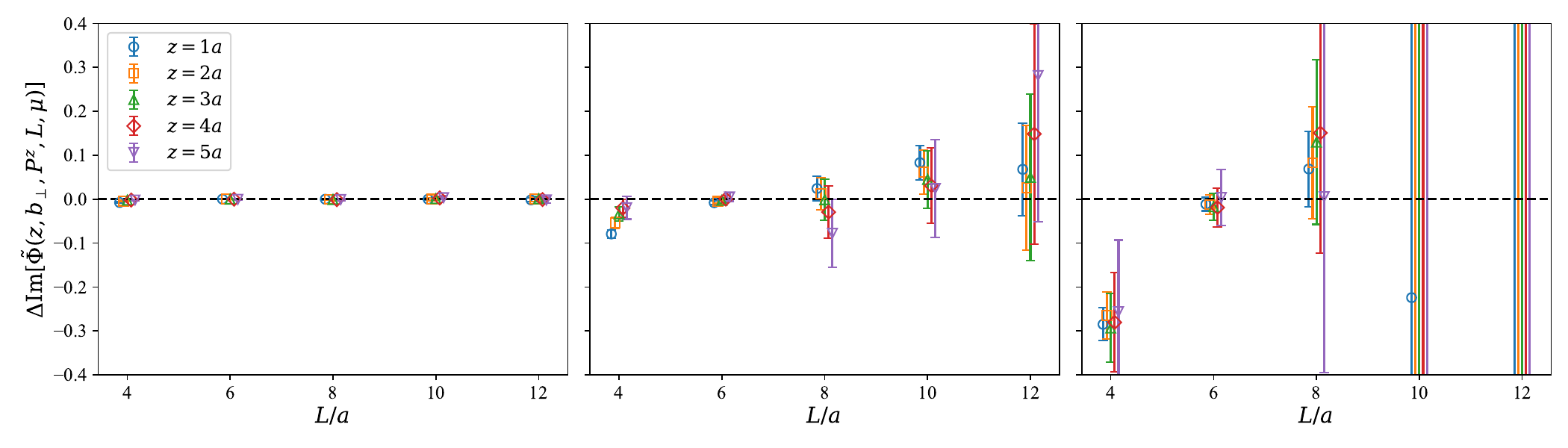}
\caption{The saturation of $L$-dependence for renormalized quasi-TMDWFs, illustrated using the results at $z\in\{1, \, 2, \, 3, \, 4, \, 5\}a$, $P^z=1.47$GeV and $b\in\{2, \, 5, \, 9\}a$ on the C32P29 ensemble. The upper panel shows the real part, and the lower panel shows the imaginary part.}
\label{fig:L dependence}
\end{figure*}

%%%%%%%%%%%%%%%%%%%%%%%%%%%%%%%%%%%%%%%%%%%%%

As mentioned earlier, to remove the linear divergence, the bare quasi-TMDWF matrix element must be divided by a Wilson line of the same shape and length. In this work, we employ the Wilson loop renormalization method ~\cite{Green:2017xeu, Chen:2016fxx, Zhang:2017bzy, Musch:2010ka, Zhang:2017zfe}, where the wilson loop defined in Eq.(\ref{eq:wilson_loop_define}) can be parametrized as proportional to the static potential $V(n_\perp)$:
\begin{align}
    Z_E(n_\perp, n_z) \propto e^{-V(n_\perp) n_z}\left(1+\mathcal{O}(e^{-n_z \Delta E})\right).
    \label{Wloop extrapolation}
\end{align}
This expression enables an extrapolation of the Wilson loop using a two-state fit, as described by the following equation:
\begin{align}
    Z_E(n_\perp, n_z) = c_0(n_\perp) e^{-E(n_\perp)n_z} \left[ 1 + c_1(n_\perp) e^{-\Delta E(n_\perp)n_z} \right].
\end{align}

Fig.~\ref{fig:Wilson loop} shows the lattice data and extrapolated results for the Wilson loop, both with and without HYP smearing, applied to the C32P29 gauge ensemble for $n_\perp = \{1, 2, 3, 4, 5\}a$ as examples. The comparison of the results from different steps of HYP smearing indicates that, for a fixed $n_\perp$, the Wilson loop without HYP smearing decreases more rapidly along $n_z$, accompanied by a more rapid deterioration of the signal-to-noise ratio. This suggests that HYP smearing effectively improves the signal quality.

Based the locality property of the gauge action, the influence of a gauge‐field disturbance decays exponentially with distance, and large Wilson loops can be factorized into approximately independent sub‑regions \cite{Luscher:2001up}. In quasi‑TMDs, one also finds that for the non‑local operator the Wilson line segment that lies far away from the quark fields is essentially identical to the corresponding segment appearing in the Wilson loop used for renormalization. Therefore, once the bare quasi‑TMDWF is divided by the Wilson loop of the same shape and length, the contributions from the distant Wilson‐link segments (those far from the quark sources) saturate and no longer depend on the further extension of the Wilson link. In other words, the long‑link contributions become independent of the Wilson‐line length beyond some moderate distance, because the additional segments lie in a region whose gauge fluctuations no longer couple strongly to the quark‐field insertion region.

For the renormalization factor $Z_O$, as mentioned earlier, it is independent of the coordinates, so it can be factored out of the Fourier transform and then canceled in the extraction of the CS kernel. However, to ensure the completeness of the quasi-TMDWF results and to verify that the renormalized distributions are independent of HYP smearing, we still extract $Z_O(1/a, \mu)$ using the self-renormalization scheme.
Tab.~\ref{tab:Z_O} presents the results for $Z_O(1/a, \mu)$ at $\mu=2\, $GeV for different lattice spacings, with a detailed analysis provided in Appendix~\ref{Ax:Self Renormalization}.

\begin{table}[tp]
    \centering
    \renewcommand{\arraystretch}{1.5}
      \setlength{\tabcolsep}{3mm}
    \begin{tabular}{c|c c c}
        \hline
         $a$/fm& 0.1053 & 0.07746 & 0.05187 \\
         \hline
        HYP=0 & 1.072(22) & 1.153(20) & 1.251(19)\\
        HYP=1 & 0.9221(62)& 0.9753(57)& 1.0397(53)\\
        HYP=2 & 0.8690(58)&   0.9230(54)   &   0.9882(50) \\
        \hline
    \end{tabular}
    \caption{The renormalization factor $Z_O$ extracted from three ensembles with HYP=0, 1, 2.}
    \label{tab:Z_O}
\end{table}

Therefore, the renormalized matrix elements can be obtained by dividing the bare matrix elements by the Wilson loop $Z_E$ as well as the renormalization factor $Z_O$, 
\begin{align}
 \tilde{\Phi}\left(z, b_\perp, P^z, L\right) = \frac{\tilde{\Phi}^{0}\left(z, b_\perp, P^z, L\right)}{Z_O(1/a, \mu)\sqrt{Z_E(2L+z, b_\perp)}}, 
 \label{renormalized_ME}
\end{align}
ensuring that the dependence on the UV divergences are properly accounted for.

For a well-defined quasi-TMDWF, the length $L$ of the Wilson link extending along the $z$-direction must be sufficiently large to ensure that the link reaches outside the hadron’s spatial extent. Note that in this work $L$ denotes the ``additional'' extension beyond the quark–antiquark separation along $z$, which convention differs from that adopted in several previous studies. Further details of this setup are provided in Appendix~\ref{Ax:diff_z_directions}.

The $L$-dependence of the bare matrix element is associated with the linear divergence of the Wilson link self-energy, which can be subtracted by the square root of the Wilson loop. Therefore, the $\tilde{\Phi}$ defined in Eq.(\ref{renormalized_ME}) should saturate to a plateau at $L$ becomes large. To extract the asymptotic behavior as $L\to\infty$, we perform a constant fit of $\tilde{\Phi}$ and define the deviation from the plateau as:
\begin{align}
&\Delta \tilde{\Phi}\left(z, b_\perp, P^z, L\right) = \notag \\
&\qquad \tilde{\Phi}\left(z, b_\perp, P^z, L\right) - \tilde{\Phi}\left(z, b_\perp, P^z, L \to \infty\right).
\end{align}
This allows us to investigate the value of $L$ at which saturation occurs. As shown in Fig.~\ref{fig:L dependence}, starting from $L=6a$ on the C32P29 ensemble, $\Delta \tilde{\Phi}$ converges to zero, indicating that the staple-shaped Wilson link along the $z$-direction is sufficiently long. The corresponding length, $L\simeq0.6\, $fm, is of the same order as the pion’s strong-interaction (matter) radius, suggesting that once the link extends beyond the hadron’s interaction region in a given direction, further increases in its length no longer affect the partonic distributions probed by the operator. Based on this, we choose $L\gtrsim0.6\, $fm on the other ensembles: for C32P23 and C48P14 we take $L=6a$, for F32P30 we take $L=8a$, and for H48P32 we take $L=12a$.

\subsubsection{Renormalized Quasi-TMDWFs}

\begin{figure*}
\centering
\includegraphics[width=0.45\textwidth]{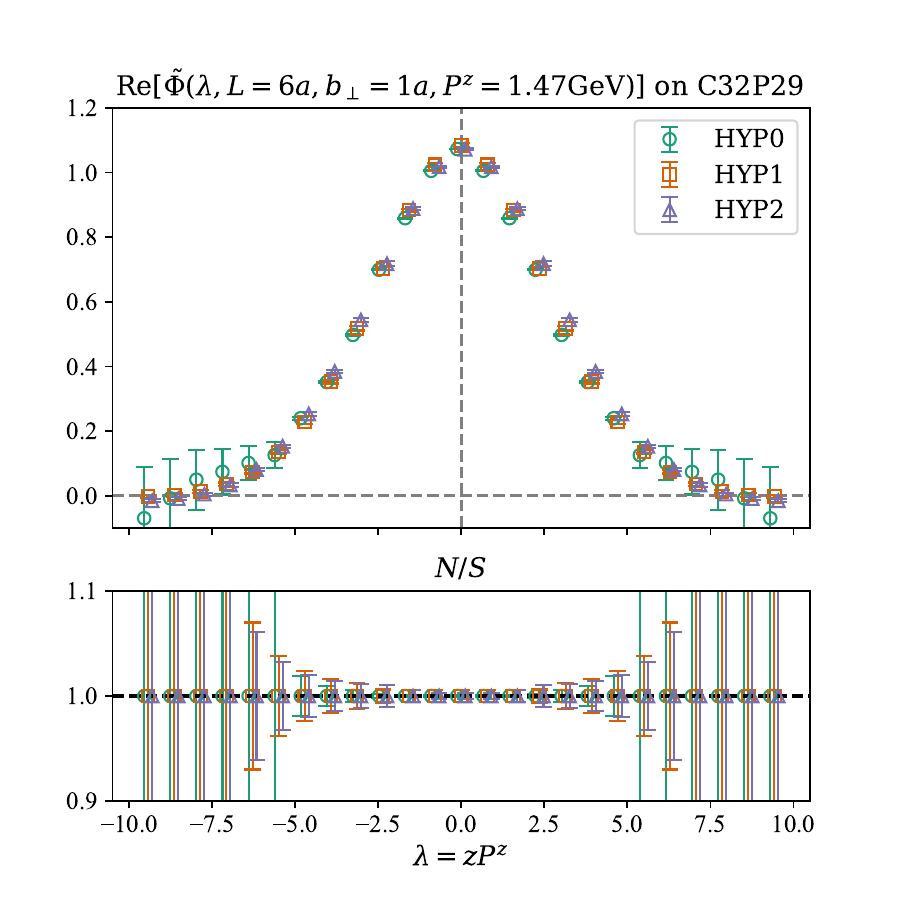}
\includegraphics[width=0.45\textwidth]{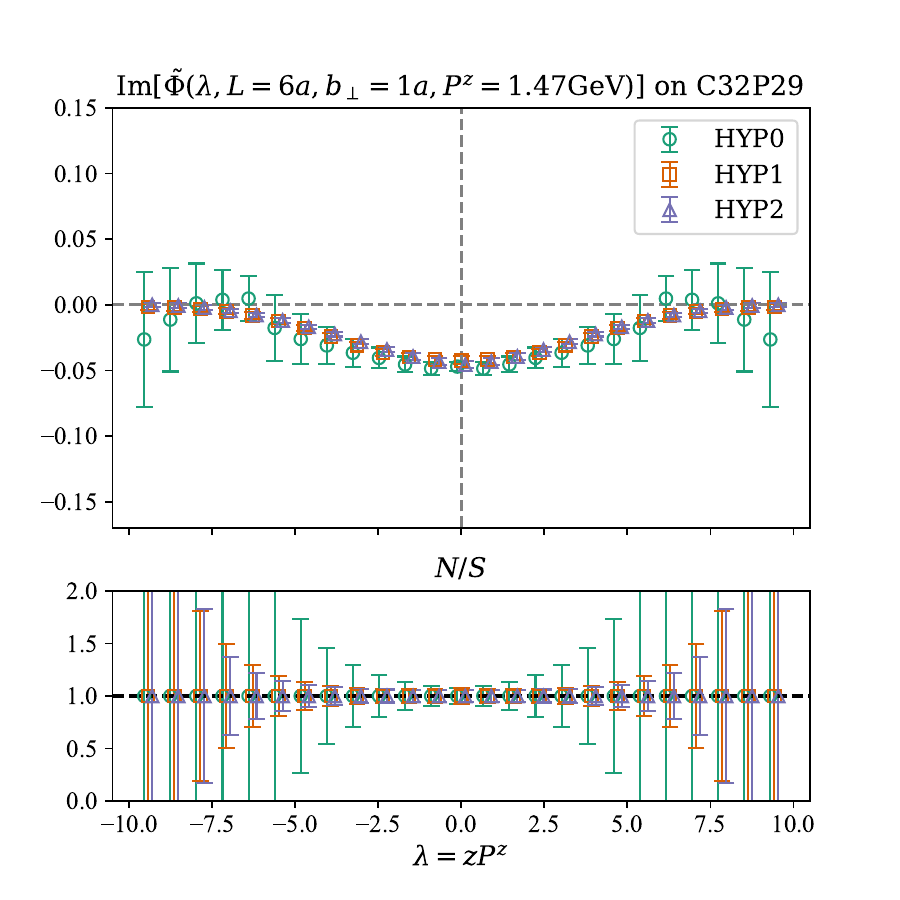}
\includegraphics[width=0.45\textwidth]{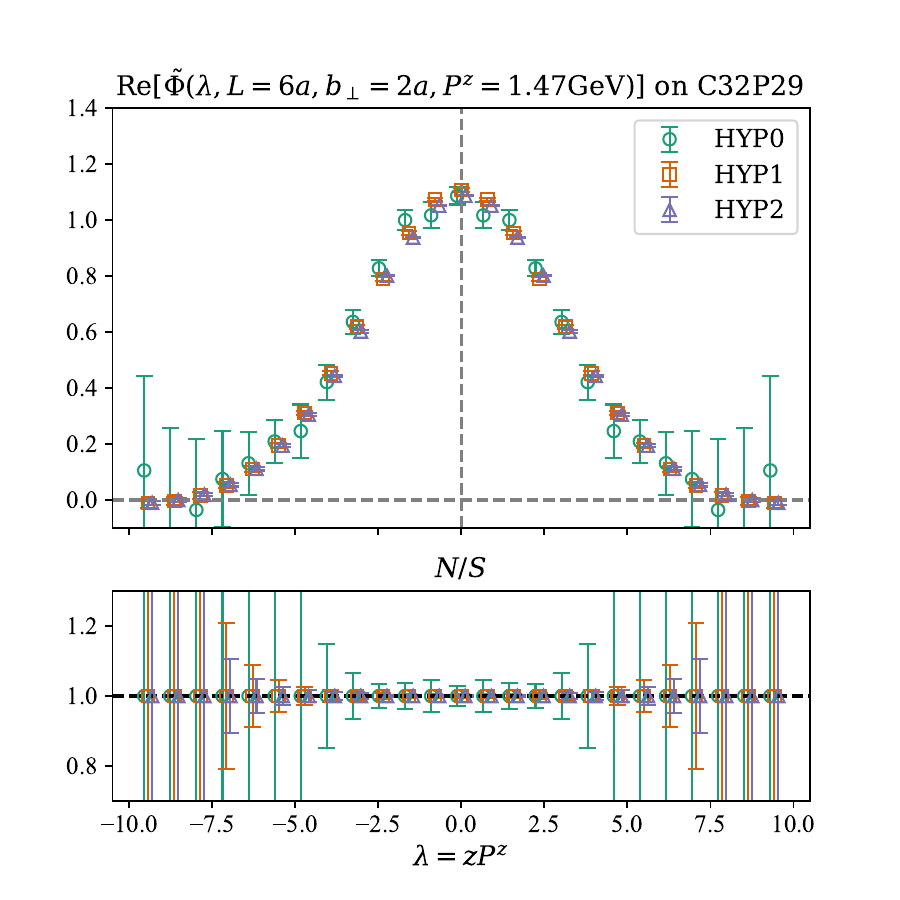}
\includegraphics[width=0.45\textwidth]{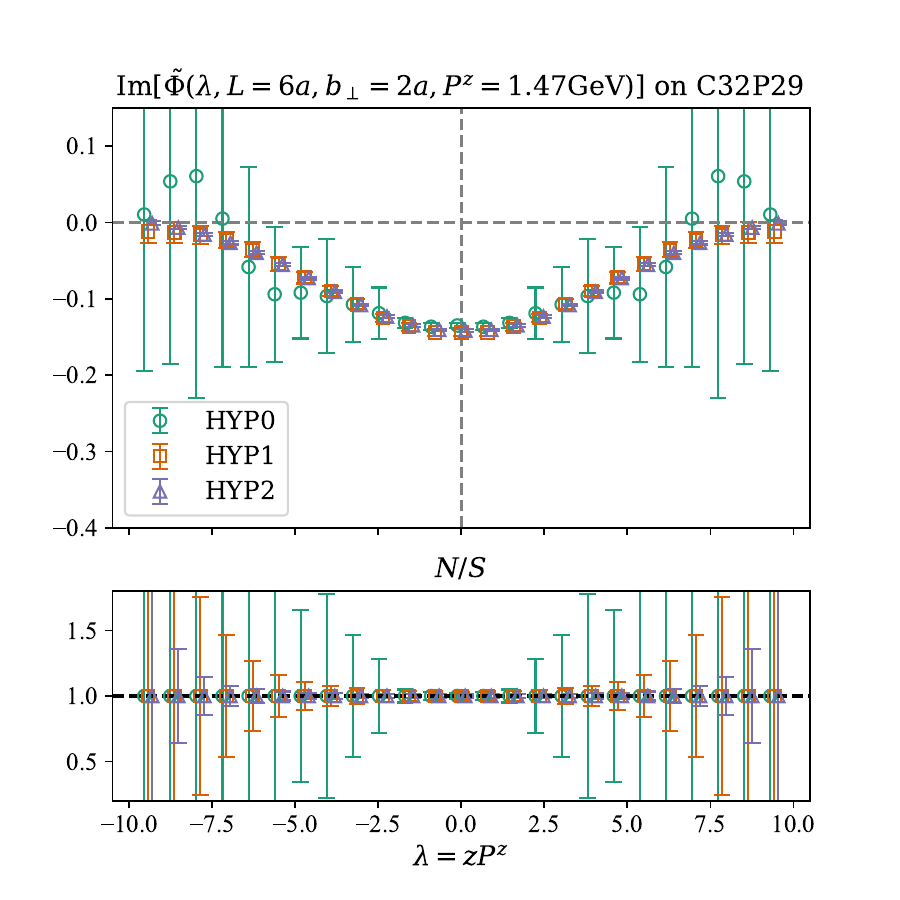}
\caption{Comparison of the real part (left panels) and imaginary part (right panels) of the renormalized matrix element from different HYP smearing steps. In this figure, we use the cases with $(L, P^z)=(6a, 1.47\mathrm{GeV})$ at $b_{\perp}=1a$ (upper panels) and $2a$ (lower panels) from the C32P29 ensemble as examples. Similar results are observed for larger $b_{\perp}$, but the results for HYP0 will suffer large errors. Below each subplot, the noisy-to-signal ratio (N/S) for the different HYP smearing steps is displayed, showing that the signal-to-noise ratio improves as the number of HYP smearing steps increases.} 
\label{fig:diff HYP in coor}
\end{figure*}

%%%%%%%%%%%%%%%%%%%%%%%%%%%%%%%
\begin{figure*}[http]
\centering
\includegraphics[width=0.48\textwidth]{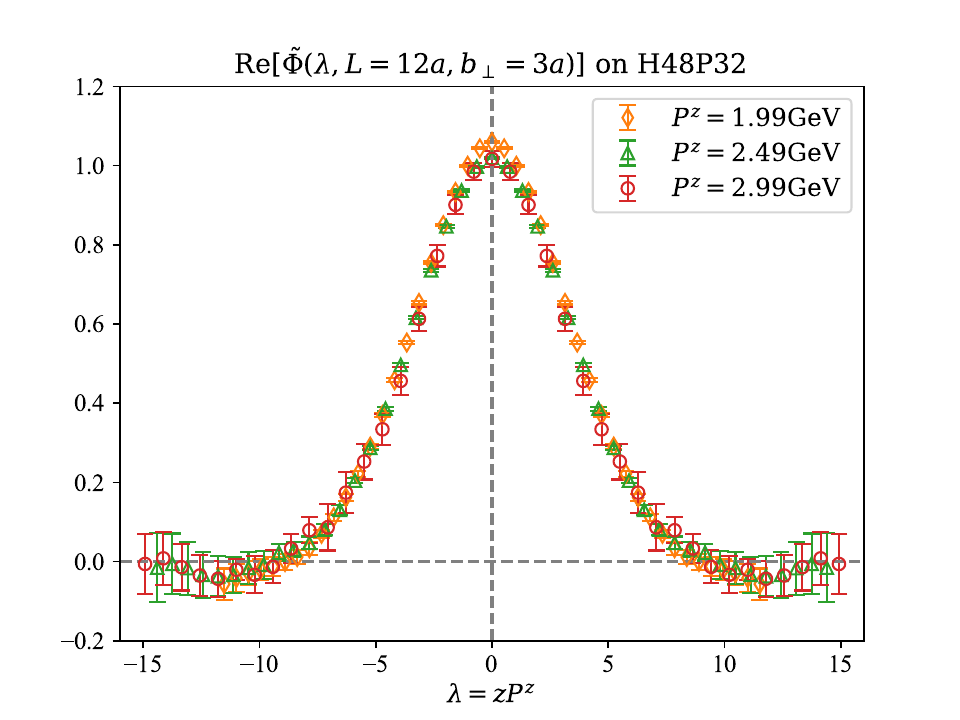} 
\includegraphics[width=0.48\textwidth]{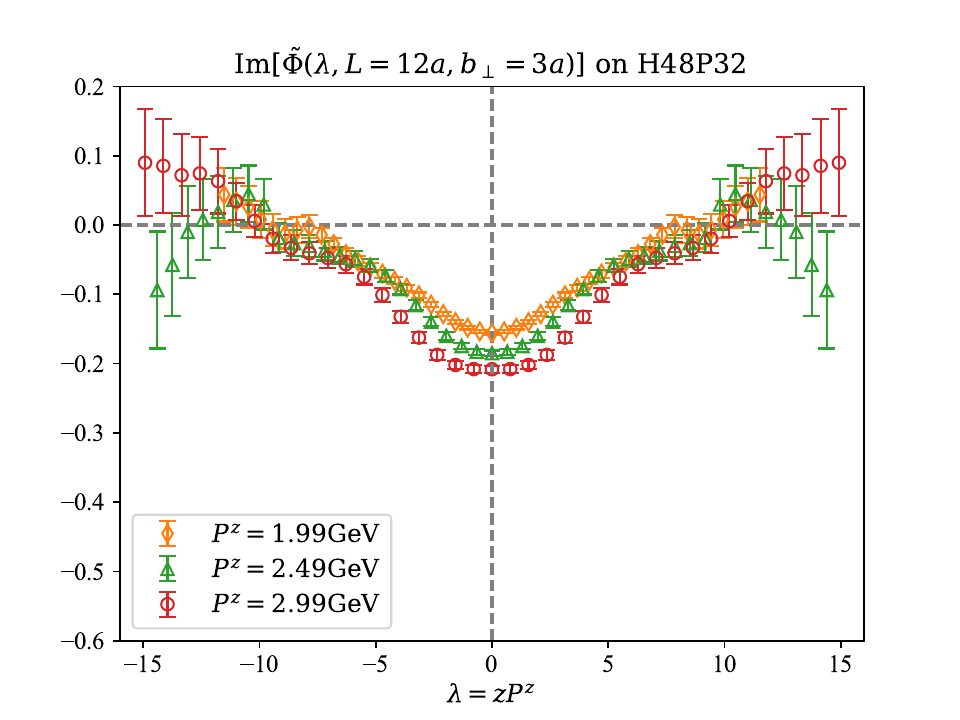}
\includegraphics[width=0.48\textwidth]{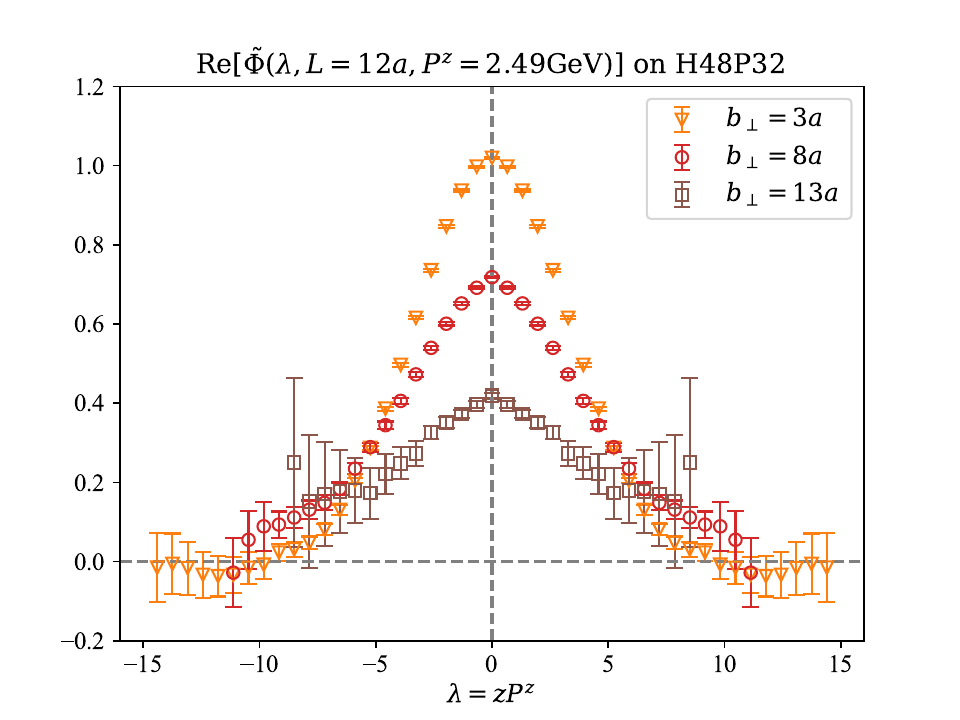} 
\includegraphics[width=0.48\textwidth]{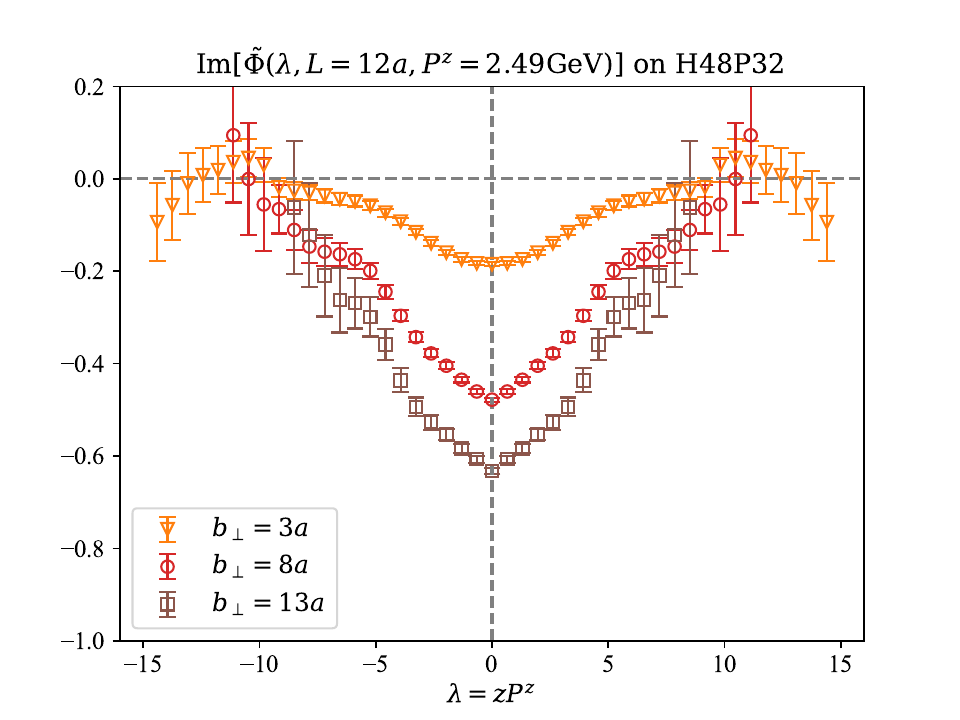}
\caption{$\lambda$-dependence of the renormalized matrix elements $\tilde{\Phi}$. Results from the H48P32 ensemble are shown as examples. The upper panel compares the results with $P^z = \{1.99, \, 2.49, \, 2.99\}\, \mathrm{GeV}$ at fixed $b_{\perp}=3a$, and the lower panel compares compares the results with $b_{\perp}=\{3, \, 8, \, 13\}a$ at fixed $P^z=2.49\, \mathrm{GeV}$. Additional cases are presented in the Appendix.
} 
\label{fig:TMDWFs in coor}
\end{figure*}
%%%%%%%%%%%%%%%%%%%%%%%%%%%%%%%

By subtracting the linear divergence in the bare matrix element from the Wilson loop $Z_E$ and removing the remaining logarithmic divergence from $Z_O$, we obtain numerical results for renormalized matrix elements as function of $\lambda=zP^z$. Fig.~\ref{fig:diff HYP in coor} compares the results of the renormalized matrix elements from different HYP smearing steps. 
From the comparisons across the cases listed in the subplots, it is clear that the results are indeed consistent across different HYP smearing steps. Additionally, the noisy-to-signal ratio (N/S) shown below each subplot demonstrates that as the Wilson link length increases, using more HYP smearing steps significantly improves the signal-to-noise ratio. For conservative estimates, we choose HYP1 as the smearing strategy in this work.

According to our numerical simulations, the renormalized coordinate-space matrix elements $\tilde{\Phi}$ as a function of $\lambda=zP^z$ are shown in Fig.~\ref{fig:TMDWFs in coor}. Here we take the results from H48P32 ensemble as examples, and additional cases are provided in the Appendix. The figure shows the dependence of the renormalized coordinate-space matrix elements on the boosted momentum $P^z$ at fixed transverse separation $b_{\perp}$, as well as their $b_{\perp}$ dependence at fixed $P^z$. Both the real and imaginary parts exhibit symmetry about $z=0$, ensuring that the renormalized quasi-TMDWF after Fourier transformation is real. Exploiting this symmetry, we average the $\pm z$ data to reduce statistical fluctuations.
The observed $P^z$ dependence at each $b_{\perp}$ encodes the rapidity evolution, which in turn enables a numerical extraction of the CS kernel.

To determine the quasi-TMDWFs $\tilde{f}(x, b_{\perp}, \mu, \zeta_z)$ in momentum space, we need to perform a Fourier transformation. Although Fig.~\ref{fig:TMDWFs in coor} shows that $\tilde{\Phi}$ converges to zero at large $\lambda$, however, a brute-force truncated Fourier transformation would introduce significant oscillations in the momentum-space distribution due to the errors \cite{Chen:2025cxr, Xiong:2025obq, Ling:2025olz}. Therefore, we extrapolate the renormalized matrix elements $\tilde{\Phi}$ by using the following form \cite{Ji:2020brr}:
\begin{align}
&\tilde{\Phi}\left(\lambda, b_{\perp}\right)=\left[\frac{c_1}{(-i\lambda)^{n_1}}+e^{i\lambda}\frac{c_2}{(i\lambda)^{n_2}}\right]e^{-\frac{\lambda}{\lambda_0}}, 
\label{Extrapolation}
\end{align}
in which all parameters $c_{1, 2}$, $n_{1, 2}$ and $\lambda_0$ depend on the transverse separation $b_{\perp}$. The terms inside the square brackets account for the algebraic behavior and is motivated by the Regge behavior \cite{Regge:1959mz} of the light-cone distributions at endpoint regions, and the exponential term is motivated by the expectation that the correlation function has a finite correlation length at finite momentum, which denoted as $\lambda_0$.

\begin{figure*}[http]
\centering
\includegraphics[width=0.45\textwidth]{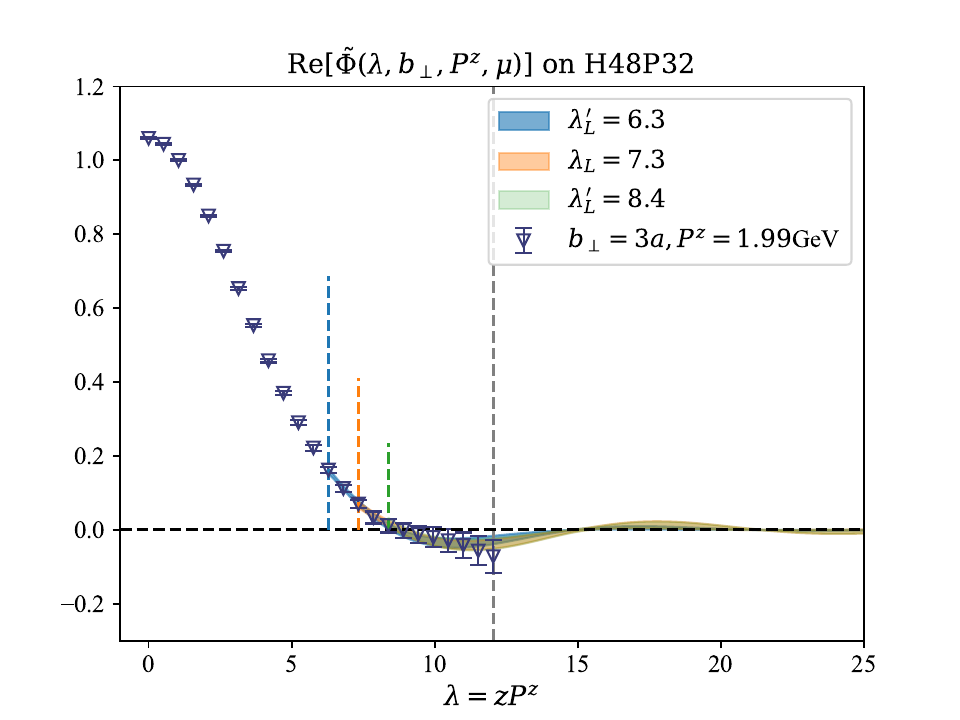} 
\includegraphics[width=0.45\textwidth]{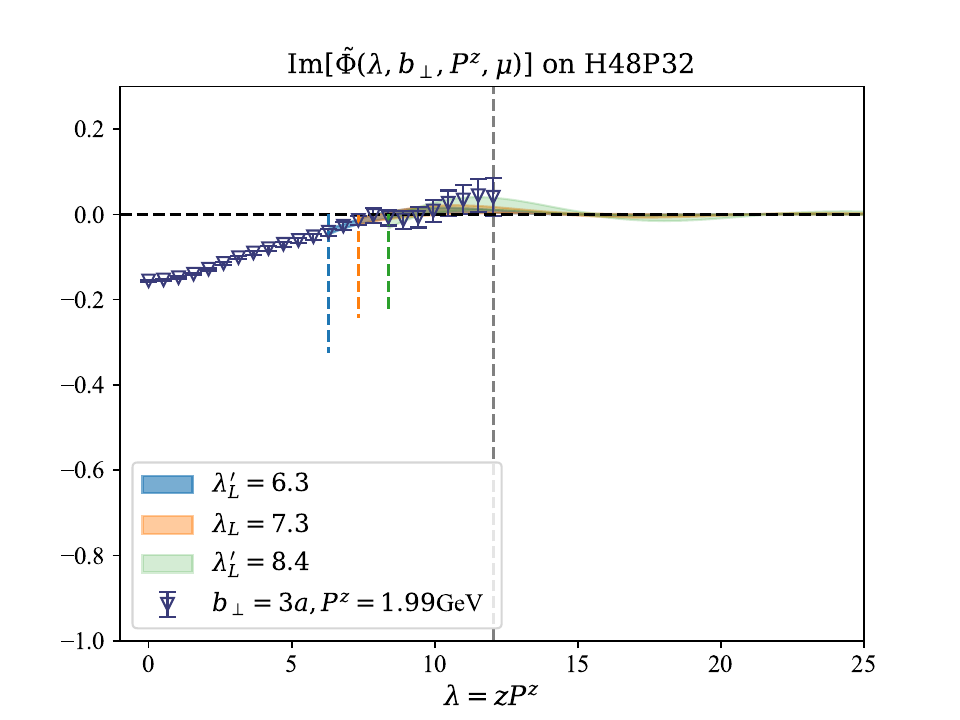} 
\includegraphics[width=0.45\textwidth]{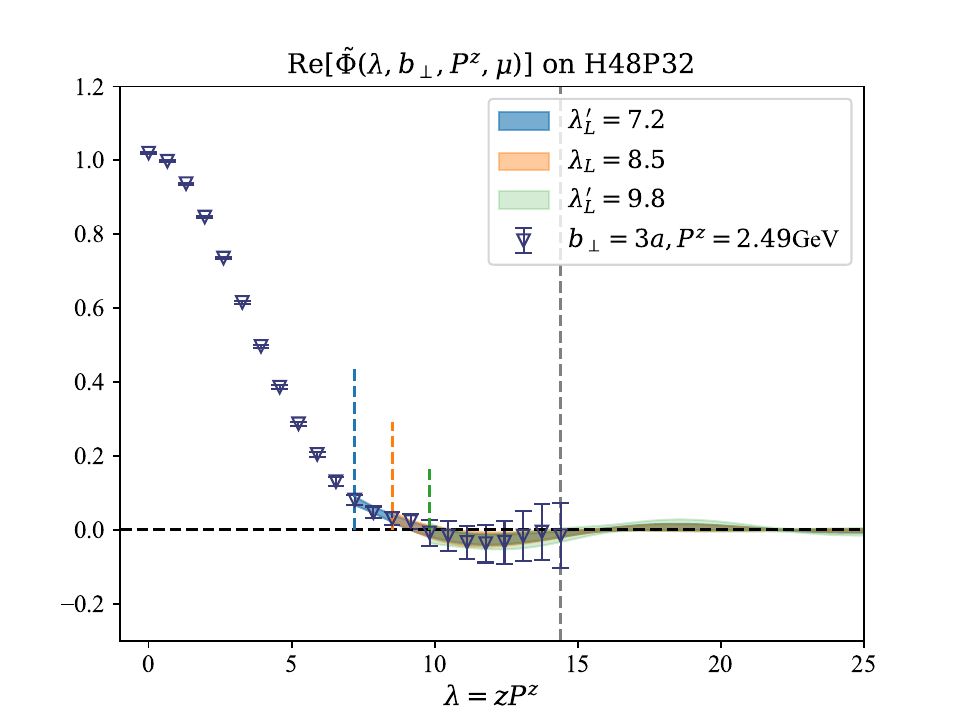} 
\includegraphics[width=0.45\textwidth]{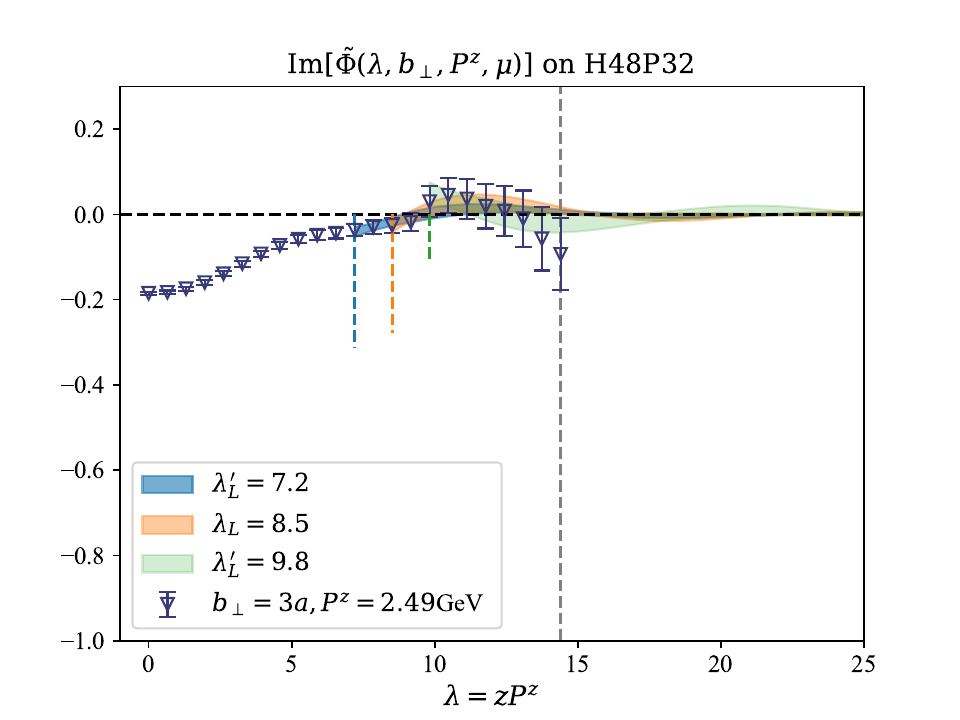} 
\includegraphics[width=0.45\textwidth]{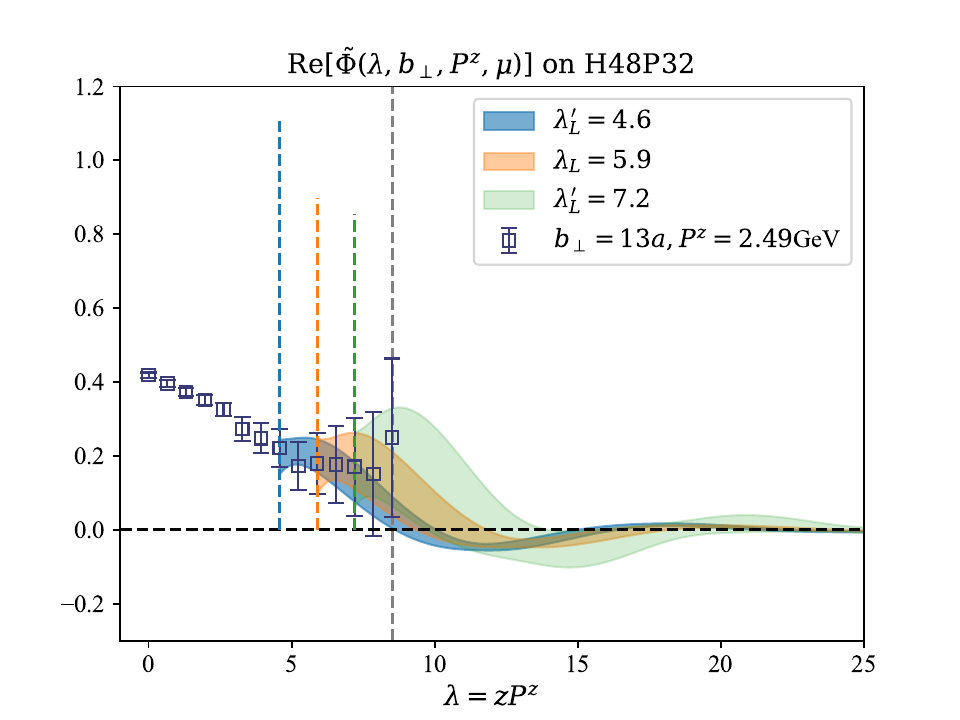} 
\includegraphics[width=0.45\textwidth]{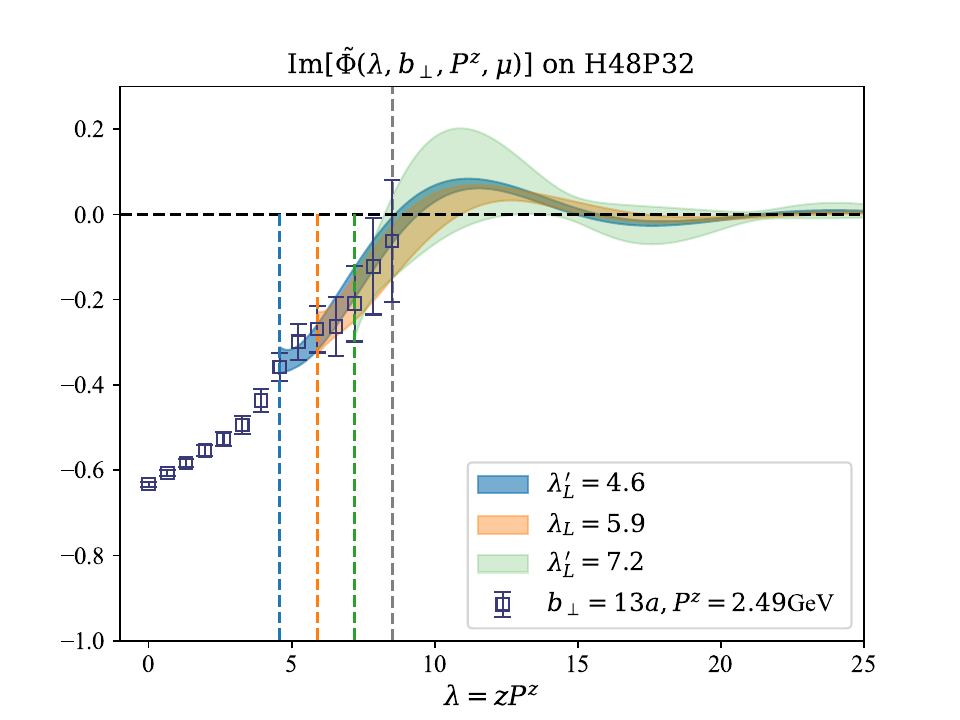} 
\caption{Comparison of the original data (points) and the extrapolated results (colored bands) at $(b_{\perp}, \, P^z)=(3a, \, 1.99\mathrm{GeV})$, $(3a, \, 2.49\mathrm{GeV})$ and $(13a, \, 2.49\mathrm{GeV})$ on the H48P32 ensemble. The renormalization scale is $\mu=2\mathrm{GeV}$. Extrapolations are performed over the fit window $\lambda\geq\lambda_L$, which gives the main results; systematic uncertainties are estimated by varying the window start to alternative values $\lambda_L^{\prime}$.
 } 
\label{fig:diff ZL}
\end{figure*}

To perform the extrapolation, a reasonable range of $\lambda$ is required to determine the parameters. 
To perform the extrapolation, a reasonable $\lambda$-range is required to constrain the parameters. In practice, we test the sensitivity of the extrapolation to the chosen fit window. The fit is carried out for $\lambda\geq\lambda_L$, where $\lambda_L$ is a truncation parameter selected in the long-tail region of the coordinate-space distribution, corresponding to the end-point region in momentum space. As shown in Fig.~\ref{fig:diff ZL}, the fitting bands obtained from different fit windows agree well with the original lattice data and are mutually consistent. This demonstrates the self-consistency of the extrapolation across fit intervals. To estimate the systematic uncertainty associated with the extrapolation form, we take the difference between the result obtained with the nominal $\lambda_L$ window and those obtained by shifting $\lambda_L$ forward and backward by two data points as the systematic error.

\begin{figure*}
\centering
\includegraphics[width=0.45\textwidth]{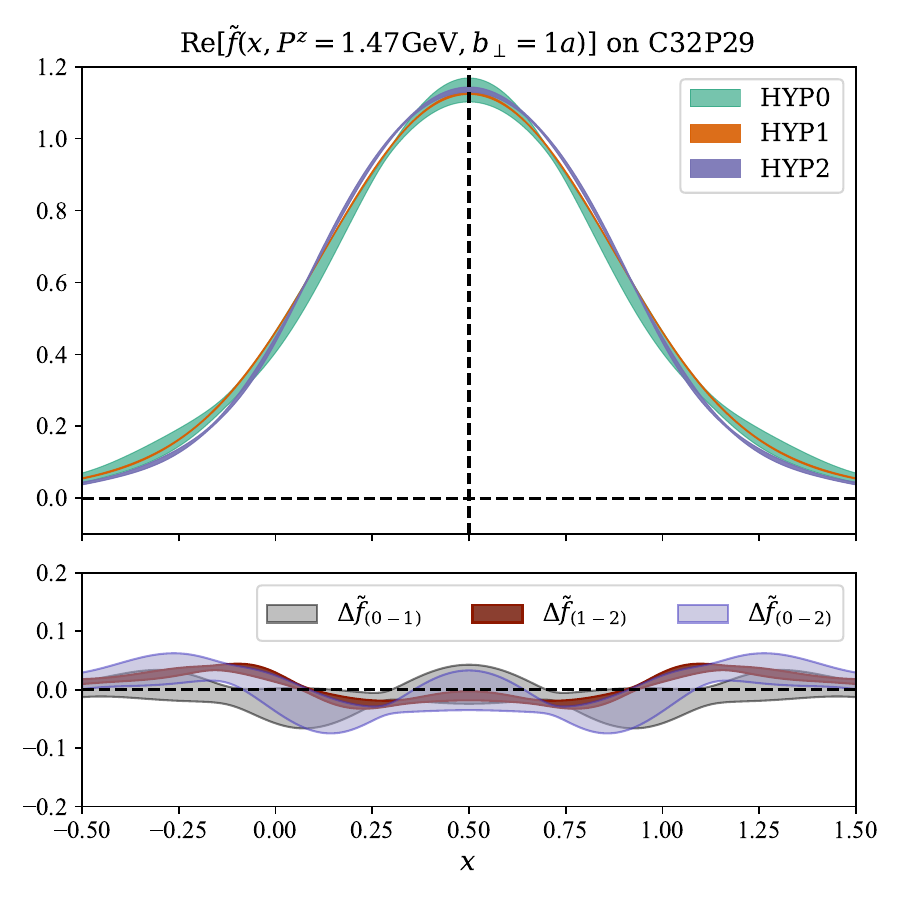}
\includegraphics[width=0.45\textwidth]{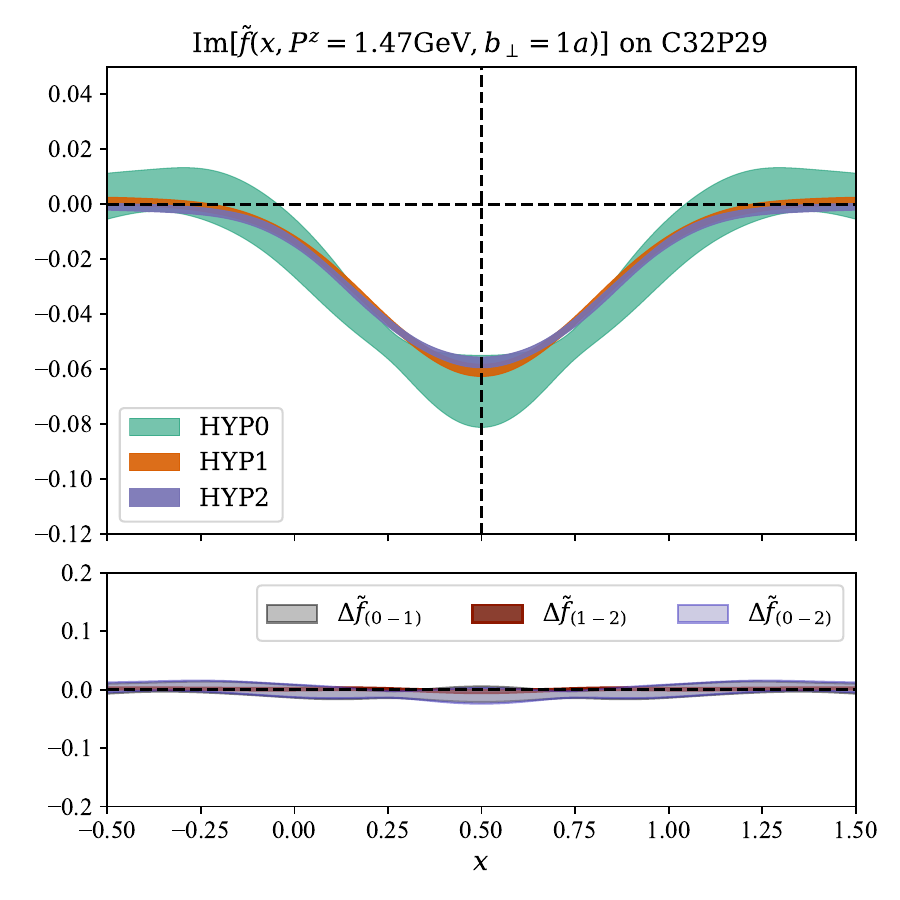}
\includegraphics[width=0.45\textwidth]{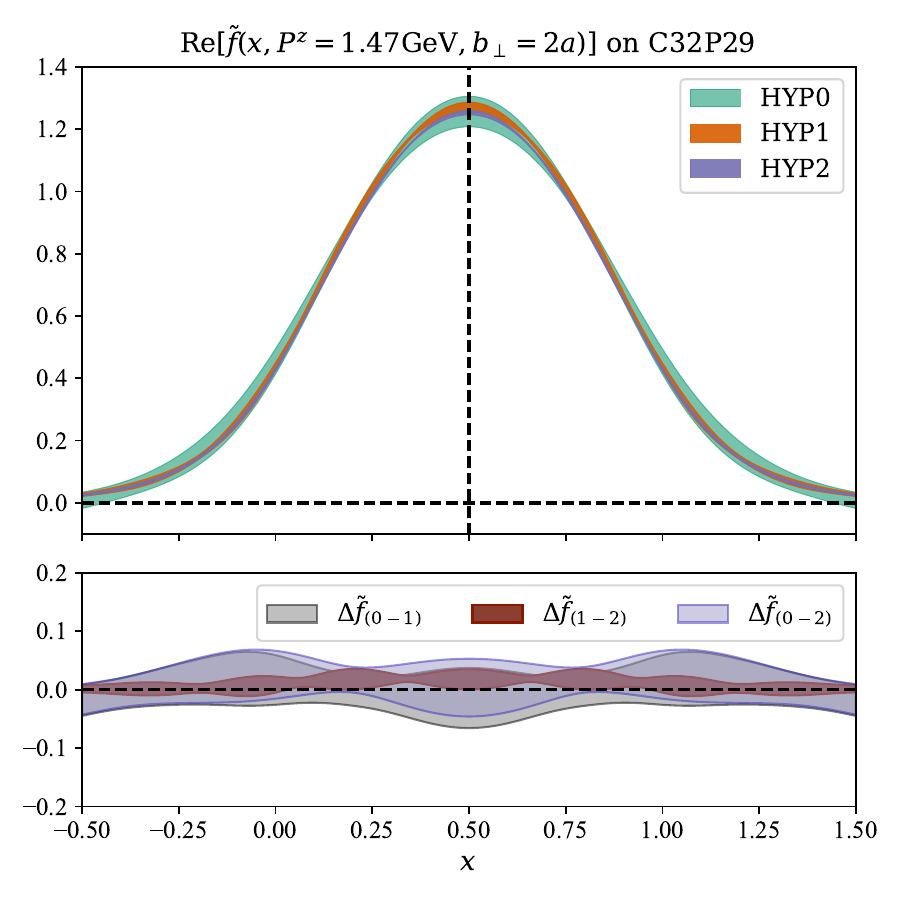}
\includegraphics[width=0.45\textwidth]{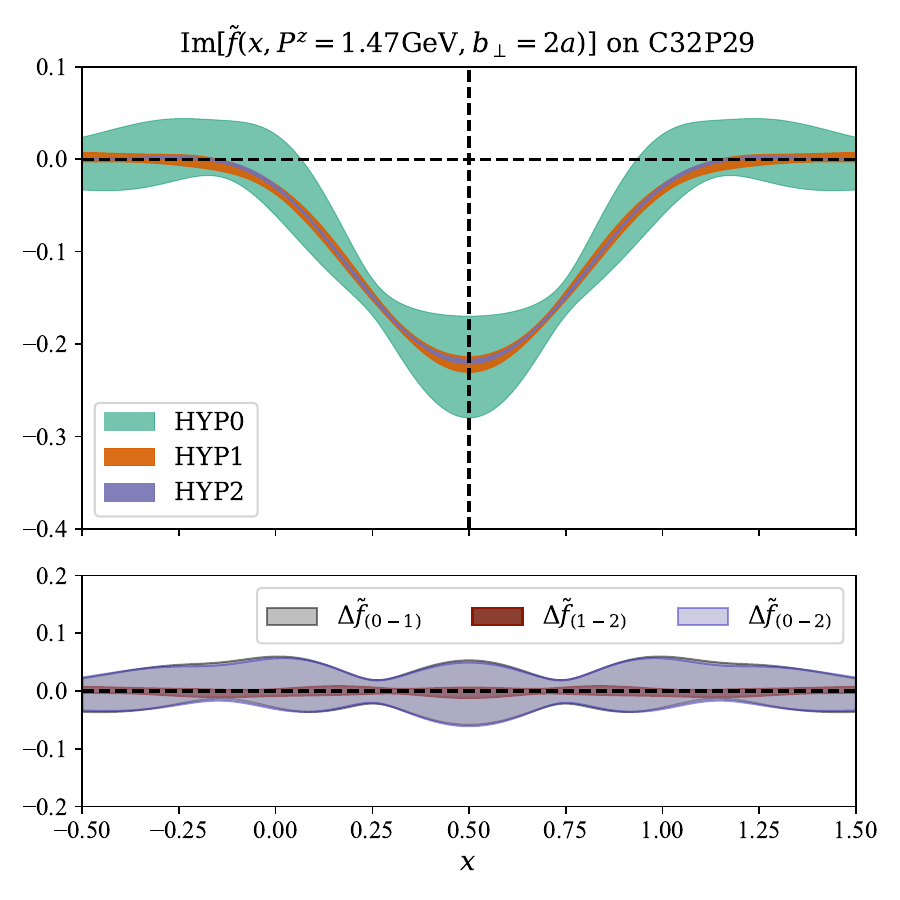}
\caption{Comparison of the real part (left panels) and imaginary part (right panels) of the renormalized quasi-TMDWF obtained with different HYP smearing steps. The settings are similar to Fig.~\ref{fig:diff HYP in coor}. The panels beneath each subplot display the differences between smearing steps.} 
\label{fig:diff HYP mom}
\end{figure*}

\begin{figure*}[http]
\centering
\includegraphics[width=0.45\textwidth]{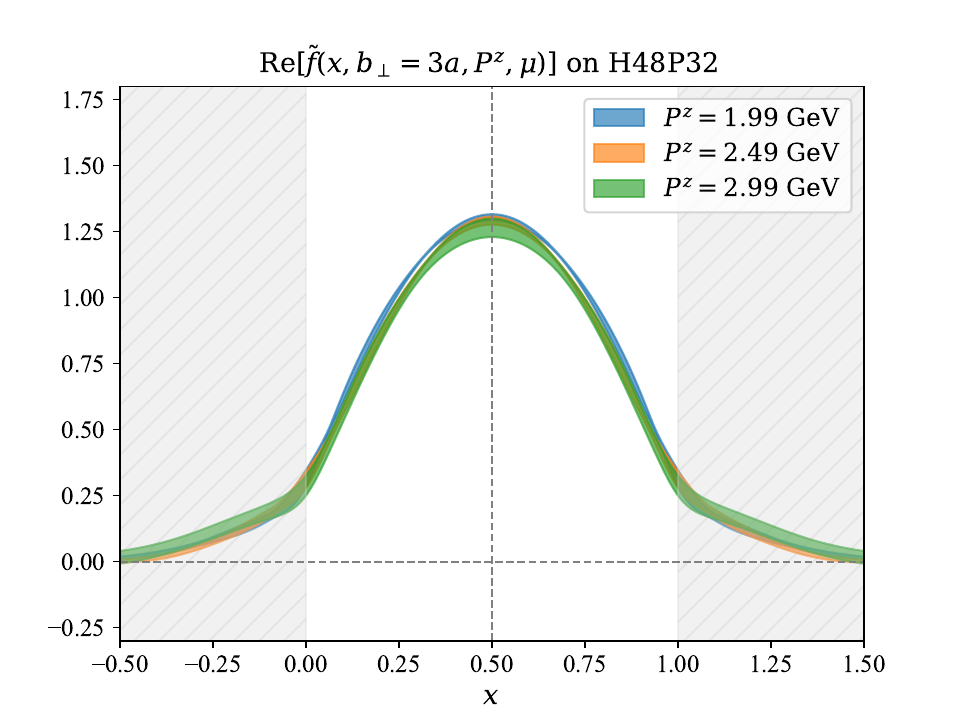} 
\includegraphics[width=0.45\textwidth]{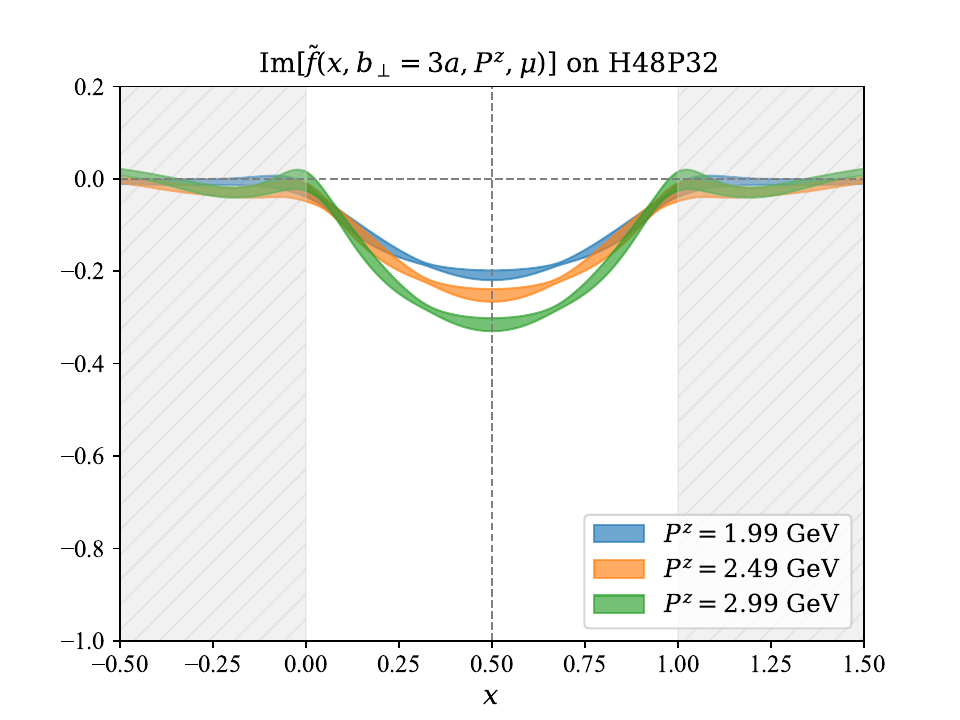}
\includegraphics[width=0.45\textwidth]{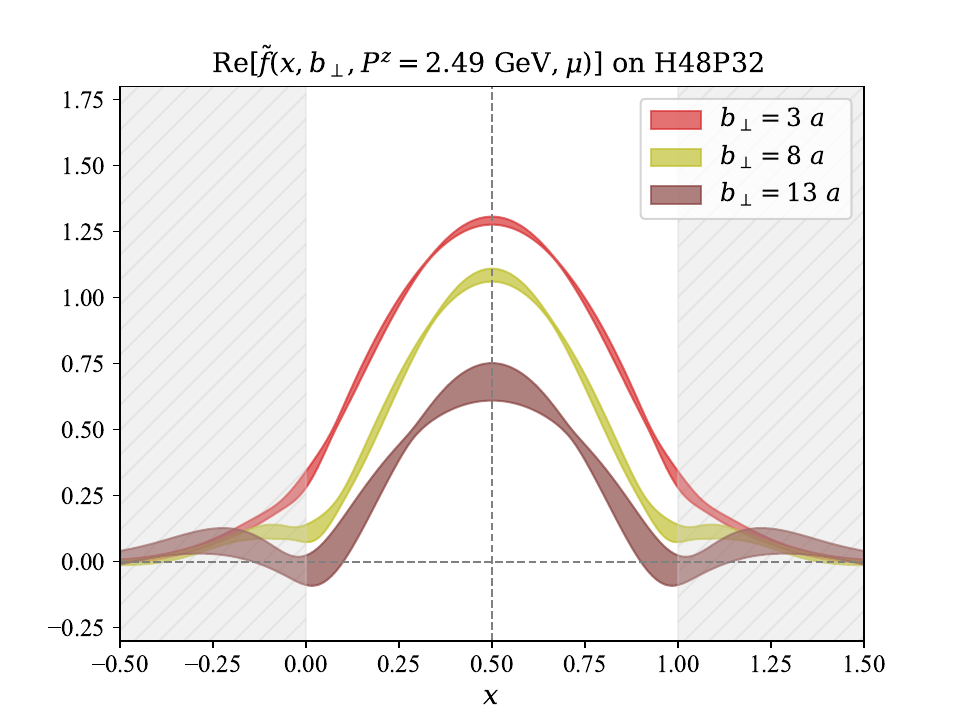} 
\includegraphics[width=0.45\textwidth]{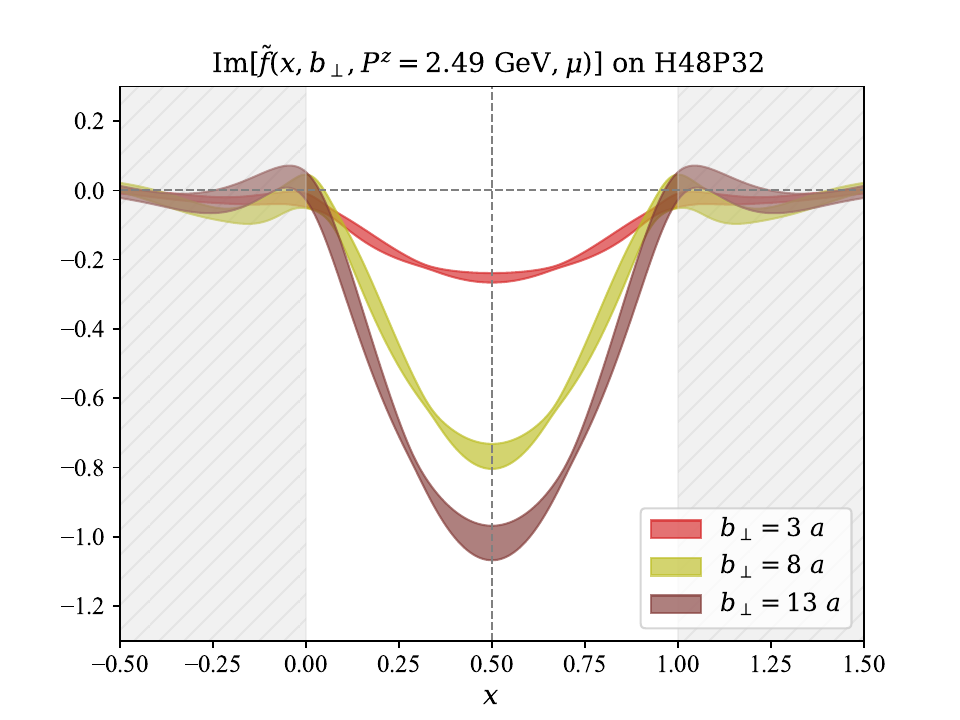}
\caption{The renormalized quasi-TMDWFs in momentum space. The upper panel shows the case of $b_\perp = 3 a$ and $P^z = \{1.99, 2.49, 2.99\}\mathrm{GeV}$ on H48P32 as examples. The lower panel compares the quasi-TMDWF in momentum space with $b_\perp\in\{3, 8, 13\}a$.
} 
\label{fig:TMDWFs in mom}
\end{figure*}

\begin{figure*}[http]
\centering
\includegraphics[width=0.45\textwidth]{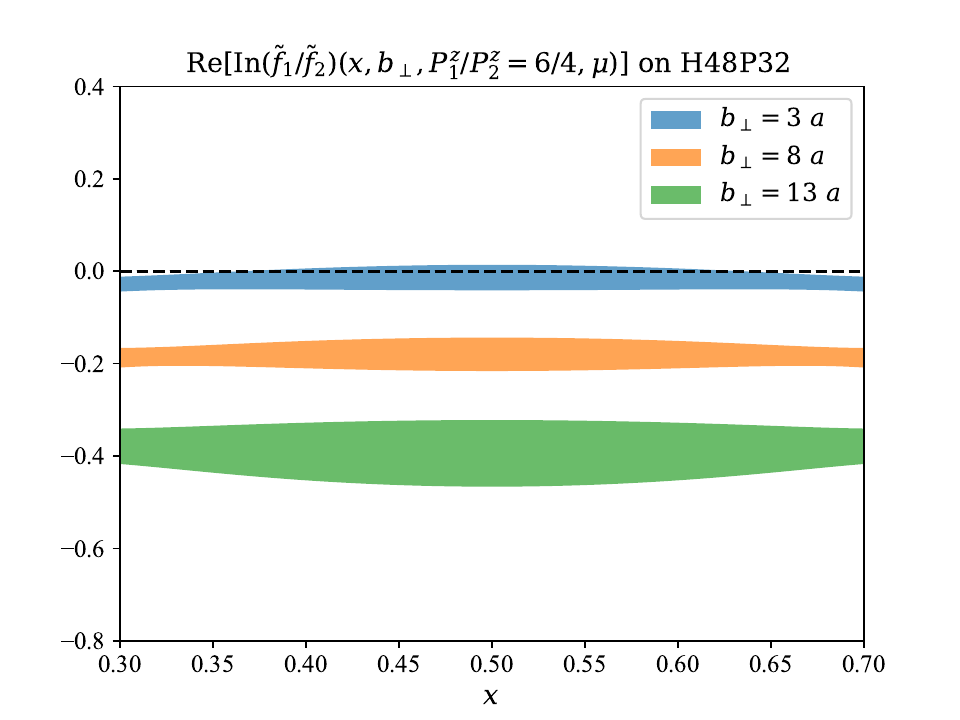} 
\includegraphics[width=0.45\textwidth]{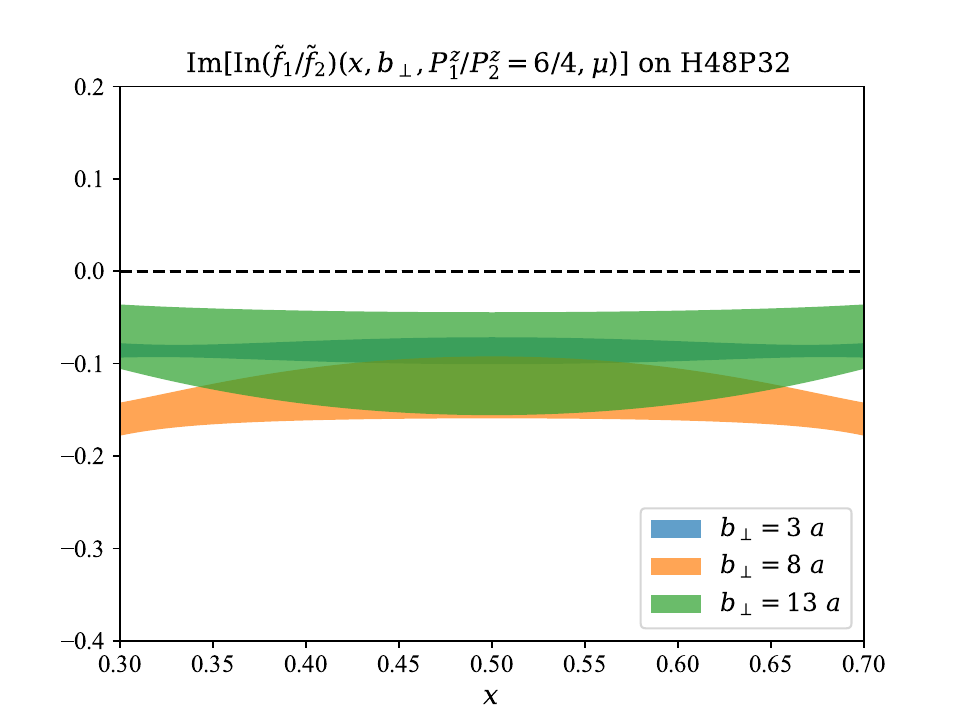}
\caption{Example of results for the real(left) and imaginary(right) part of the ratio for quasi-TMDWFs at different $b_\perp$, 
$\operatorname{ln}\dfrac{\tilde{f}(x, b_\perp, P_1^z, \mu)}{\tilde{f}(x, b_\perp, P_2^z, \mu)}$, 
with $P^z_1/P^z_2=6/4$ as a function of momentum fraction $x$. } 
\label{fig:TMDWFs ratio}
\end{figure*}

After Fourier transforming the coordinate-space matrix elements to momentum space, the quasi-TMDWF can be obtained as
\begin{align}
    \tilde{f}(x, b_\perp, P^z) =\int \frac{d\lambda}{2\pi}e^{i(x-\frac{1}{2})\lambda}\tilde{\Phi}(\lambda, b_\perp).
\label{Fourier transformation of extracted TMDWFs}
\end{align}
In Fig.~\ref{fig:diff HYP mom}, we present the results of the renormalized quasi-TMDWF in momentum space. Corresponding to Fig.~\ref{fig:TMDWFs in coor}, we compare the results for different HYP smearing steps, and the differences between the results obtained with different smearing steps are shown in the panels below each subplot. 
The difference, $\Delta \tilde{f}_(i-j)=\tilde{f}_{\rm HYPi}-\tilde{f}_{\rm HYPj}$, remains consistent with zero across the region to extract CS kernel $x \in [-0.5, 1.5]$, indicating that while HYP smearing substantially improves statistical precision, its impact on the central values is statistically insignificant. Potential systematic effects induced by HYP smearing are well below the statistical uncertainties at our current precision and are therefore not considered in the subsequent analysis.

In Fig.~\ref{fig:TMDWFs in mom} we present the renormalized quasi-TMDWF $\tilde{f}(x, P^z, b_\perp, \mu)$ as a function of the momentum fraction $x$, together with its dependence on the longitudinal momentum $P^z$ and the transverse separation $b_{\perp}$. The upper panels show the $P^z$ dependence at fixed $b_{\perp}$, which encodes the evolution in the rapidity scale $\zeta_z=(xP^z)^2$, and enables the extraction of the CS kernel. The lower panels display the $b_{\perp}$ dependence at fixed $P^z$, one can see that as $b_{\perp}$ increases, the real part of $\tilde{f}$ decreases while the imaginary part grows, indicating that the $b_{\perp}$ dependence predominantly enters through a phase in $\tilde{f}$.

Furthermore, we compare ratios of renormalized quasi-TMDWFs at different $P^z$, which isolate the $\zeta_z$ evolution. As indicated by Eq.(\ref{cs kernel of ratio}), the connection between the rapidity evolution of $\tilde{f}$ and that of the physical TMD $f$, that is the CS kernel, is perturbative. Notably, the ratios shown in Fig.~\ref{fig:TMDWFs ratio} show a strong $b_{\perp}$ dependence in the real part but hardly any $b_{\perp}$ dependence in the imaginary part, suggesting (within our current precision) that changing $P^z$ mainly rescales the magnitude through $\zeta_z$, while the phase is nearly $P^z$-independent and cancels in the ratio. Consequently, the real part is driven by the rapidity evolution kernel of $\tilde{f}$, producing the significant $b_{\perp}$ dependence, whereas the imaginary part reflects only a small, almost $b_{\perp}$- and $x$ independent global phase offset. In the large-$P^z$ limit this offset should vanish, and the residual imaginary signal is consistent with $1/P^z$ power corrections. These features make momentum ratios a clean, $x$-insensitive handle for extracting the CS kernel $K(b_\perp, \mu)$.

%%%%%%%%%%%%%%%%%%%%%%%%%%%%%%%%%%%%%

\subsection{CS Kernel from Quasi-TMDWFs}
\label{subsec:lattice CS kernel}

\begin{figure*}
\centering
\includegraphics[width=0.4\textwidth]{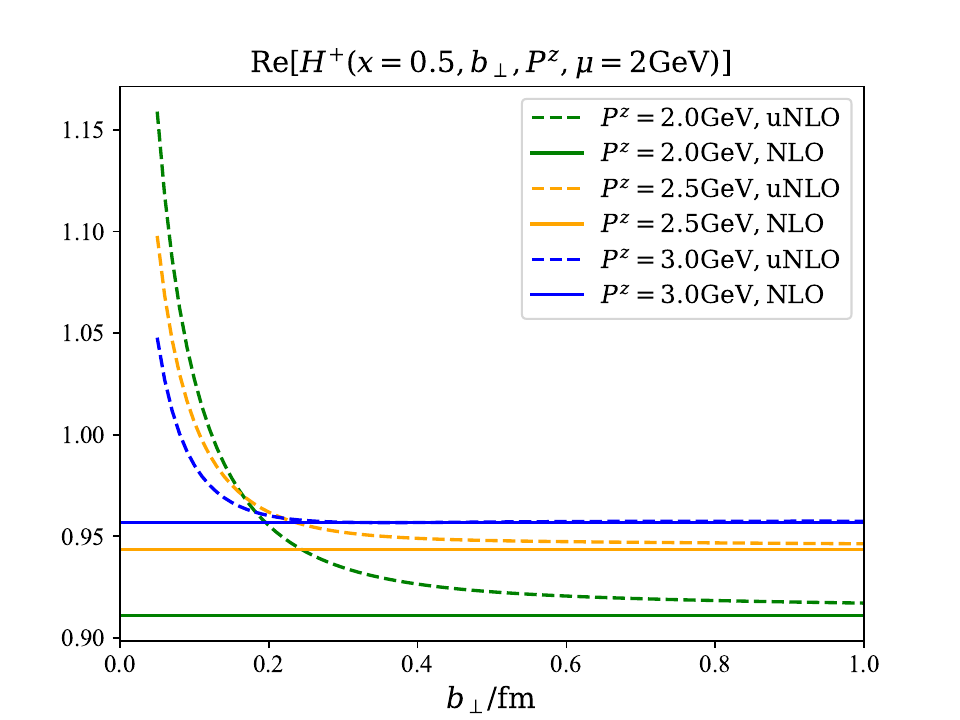} 
\includegraphics[width=0.4\textwidth]{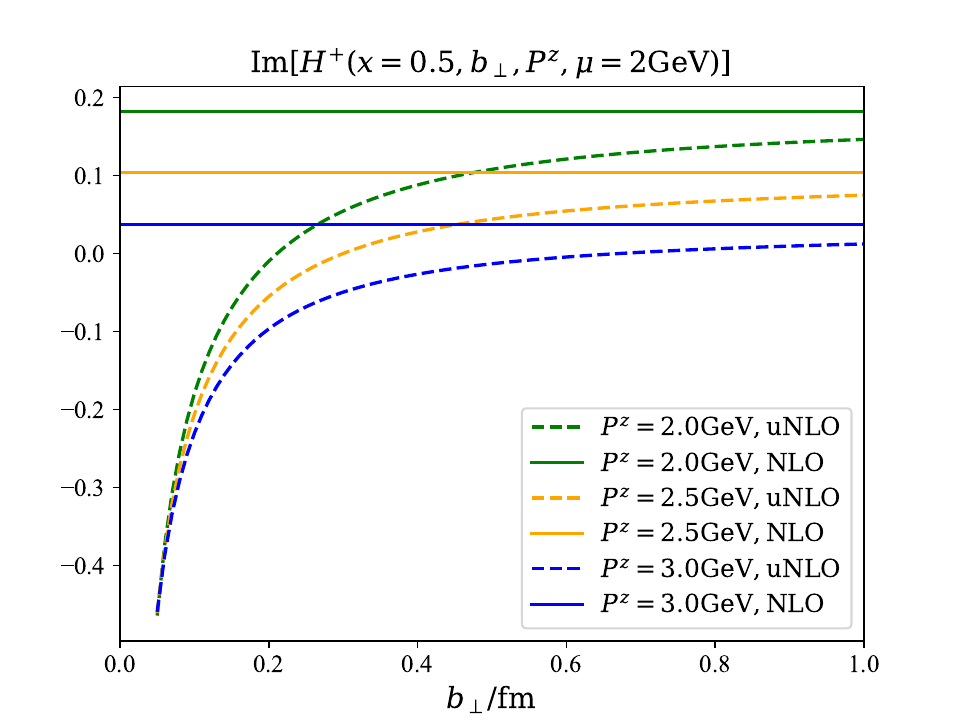}
\caption{Comparison of matching kernels at NLO (solid) and $b_\perp$-unexpanded NLO (uNLO, dotted) for fixed $x = 0.5$ and $\mu = 2$ GeV.
} 
\label{fig:uNLO C}
\end{figure*}

\begin{figure*}[http]
\centering
\includegraphics[width=0.45\textwidth]{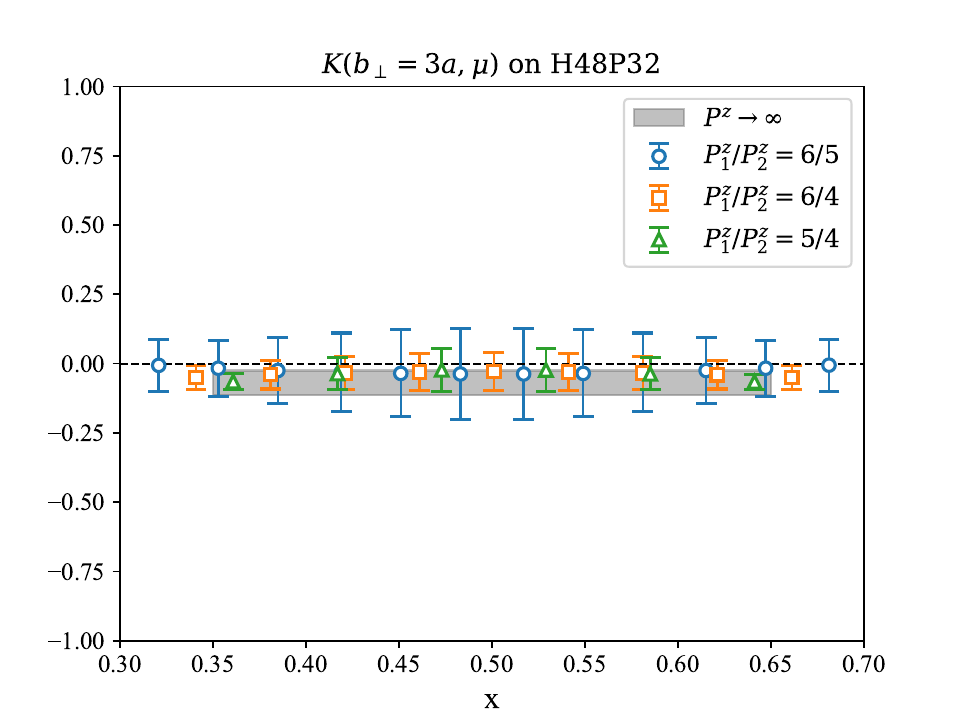} 
\includegraphics[width=0.45\textwidth]{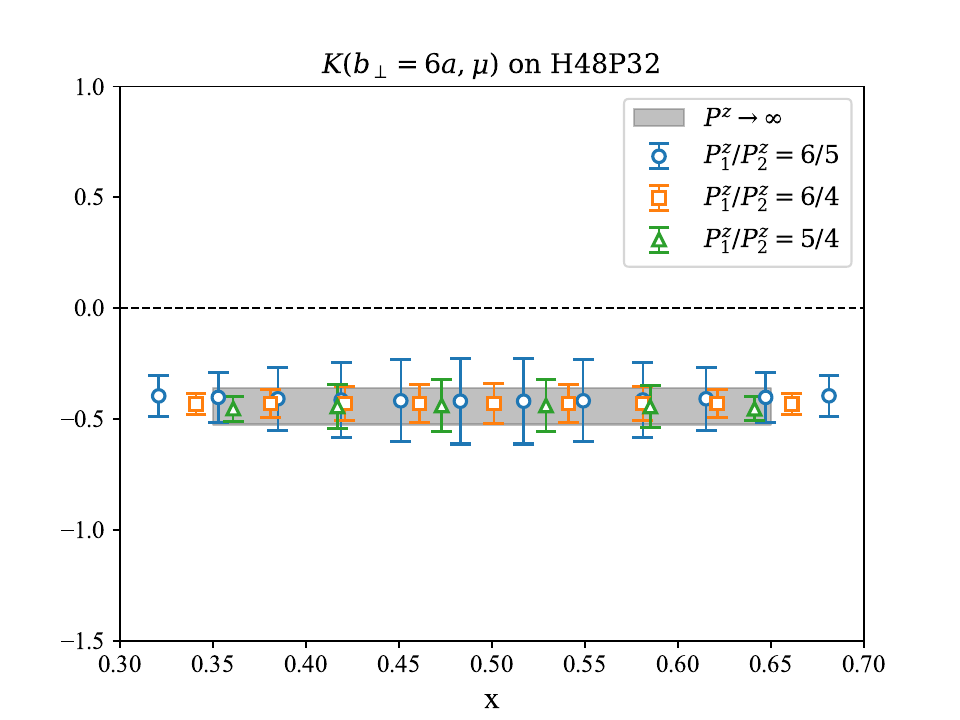}
\includegraphics[width=0.45\textwidth]{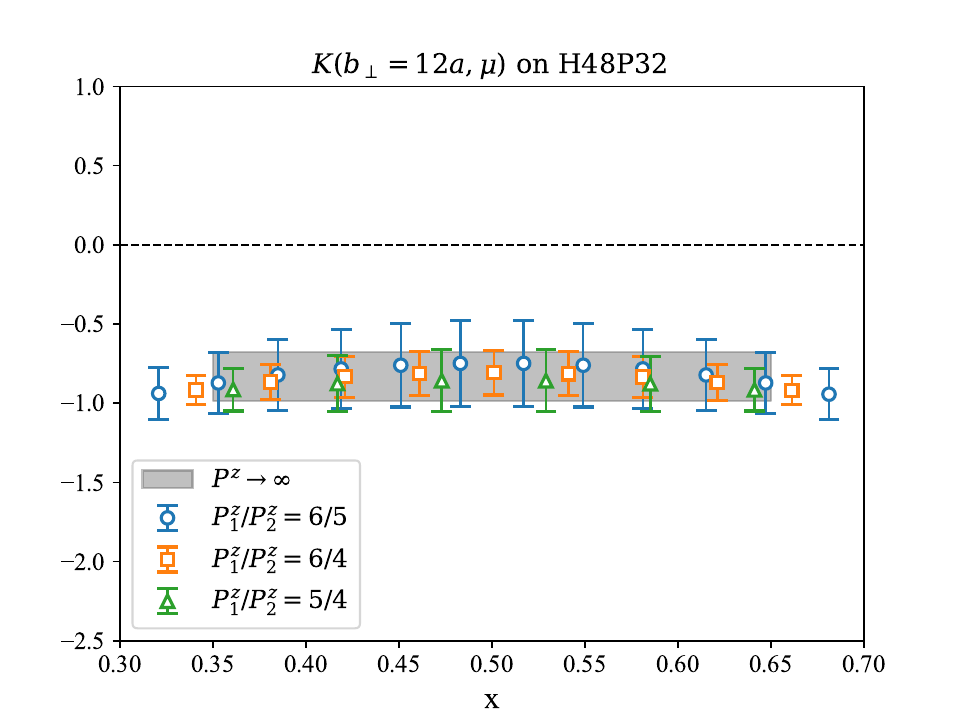} 
\includegraphics[width=0.45\textwidth]{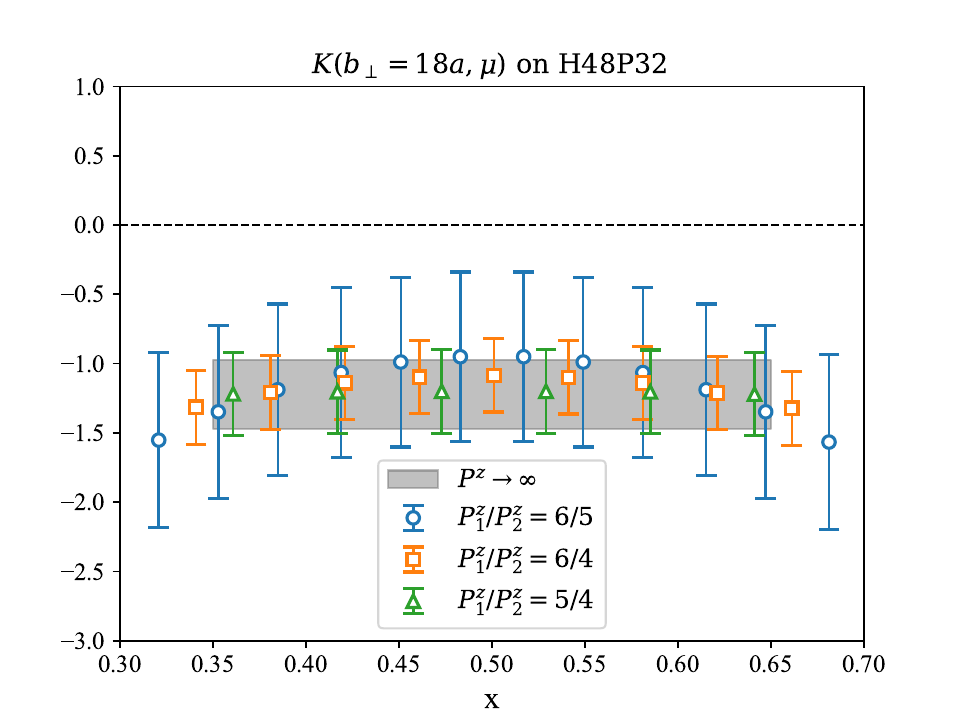}
\caption{CS kernel extracted from quasi-TMDWFs and its extrapolation to the large-momentum limit. Results are shown for $b_{\perp}={3, 6, 12, 18}a$ on the H48P32 ensemble, using three distinct momentum pairs $P_1^z/P_2^z$. Data points with errors denote $K(b_\perp, \mu;x, P_1^z, P_2^z)$ from individual momentum combinations, and the band indicates the large-momentum extrapolated result of  $K(b_\perp, \mu)$.
} 
\label{fig:CS kernel examples}
\end{figure*}

As indicated in Eq.~(\ref{factorizaton of quasi-TMDWFs}), the factorization of the renormalized quasi-TMDWF requires sufficiently large hadron momenta, such that either $x P^z b_\perp \gg 1 $ or $(1-x) P^z b_\perp \gg 1 $. However, in practice these conditions are not always satisfied, especially in the small-$b_\perp$ region. Therefore, following Ref.~\cite{Avkhadiev:2023poz}, we adopt the $b_\perp$-unexpanded matching kernel, in which the full $b_\perp$ dependence of the perturbative kernel is retained rather than expanded in powers of $b_\perp$. This treatment avoids relying on the small-$b_\perp$ expansion and thus provides a more reliable description over a broader range of transverse separations.

Fig.~\ref{fig:uNLO C} compares the results of $b_\perp$-unexpanded NLO (uNLO) matching kernel with the conventional NLO results~\cite{Deng:2022gzi}. A significant discrepancy is observed at small $b_\perp$, which diminishes as $b_\perp$ increases. The uNLO treatment improves the systematics in the small-$b_\perp$ region and helps suppress unphysical imaginary parts in the extracted CS kernel. This improvement stems from avoiding the fixed-order expansion in $b_\perp$, which implicitly assumes the hierarchy $1/b_\perp \gg P^z$ and is not strictly valid in typical lattice kinematics. Retaining the full $b_\perp$ dependence preserves a more faithful analytic continuation between coordinate and momentum space, leading to a more stable CS kernel extraction and better perturbative behavior in the nonperturbative regime.

Based on Eq.~\eqref{cs kernel of ratio}, we extract the CS kernel from the renormalized quasi-TMDWFs using the $b_{\perp}$-unexpanded matching kernel at different hadron momenta. As illustrated by the data points with uncertainties in Fig.~\ref{fig:CS kernel examples}, we obtain $K(b_{\perp}, \mu; x, P_1^z, P_2^z)$ from ratios formed at various large-momentum pairs $(P_1^z, P_2^z)$, and then perform an infinite-momentum extrapolation using
\begin{align}
&K(b_\perp, \mu;x, P_1^z, P_2^z,a,m_\pi) = K(b_\perp, \mu;a,m_\pi)\notag \\
& \qquad  + A(b_\perp, \mu;x,a,m_\pi)\left[\frac{1}{(P_1^z)^2}-\frac{1}{(P_2^z)^2}\right] 
\label{cs kernel for extrapolation}, 
\end{align}
which is motivated by factorization analysis of the TMDWF in Eq.~\eqref{factorizaton of quasi-TMDWFs}. The coefficient $A(b_\perp,\mu;x,a,m_\pi)$ parametrizes next-to-leading power effects, which can be numerically enhanced at the endpoint regions in $x$.

The extractions of $K(b_{\perp},\mu;a,m_{\pi})$ in the infinite-momentum limit are shown in Fig.~\ref{fig:CS kernel examples}. We find that the results obtained from different momentum pairs are mutually consistent, indicating that the chosen hadron momenta are sufficiently large and that residual higher-power contributions are well suppressed. Based on Eq.~(\ref{cs kernel for extrapolation}), we therefore perform the infinite-momentum extrapolation in the intermediate-$x$ window, $x\in[0.3,0.7]$, where power corrections are expected to be minimized. The resulting leading power CS kernel $K(b_\perp, \mu;a,m_\pi)$ are presented as the gray band in Fig.~\ref{fig:CS kernel examples}.

\begin{figure*}[http]
\centering
\includegraphics[width=0.48\textwidth]{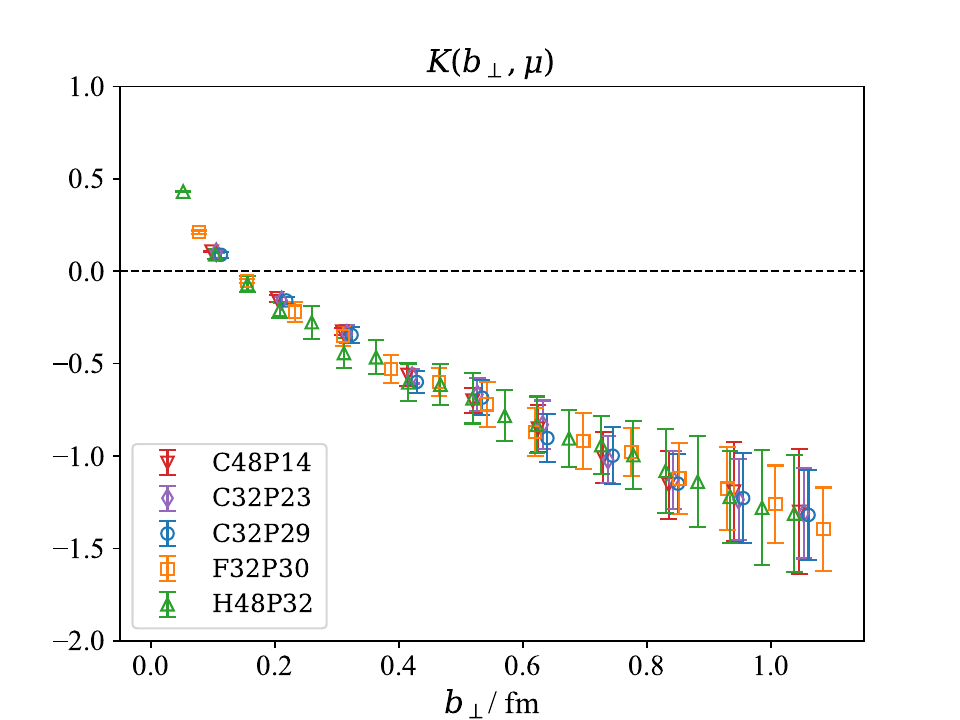} 
\includegraphics[width=0.48\textwidth]{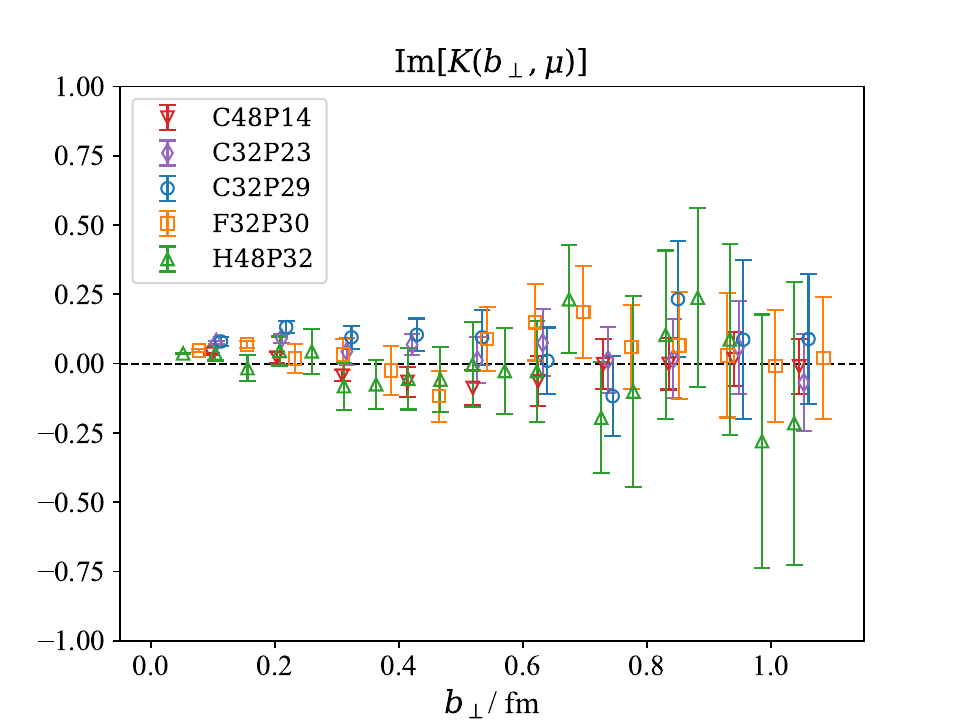} 
 \caption{Real (left) and imaginary (right) parts of the CS kernel extracted from five ensembles. Error bars denote statistical uncertainties.
} 
\label{fig:CS kernel real and imag part}
\end{figure*}

Following the procedure outlined above, we extract the CS kernel on the ensembles with different lattice spacings and pion masses, as shown in Fig.~\ref{fig:CS kernel real and imag part}. The real part of the CS kernels, illustrated in the left panel, exhibit remarkable consistency across the five ensembles, indicating that the discretization effects and mass corrections have been well controlled.

Theoretically, the CS kernel is purely real, and this expectation is borne out by the right panel of Fig.~\ref{fig:CS kernel real and imag part}. The imaginary parts from all ensembles are basically consistent with zero, especially for large $b_{\perp}$ cases. The residual unphysical imaginary contributions primarily arise from the perturbative matching kernel, which will mainly contribute to the small-$b_{\perp}$ region~\cite{LatticePartonLPC:2022eev}. 
To estimate the impact of this effect on the final results, we quantify the systematic uncertainty associated with the residual imaginary part as
\begin{align}
    \delta^{\mathrm{Im}}_{\mathrm{sys}} = 
    \sqrt{K(b_\perp, \mu)^2 + \operatorname{Im}[K(b_\perp, \mu)^2]} - |K(b_\perp, \mu)|.
    \label{eq:Imag systmatic error}
\end{align}
which measures the deviation between the complex magnitude and the real part of $K(b_{\perp}, \mu)$, and thus isolates the contribution from the imaginary part under the expectation that the physical CS kernel is real. A detailed discussion of this systematic appears in {Sec.~\ref{subsec:systematic uncertainties}.}

\begin{figure*}[http]
\centering
\includegraphics[width=0.3\textwidth]{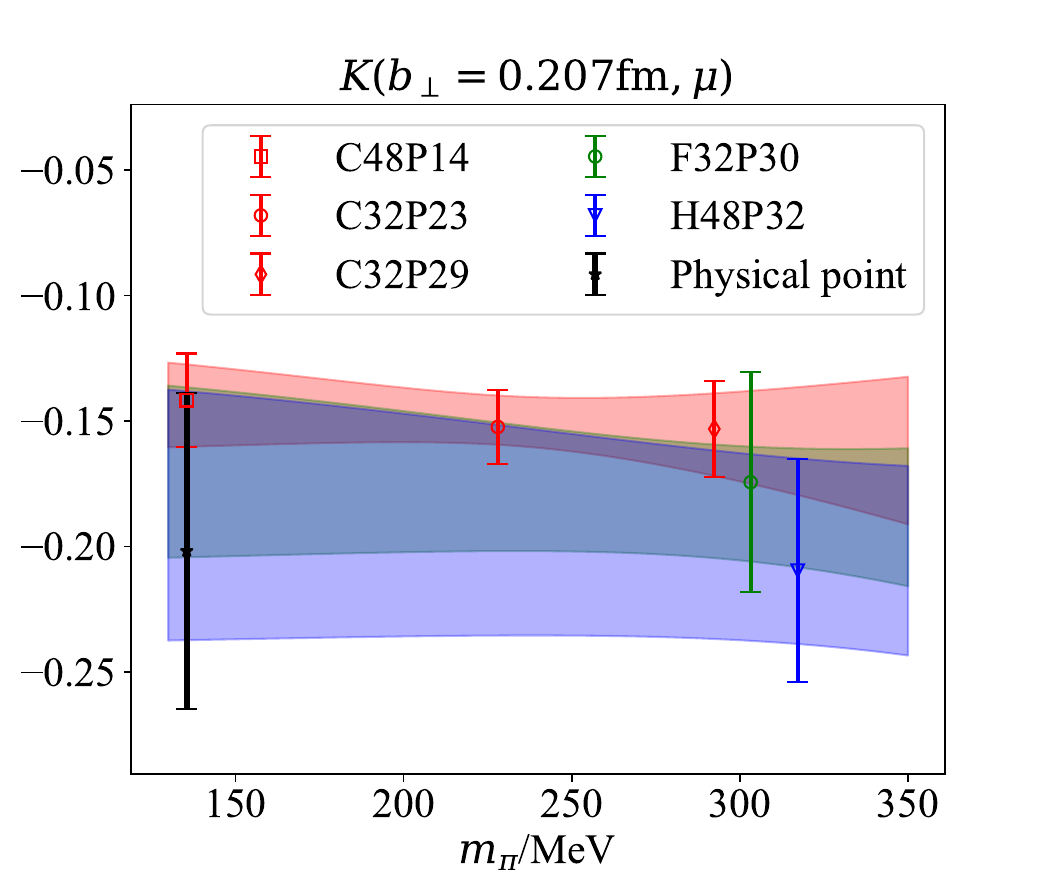} 
\includegraphics[width=0.3\textwidth]{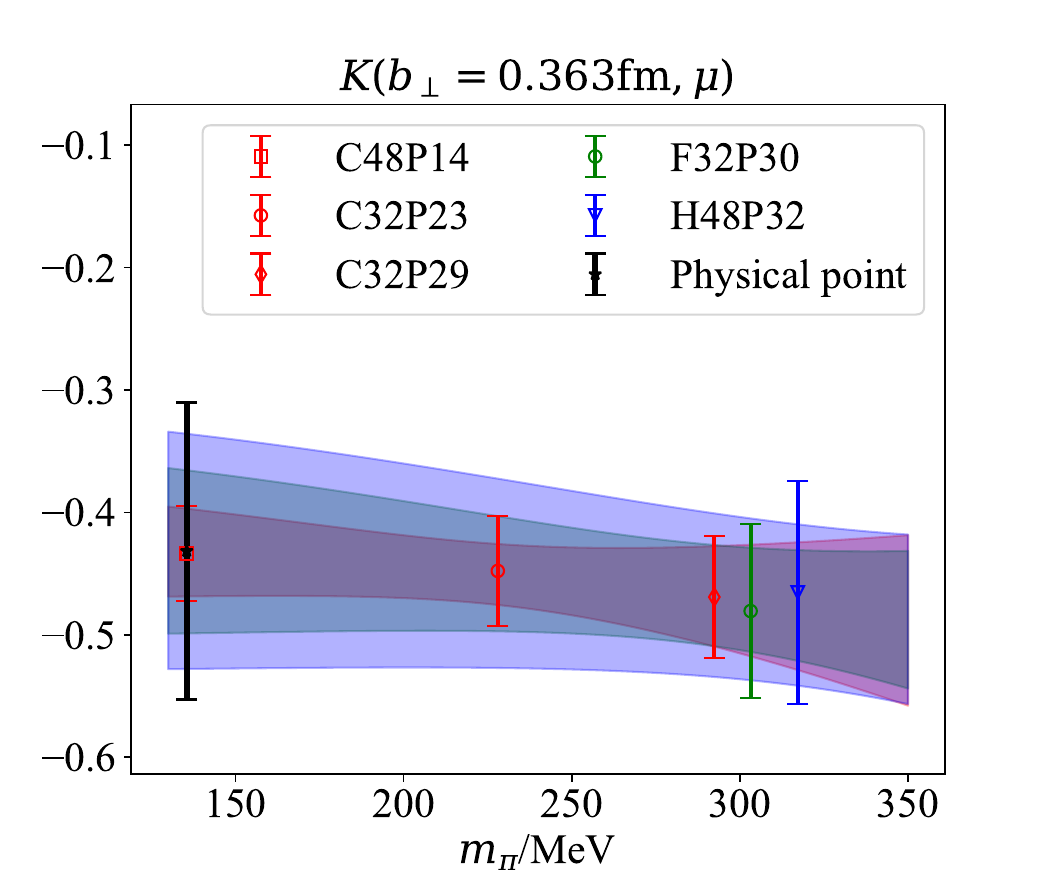}
\includegraphics[width=0.3\textwidth]{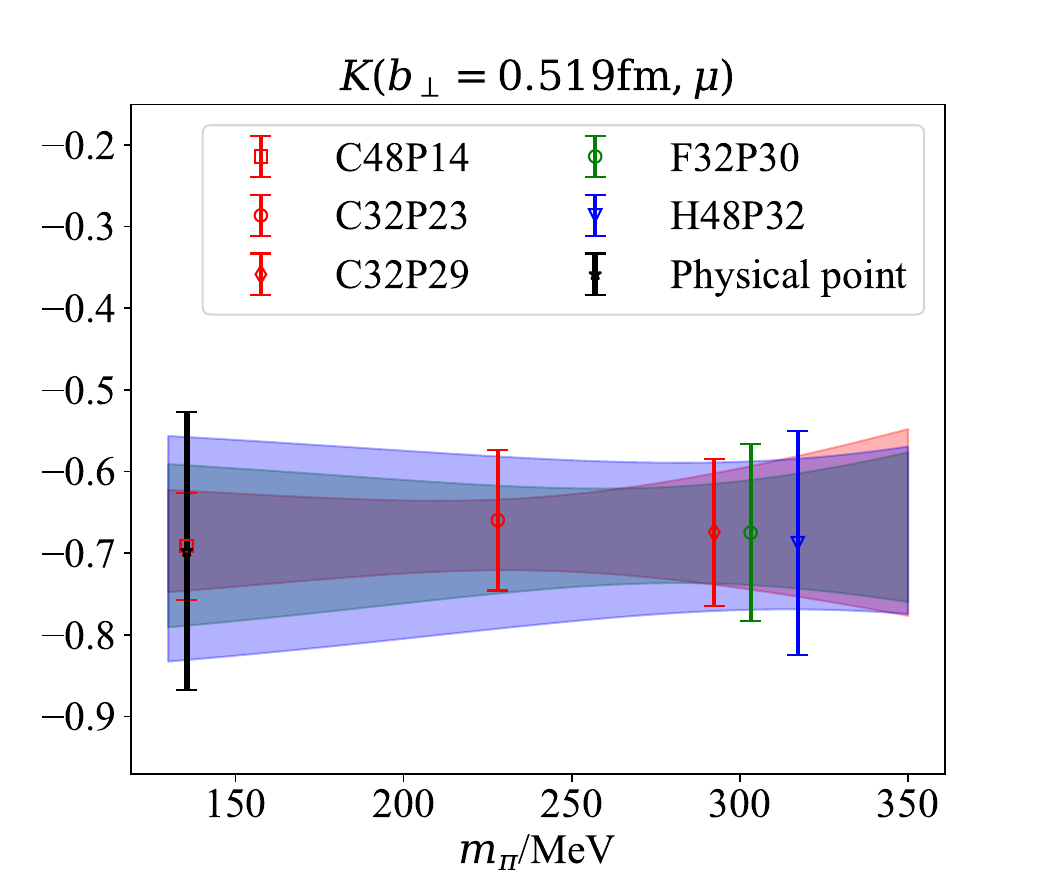}
\includegraphics[width=0.3\textwidth]{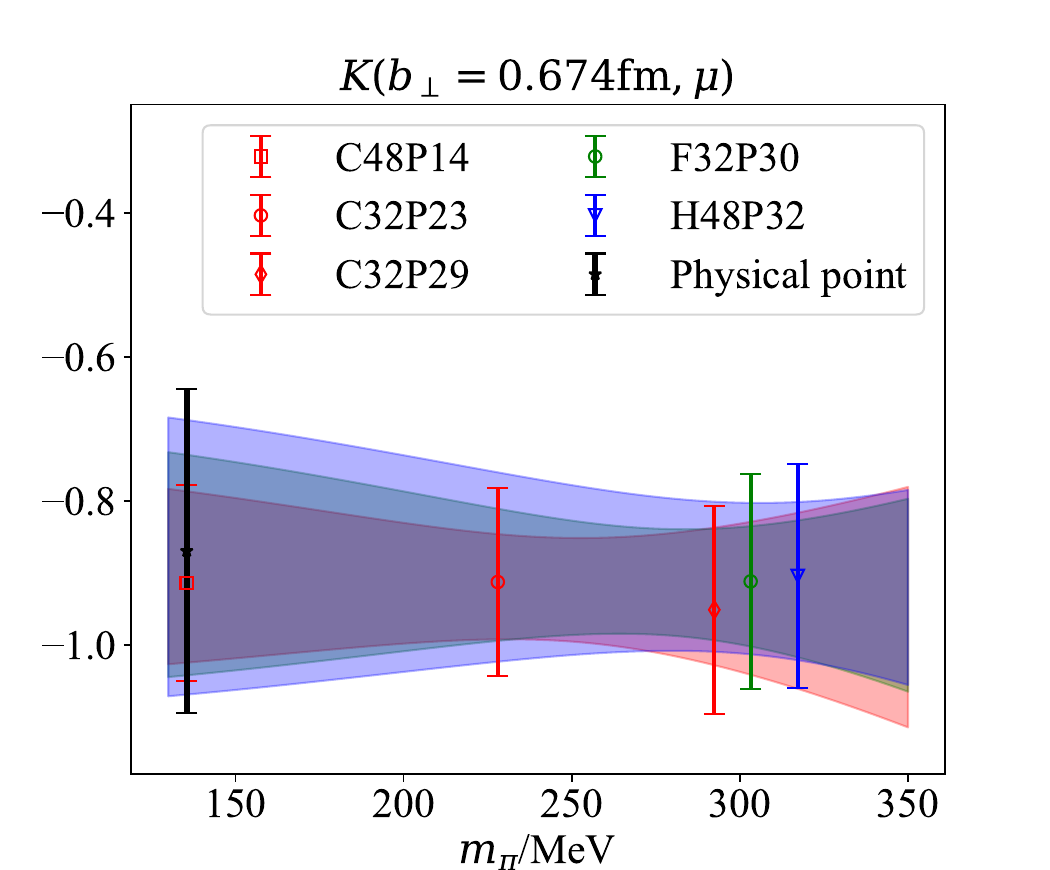}
\includegraphics[width=0.3\textwidth]{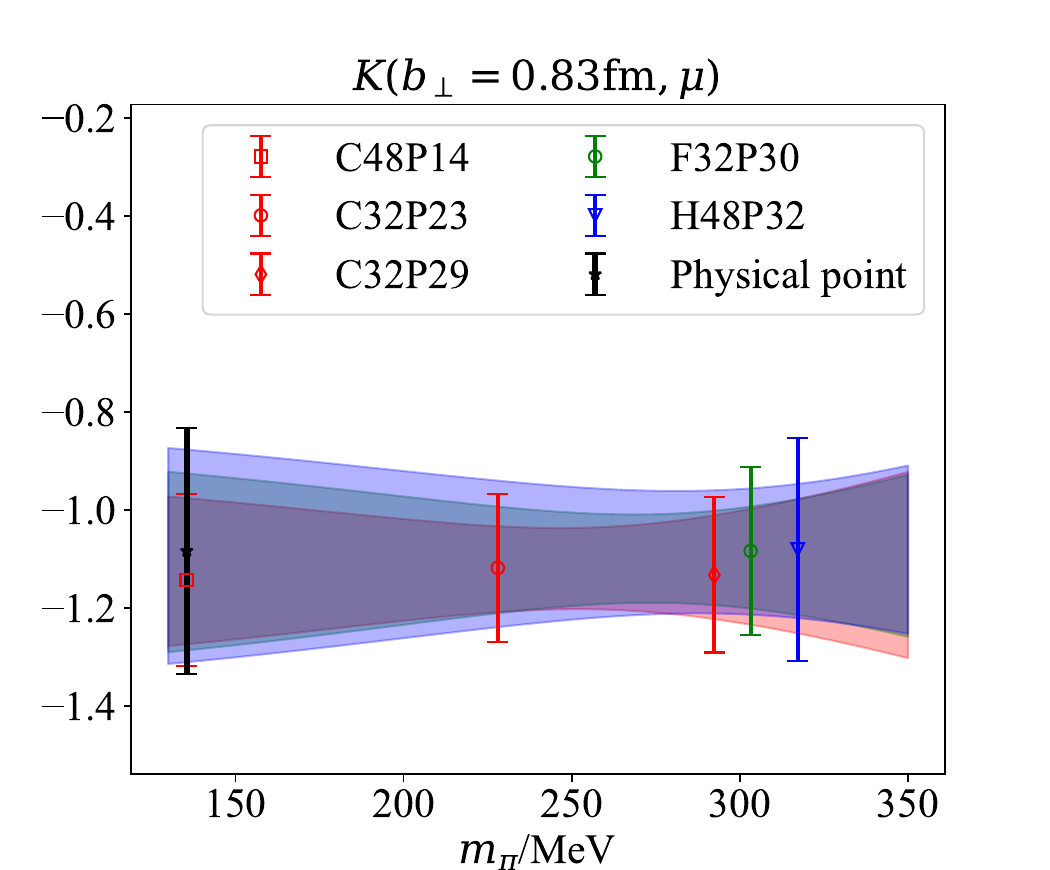}
\includegraphics[width=0.3\textwidth]{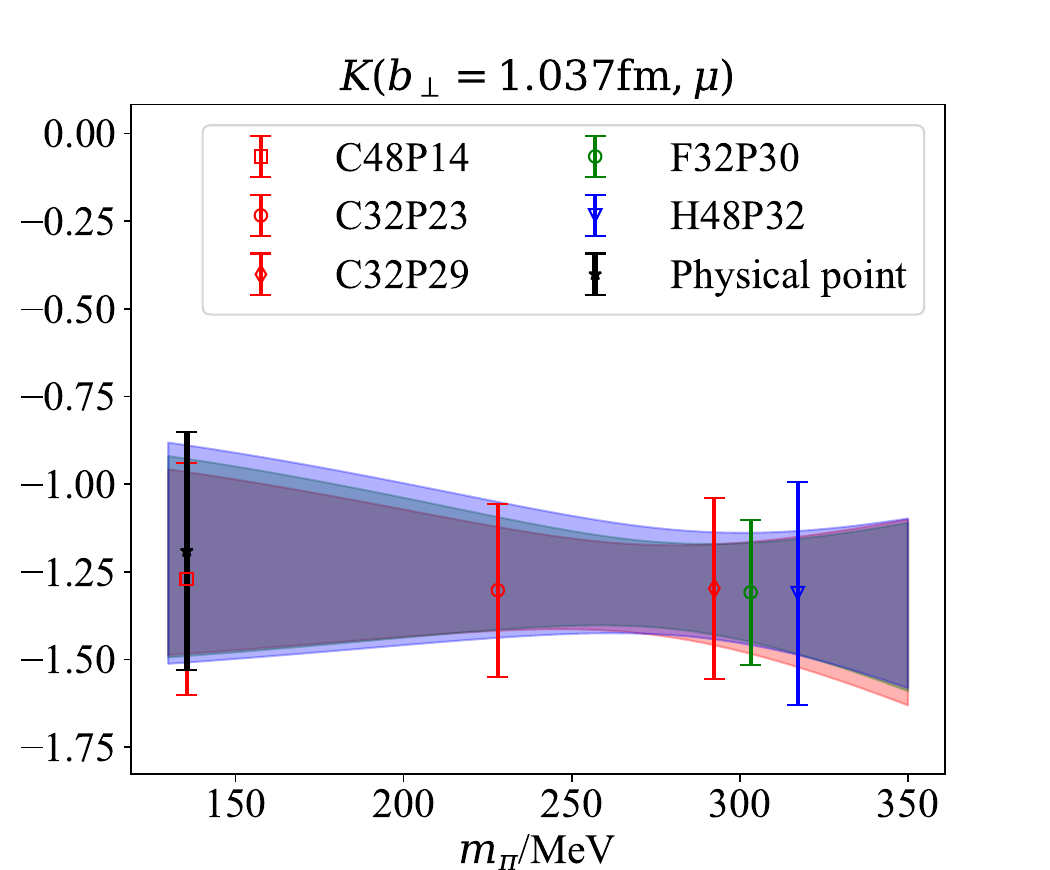}
\caption{Continuum and chiral extrapolations of the CS kernel. Colored points denote results from each ensembles, and black points indicate the extrapolated values. Error bars reflect statistical uncertainties only.
} 
\label{fig:physical extrap}
\end{figure*}

To perform the continuum extrapolation, results obtained at different lattice spacings are first interpolated to a common spacing $a'=0.05187$~fm. We then carry out, for each fixed $b_{\perp}$, a combined continuum and chiral extrapolation using
\begin{align}
    K(b_\perp, \mu;a, m_\pi) = &K(b_\perp, \mu) + a^2B(b_\perp, \mu) \notag\\
    &+ (m_\pi^2-m^2_{\pi, \mathrm{phy}})C(b_\perp, \mu)
\label{Physical Extra}
\end{align}
where the term $B$ encodes discretization effects and $C$ parameterizes deviations from unphysical pion masses. The fits are illustrated in Fig.~\ref{fig:physical extrap}, which compares data from all ensembles with the extrapolated curves at several $b_{\perp}$.
Across all shown $b_{\perp}$ values, the dependence on the pion mass is negligible, consistent with the expected universality of the CS kernel at sufficiently large boost factors $P^z/m_{\pi}$. The lattice spacing dependence is likewise modest, indicating that discretization effects are under good control. An exception occurs at very small transverse separations, $b_{\perp}\lesssim0.1$fm, where $b_{\perp}$ approaches the lattice cutoff ($b_{\perp}\sim a$) and discretization artifacts become significant. Since this region is beyond the reliable range of LaMET, we omit these points from the final predictions.

Beyond the leading $1/(P^z)^2$ corrections included in Eq.~(\ref{cs kernel for extrapolation}), there also may exist potential lattice-spacing–dependent discretization corrections proportional to $aP^z$~\cite{Avkhadiev:2024mgd}. Such contributions may originate from discretization artifacts at large hadron momentum and could, in principle, affect the infinite-momentum extrapolation. 

Beyond the leading $1/(P^z)^2$ corrections included in Eq.~(\ref{cs kernel for extrapolation}), additional lattice-spacing–dependent power corrections proportional to $aP^z$ may also arise, as suggested in Ref.~\cite{Avkhadiev:2024mgd}. Such terms can originate from discretization artifacts at large hadron momenta and could, in principle, affect the infinite-momentum extrapolation in Eq.~\eqref{cs kernel for extrapolation}. 
In the present analysis, we observe that these   corrections are small for two reasons. First, as shown in Fig.~\ref{fig:dispersion}, the boosted hadron energies on all ensembles are consistent with the continuum dispersion relation over the range of momenta used, indicating that discretization effects that grow with $P^z$ are well under control. Secondly, as illustrated in the left panel of Fig.~\ref{fig:CS kernel real and imag part}, the CS kernels extracted at the same momentum combinations but at different lattice spacings $a$ agree within uncertainties, providing direct evidence that the residual $aP^z$-dependent effects are swamped by the statistical errors.

To further validate this assumption, we combine the extrapolation ans\"atze in Eqs.~\eqref{cs kernel for extrapolation} and \eqref{Physical Extra}, and explicitly include a mixed term proportional to $aP^z$ to parameterize lattice-spacing–dependent power corrections suggested in Ref.~\cite{Avkhadiev:2024mgd},
\begin{align}
 &K(b_\perp,\mu;x,P^z_1,P^z_2,a,m_\pi)= K(b_\perp,\mu) \notag\\
&\qquad+ A(x,b_\perp,\mu)\!\left[\frac{1}{(P^z_1)^2}-\frac{1} {(P^z_2)^2}\right] \notag\\
&\qquad + a^2 B(b_\perp,\mu)
 + (m_\pi^2-m_{\pi,\mathrm{phy}}^2)C(b_\perp,\mu) \notag\\
&\qquad+ D(b_\perp,\mu)\, a(P^z_1-P^z_2),
\label{eq:simultaneous_fit}
\end{align}
where the term proportional to $D(b_\perp,\mu)$ is introduced to capture potential discretization artifacts that grow linearly with the hadron momentum.

\begin{figure}[http]
\centering
\includegraphics[width=0.95\textwidth]{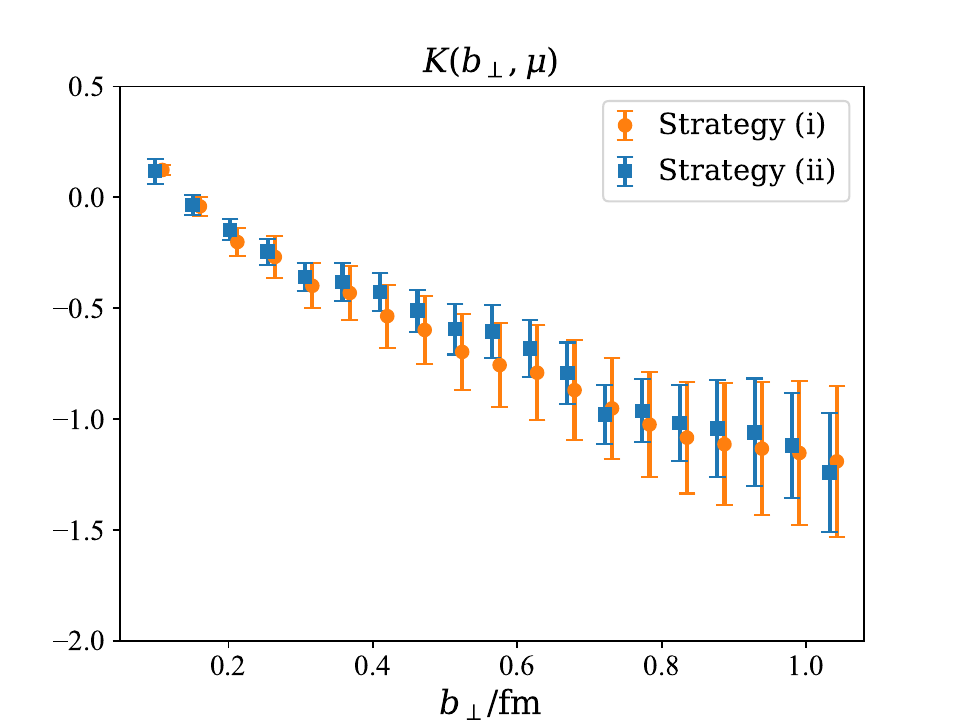} 
\caption{Comparison of the extrapolated CS kernel $K(b_\perp,\mu)$ obtained with two fitting strategies: (i) a chained fit based on Eqs.~(\ref{cs kernel for extrapolation}) and (\ref{Physical Extra}) and (ii) a joint fit using Eq.~(\ref{eq:simultaneous_fit}). Only statistical uncertainties are shown.}
\label{fig:aPz joint fit}
\end{figure}

The comparison of the extrapolated result of $K(b_\perp,\mu)$ from the fitting strategies i) a chained fit based on Eqs.~(\ref{cs kernel for extrapolation}) and (\ref{Physical Extra}) and ii) a joint fit from Eq.~(\ref{eq:simultaneous_fit}) are displayed in Fig.~\ref{fig:aPz joint fit}. One can see that within the current statistical precision, the results for two strategies are consistent within uncertainties. In particular, no statistically significant $aP^z$ term can be resolved, indicating that its contribution cannot be reliably disentangled from the leading power corrections with the present data. For the remainder of this work, we adopt the results obtained from the chained fitting strategy for our final analysis.

\subsection{Estimation of Systematic Uncertainties and Final Results}
\label{subsec:systematic uncertainties}

\begin{figure*}[http]
\centering
\includegraphics[width=0.8\textwidth]{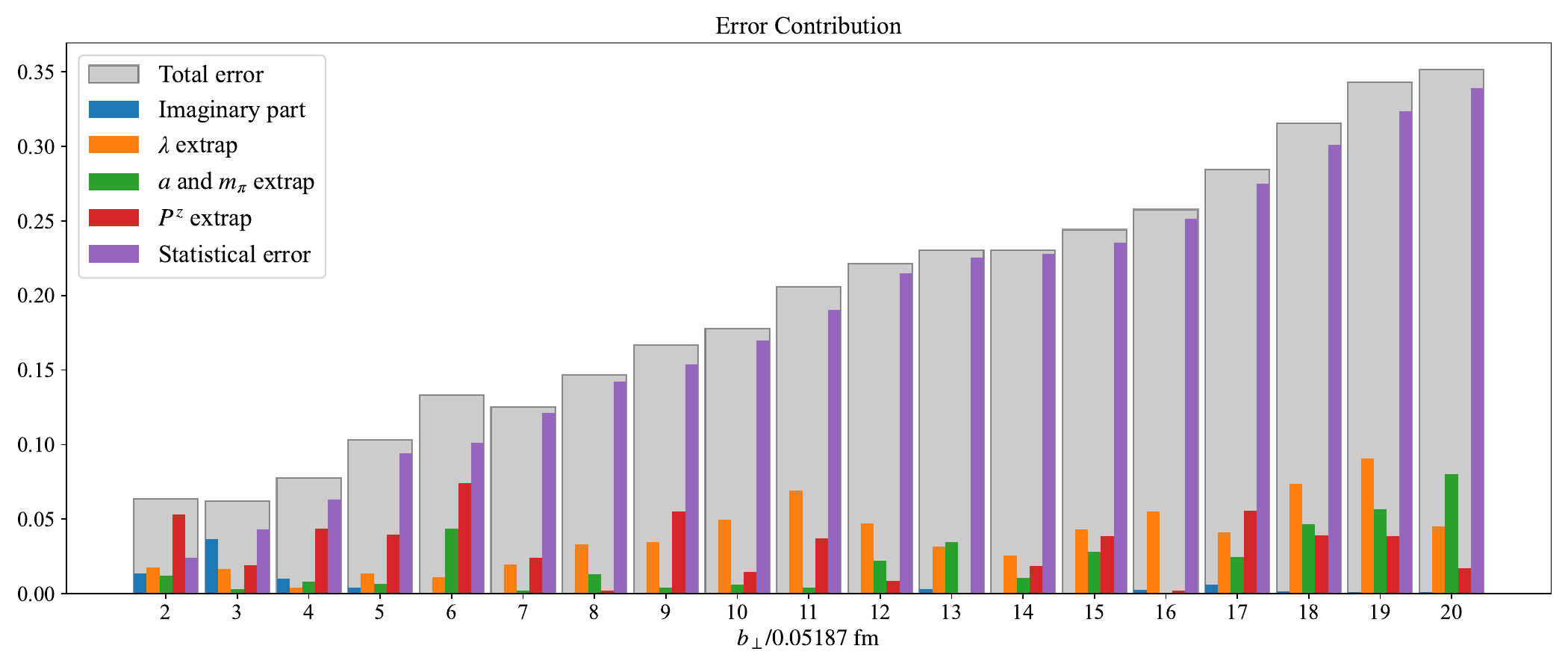} 
\caption{Error budget for the CS kernel, showing statistical uncertainties and systematic ones from different sources. Total uncertainties are obtained by adding these contributions in quadrature.
} 
\label{fig:System Error Contribution}
\end{figure*}

\begin{figure*}[http]
  \centering
     {\includegraphics[width=0.45\textwidth]{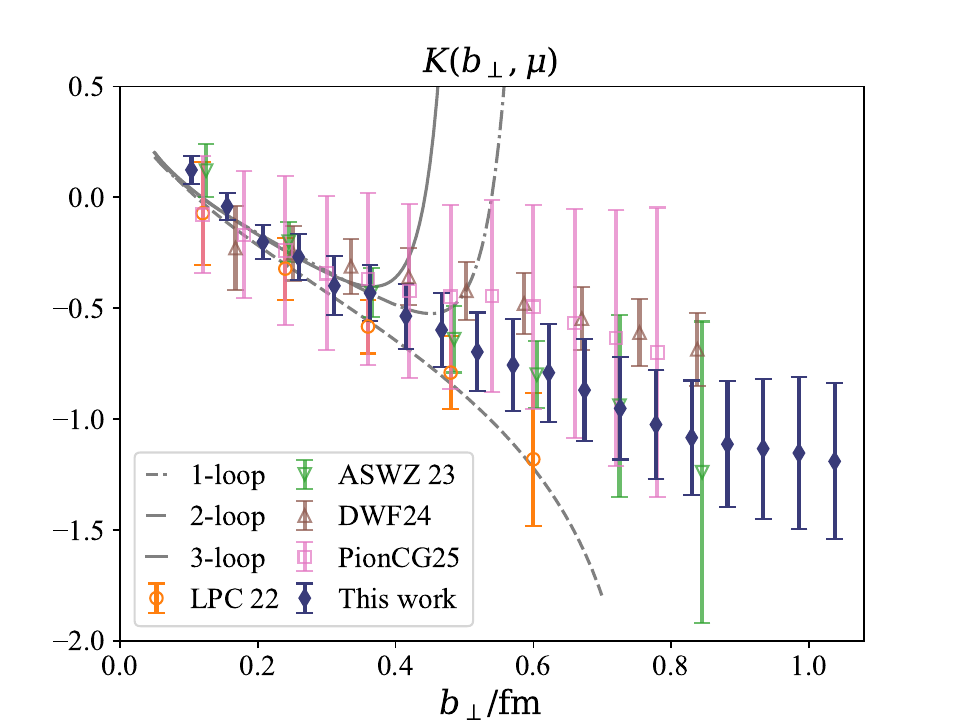}}
  {\includegraphics[width=0.45\textwidth]{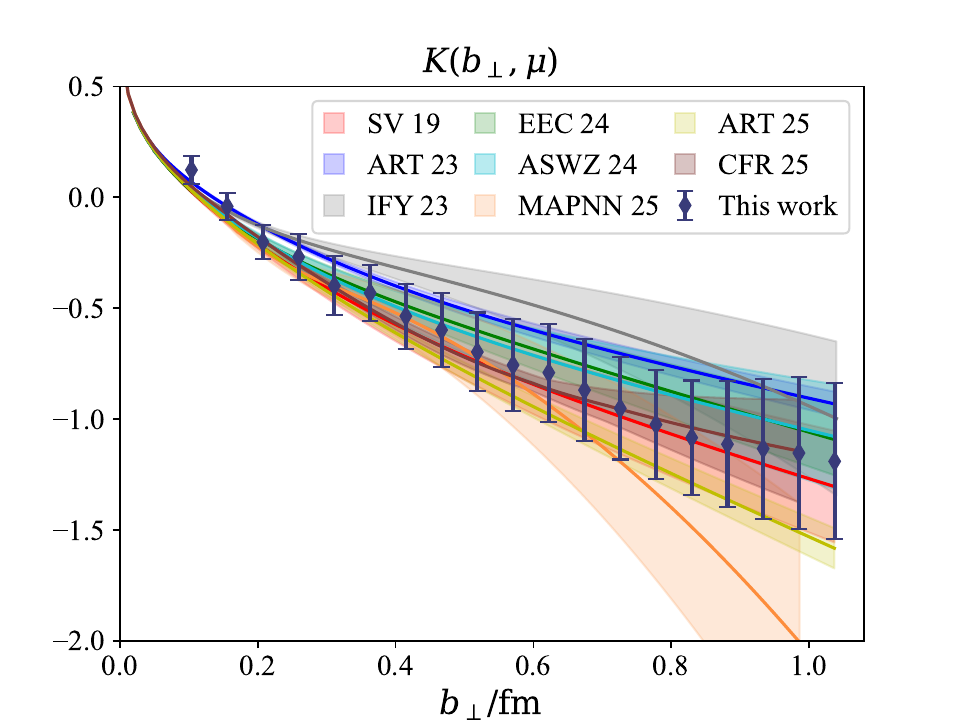}}
  \caption{Left panel: comparison of our result $K(b_\perp, \mu)$ the lattice calculations by LPC22~\cite{LatticePartonLPC:2022eev}, ASWZ23~\cite{Avkhadiev:2023poz}, DWF24~\cite{Bollweg:2024zet}, PionCG25~\cite{Bollweg:2025iol}, as well as the perturbative calculations up to three loops~\cite{Li:2016ctv, Moch:2017uml}.
  Right panel: comparison of our result with phenomenological extractions from SV19~\cite{Scimemi:2019cmh}, ART23~\cite{Moos:2023yfa}, IFY23~\cite{Isaacson:2023iui}, EEC24~\cite{Kang:2024dja}, ASWZ24~\cite{Avkhadiev:2024mgd}, MAPNN25~\cite{Bacchetta:2025ara}, ART25~\cite{Moos:2025sal}, CFR25~\cite{Camarda:2025lbt}.
}
  \label{fig:CS kernel result}
\end{figure*}    

For a  reliable prediction, we evaluate the main sources of systematic uncertainties in lattice QCD calculation and analysis as follows
\begin{enumerate}
    \item {\it $\delta^{\mathrm{Im}}_{\mathrm{sys}}$: from the imaginary part.} \\
    As mentioned above, physical CS kernel is purely real, and the nonzero imaginary component primarily originates from truncation effects in the perturbative matching kernel (particularly in the small-$b_{\perp}$ region). Although adopting the $b_{\perp}$-unexpanded matching kernel substantially reduces this artifact (compared with our earlier analysis \cite{LatticePartonLPC:2022eev, LatticePartonLPC:2023pdv}), the residual imaginary part, which now consistent with zero within uncertainties as illustrated in Fig.~\ref{fig:CS kernel real and imag part}, can still inflate statistical fluctuations and thus affect the final precision. To quantify the associated systematic, we evaluate $\delta^{\mathrm{Im}}_{\mathrm{sys}}$ by taking $\mathrm{Im}[K]$ as he ensemble-averaged imaginary part of the CS kernel over the five ensembles and  into Eq.(\ref{eq:Imag systmatic error}) to quantify the systematic uncertainty from the unphysical imaginary component.

    \item {\it $\delta^{\lambda}_{\mathrm{sys}}$: from $\lambda$ extrapolation.} \\
    On the lattice, matrix elements can only be computed up to a finite spatial separation, whereas the Fourier transform to momentum space requires integration over $|\lambda|=|zP^z|\to\infty$. To solve this, we employ the extrapolation form in Eq.\ref{Extrapolation} for the dimensionless variable $\lambda$. Physically, a reliable extrapolation should be performed in the asymptotic region where the renormalized coordinate–space matrix elements have entered the exponential–decay region at large $\lambda$. The residual ambiguity in choosing the fit window induces a systematic uncertainty.
    In practice, we assess this by varying the truncation parameter $\lambda_L$. By shifting $\lambda_L$ forward and backward by $\sim0.1$fm in separation (corresponding to 1–2 lattice spacings), repeat the fits, and take the discrepancy of the resulting momentum–space observables as $\delta^{\lambda}_{\mathrm{sys}}$. 

    \item {\it $\delta^{a, m_\pi}_{\mathrm{sys}}$: from continuum and chiral extrapolations.}
    Lattice QCD is formulated at finite lattice spacings, and the results must be extrapolated to the continuum. Similarly, ensembles with unphysical quark masses require a chiral extrapolation to the physical point. In this work, we employ five ensembles spanning three lattice spacings $a\simeq(0.105, ~0.077, ~0.052)$fm and a range of pion masses including the physical one. For each fixed $b_{\perp}$, we fit the data using Eq.(\ref{cs kernel for extrapolation}) and obtain the physical result $K(b_{\perp}, \mu)$. To estimate the associated systematic uncertainty, we take the single data point (from any ensemble) whose central value lies closest to the extrapolated result and use the absolute difference between this point and the extrapolated value as $\delta^{a, m_\pi}_{\mathrm{sys}}$. This choice provides a conservative estimate of the residual model dependence of the extrapolation.

    \item {\it $\delta^{P^z}_{\mathrm{sys}}$: from infinite–momentum extrapolation. } \\
    LaMET requires sufficiently large hadron boosts such that power corrections of order $1/(P^z)^2$ are under control. In practice, only finite momenta are accessible, and we therefore extrapolate to $P^z\to\infty$ using Eq.~(\ref{cs kernel for extrapolation}) to remove the leading $1/(P^z)^2$ effects. We estimate the associated systematic uncertainty by comparing the extrapolated value $K(b_{\perp},\mu)$ with the result obtained at the largest available lattice momentum, and take the absolute difference as $\delta^{P^z}_{\mathrm{sys}}$.

    \item {\it $\delta^{aP^z}_{\mathrm{sys}}$: from lattice-spacing-dependent power corrections. } \\
    As discussed in Sec.~\ref{subsec:lattice CS kernel},  results from the joint-fit strategy based on Eq.~\eqref{eq:simultaneous_fit} are consistent with those from the chained-fit strategy. For a conservative error budget, we adopt the latter one in the final analysis, which allows us to estimate the systematic uncertainties from the continuum extrapolation and the infinite-momentum extrapolation separately. Accordingly, we do not introduce an independent uncertainty $\delta^{aP^z}_{\mathrm{sys}}$; instead, any residual $aP^z$-dependent effects are treated as subleading and are effectively covered by the systematic variations already included in $\delta^{P^z}_{\mathrm{sys}}$ and $\delta^{a}_{\mathrm{sys}}$.

\end{enumerate}

In Fig.~\ref{fig:System Error Contribution}, we compare the statistical uncertainty of the final CS kernel with the systematic ones discussed above. As noted above, our predictions for the CS kernel are quoted for $b_{\perp}\gtrsim0.1$fm. The comparison shows that at small $b_{\perp}$ various systematics remain non-negligible, whereas at larger $b_{\perp}$ the statistical error becomes dominant. We combine uncertainties in quadrature, 
\begin{align}
    \delta_{\mathrm{tot}} &= \sqrt{\delta^2_{\mathrm{stat}}+\delta^2_{\mathrm{sys}}} \notag \\
    &=\sqrt{\delta^2_{\mathrm{stat}}+(\delta^{\mathrm{Im}}_{\mathrm{sys}})^2+(\delta^{\lambda}_{\mathrm{sys}})^2+(\delta^{P^z}_{\mathrm{sys}})^2+(\delta^{a, m_\pi}_{\mathrm{sys}})^2}, 
\end{align} 
where $\delta_{\mathrm{stat}}$ denotes the statistical uncertainty. The statistical uncertainty dominates at large $b_{\perp}$ because the signal of quasi-TMDWF decays exponentially with increasing of spatial separation, while the noise, set by correlators of local operators, remains approximately constant, leading to a rapid deterioration of the signal-to-noise ratio.
Overall, the results shown indicate that the uncertainties remain under control up to $b_{\perp}\sim1$fm, this allows us to make reliable predictions for the CS kernel in the large-$b_{\perp}$ region.

Combining statistical and systematic uncertainties, Fig.~\ref{fig:CS kernel result} presents our results (black data points) for the CS kernel at $\mu=2$GeV. 
As shown in the left panel, in the perturbative region where $1/b_{\perp}\gtrsim1$GeV, our results agree well with perturbation theory up to three-loop accuracy~\cite{Li:2016ctv, Moch:2017uml}. In the nonperturbative region, we  compare with previous first-principles lattice determinations, including LPC22~\cite{LatticePartonLPC:2022eev}, ASWZ23~\cite{Avkhadiev:2023poz}, ASWZ24~\cite{Avkhadiev:2024mgd}, DWF24~\cite{Bollweg:2024zet}, and PionCG25~\cite{Bollweg:2025iol}.
Notably, our results provide improved precision, even for $b_{\perp}\gtrsim1$fm, the uncertainties remain under control.

In addition, we compare our results with those from global analyses, as shown in the right panel of Fig.~\ref{fig:CS kernel result}. Specifically, we include SV19~\cite{Scimemi:2019cmh}, ART23~\cite{Moos:2023yfa}, IFY23~\cite{Isaacson:2023iui}, EEC24~\cite{Kang:2024dja}, MAPNN25~\cite{Bacchetta:2025ara}, ART25~\cite{Moos:2025sal}, and CFR25~\cite{Camarda:2025lbt}. Usually, global analyses typically achieve smaller uncertainties than direct lattice determinations by imposing explicit model parametrizations, in particular for the $b_\perp$ dependence (see, e.g., Ref.~\cite{Avkhadiev:2024mgd}). 
Remarkably, our determination reaches a comparable precision, especially at large $b_\perp$, without introducing a dedicated model parametrization for the $b_\perp$ dependence in the continuum-extrapolated CS kernel. In our analysis, the independence of a $b_\perp$ ansatz is enabled by the infinite-momentum extrapolation in Eq.~\eqref{cs kernel for extrapolation}, where we find no statistically significant sensitivity to the mixed discretization term proportional to $aP^z$ and therefore neglect it at the current level of precision.
The resulting agreement between our CS kernel and the global-fit determinations demonstrates full compatibility with phenomenological extractions within uncertainties.

\subsection{Some attempted global Fits using Phenomenology and Lattice Inputs}

%%%%%%%%%%%%%%%%%%%%%%%%%%%%%%%%
\begin{table*}[t]
\centering
    \renewcommand{\arraystretch}{1.5}
      \setlength{\tabcolsep}{3mm}
\begin{tabular}{@{} cc ccc @{}} 
\hline\hline
 & Fit scenario & $g_2$ & $\chi^2/\mathrm{dof}$ \\  
\hline
& Only Phenomenological extractions       & $0.0702 (28)$  & $2.12$ \\
Model 1 & Lattice QCD   & $0.0711 (36)$ & $0.80$ \\
& Joint Analysis &    $0.0705 (22) $ & $1.46$ \\
\hline
& Only Phenomenological extractions      & $0.1570 (33) $ & $3.34$ \\
Model 2 & Lattice QCD   & $ 0.1435 (46)$ & $1.02$ \\
& Joint Analysis &   $0.1521 (27)$ & $2.26$\\  
\hline
& Only Phenomenological extractions     & 0.1972 (81)  & 2.73 \\
Model 3 & Lattice QCD   &  0.210 (10) & 0.97 \\
& Joint Analysis &   0.1996 (64) & 1.86 \\  
\hline
\end{tabular}
\caption{Summary of the global one-parameter fits to the Collins--Soper kernel.
The fitted parameter $g_2$ and the corresponding $\chi^2/\mathrm{dof}$ are also listed.}
\label{tab:global_fit_summary}
\end{table*}
%%%%%%%%%%%%%%%%%%%%%%%%%%%%%%%%

To quantitatively assess the potential influence of lattice data on phenomenological determinations of the CS kernel, 
we perform a global one-parameter fit combining phenomenological and lattice results. 
The analysis follows the standard perturbative–nonperturbative decomposition of the CS kernel~\cite{Scimemi:2019cmh, Kang:2024dja, Moos:2023yfa, Moos:2025sal}, 
\begin{align}
K(b_\perp, \mu) = -2 \int_{\mu_{*}}^{\mu} \frac{d\mu'}{\mu'} \Gamma_{\mathrm{cusp}}\!\left[\alpha_s(\mu')\right]
\notag\\- 2 D_{\mathrm{pert}}(b_*, \mu_*) - 2 D_{\mathrm{NP}}(b_\perp), 
\label{eq:CSkernel_model}
\end{align}
where $ \Gamma_{\mathrm{cusp}} $ is the cusp anomalous dimension, and $ D_{\mathrm{pert}} $ and $ D_{\mathrm{NP}} $ denote the perturbative and nonperturbative contributions, respectively.
The variable $ b_*(b_\perp) = b_\perp/\sqrt{1+b_\perp^2/b_{\max}^2} $ and  $\mu_*(b_\perp)=2e^{-\gamma_E}/b_*(b_\perp)$ are introduced to regulate the Landau pole~\cite{Collins:1984kg, Collins:2014jpa}, 
and we choose $ b_{\max}=2 e^{-\gamma_E}\, \mathrm{GeV}^{-1} $.
In our analysis, $ D_{\mathrm{pert}}(b_*, \mu_{*}) $ is evaluated at 2-loop level~\cite{Scimemi:2018xaf, Echevarria:2012pw}, 
and the 3-loop cusp anomalous dimension~\cite{Henn:2019swt, Moch:2004pa, Vogt:2004mw, vonManteuffel:2020vjv} is used in the evolution kernel to achieve a consistent perturbative accuracy.

For the nonperturbative part, we adopt the following models~\cite{Scimemi:2019cmh, Scimemi:2019cmh, Avkhadiev:2025wps}, 
\begin{align}
\mathrm{Model~1: \quad}& D_{\mathrm{NP}}(b_\perp) = g_2\, b_\perp\, b_* , \\
\mathrm{Model~2: \quad}& D_{\mathrm{NP}}(b_\perp) = g_2 \, b_\perp^2 , \\
\mathrm{Model~3: \quad}& D_{\mathrm{NP}}(b_\perp) = g_2\, \operatorname{ln}(b_\perp/b_*), 
\label{eq:DNP_model}
\end{align}
which introduces a single free parameter $ g_2 $ that governs the long-distance behavior of the CS kernel. 
This parametrization effectively captures the mild $b_\perp$-dependence of nonperturbative effects at large $b_\perp$, 
and has been widely used in recent global TMD analyses.

Using this model, we perform three fitting scenarios to assess how lattice inputs constrain the CS kernel:
\begin{enumerate}
    \item \textbf{Exp only}: fit using phenomenological extractions from global TMD analyses (SV19~\cite{Scimemi:2019cmh}, IFY23~\cite{Isaacson:2023iui}, EEC24~\cite{Kang:2024dja}, MAPNN25~\cite{Bacchetta:2025ara});
    \item \textbf{Lattice only}: fit using recent lattice calculations (ASWZ23~\cite{Avkhadiev:2023poz}, DWF24~\cite{Bollweg:2024zet}, PionCG25~\cite{Bollweg:2025iol}, LPC25);
    \item \textbf{Combined}: simultaneous fit including both phenomenological and lattice data.
\end{enumerate}

To ensure a uniform comparison, all available determinations of the CS kernel are first interpolated with a fixed spacing of $0.1\, \mathrm{fm}$ over the range 
$[0.2, 1.0]\, \mathrm{fm}$, and the global fits are then performed on the interpolated dataset.
Each fitting scenario contains only one free parameter, $ g_2 $, and the quality of fit is characterized by the resulting $\chi^2/\mathrm{dof}$.

As summarized in Table~\ref{tab:global_fit_summary}, 
the inclusion of lattice results in the combined analysis produces a visible reduction in the statistical uncertainty of $ g_2 $, 
while the central value shifts moderately compared with the phenomenology-only fit.
This demonstrates that lattice determinations provide complementary constraints on the long-distance behavior of the CS kernel, 
particularly in the large-$b_\perp$ region where phenomenological extractions alone tend to be less precise.

It should be pointed out that this study is only  intended to illustrate the potential of fits that combine lattice data and phenomenological input. We fully acknowledge that doing so correctly is a highly subtle endeavor that will require a comprehensive joint analysis, far beyond the simple combination of individual results we have presented here. At present, we lack the capability to conduct such a rigorous analysis, which would also need to account for correlations among experimental datasets (such as in Ref.~\cite{Avkhadiev:2025wps}). Our primary aim here is merely to highlight that incorporating lattice QCD input into global TMD analyses could represent a promising direction for future research.

\section{Summary and outlook}
\label{sec:Summary and outlook}

In order to achieve a first-principles determination of the Collins--Soper (CS) kernel on the lattice, 
we have carried out a comprehensive calculation within the framework of LaMET using five $(2+1)$-flavor QCD ensembles from CLQCD, 
with three different lattice spacings and four valence pion masses, and boosted the hadron momentum up to $3~\mathrm{GeV}$. 
The quasi-TMDWF matrix elements were extracted from carefully constructed nonlocal two-point correlation functions. 
The Wilson-loop renormalization procedure was employed to remove the linear divergence of the matrix elements. We have implemented a renormalization approach inspired by the self-renormalization, 
which allowed us to investigate in detail the logarithmic divergences of the quasi-TMDWF matrix elements 
and successfully match the lattice data to perturbative results in the $\overline{\mathrm{MS}}$ scheme. 

To optimize the statistical precision of our simulations, we adopted HYP smearing for the gauge fields, 
and explicitly demonstrated that the renormalized matrix elements remain unaffected by the smearing procedure. 
By performing large-$\lambda$ extrapolations and Fourier transformations, we obtained the quasi-TMDWFs in momentum space. 
To further suppress systematic uncertainties from the imaginary part of the CS kernel, 
we adopted the $b_\perp$-unexpanded matching kernel at NLO. 
In addition, the large-momentum extrapolation of the CS kernel under different momentum combinations 
was used to systematically eliminate power corrections. 

Finally, through a combined extrapolation across all four ensembles, we achieved, for the first time, a determination of the CS kernel up to $1~\mathrm{fm}$ in the continuum limit and at the physical pion mass without introducing any explicit parametrization of its $b_{\perp}$ dependence. 
All major sources of systematic uncertainty have been carefully analyzed and quantified, 
with their contributions explicitly presented. 
Our results are in good agreement with both previous lattice QCD calculations and phenomenological extractions from experimental data, 
demonstrating the reliability of this first-principles approach. 

Looking ahead, the methods established in this work provide a solid foundation 
for further lattice QCD studies of TMD observables. 
With increasing computational resources and future extensions to higher statistics, finer lattice spacings, 
and larger hadron momenta, 
it will become possible to achieve even more precise determinations of the CS kernel and related quantities. 
Such efforts will play an essential role in connecting lattice QCD, perturbative QCD, and experimental measurements, 
thereby deepening our understanding of the three-dimensional structure of hadrons.

\section*{Acknowledgement}
We thank Huey-Wen Lin, Tian-Bo Liu, Yong Zhao for useful discussions. We thank the CLQCD collaborations for providing us the gauge configurations with dynamical fermions~\cite{CLQCD:2023sdb}, which are generated on the HPC Cluster of ITP-CAS, the Southern Nuclear Science Computing Center(SNSC), the Siyuan-1 cluster supported by the Center for High Performance Computing at Shanghai Jiao Tong University and the Dongjiang Yuan Intelligent Computing Center. 
This work is supported in part by  Natural Science Foundation of China under grant No. 12375069, 12125503, 12305103,  12565014, 12405101, 12293060, 12205106, 12575084, 12293062, 12525504, 12435002 and 12447101 by the Fundamental Research Funds for the Central Universities. 
Q.A.Z is also supported by the Fundamental Research Funds for the Central Universities.  
J.H is also supported by Guang-dong Major Project of Basic and Applied Basic Research No. 2025A1515012199. 
H.L is also supported by the GHfund B (202407028788).  
Y.Y is also supported by supported in part by National Key R\&D Program of China No.2024YFE0109800. 
J.L.Z is also supported by T.D. Lee scholarship.
J.Z. is also supported by the Talent Research Startup Foundation of Hainan Normal University HSZK-KYQD-202523, and Opening Foundation of Shanghai Key Laboratory of Particle Physics and Cosmology under Grant No. 22DZ2229013-5.
The computations in this paper were run on the Siyuan-1 cluster supported by the Center for High Performance Computing at Shanghai Jiao Tong University, and Advanced Computing East China Sub-center.   The LQCD simulations were performed using the PyQUDA software suite~\cite{Jiang:2024lto} and QUDA~\cite{Clark:2009wm, Babich:2011np, Clark:2016rdz} through HIP programming model~\cite{Bi:2020wpt}.

\begin{appendix}
\begin{widetext}

\section{Self Renormalization}
\label{Ax:Self Renormalization}

The renormalization group (RG) equations allow us to establish the factorization between the ultraviolet (UV) divergence scale (represented by $\mu$ or $1/a$) and the relevant physical scales. This leads to the definition of a multiplicative renormalization factor $Z_O$, which serves to convert the lattice results into the $\overline{\rm MS}$ scheme~\cite{Ji:1995vv}. Building upon these first-principle insights, phenomenologically motivated renormalization strategies are often employed in practical lattice computations to effectively incorporate such conversions.

We need to eliminate the $a$-dependence of the subtracted quasi-TMDWF matrix elements as defined in Eq.~(\ref{renormalized_ME}), and match the results to those in the $\overline{\text{MS}}$ scheme. Motivated by the idea of self-renormalization, we propose the following new scheme to extract the renormalization factor $Z_O$~\cite{LatticePartonLPC:2021gpi}.

When the external hadron momentum is zero, the renormalization condition for the subtracted quasi-TMDWFs can be expressed as~\cite{Deng:2022gzi, Zhang:2022xuw}:
\begin{align}
    \tilde{\Phi}(z, P^z=0, b_\perp, &1/a, m) = Z_O(1/a, \mu)   \times \tilde{\Phi}^{\rm \overline{MS}}(\mu, z, P^z=0, b_{\perp}, m), 
\end{align}
where the finite quark mass $m$ is introduced to regulate the IR divergence. The perturbative result up to one loop for zero momentum matrix element can be found in Ref.~\cite{Zhang:2022xuw}, 
\begin{align}
&\tilde{\Phi}^{\rm \overline{MS}}(\mu, z, P^z=0, b_{\perp}, m) \equiv \tilde{\Phi}_{0}^{\rm \overline{MS}}(\mu, z, b_{\perp}) \notag\\
&=  1 + \frac{\alpha_{\overline{\rm MS}}(\mu) C_F}{2\pi}\left[ \frac{3}{2}\ln\frac{\mu^2 (z^2+b_{\perp}^2)}{4 e^{-2\gamma_E}} + \frac{1}{2} - \frac{2z}{b_{\perp}}\arctan{\frac{z}{b_{\perp}}}\right].
\end{align}

The parametrization of the matrix element $\tilde{\Phi}$ is written as\cite{LatticePartonLPC:2021gpi}:
\begin{align}
     \operatorname{ln} \tilde{\Phi}(z, P^z=0, b_\perp, &1/a)  = \frac{\gamma_0}{\beta_0}\operatorname{ln} \biggl[ \operatorname{ln}[1/(a\Lambda_{\mathrm{QCD}} )]   \biggr]  + \frac{c_1}{\operatorname{ln}[1/(a\Lambda_{\mathrm{QCD}} )]} + g(z, b_\perp), 
\end{align}
where the first two terms correspond to logarithmic divergences, while $g(z, b_\perp)$ denotes the intrinsic non-perturbative physics. 

It should be noted that, for clarity of presentation, the lattice data and fit results shown below are based on data with HYP smearing once(HYP=1). Results for HYP=0, 2 and others for HYP=1 are presented in the Appendix.~\ref{ax:Results_for_self_renormalization}. After interpolating the lattice data at three different lattice spacings to a common spacing of $a' = 0.05~\mathrm{fm}$, a joint fit allows us to constrain the form of the logarithmic divergence. In the fit, we fix $\mu = 2~\mathrm{GeV}$ and $\Lambda_{\mathrm{QCD}} = 0.2~\mathrm{GeV}$, and we introduce the associated systematic uncertainty:
\begin{align}
    \delta_{\rm all}^{\operatorname{ln}\tilde{\Phi}} = \sqrt{( \delta_{\rm stat}^{\operatorname{ln}\tilde{\Phi}})^2+ (\delta_{\rm sys} a\mu)^2}, 
\end{align}
where we set $\delta_{\rm sys} = 0.007$, and from the joint fit for all lattice matrix elements ($b_\perp \geq 0.1~\mathrm{fm}$), we obtain $c_1 = 0.020(15)$ with $\chi^2/\mathrm{dof} = 1.0$.

\begin{figure}[http]
\centering
\includegraphics[width=0.47\textwidth]{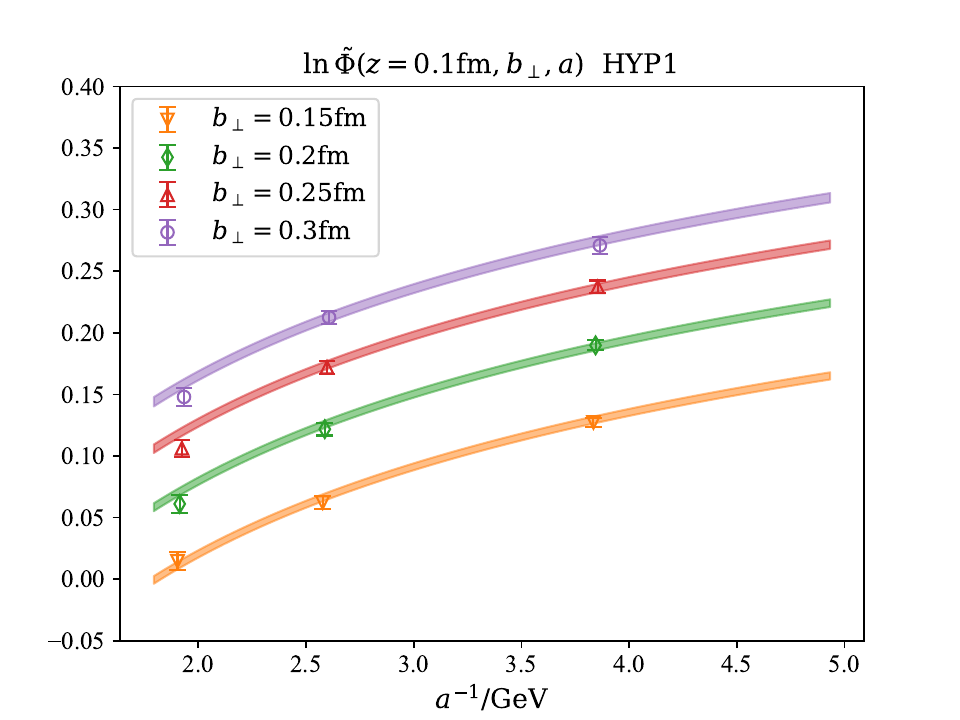} 
\includegraphics[width=0.43\textwidth]{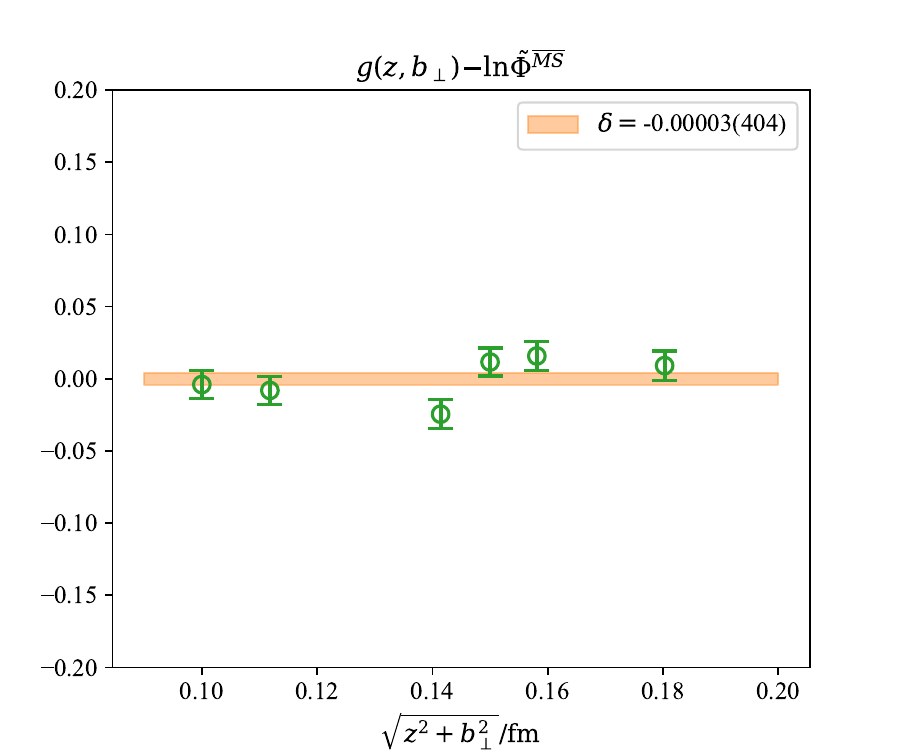}
\caption{The result of joint fit for the self renormalization, taking $b_\perp = \{0.15, 0.2, 0.25, 0.3\}~\mathrm{fm}$ and $z = 0.1~\mathrm{fm}$ as an example(left panel). Matching result of the $g(z, b_\perp)$ to the $\overline{\text{MS}}$ scheme with HYP=1(right panel).} 
\label{fig:Self renormalization example and renormalization match}
\end{figure}

Besides taking into account that discretization effects may play a role in the region of small $b_\perp$ and $z$, we fit the lattice data in the perturbative range using the following formula $(\sqrt{z^2+b_\perp^2} \leq 0.2~\mathrm{fm})$ :
\begin{align}
     \operatorname{ln} \tilde{\Phi}&(z, P^z=0, b_\perp, 1/a)  = \frac{\gamma_0}{\beta_0}\operatorname{ln} \biggl[ \operatorname{ln}[1/(a\Lambda_{\mathrm{QCD}} )]   \biggr]  + \frac{c_1}{\operatorname{ln}[1/(a\Lambda_{\mathrm{QCD}} )]} + g'(z, b_\perp) + f(z, b_\perp)a^2
\end{align}
where the $g'(z, b_\perp)$ term will correspond to the result of the scheme $\overline{\rm MS}$, and the $f(z, b_\perp)$ term denotes the contribution of the discretization error. In addition, considering the renormalization factor $Z_O(1/a, \mu)$ can be parameterized as:
\begin{align}
    \operatorname{ln}Z_O(1/a, &\mu)  = \frac{\gamma_0}{\beta_0}\operatorname{ln} \biggl[ \operatorname{ln}[1/(a\Lambda_{\mathrm{QCD}} )]   \biggr]  + \frac{c_1}{\operatorname{ln}[1/(a\Lambda_{\mathrm{QCD}} )]} + d'(\mu), 
\end{align}
where the $d'(\mu)$ term absorbs all the scale dependence in $Z_O$. In order to match $g'$ to the zero-momentum matrix element calculated in the $\overline{\text{MS}}$ scheme, we include the parameter $d'$ in the fit within the perturbative regime and tune its value such that:
\begin{align}
    \operatorname{exp}[g'(z, b_\perp)-d'(\mu)] = \tilde{\Phi}^{\rm \overline{MS}}(\mu, z, P^z=0, b_{\perp}, m)
\end{align}

As shown in the right panel of  Fig.~\ref{fig:Self renormalization example and renormalization match}, when $d'(\mu) = -0.448$, the average deviation $\delta$ between the $g'(z, b_\perp)$ term and the $\overline{\text{MS}}$ result is consistent with zero within uncertainties. This allows us to determine the form of the logarithmic divergence, eliminate discretization effects, and ensure consistency between the lattice results and the $\overline{\text{MS}}$ results in the perturbative region.

Finally, after determining the form of the logarithmic divergence and the renormalization scale dependence, the results for different lattice spacings and HYP smearing steps are obtained, as shown in Tab.~\ref{tab:Z_O}.

% ====================================================================================

\section{Convention on the Length of Wilson line  in lattice simulation}
\label{Ax:diff_z_directions}

In lattice simulation, we adopt the convention on the length of Wilson line shown in Fig.~\ref{fig:diff_z_directions_wilson_line}, 
where the direction of the relative displacement between the quark and antiquark is taken as the
$z$-direction, while maintaining the requirement of a large Wilson link length $L$. 
The left panel illustrates the convention when $z>0$, which follows directly
from Eq.~(\ref{eq:Wilson_link_define}).  
The right panel shows  the convention for $z<0$. To ensure that the Wilson
link remains sufficiently long in this case, one can redefine the length as $L' = L + |z| \ge L$, which is also consistent with Eq.~(\ref{eq:Wilson_link_define}). 
With this convention, the Wilson loop has a total longitudinal extent of $(2L + |z|)$ for both
positive and negative $z$, and a transverse width of $b_\perp$. 
This convention guarantees that the Wilson line is sufficiently long to ensure the saturation of the final results for quasi TMDWFs.

\begin{figure}[http]
\centering
\includegraphics[width=0.9\textwidth]{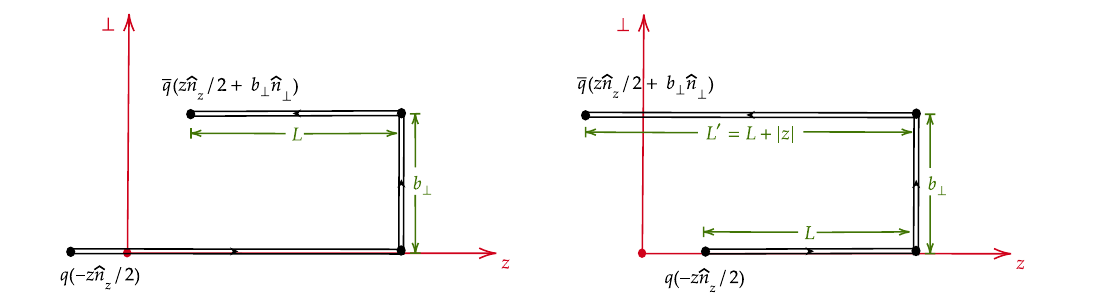} 
\caption{Construction of the Wilson link in the lattice simulation for $z>0$(left panel) and $z<0$(right panel).} 
\label{fig:diff_z_directions_wilson_line}
\end{figure}
% ====================================================================================

\section{Symmetry in quasi-TMDWFs}
\label{ax:The symmetry of quasi-TMDWFs}
In this appendix, we will show that the quasi-TMDWF of the pion is symmetric around $x=1/2$ under the assumption of isospin symmetry. To make it transparent, we label the color and spinor indices explicitly, and the staple-shaped Wilson link is denoted as $U_{ab}(z\hat{n}_z/2+b_\perp\hat n_\perp, -z\hat{n}_z/2)$.
So the bare matrix element is now expressed as
\begin{equation}
\label{app-def}
\tilde\Phi^0(z, b_\perp, P^z, L)=\langle 0|\bar{q}_{i, b}(z\hat{n}_z/2+b_\perp\hat n_\perp)\Gamma_{ij}U_{ab}(z\hat{n}_z/2+b_\perp\hat n_\perp, -z\hat{n}_z/2)\psi_{j, b}(-z\hat{n}_z/2)|P^z\rangle, 
\end{equation}
where $a, b$ are color indices in the fundamental representation and $i, j$ are spinor indices. 
To proceed, we need the charge conjugation of the Wilson line
\begin{equation}
\begin{aligned}
		\mathcal{C}U_{ab}(x, 0)\mathcal{C}^{-1}&= (e^{ig\mathcal{P}\int_{0}^xn\cdot \mathcal{C}A(nt)\mathcal{C}^{-1}dt})_{ab}= (e^{-ig\mathcal{P}\int_{0}^xn\cdot A^T(nt)dt})_{ab}\\&= (e^{ig\bar{\mathcal{P}}\int_{0}^x(-n)\cdot A^T(x-nt)dt})_{ab}=(e^{ig\mathcal{P}\int_{0}^x(-n)\cdot A(x-nt)dt})_{ba}=U_{ba}(0, x), 
\end{aligned}
\end{equation}
where $\mathcal{C}$ is the charge conjugation operator. The effect of charge conjugation is basically to reverse the direction of the Wilson line.

Then, we insert $I=\mathcal{C}^{-1}\mathcal{C}$ into Eq.~(\ref{app-def}) and obtain 
\begin{equation}
\label{app_pv}
\begin{aligned}
\tilde\Phi^0&(z, b_\perp, P^z, L)\\
&=\langle 0|\mathcal{C}^{-1}\mathcal{C}\bar{q}_{i, b}(z\hat{n}_z/2+b_\perp\hat n_\perp)\mathcal{C}^{-1}\Gamma_{ij}\mathcal{C}U_{ab}(z\hat{n}_z/2+b_\perp\hat n_\perp, -z\hat{n}_z/2)\mathcal{C}^{-1}\mathcal{C}q_{j, b}(-z\hat{n}_z/2)\mathcal{C}^{-1}\mathcal{C}|P^z\rangle, \\
&=-\langle 0|q_{i, a}(z\hat{n}_z/2+b_\perp \hat n_\perp)C_{ij}\Gamma_{jk}U_{ba}(-z\hat{n}_z/2, z\hat{n}_z/2+b_\perp\hat n_\perp)C^{-1}_{kl}\bar q_{l, b}(-z\hat{n}_z/2)|P^z\rangle, \\
&=\langle 0|\bar q_{l, b}(-z\hat{n}_z/2)C_{ij}\Gamma_{jk}C^{-1}_{kl}U_{ba}(-z\hat{n}_z/2, z\hat{n}_z/2+b_\perp\hat{n}_\perp){q}_{i, a}(z\hat{n}_z/2+b_\perp \hat n_\perp)|P^z\rangle, \\
&=\langle 0|\bar q_{l, b}(-z\hat{n}_z/2)\Gamma_{li}'U_{ba}(-z\hat{n}_z/2, z\hat{n}_z/2+b_\perp\hat{n}_\perp){q}_{i, a}(z\hat{n}_z/2+b_\perp \hat n_\perp)|P^z\rangle, 
\end{aligned}
\end{equation}
where 
\begin{equation}
	\Gamma'=(C\Gamma C^{-1})^T=(C\gamma^tC^{-1}C\gamma_5 C^{-1})^T=-({\gamma^t}^T\gamma_5^T)^T=\gamma^t\gamma_5=\Gamma.
\end{equation}
In the third line of Eq.~(\ref{app_pv}), we use the fact that 
\begin{equation}
	\mathcal{C}|\pi^0\rangle=|\pi^0\rangle, 	\quad \mathcal{C}|\pi^-\rangle=|\pi^+\rangle, \quad 	\mathcal{C}|\pi^+\rangle=|\pi^-\rangle.
\end{equation} 
For $\pi^+$ and $\pi^-$, further using the isospin symmetry brings us back to the original meson. 

Furthermore, we exploit the translational and rotational invariance in the $\hat{n}_\perp$ plane
\begin{equation}
\label{app_pv}
\begin{aligned}
\tilde\Phi^0&(z, b_\perp, P^z, L)\\
&=\langle 0|\bar q_{l, b}(-z\hat{n}_z/2-b_\perp \hat{n}_\perp)\Gamma_{li}'U_{ba}(-z\hat{n}_z/2-b_\perp \hat{n}, z\hat{n}_z/2){q}_{i, a}(z\hat{n}_z/2)|P^z\rangle, \\
&=\langle 0|\bar q_{l, b}(-z\hat{n}_z/2+b_\perp \hat{n}_\perp)\Gamma_{li}'U_{ba}(-z\hat{n}_z/2+b_\perp \hat{n}, z\hat{n}_z/2){q}_{i, a}(z\hat{n}_z/2)|P^z\rangle, \\
&=\tilde\Phi^0(-z, b_\perp, P^z, L).
\end{aligned}
\end{equation}
The renormalized matrix elements should have the same symmetry property as the bare matrix element, since the renormalization factor $Z_O(1/a, \mu)$ and the Wilson loop $Z_E(2L+z, b_\perp)$ are the same under $z\leftrightarrow -z$. Therefore, we arrive at
\begin{equation}
\tilde\Phi(z, b_\perp, P^z, L)=\tilde\Phi(-z, b_\perp, P^z, L).
\end{equation}  

Transforming to momentum space, we have
\begin{equation}
\begin{aligned}
\tilde f(x, b_\perp, \mu, \zeta_z)&=\lim_{L\to \infty}\int \frac{dz}{2\pi}\, e^{i(x-1/2)zP^z}\tilde\Phi(z, b_\perp, P^z, L), \\
&=\lim_{L\to \infty}\int \frac{dz}{2\pi}\, e^{i(x-1/2)zP^z}(\tilde\Phi(z, b_\perp, P^z, L)+\tilde\Phi(-z, b_\perp, P^z, L))/2, \\
&=\lim_{L\to \infty}\int \frac{dz}{2\pi}\, e^{i(1/2-x)zP^z}(\tilde\Phi(-z, b_\perp, P^z, L)+\tilde\Phi(z, b_\perp, P^z, L))/2, \\
&=\lim_{L\to \infty}\int \frac{dz}{2\pi}\, e^{i(1-x-1/2)zP^z}\tilde\Phi(z, b_\perp, P^z, L), \\
&=\tilde f(1-x, b_\perp, \mu, \zeta_z), 
\end{aligned}
\end{equation}
which implies that the quasi-TMDWF is symmetric around $x=1/2$ under isospin symmetry.

%===================================================================================
\section{More results for Section~\ref{sec:Lattice Simulations and Results}}

\subsection{Results for Wilson loop}

More results for the Wilson loop on different Lattice ensembles are given in Fig.~\ref{fig:Wilson_loop_more_1}.

%%%%%%%%%%%%%%%%%%%%%%%%%%%%%
\begin{figure}[http]
\centering
\includegraphics[width=0.45\textwidth]{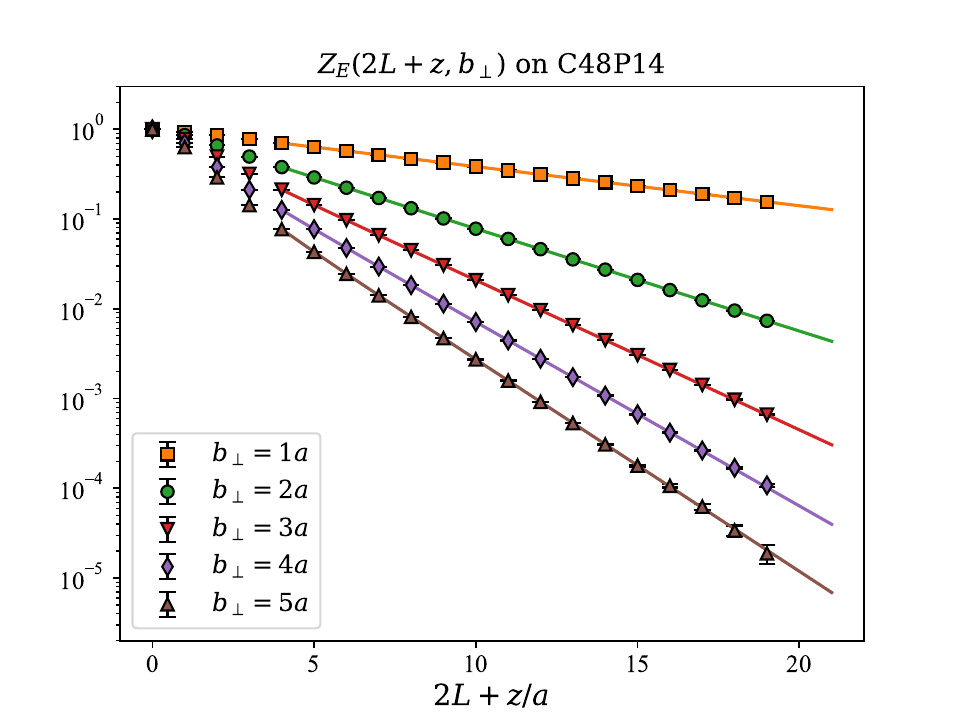} 
\includegraphics[width=0.45\textwidth]{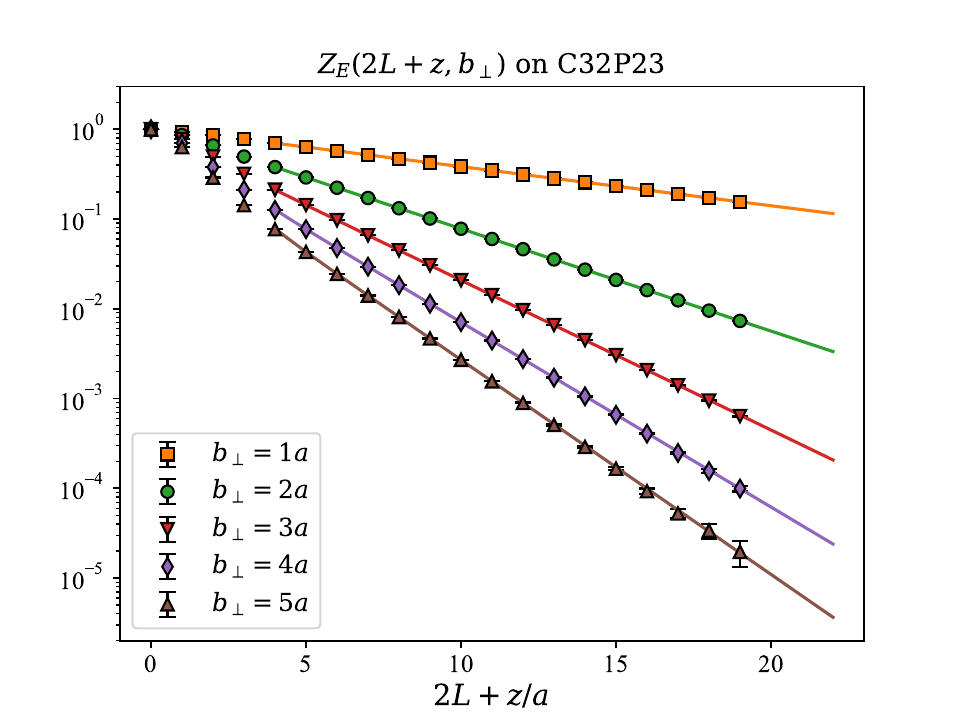} 
\includegraphics[width=0.45\textwidth]{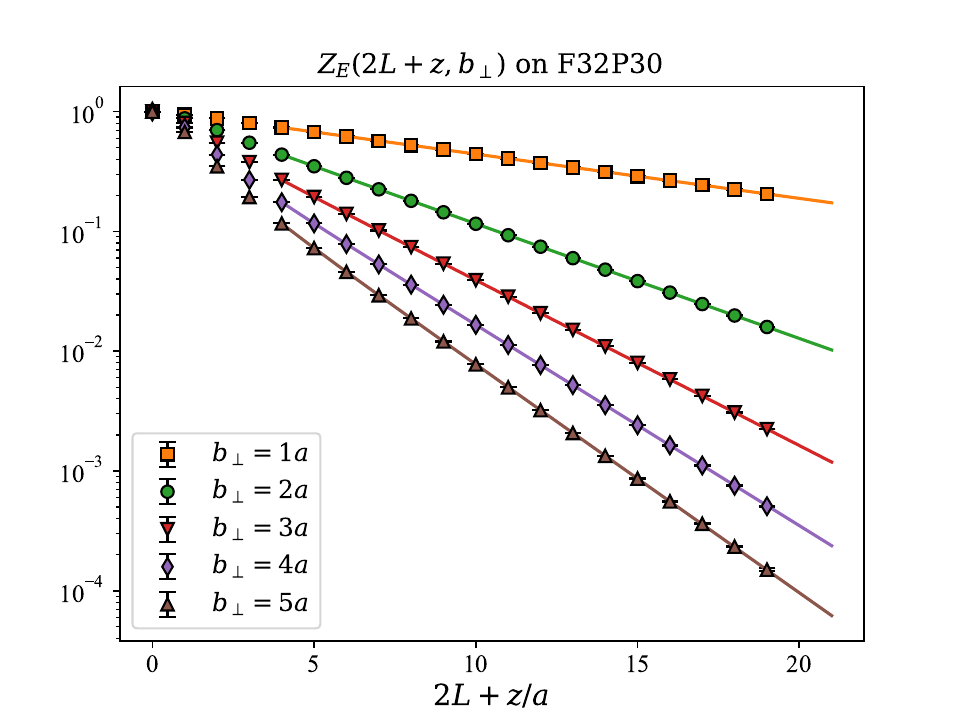} 
\includegraphics[width=0.45\textwidth]{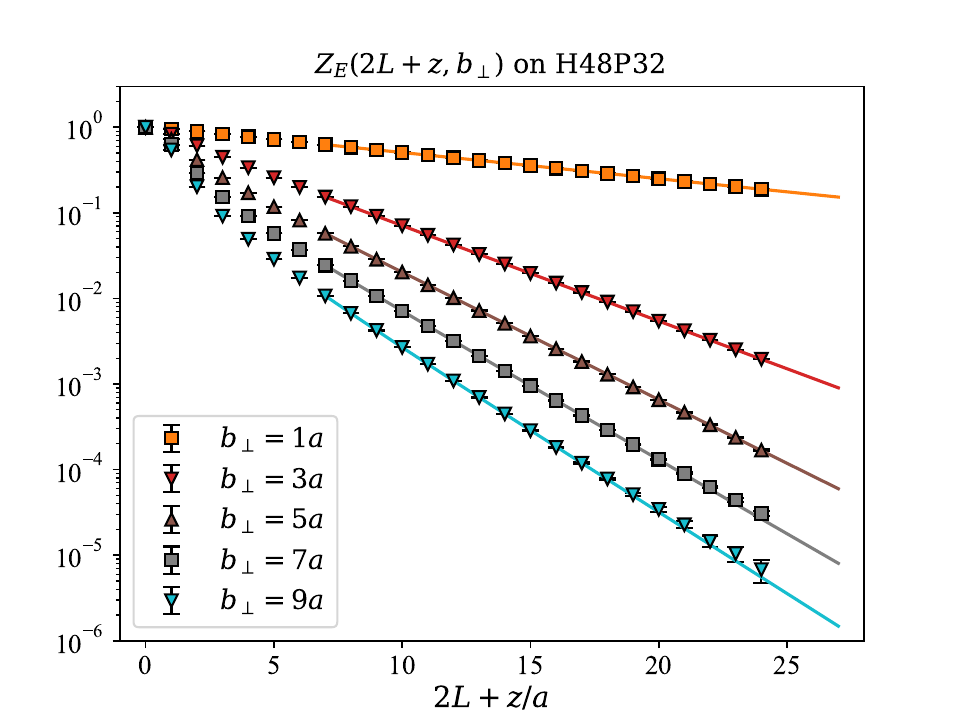} 
\caption{Results of the Wilson loop for different shapes $(2L+z)\times b_\perp$ on the different ensembles, with three panels showing a comparison of the results from HYP=1. The solid lines represent the extrapolated results based on Eq.(\ref{Wloop extrapolation}).} 
\label{fig:Wilson_loop_more_1}
\end{figure}
%%%%%%%%%%%%%%%%%%%%%%%%%%%%%

\subsection{Results for self renormalization }
\label{ax:Results_for_self_renormalization}

More results for the self-renormalization on different HYP smearing steps are given in Fig.~\ref{fig:self_renormalization_more_1}, \ref{fig:self_renormalization_more_2} , \ref{fig:self_renormalization_more_3} and \ref{fig:self_renormalization_more_4}. 

\begin{figure}[http]
\centering
\includegraphics[width=0.32\textwidth]{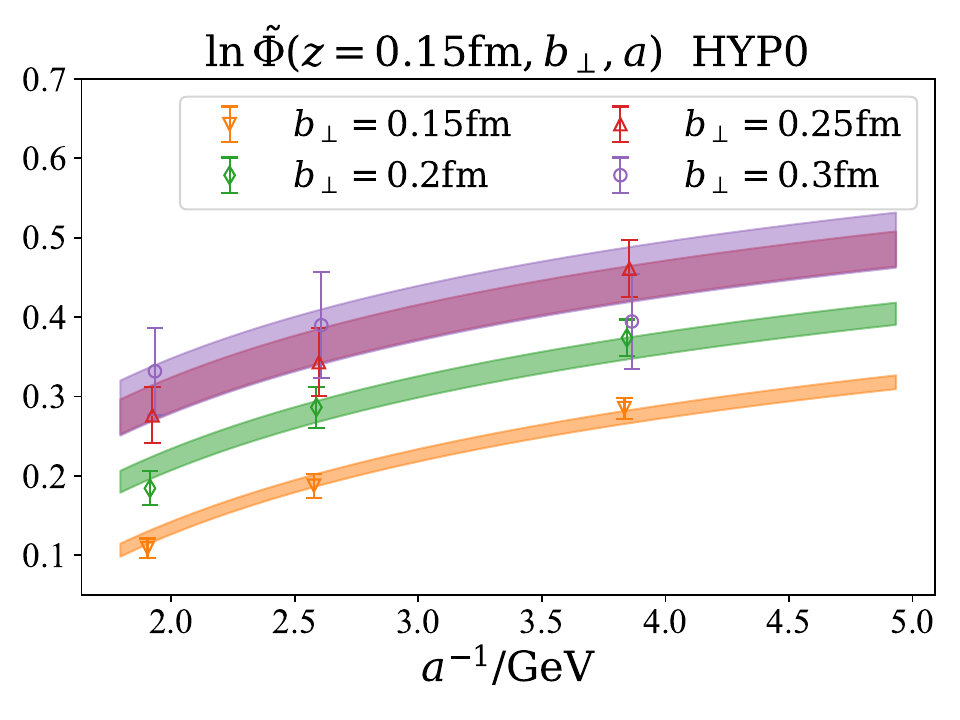} 
\includegraphics[width=0.32\textwidth]{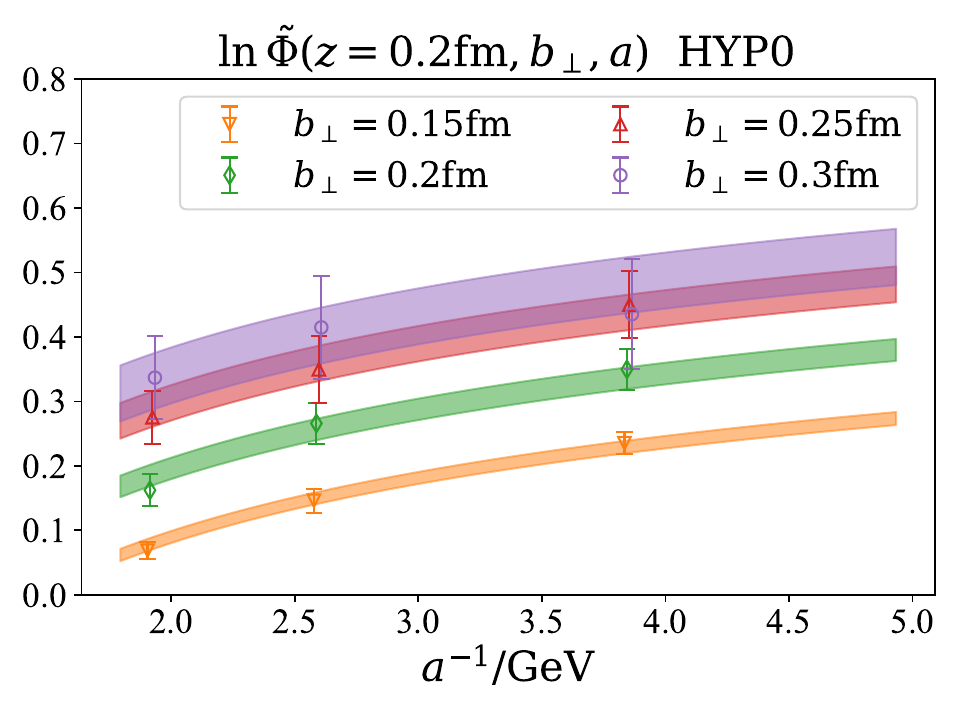} 
\includegraphics[width=0.32\textwidth]{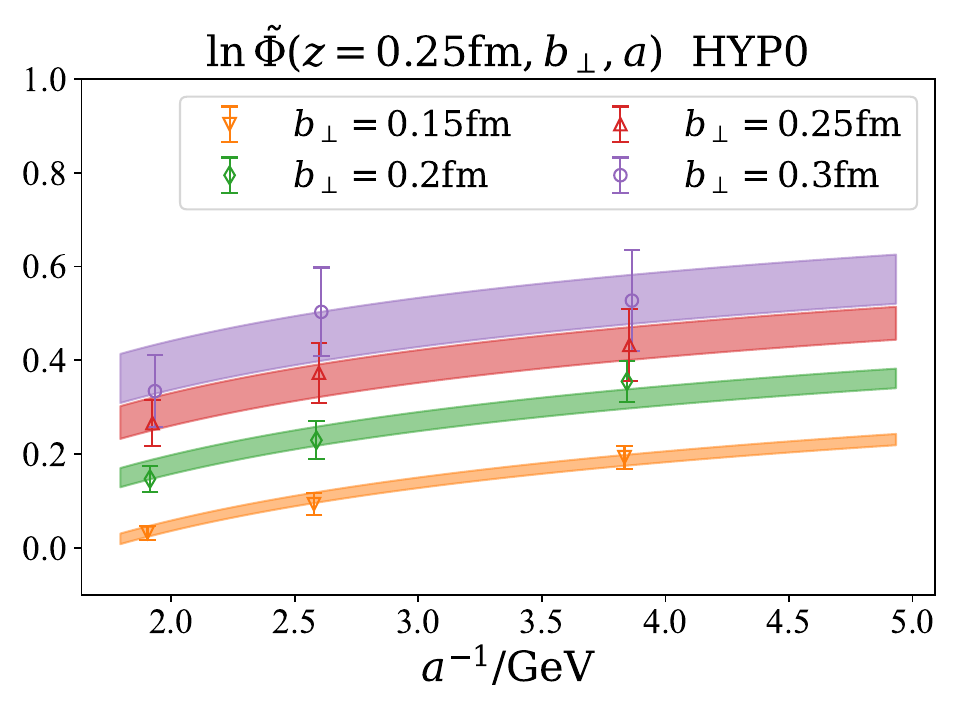} 
\caption{The result of joint fit for the self renormalization with HYP=0, $b_\perp = \{0.15, 0.2, 0.25, 0.3\}~\mathrm{fm}$ and $z = \{0.15, 0.2, 0.25\}~\mathrm{fm}$ .} 
\label{fig:self_renormalization_more_1}
\end{figure}

\begin{figure}[http]
\centering
\includegraphics[width=0.32\textwidth]{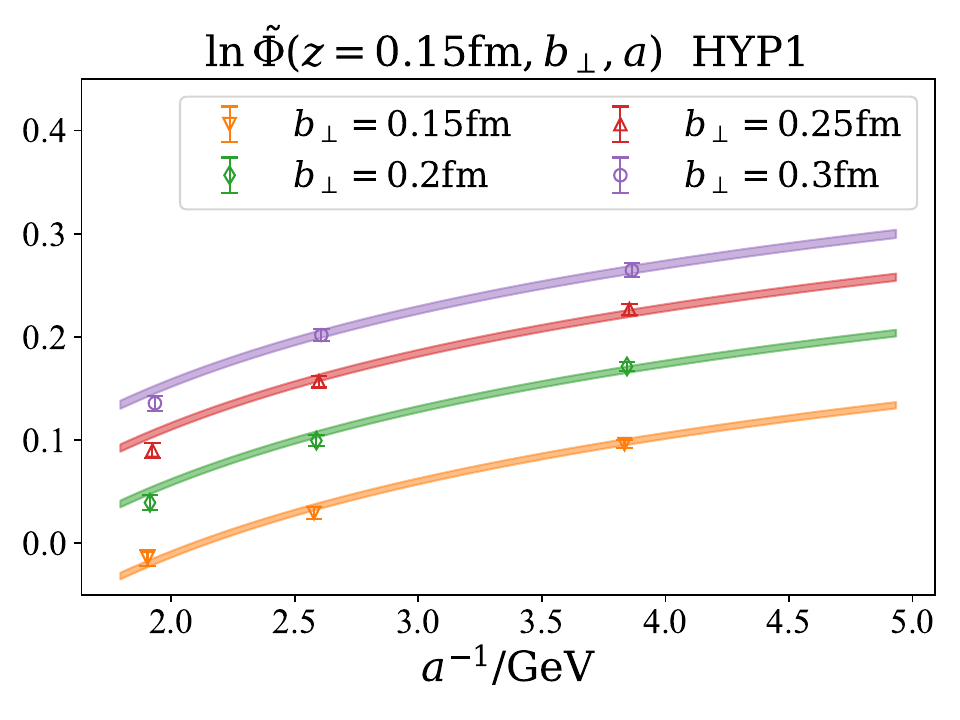} 
\includegraphics[width=0.32\textwidth]{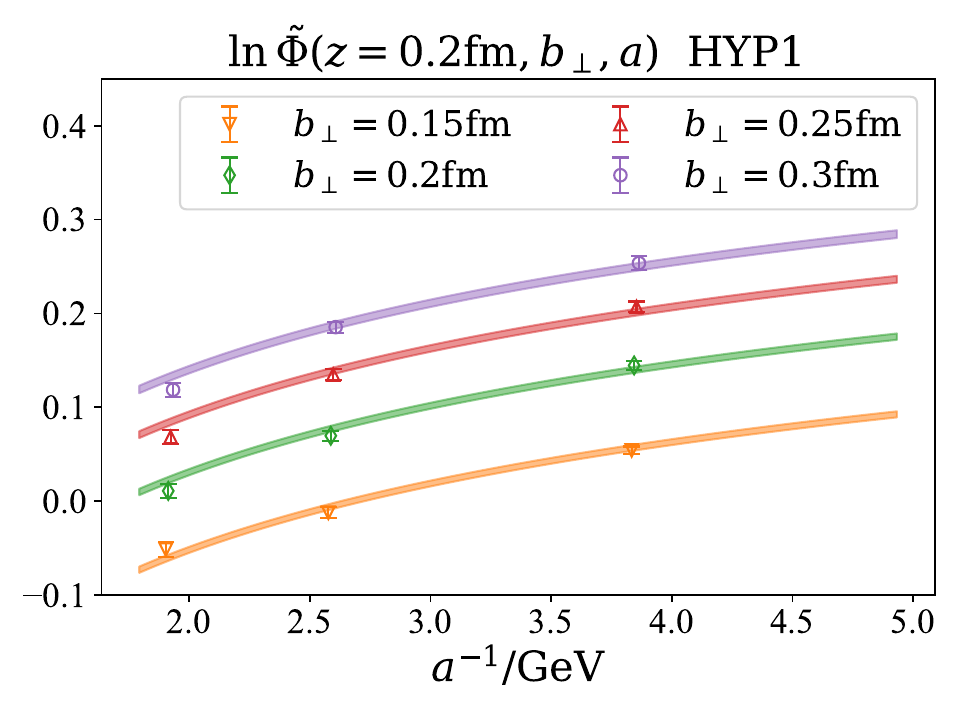} 
\includegraphics[width=0.32\textwidth]{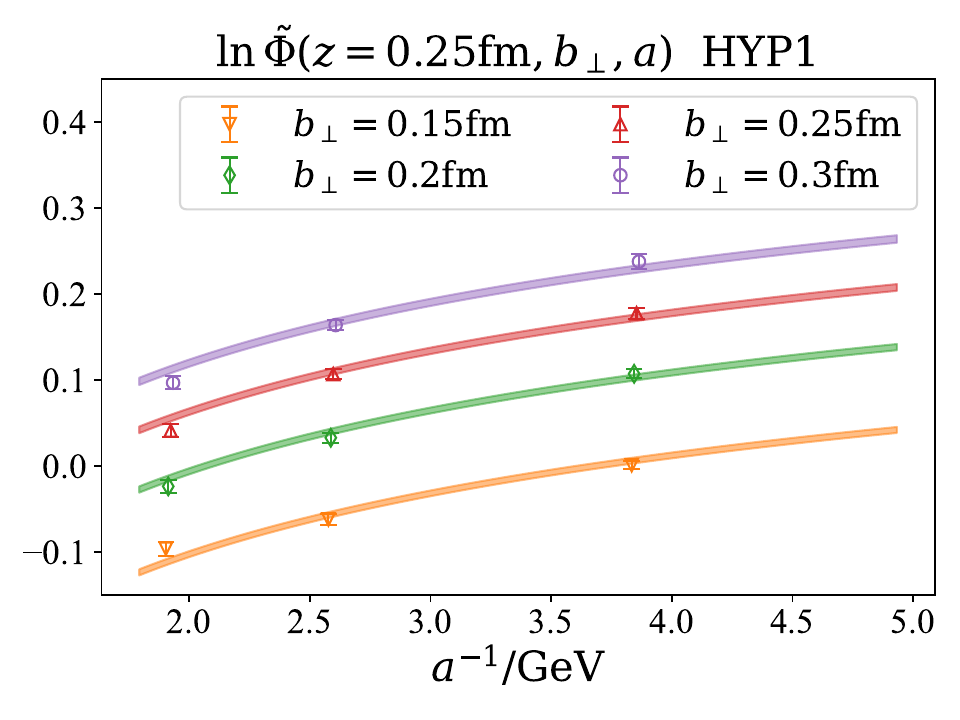} 
\caption{The result of joint fit for the self renormalization with HYP=1, $b_\perp = \{0.15, 0.2, 0.25, 0.3\}~\mathrm{fm}$ and $z = \{0.15, 0.2, 0.25\}~\mathrm{fm}$ .} 
\label{fig:self_renormalization_more_2}
\end{figure}

\begin{figure}[http]
\centering
\includegraphics[width=0.32\textwidth]{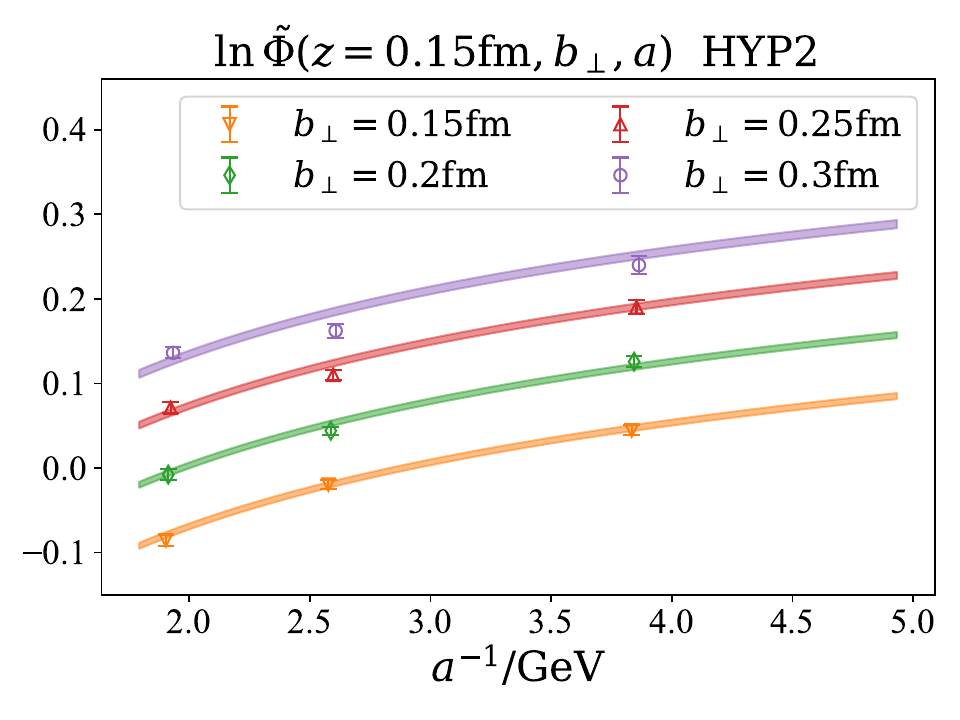} 
\includegraphics[width=0.32\textwidth]{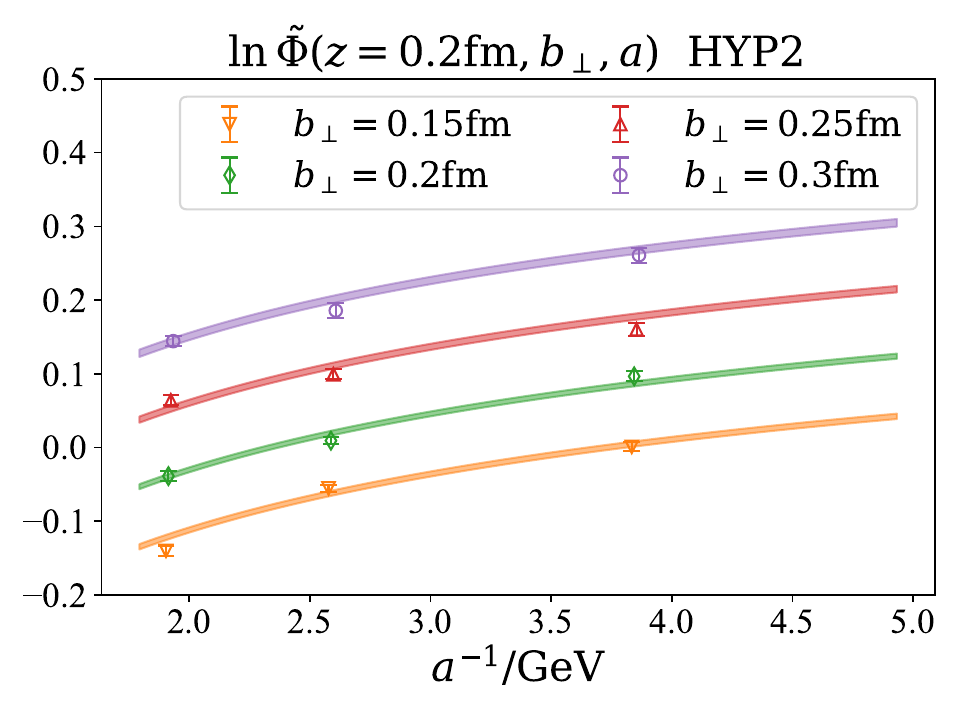} 
\includegraphics[width=0.32\textwidth]{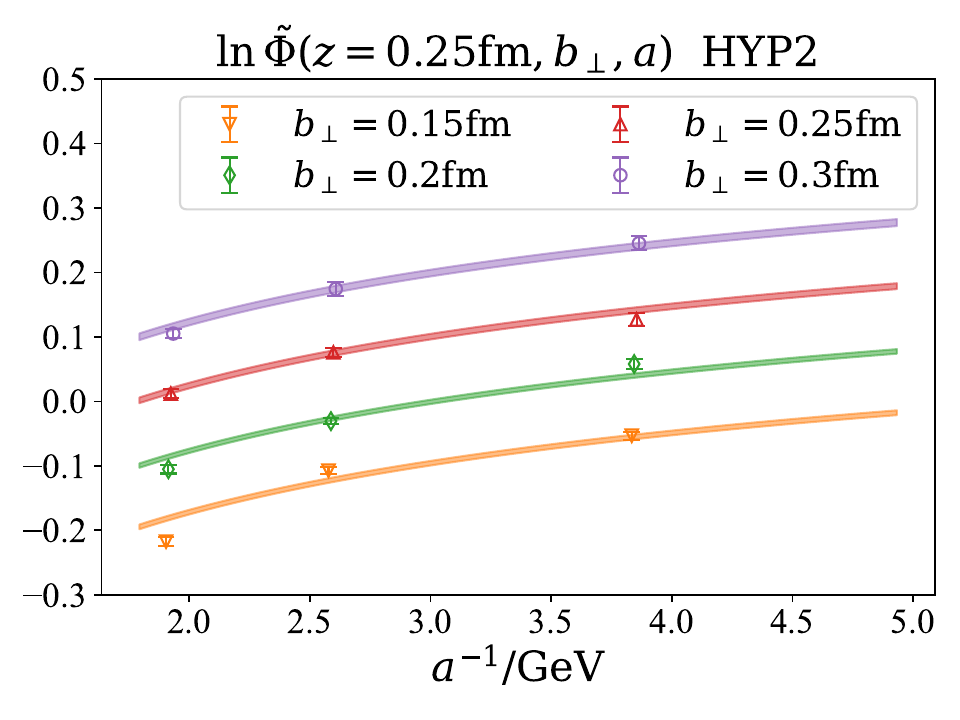} 
\caption{The result of joint fit for the self renormalization with HYP=2, $b_\perp = \{0.15, 0.2, 0.25, 0.3\}~\mathrm{fm}$ and $z = \{0.15, 0.2, 0.25\}~\mathrm{fm}$ .} 
\label{fig:self_renormalization_more_3}
\end{figure}

\begin{figure}[http]
\centering
\includegraphics[width=0.4\textwidth]{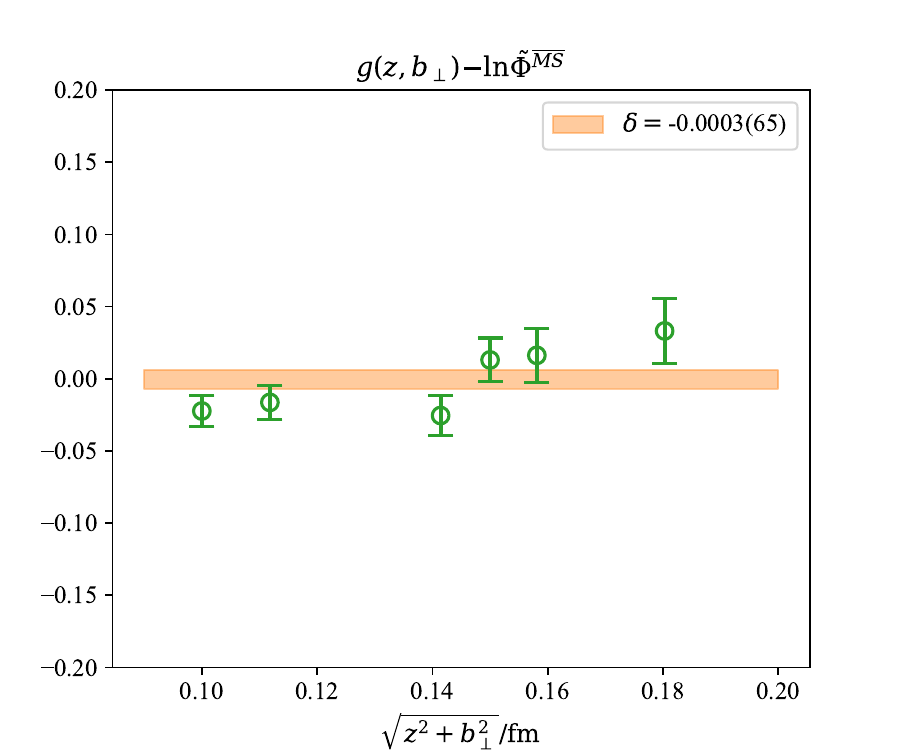} 
\includegraphics[width=0.4\textwidth]{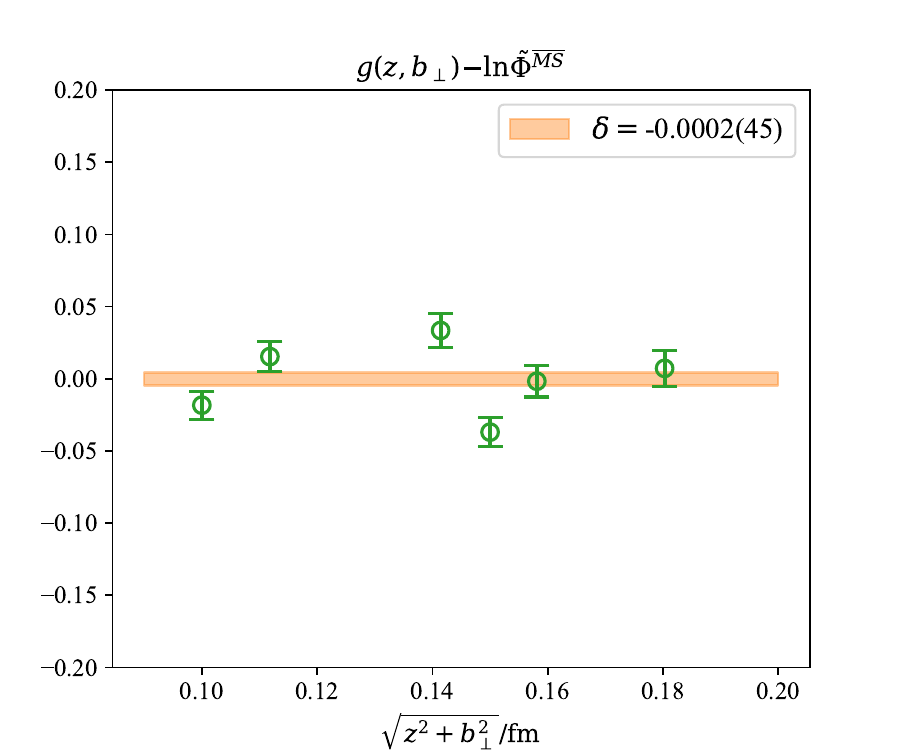} 
\caption{The result of matching to $\bar{\rm MS}$ scheme for the self renormalization with HYP=0(left panel), 2(right panel).} 
\label{fig:self_renormalization_more_4}
\end{figure}

\subsection{Results for fits to the bare quasi-TMDWF matrix elements }
More results for the fits to the bare quasi-TMDWF matrix elements on different ensembles are given in Fig.~\ref{fig:more_fit_1}, ~\ref{fig:more_fit_2}, ~\ref{fig:more_fit_3} and ~\ref{fig:more_fit_4}.

\begin{figure}[http]
\centering
\includegraphics[width=0.45\textwidth]{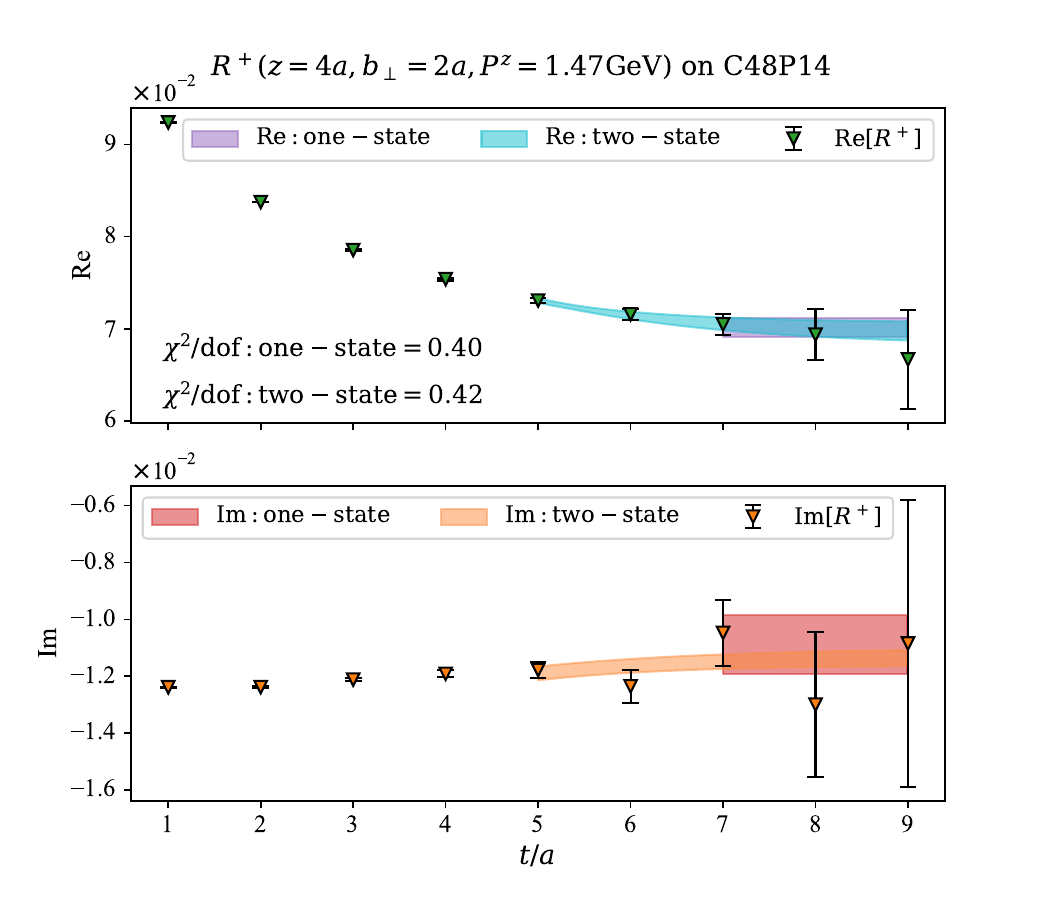}
\includegraphics[width=0.45\textwidth]{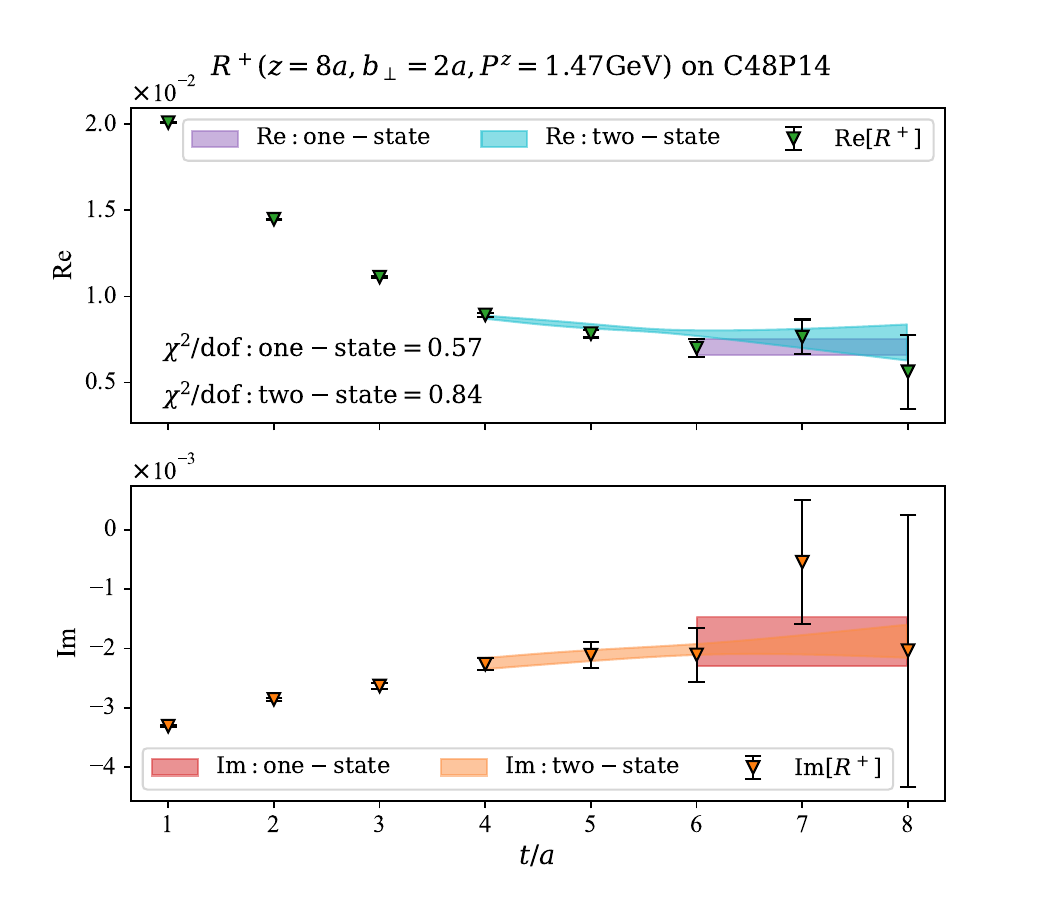}
\includegraphics[width=0.45\textwidth]{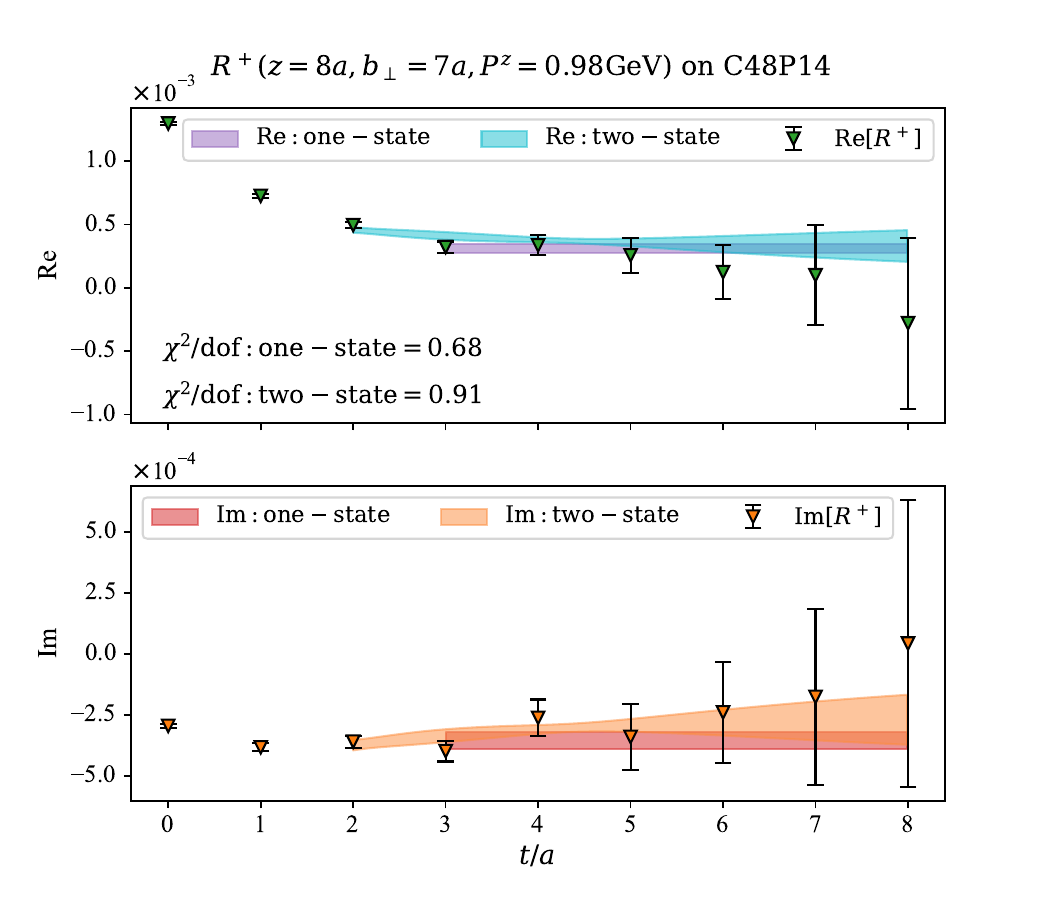}
\includegraphics[width=0.45\textwidth]{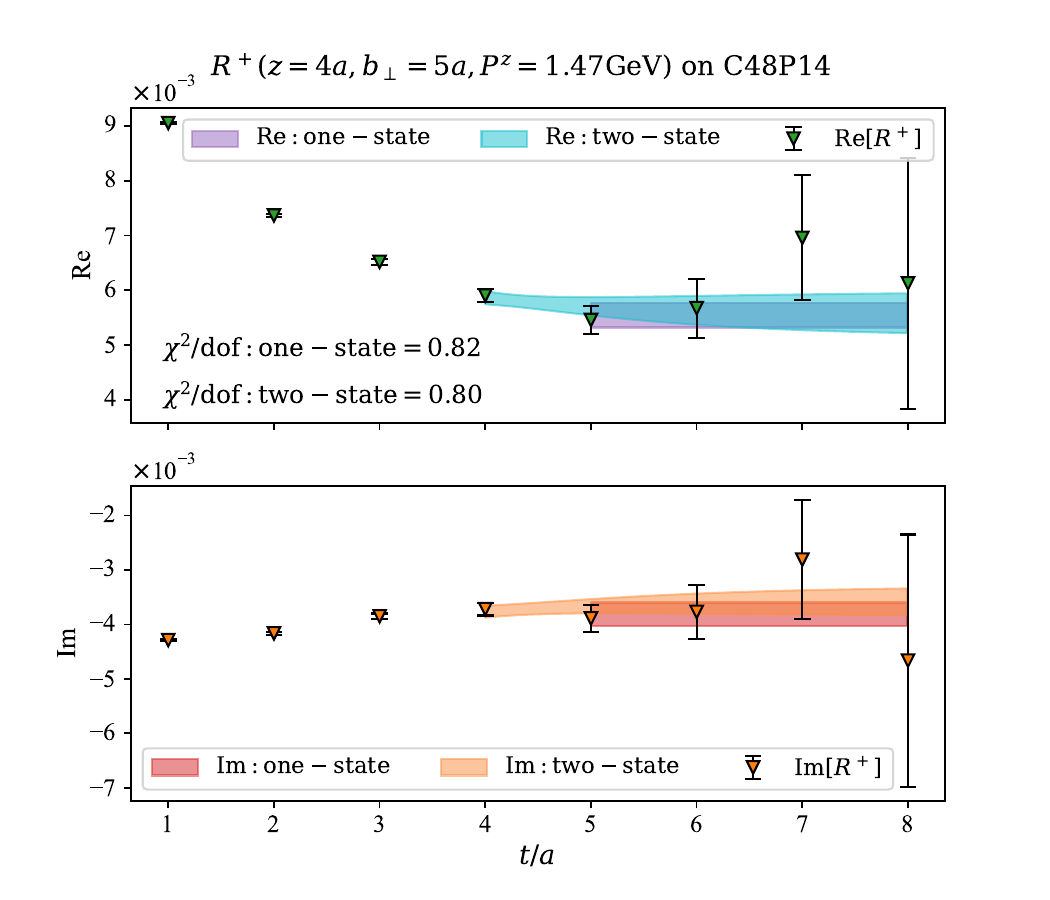}
\includegraphics[width=0.45\textwidth]{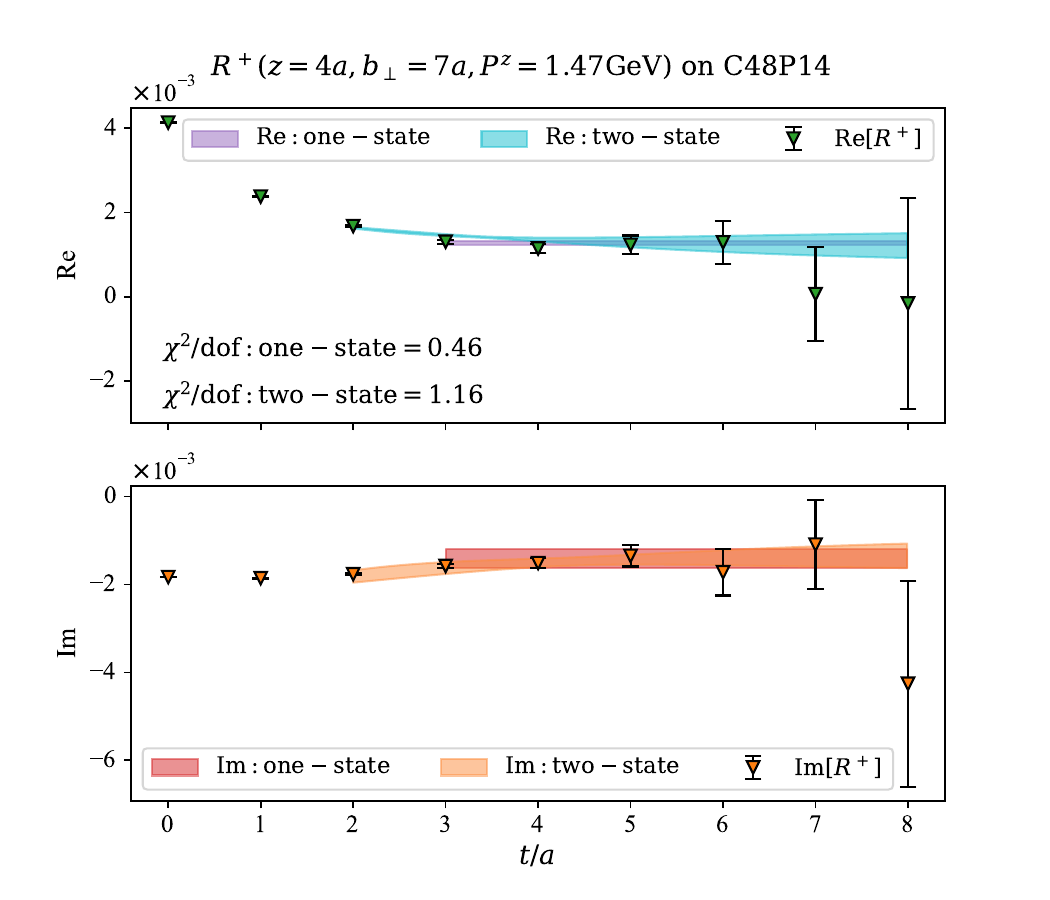}
\includegraphics[width=0.45\textwidth]{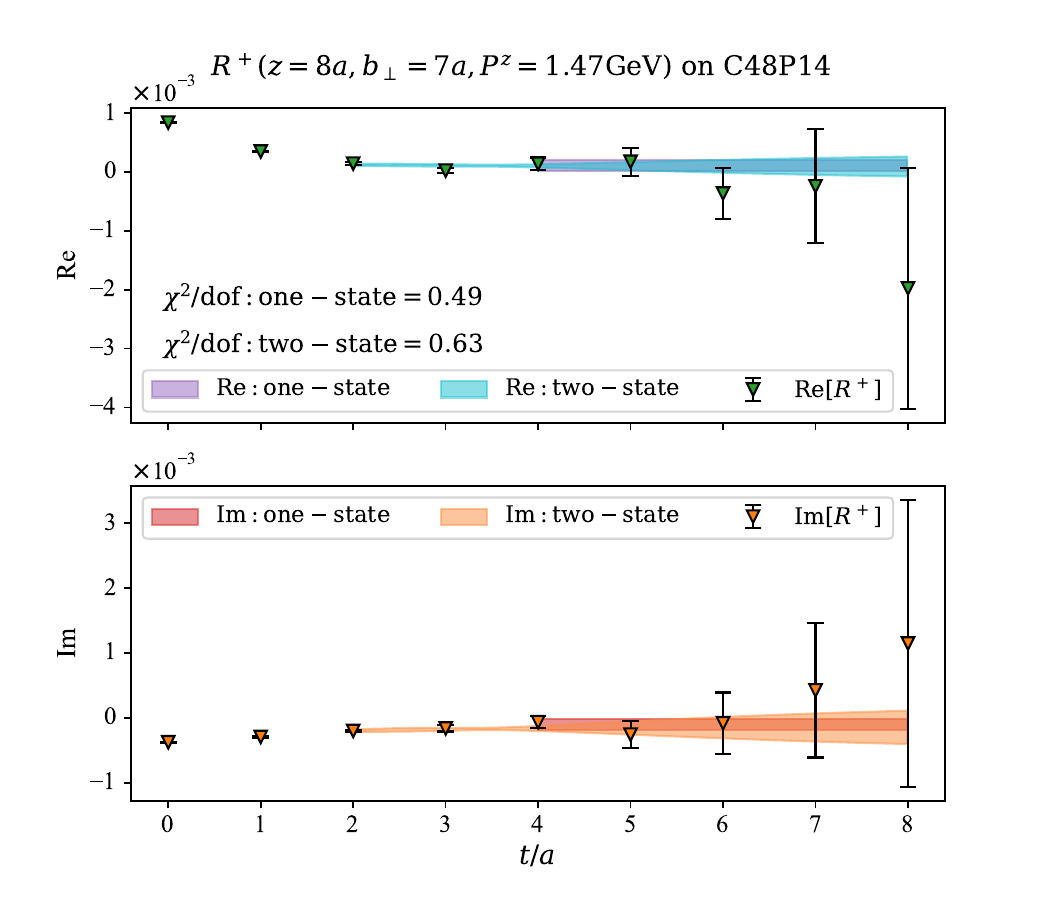}
\caption{The figures present results from the C48P14 ensemble with 1-step HYP smearing, and compare the combinations for $\{z, b_\perp, P^z\}=\{4a, 2a, 1.47\mathrm{GeV}\}$, $\{8a, 2a, 1.47\mathrm{GeV}\}$, $\{8a, 7a, 0.98\mathrm{GeV}\}$ , $\{4a, 5a, 1.47\mathrm{GeV}\}$, $\{4a, 7a, 1.47\mathrm{GeV}\}$and$\{8a, 7a, 1.47\mathrm{GeV}\}$} 
\label{fig:more_fit_1}
\end{figure}

\begin{figure}[http]
\centering
\includegraphics[width=0.45\textwidth]{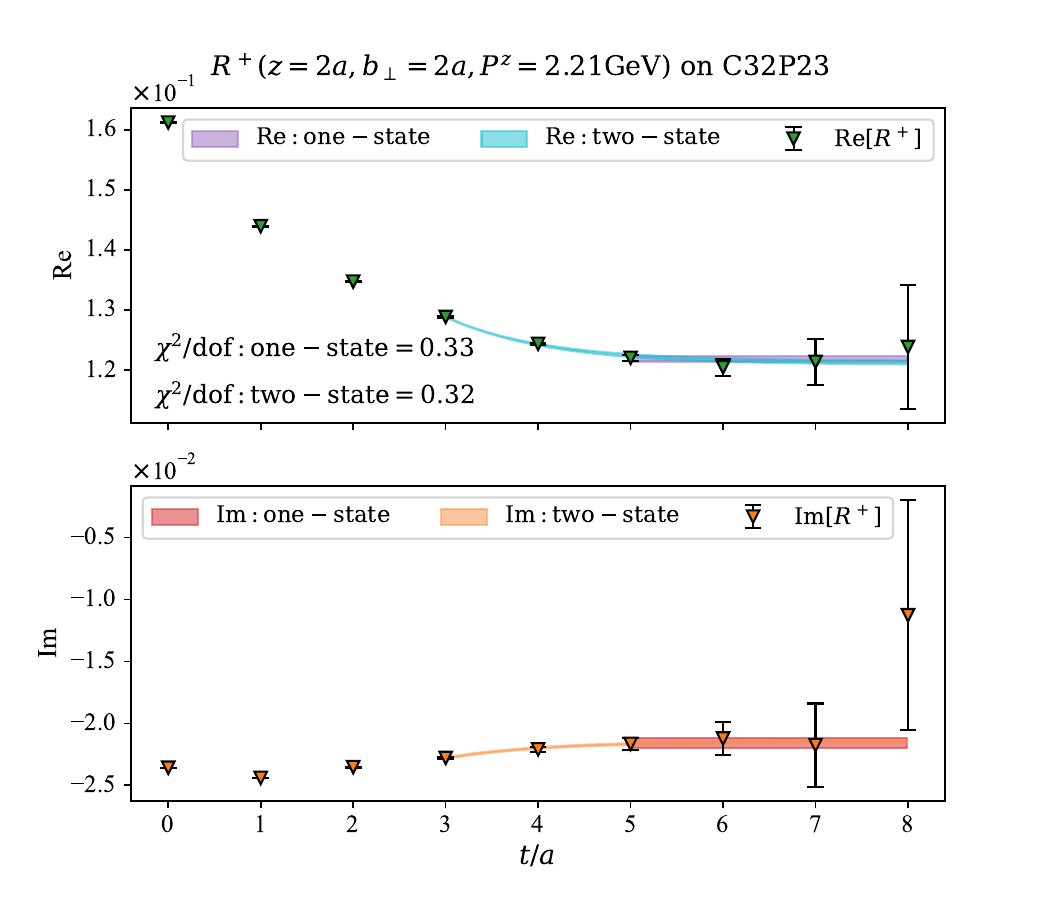}
\includegraphics[width=0.45\textwidth]{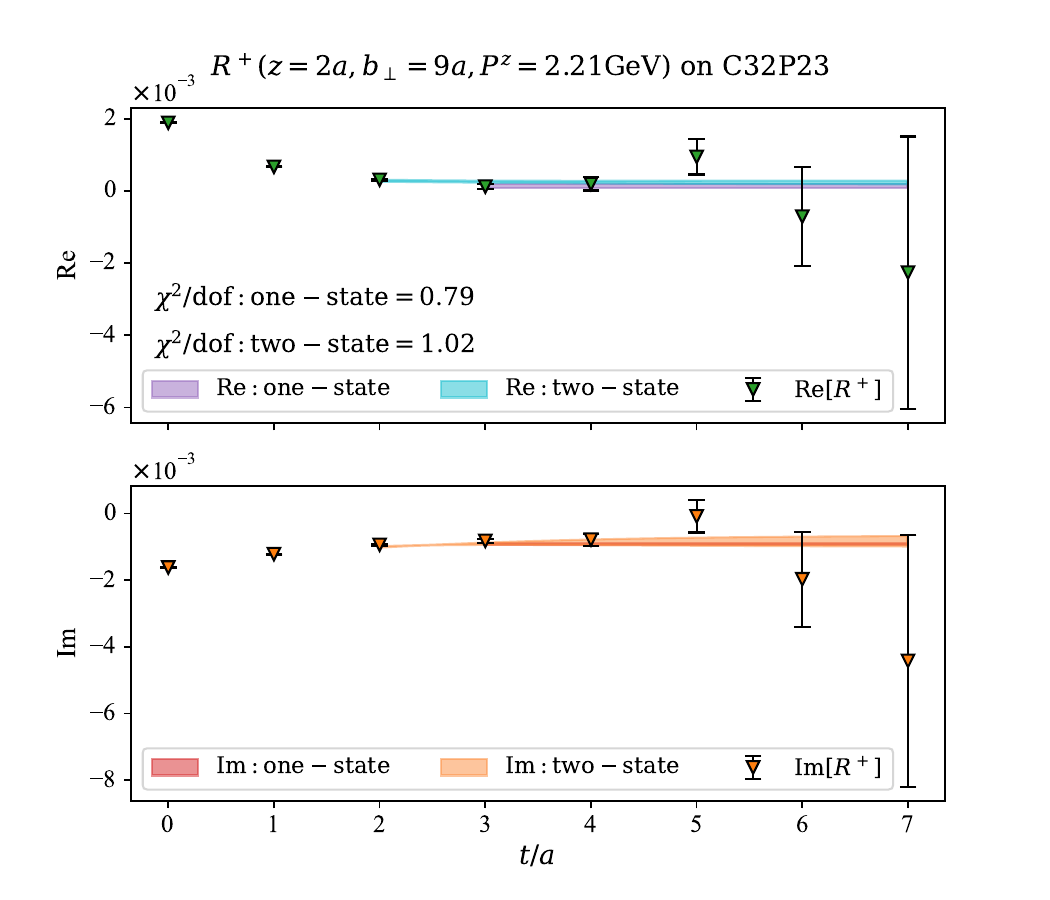}
\includegraphics[width=0.45\textwidth]{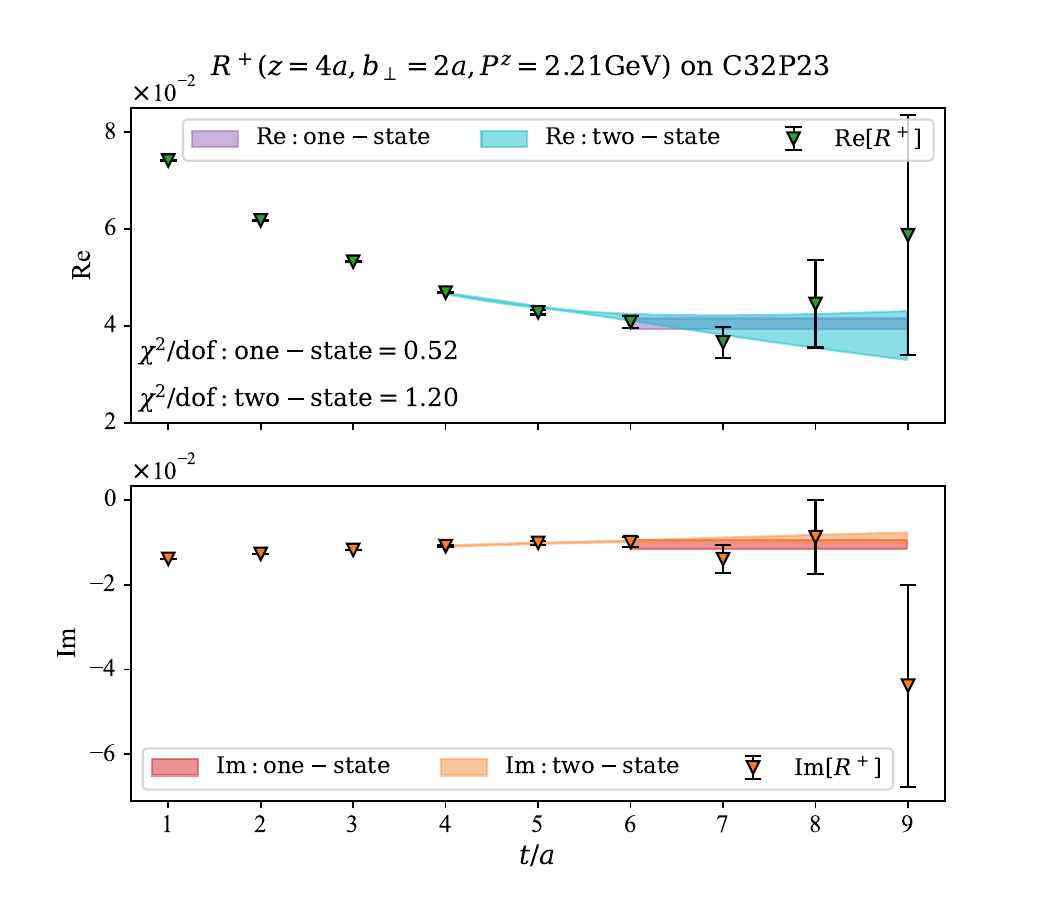}
\includegraphics[width=0.45\textwidth]{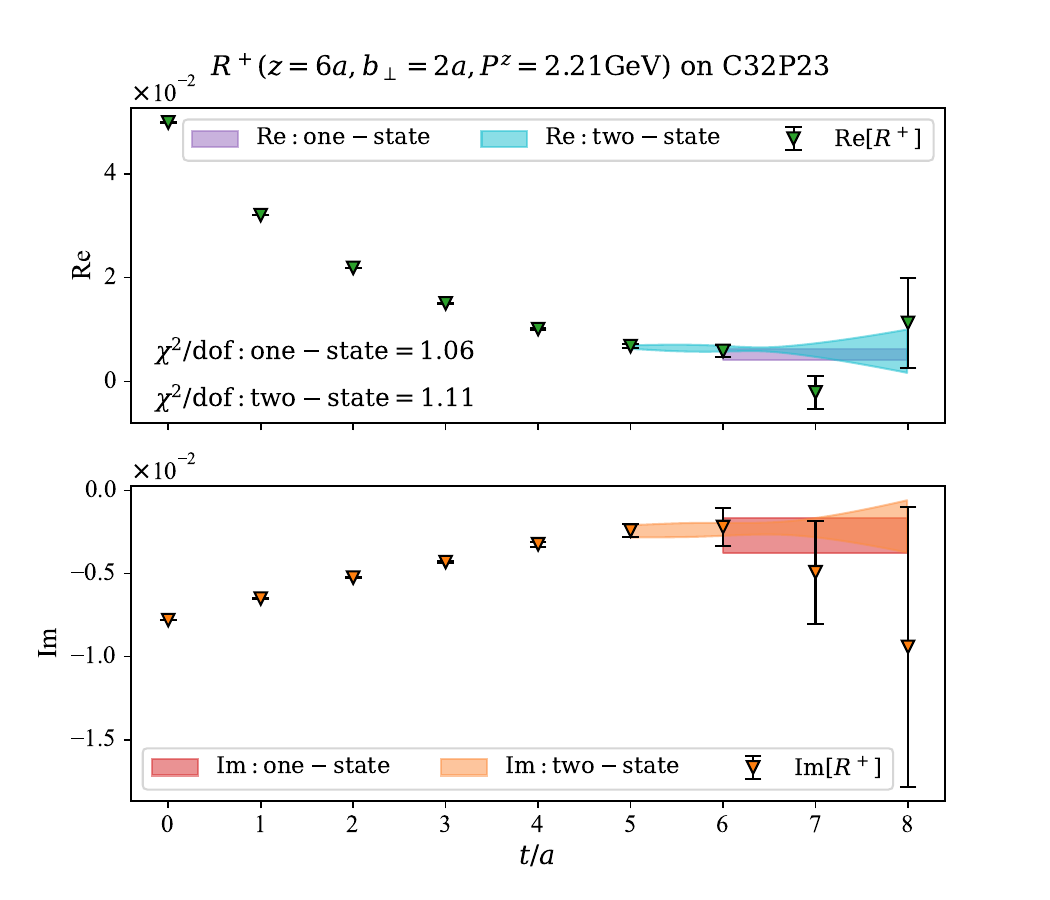}
\includegraphics[width=0.45\textwidth]{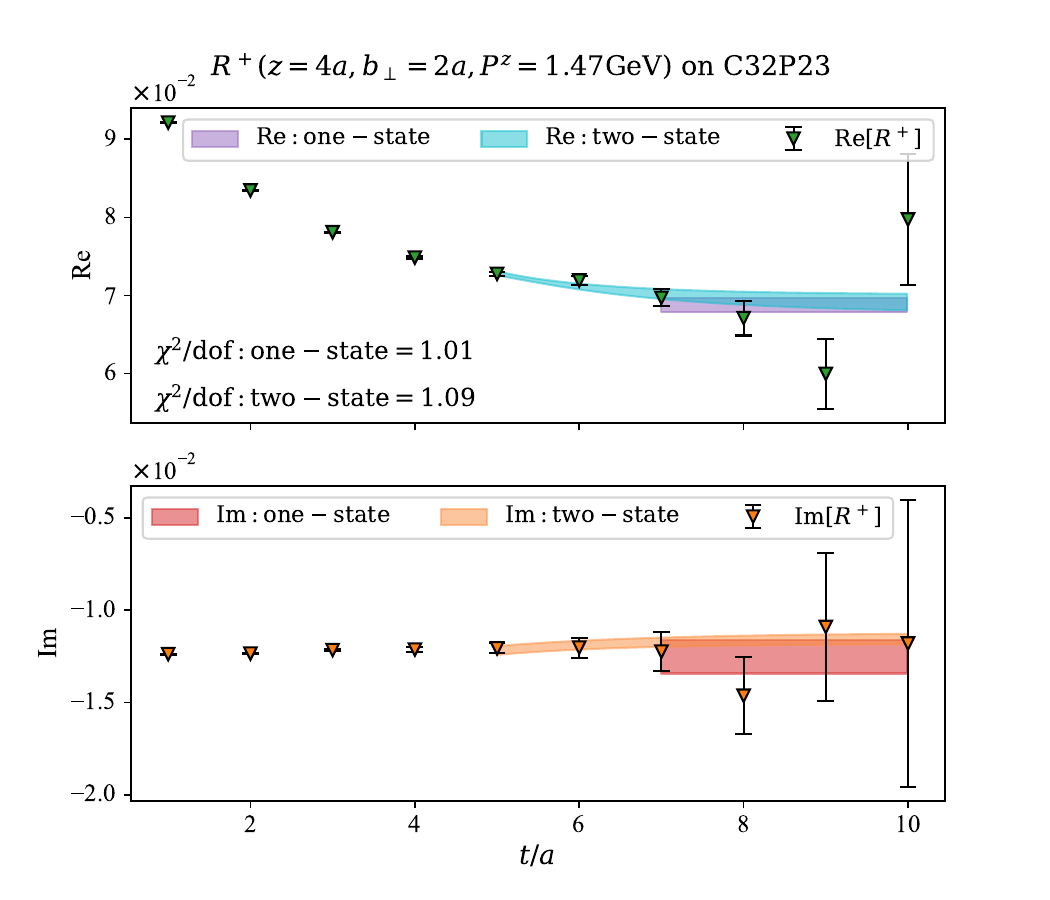}
\includegraphics[width=0.45\textwidth]{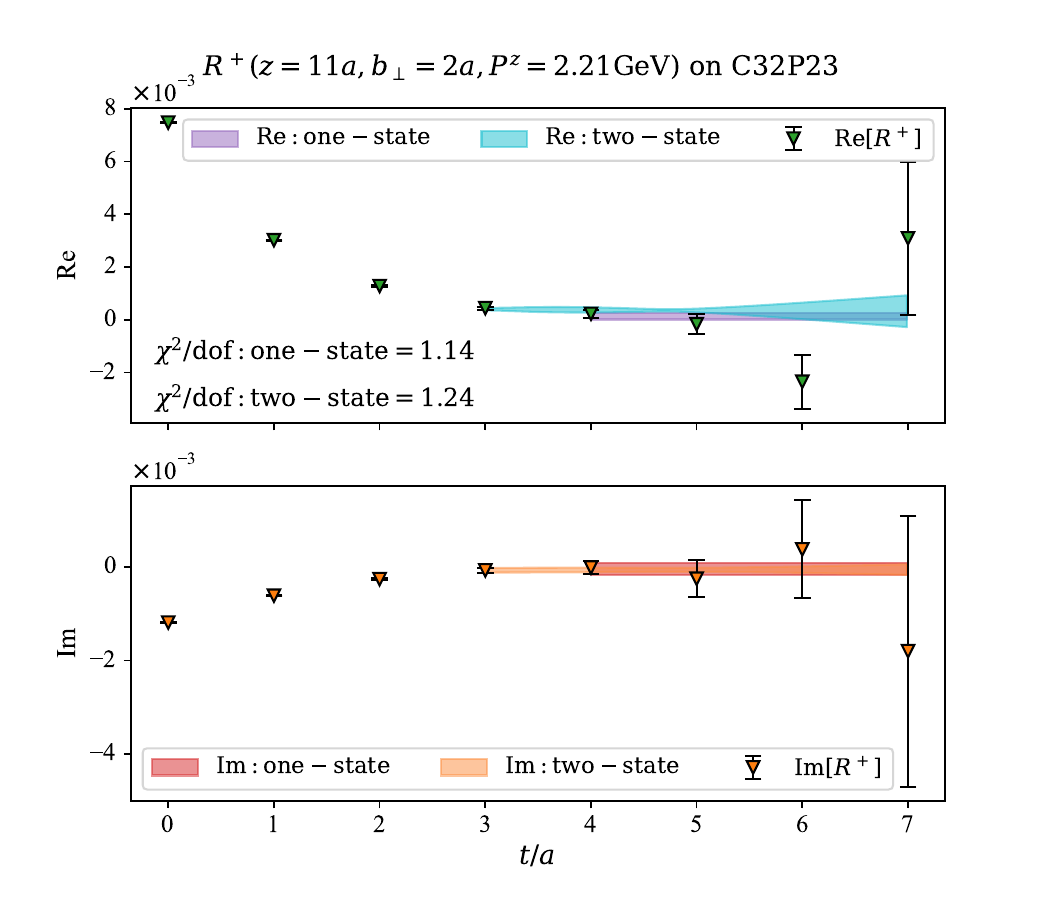}
\caption{The figures present results from the C32P23 ensemble with 1-step HYP smearing, and compare the combinations for $\{z, b_\perp, P^z\}=\{2a, 2a, 2.21\mathrm{GeV}\}$, $\{2a, 9a, 2.21\mathrm{GeV}\}$, $\{4a, 2a, 2.21\mathrm{GeV}\}$ , $\{6a, 2a, 2.21\mathrm{GeV}\}$, $\{4a, 2a, 1.47\mathrm{GeV}\}$and$\{11a, 2a, 2.21\mathrm{GeV}\}$} 
\label{fig:more_fit_2}
\end{figure}

\begin{figure}[http]
\centering
\includegraphics[width=0.45\textwidth]{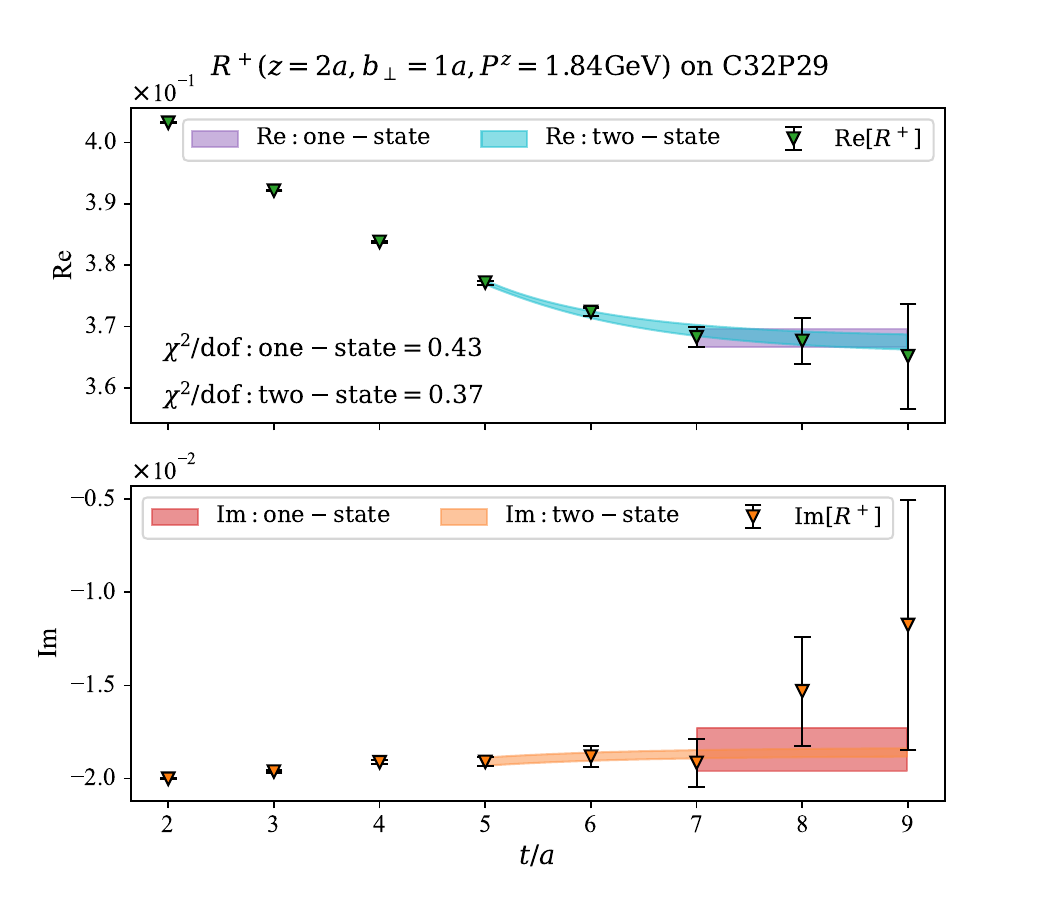}
\includegraphics[width=0.45\textwidth]{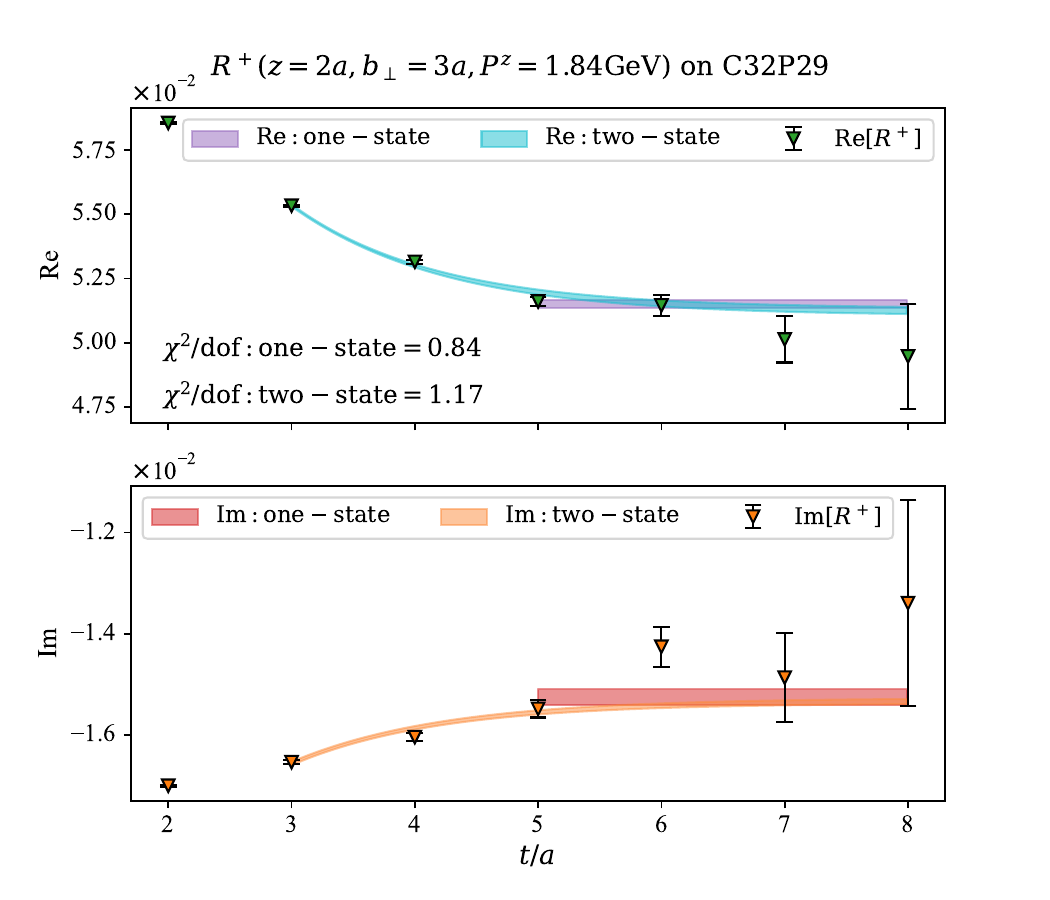}
\includegraphics[width=0.45\textwidth]{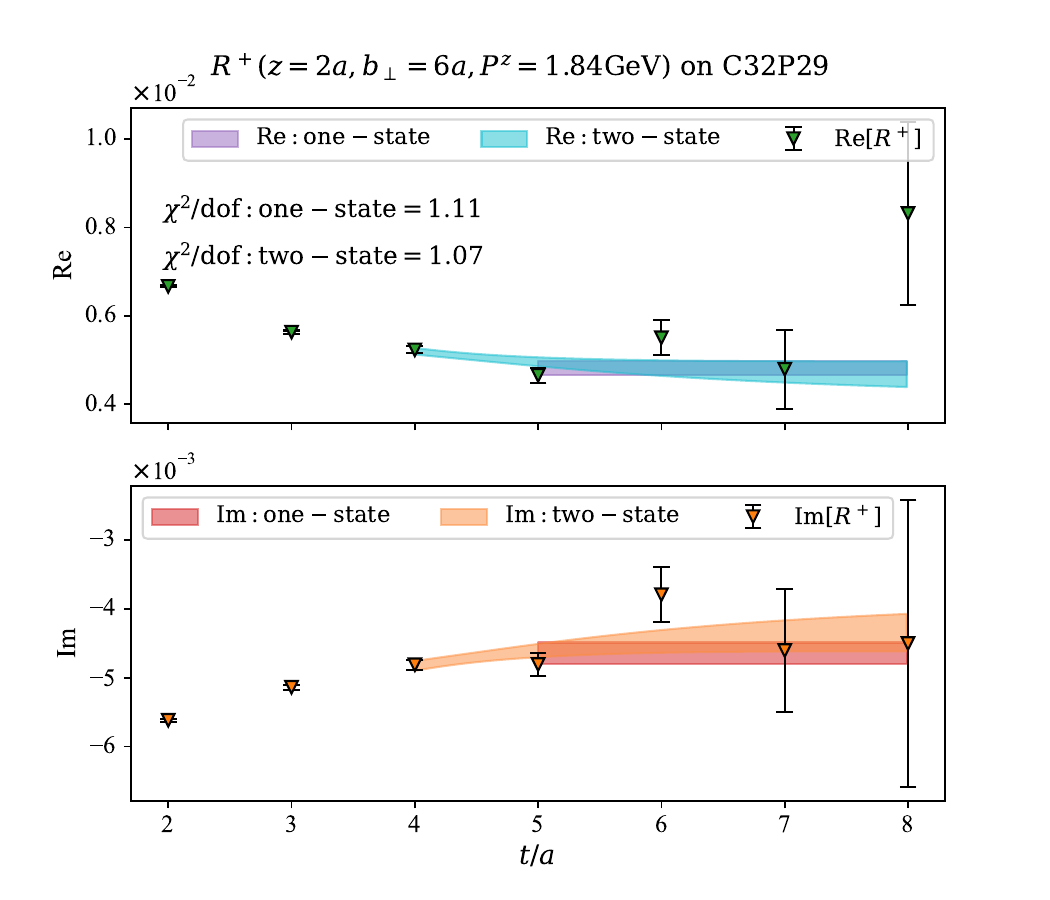}
\includegraphics[width=0.45\textwidth]{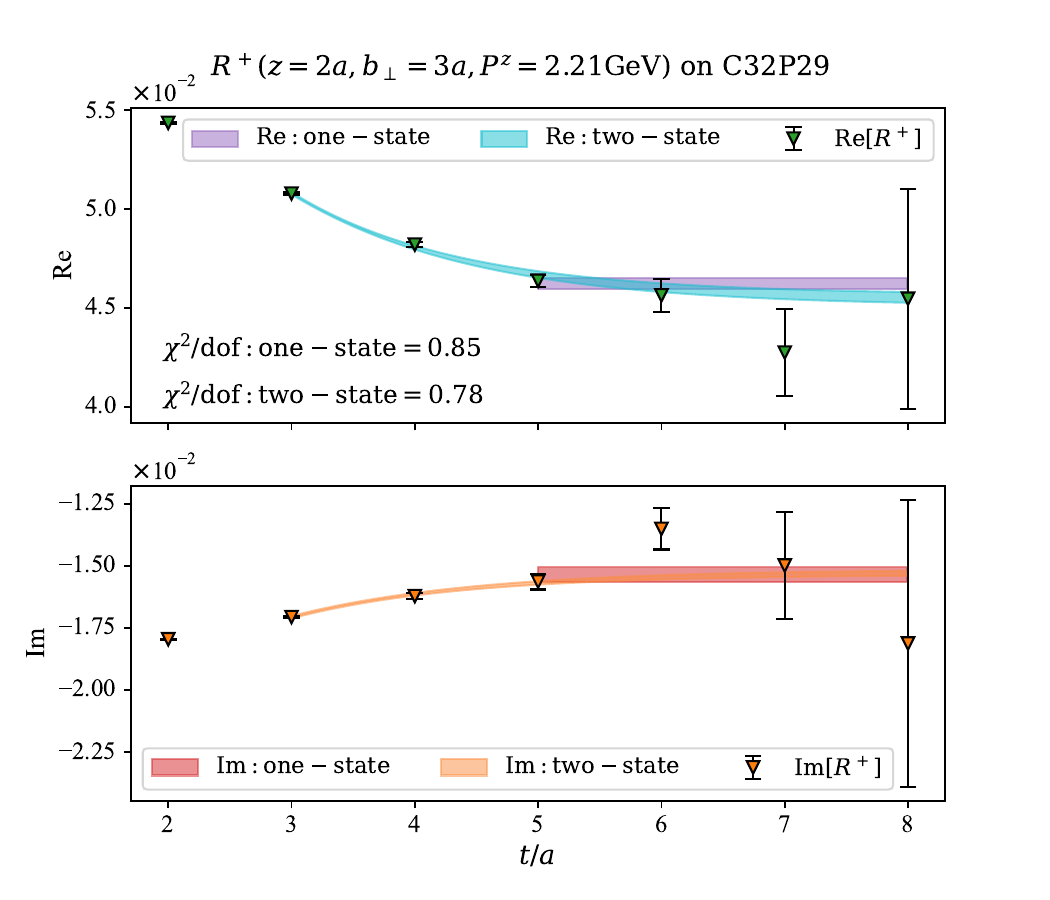}
\includegraphics[width=0.45\textwidth]{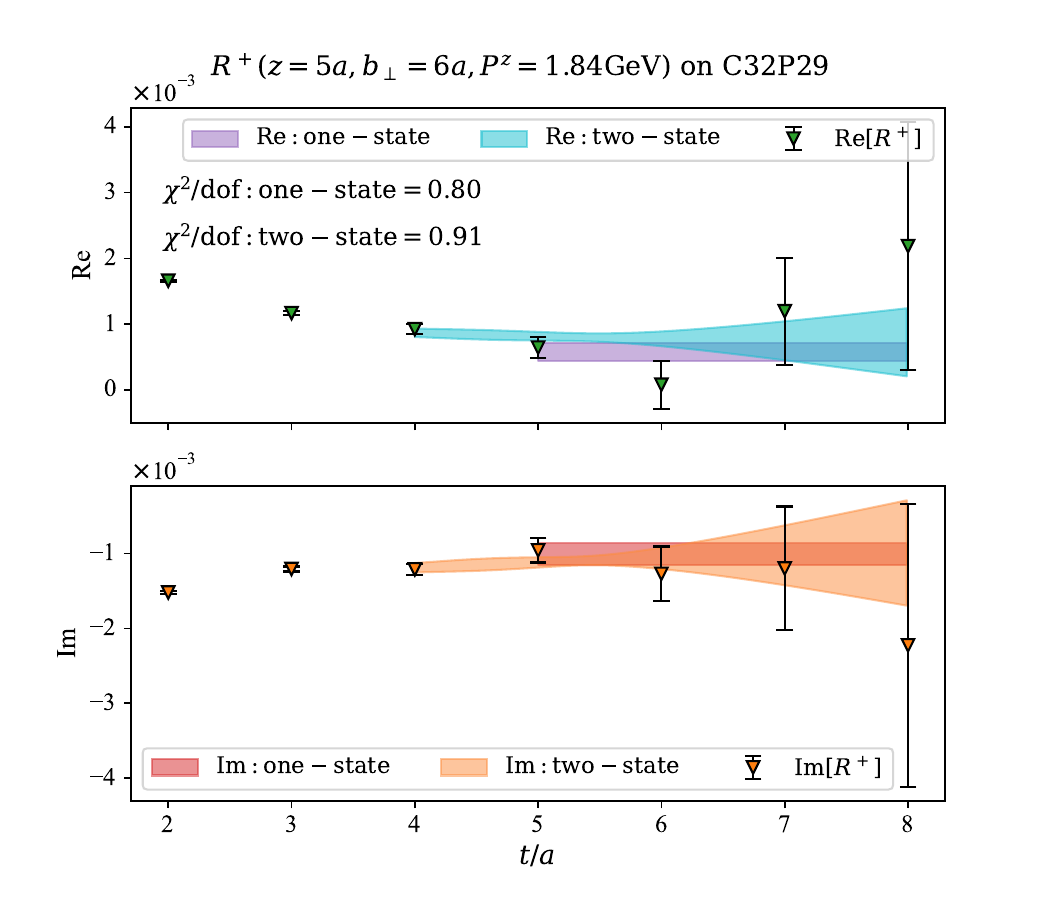}
\includegraphics[width=0.45\textwidth]{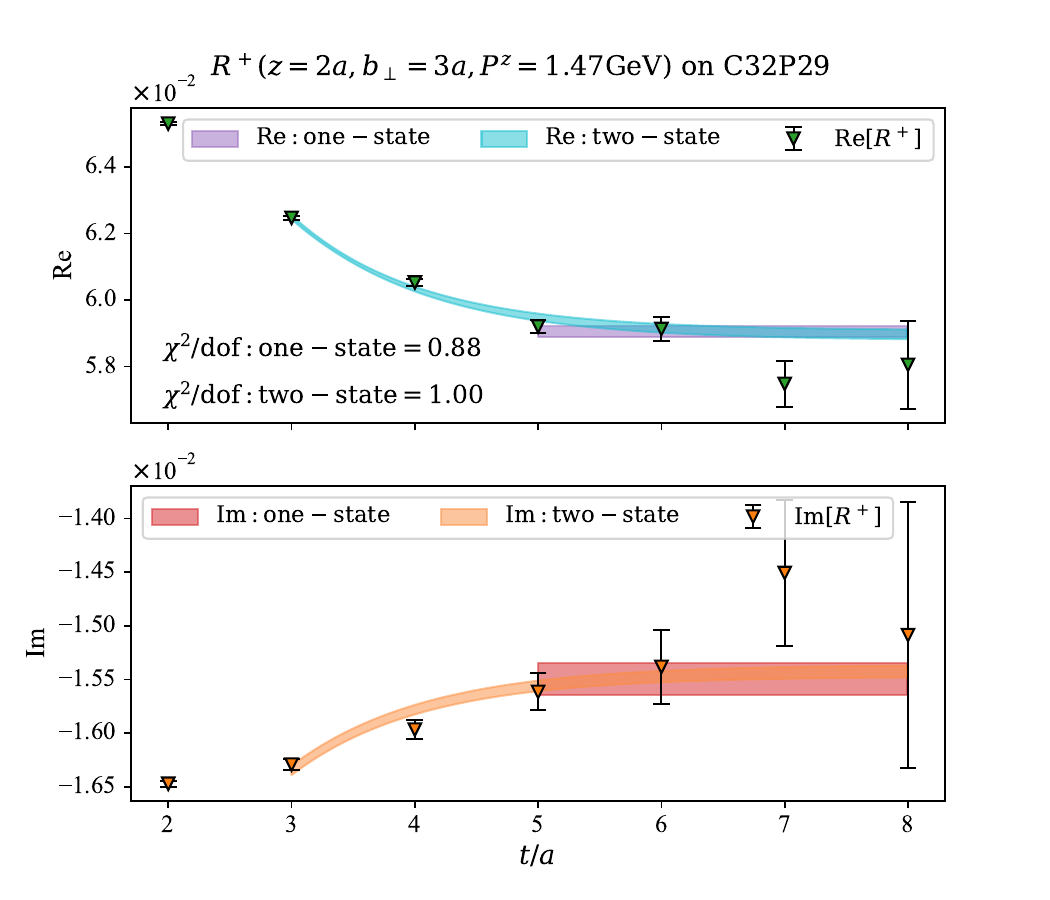}
\caption{The figures present results from the C32P29 ensemble with 1-step HYP smearing, and compare the combinations for $\{z, b_\perp, P^z\}=\{2a, a, 1.84\mathrm{GeV}\}$, $\{2a, 3a, 1.84\mathrm{GeV}\}$, $\{2a, 6a, 1.84\mathrm{GeV}\}$ , $\{2a, 3a, 2.21\mathrm{GeV}\}$, $\{5a, 6a, 1.84\mathrm{GeV}\}$and$\{2a, 3a, 1.47\mathrm{GeV}\}$} 
\label{fig:more_fit_3}
\end{figure}

\begin{figure}[http]
\centering
\includegraphics[width=0.45\textwidth]{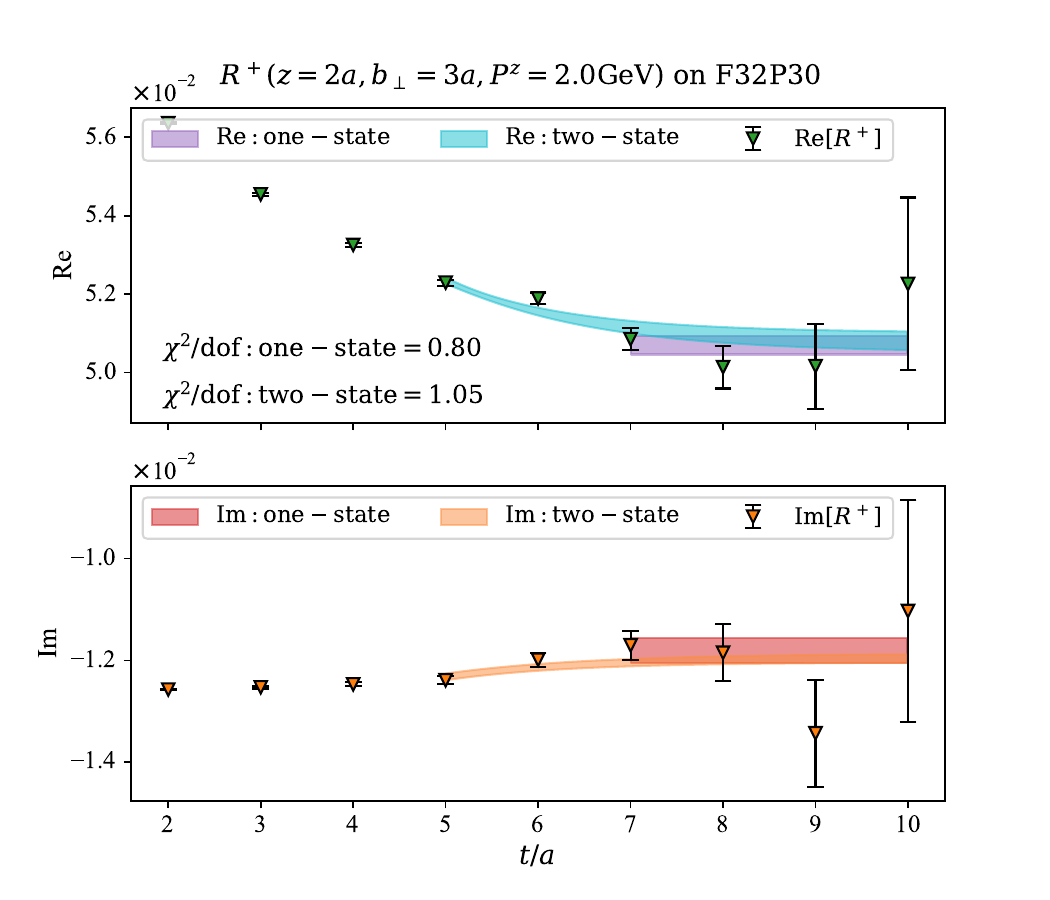}
\includegraphics[width=0.45\textwidth]{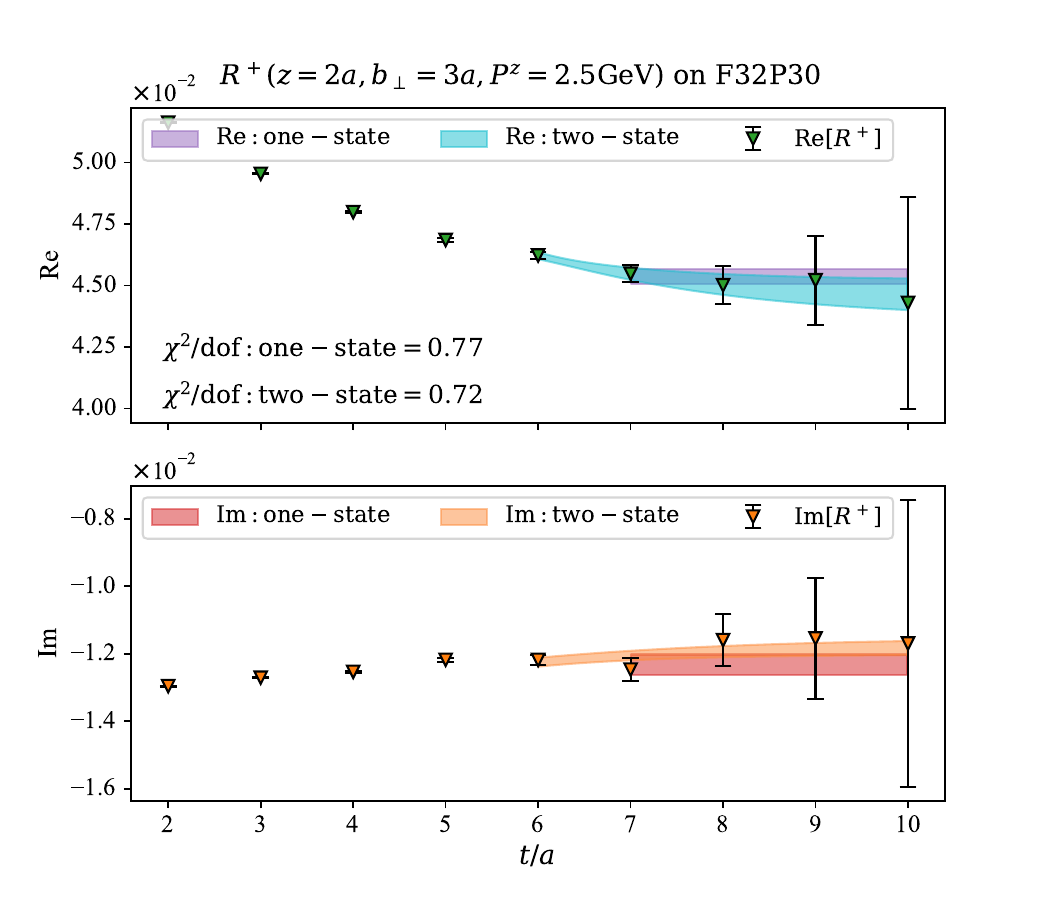}
\includegraphics[width=0.45\textwidth]{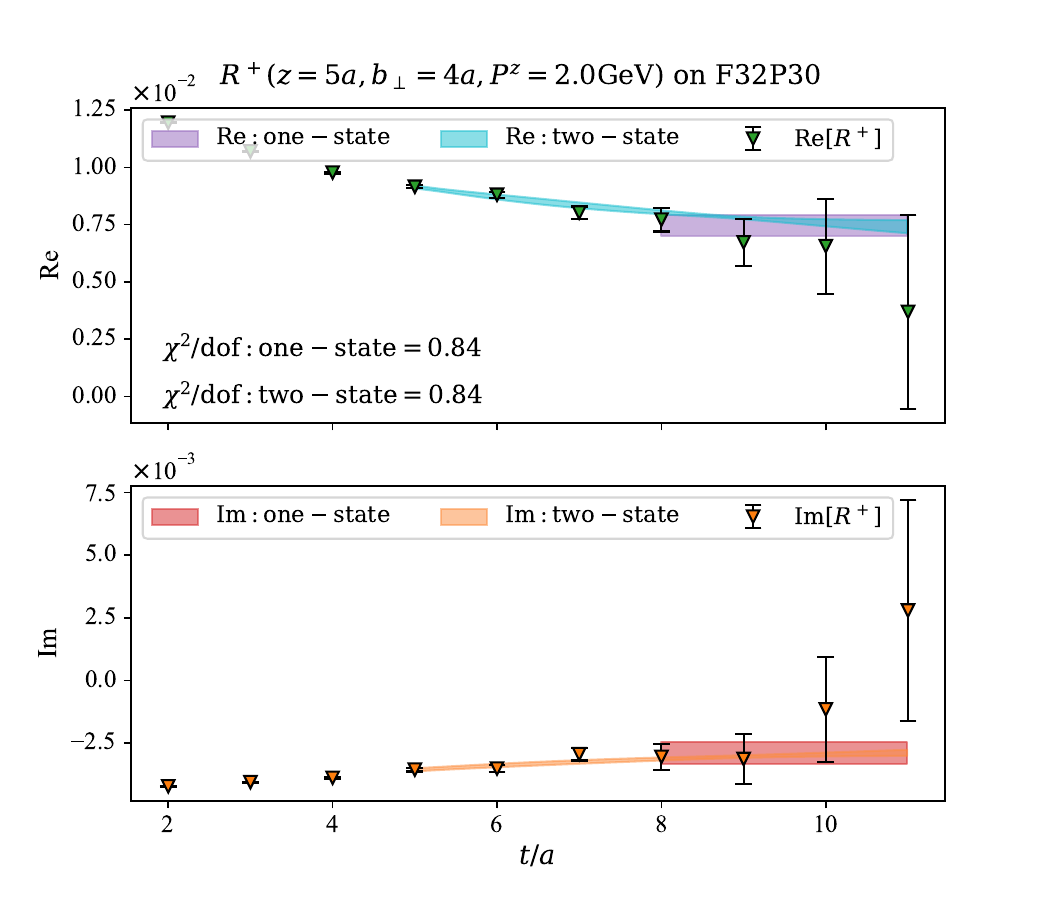}
\includegraphics[width=0.45\textwidth]{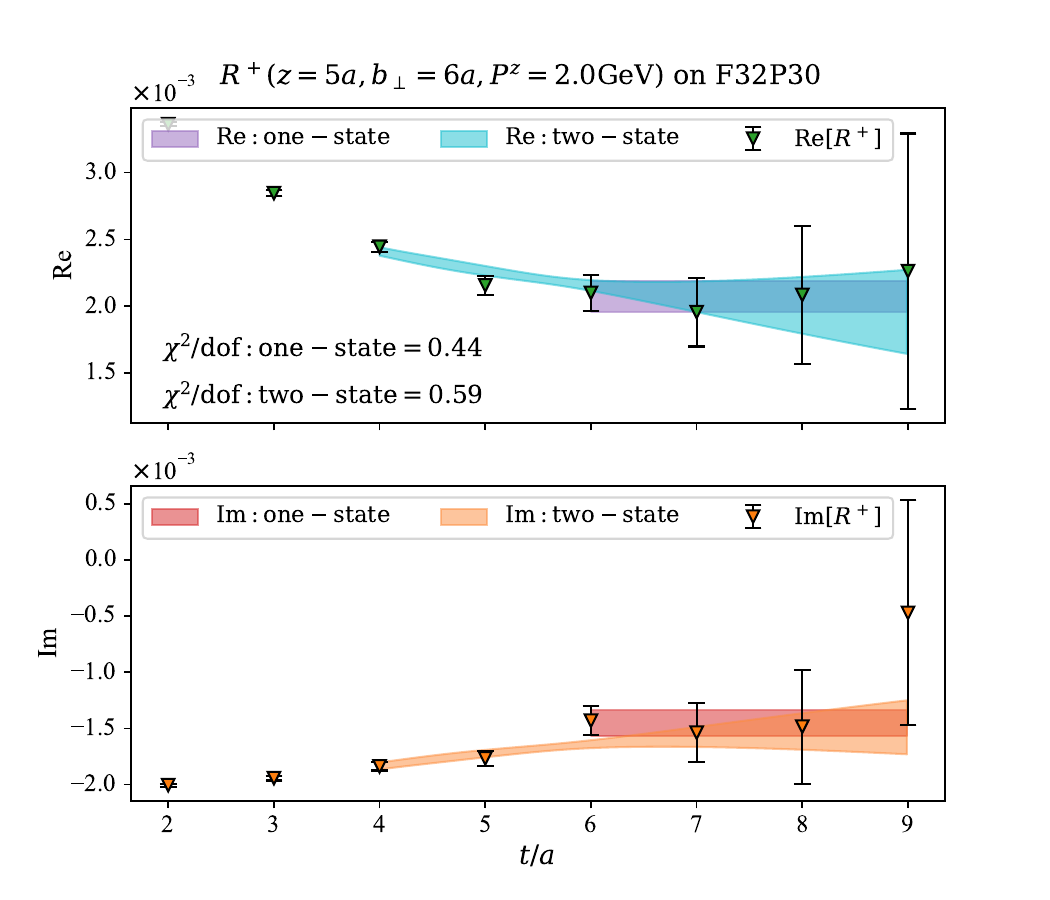}
\includegraphics[width=0.45\textwidth]{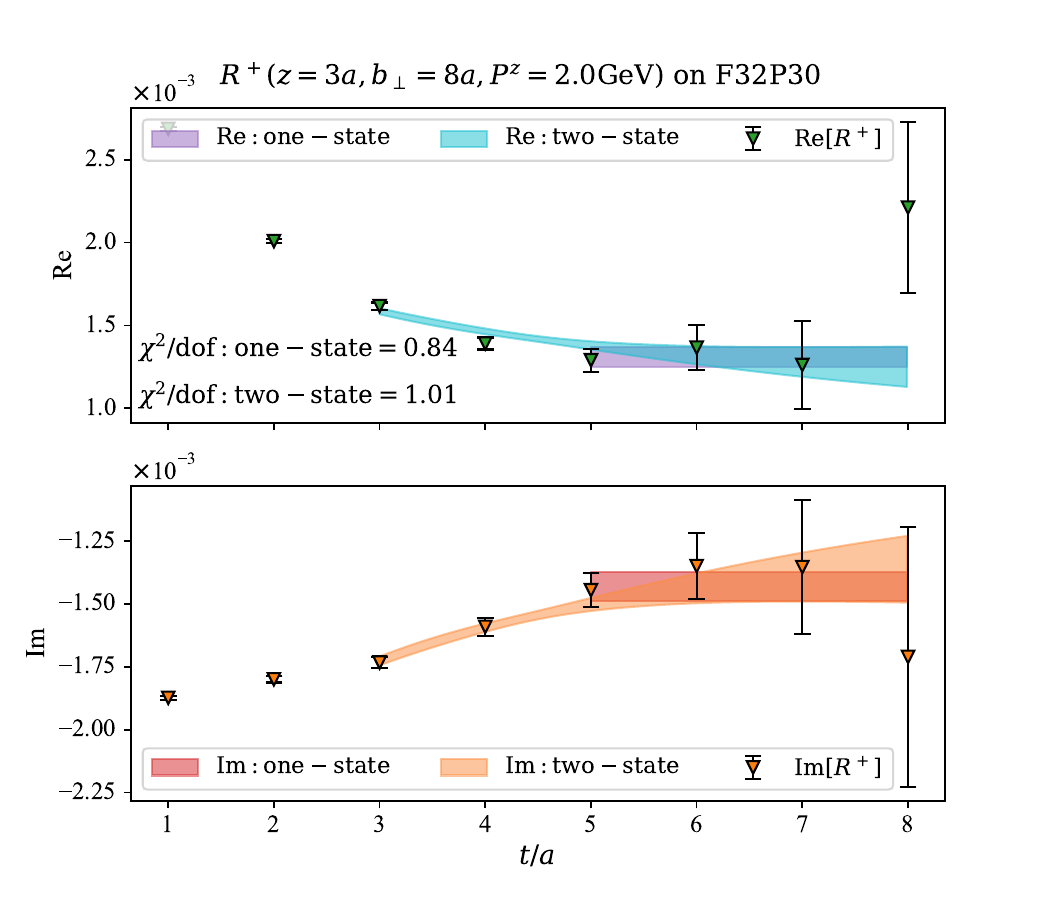}
\includegraphics[width=0.45\textwidth]{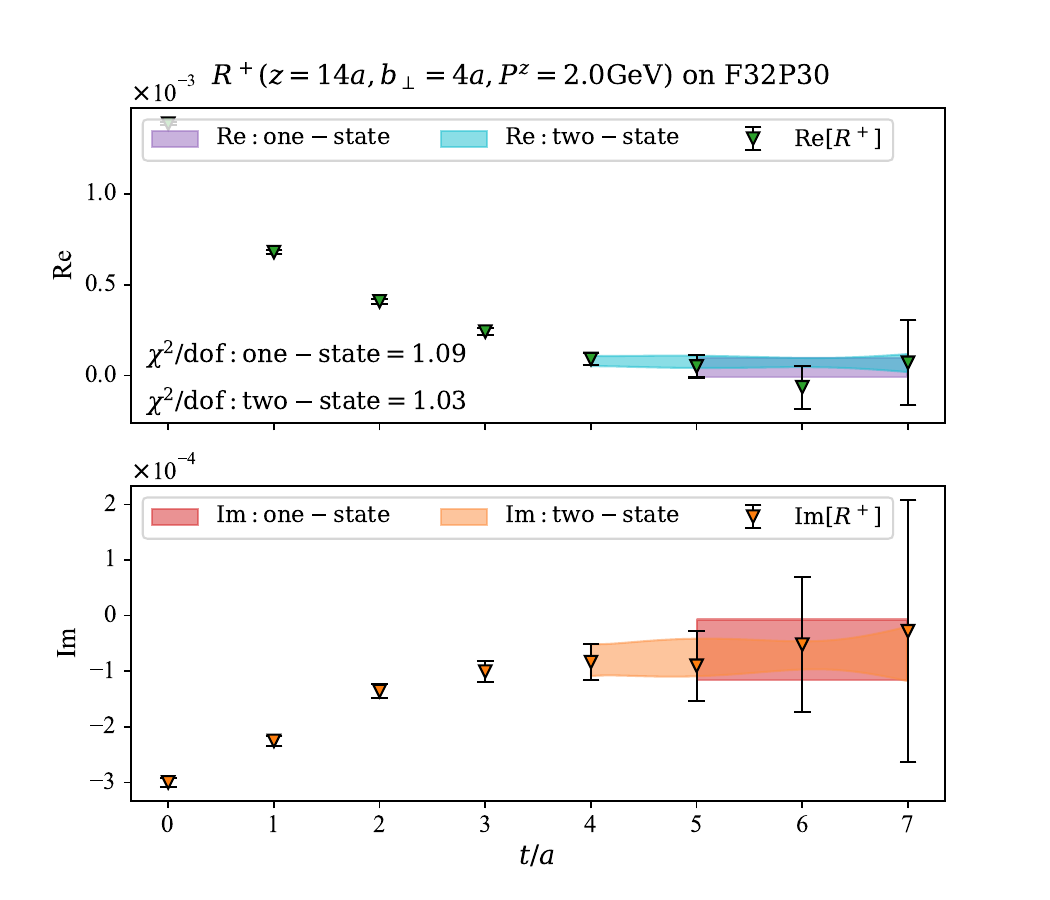}
\caption{The figures present results from the F32P30 ensemble with 1-step HYP smearing, and compare the combinations for $\{z, b_\perp, P^z\}=\{2a, 3a, 2.0\mathrm{GeV}\}$, $\{2a, 3a, 2.5\mathrm{GeV}\}$, $\{5a, 4a, 2.0\mathrm{GeV}\}$ , $\{5a, 6a, 2.0\mathrm{GeV}\}$, $\{3a, 8a, 2.0\mathrm{GeV}\}$and$\{14a, 4a, 2.0\mathrm{GeV}\}$} 
\label{fig:more_fit_4}
\end{figure}

\subsection{Results for renormalized quasi-TMDWF matrix elements }
More results for the renormalized quasi-TMDWF matrix elements on different ensembles are given in Fig.~\ref{fig:more_ME_1}, ~\ref{fig:more_ME_2}, ~\ref{fig:more_ME_3} and ~\ref{fig:more_ME_4}.

\begin{figure}[http]
\centering
\includegraphics[width=0.45\textwidth]{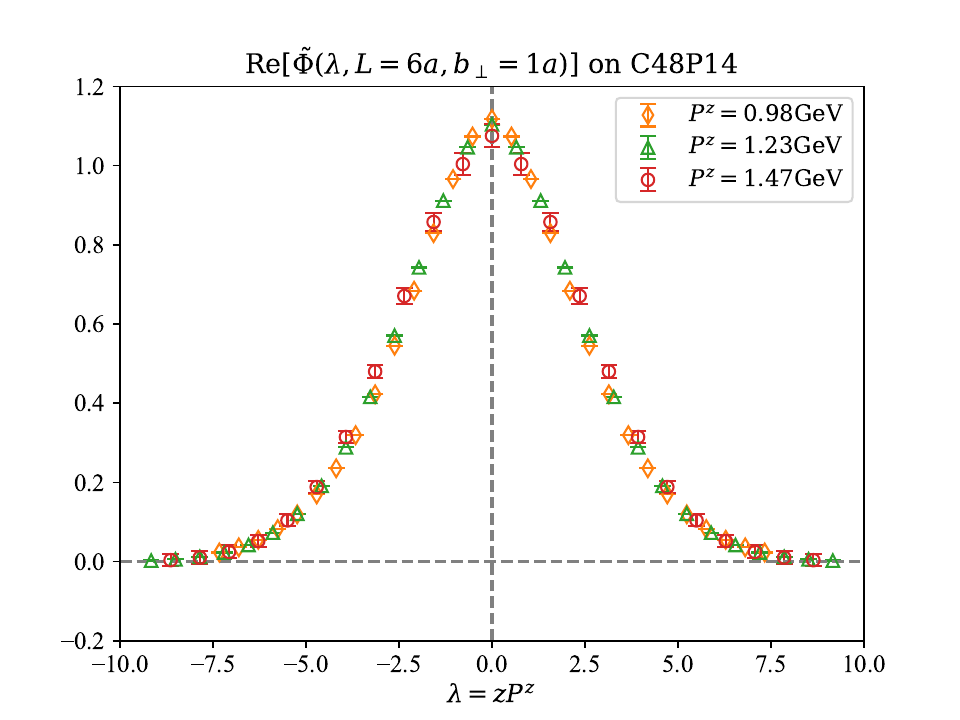}
\includegraphics[width=0.45\textwidth]{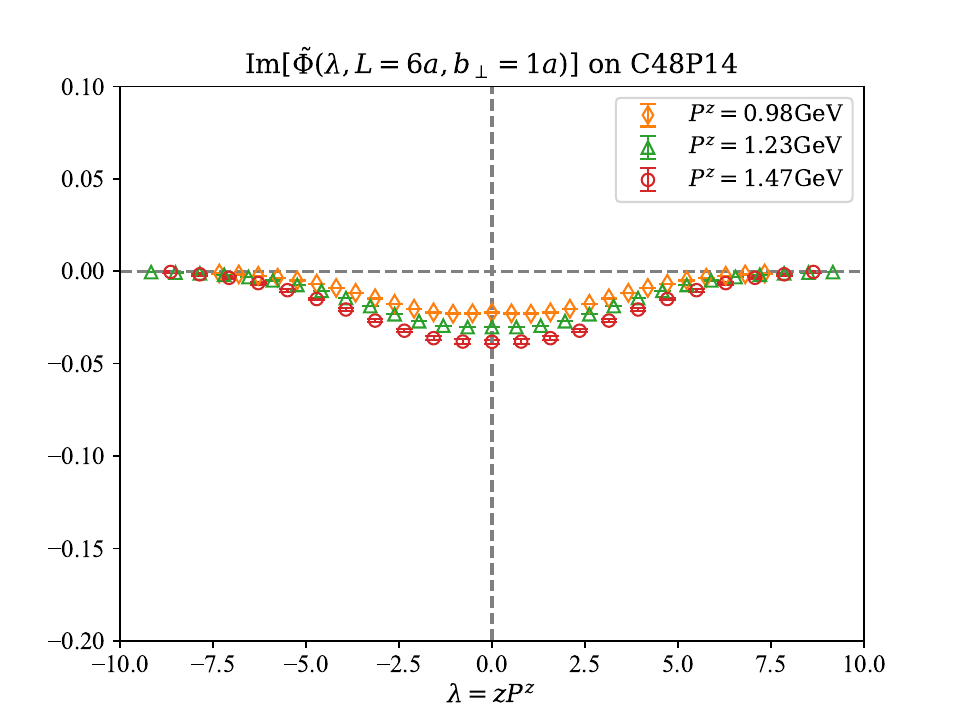}
\includegraphics[width=0.45\textwidth]{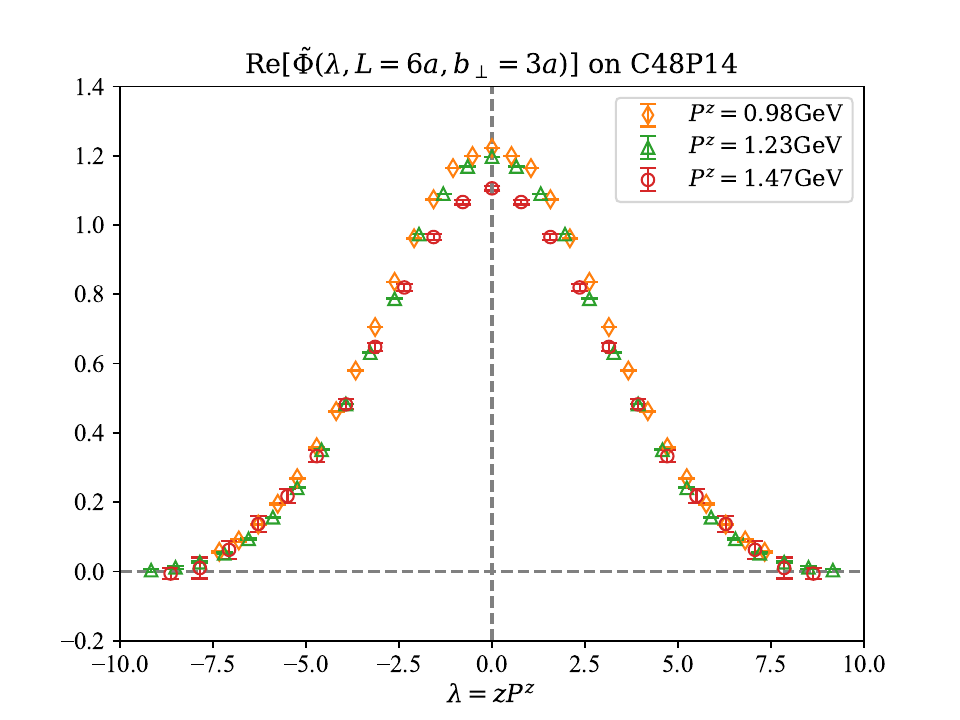}
\includegraphics[width=0.45\textwidth]{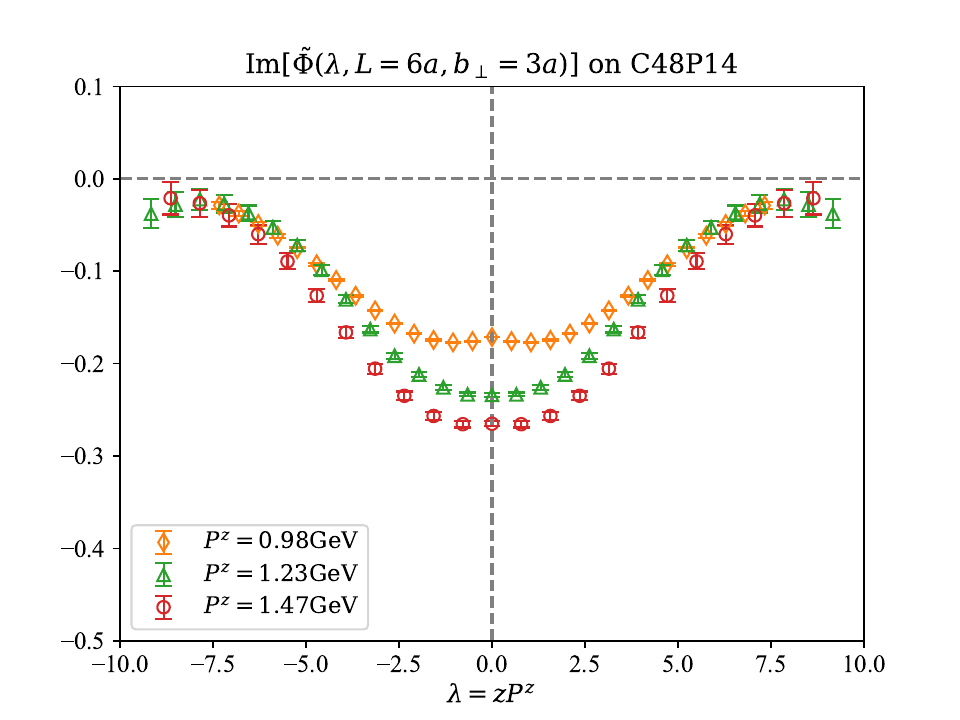}
\includegraphics[width=0.45\textwidth]{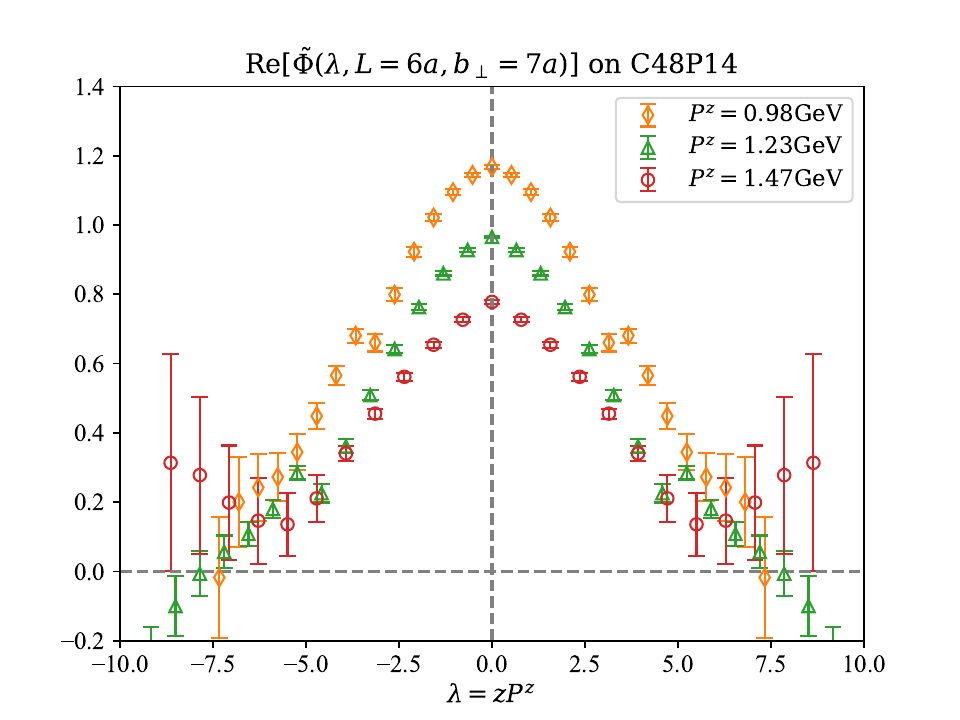}
\includegraphics[width=0.45\textwidth]{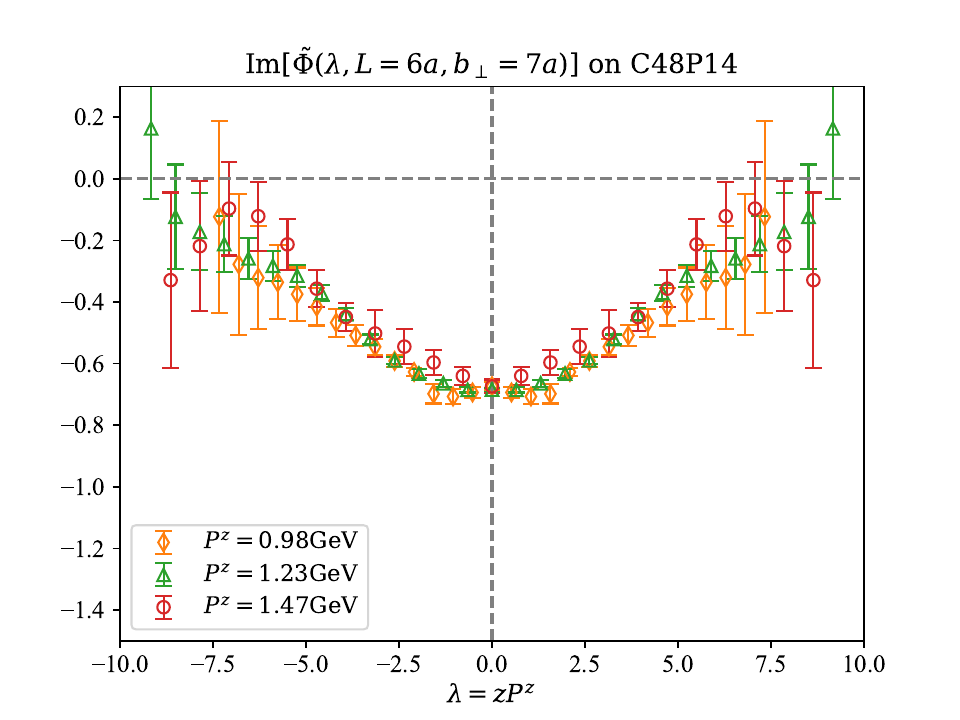}
\caption{Results from the C48P14 ensemble . A comparison of the results for $P^z = \lbrace 0.98, 1.23, 1.47 \rbrace \, \mathrm{GeV}$ at different $b_{\perp}$ values is presented in the three panels.} 
\label{fig:more_ME_1}
\end{figure}

\begin{figure}[http]
\centering
\includegraphics[width=0.45\textwidth]{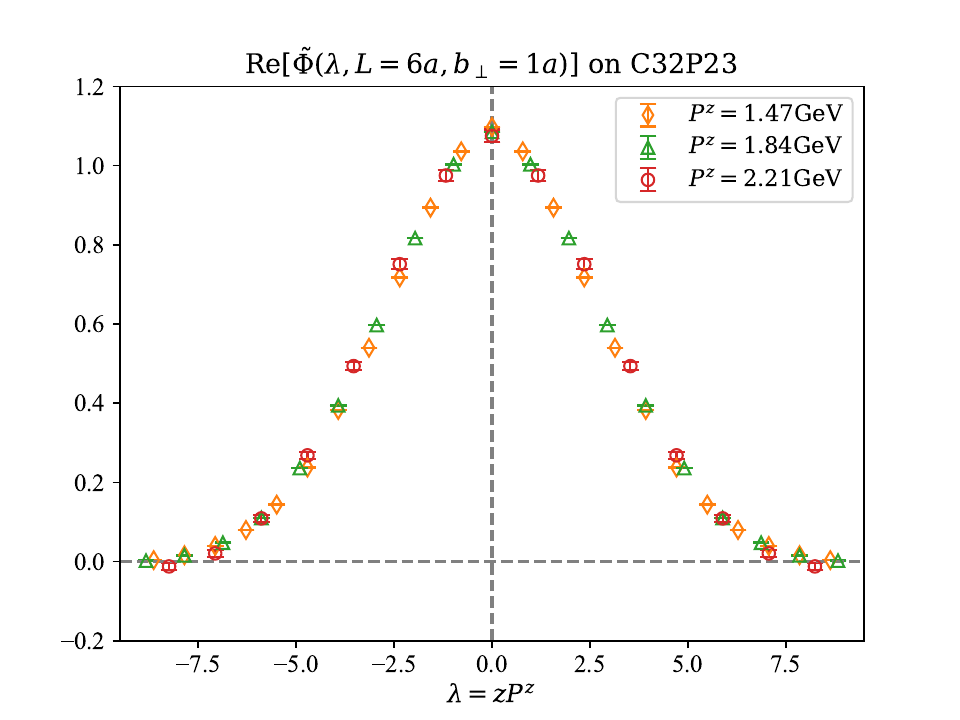}
\includegraphics[width=0.45\textwidth]{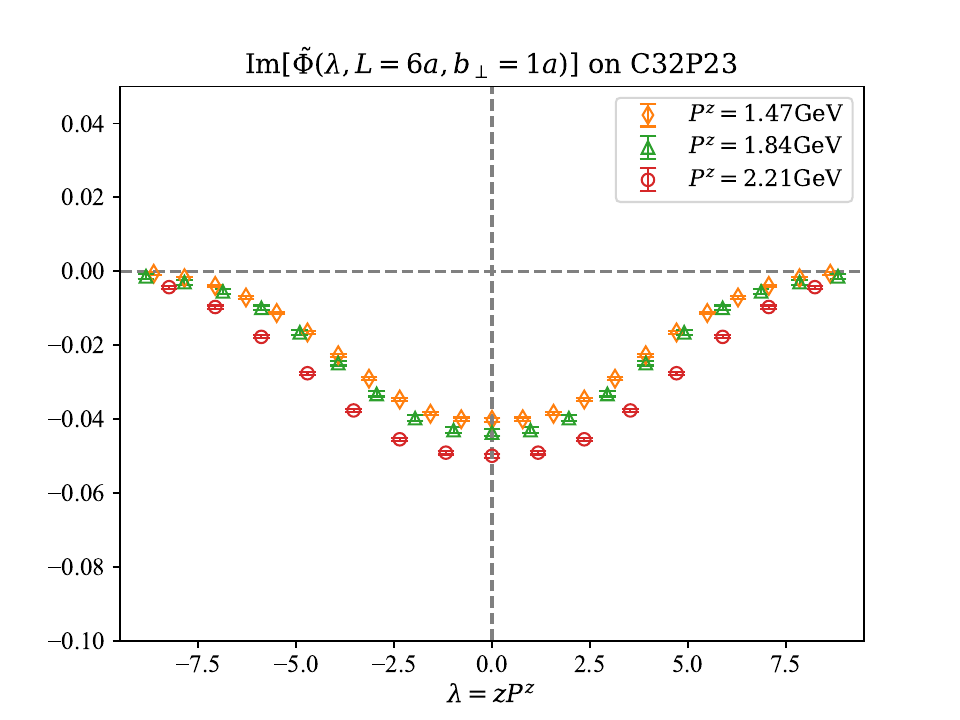}
\includegraphics[width=0.45\textwidth]{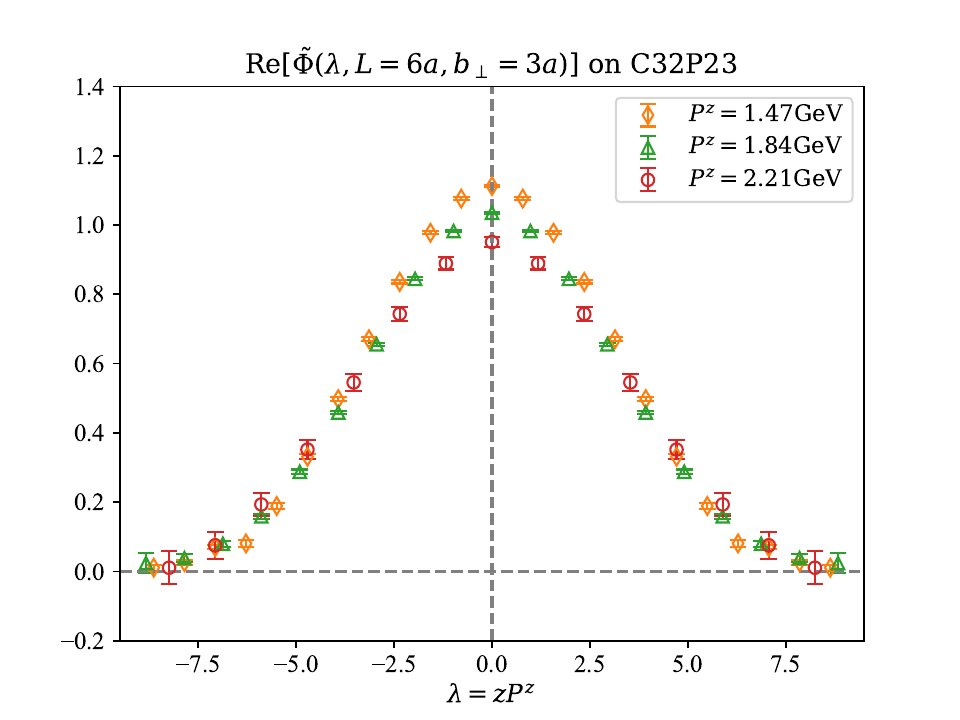}
\includegraphics[width=0.45\textwidth]{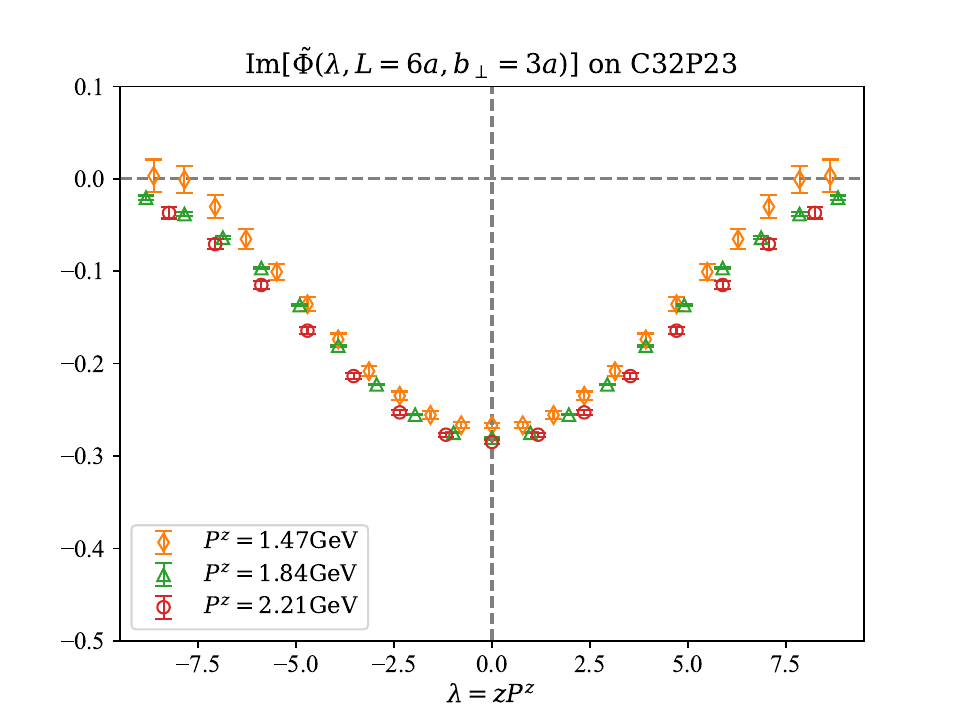}
\includegraphics[width=0.45\textwidth]{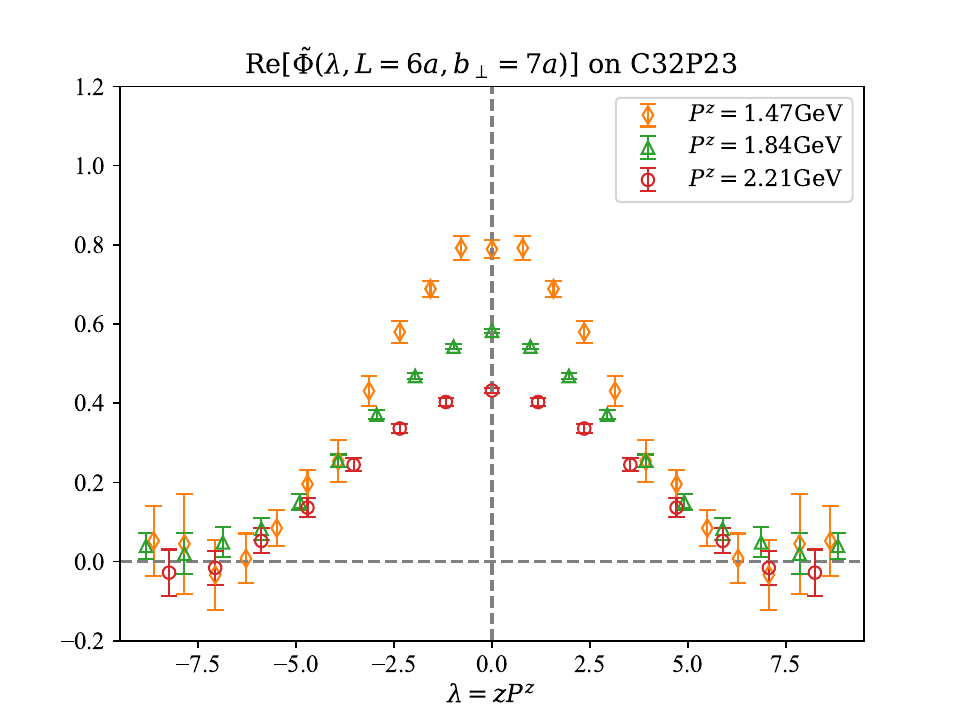}
\includegraphics[width=0.45\textwidth]{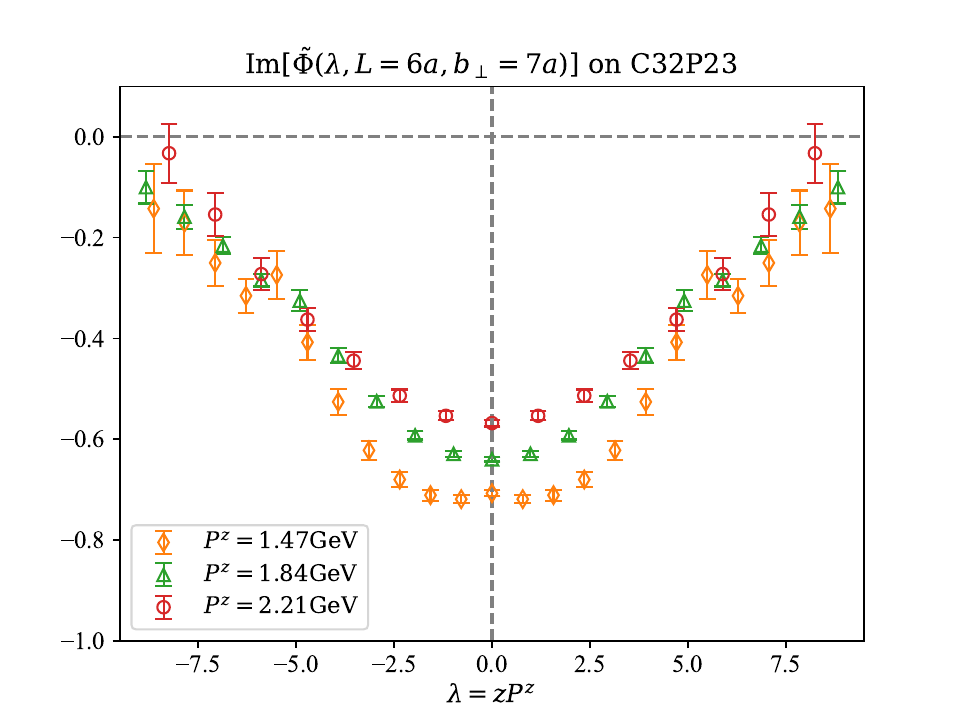}
\caption{ Results from the C32P23 ensemble . A comparison of the results for $P^z = \lbrace  1.47, 1.84, 2.21 \rbrace \, \mathrm{GeV}$ at different $b_{\perp}$ values is presented in the three panels.} 
\label{fig:more_ME_2}
\end{figure}

\begin{figure}[http]
\centering
\includegraphics[width=0.45\textwidth]{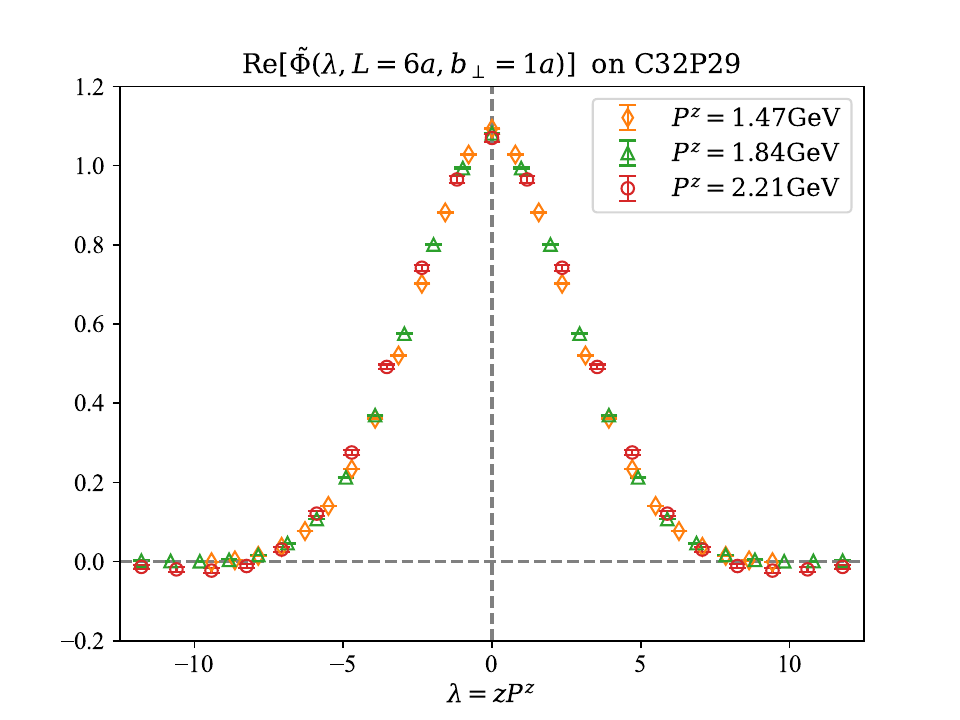}
\includegraphics[width=0.45\textwidth]{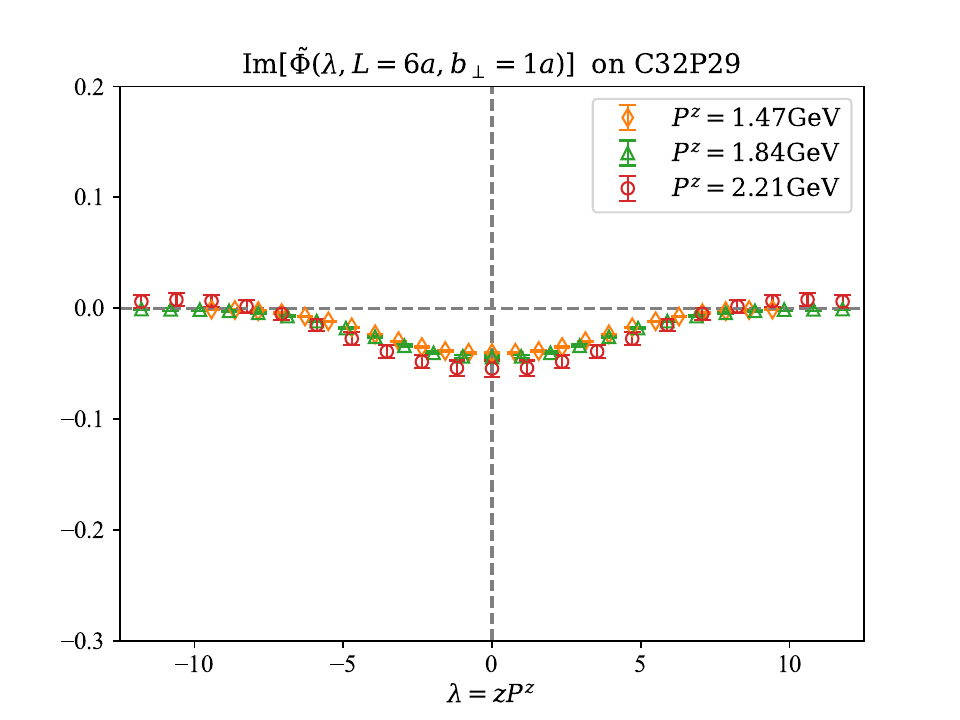}
\includegraphics[width=0.45\textwidth]{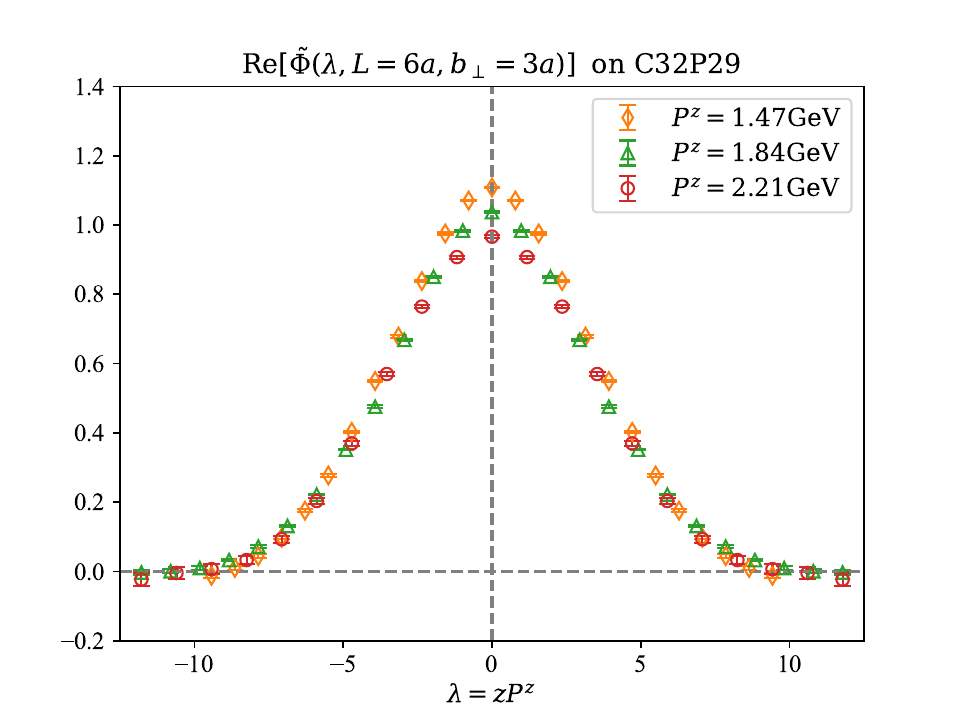}
\includegraphics[width=0.45\textwidth]{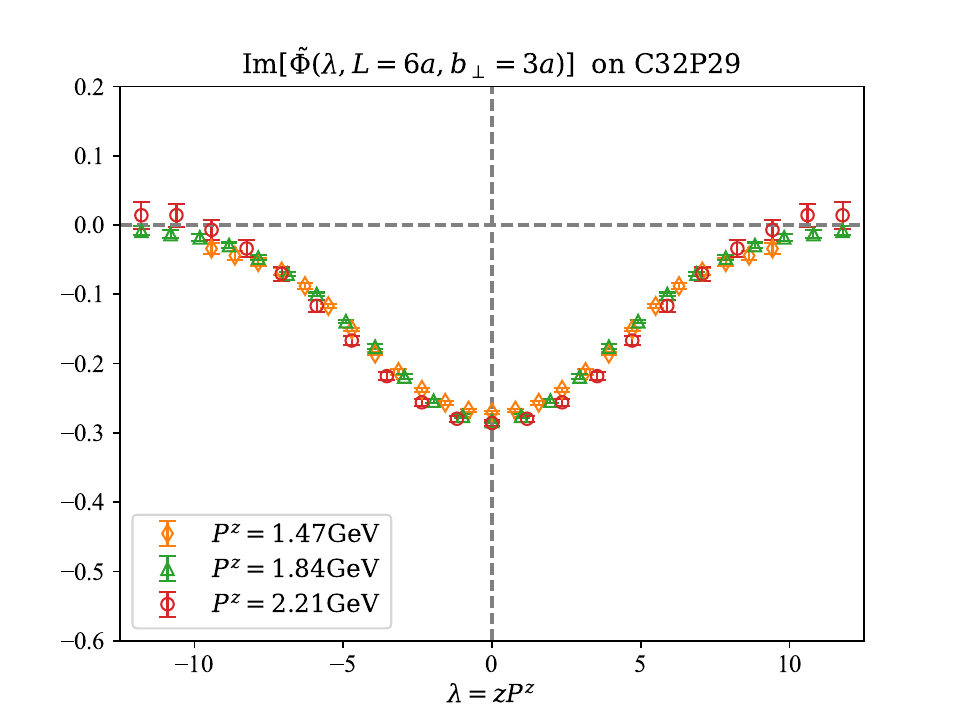}
\includegraphics[width=0.45\textwidth]{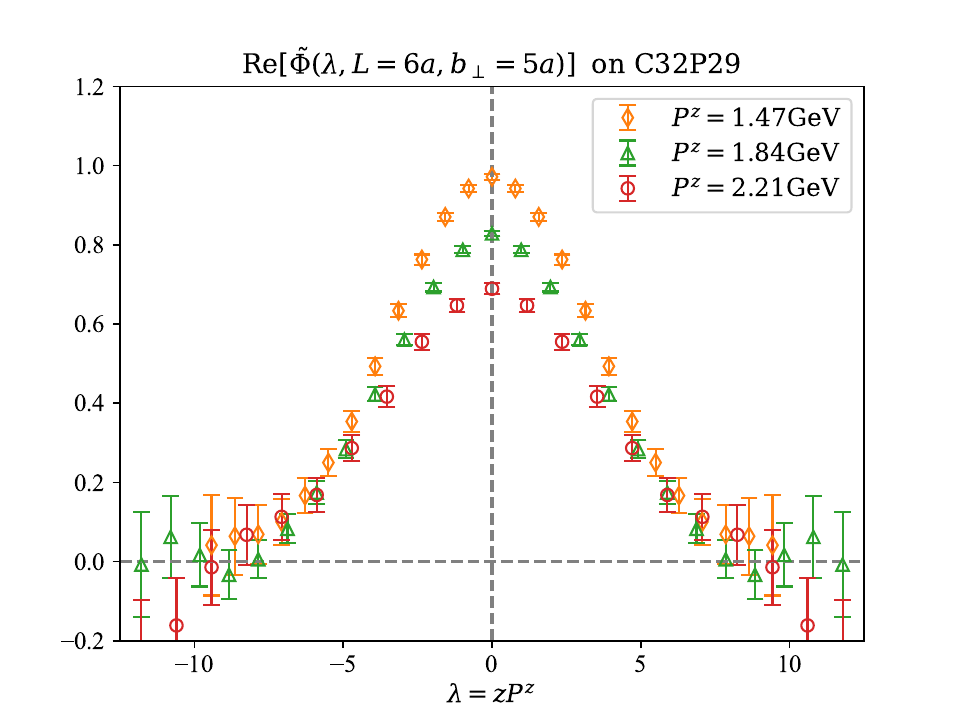}
\includegraphics[width=0.45\textwidth]{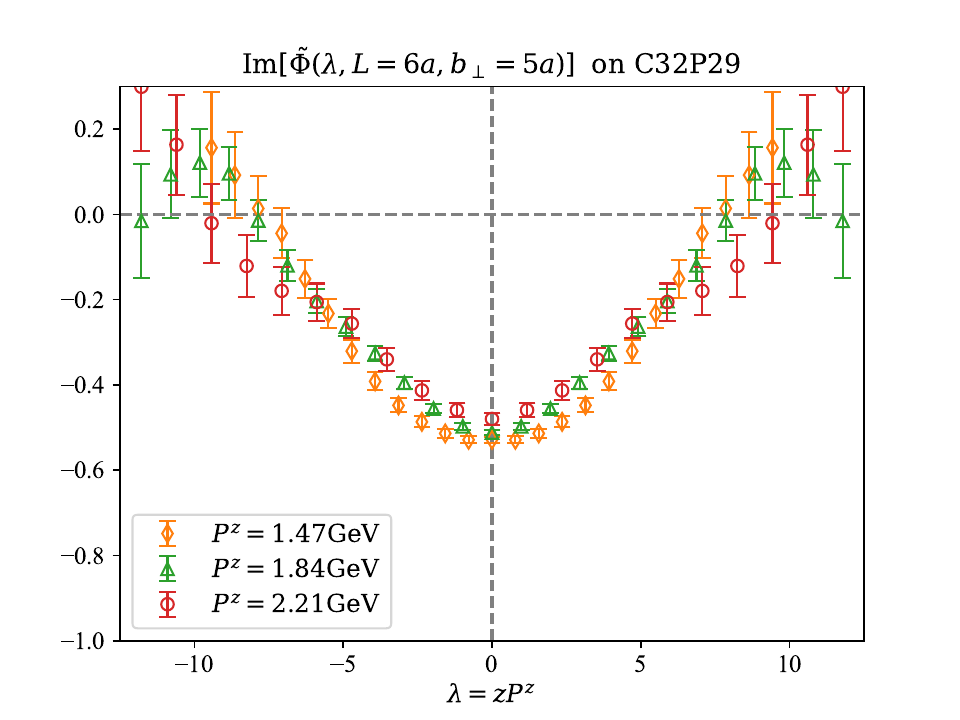}
\caption{Results from the C32P29 ensemble . A comparison of the results for $P^z = \lbrace 1.47, 1.84, 2.21 \rbrace \, \mathrm{GeV}$ at different $b_{\perp}$ values is presented in the three panels.} 
\label{fig:more_ME_3}
\end{figure}

\begin{figure}
\centering
\includegraphics[width=0.45\textwidth]{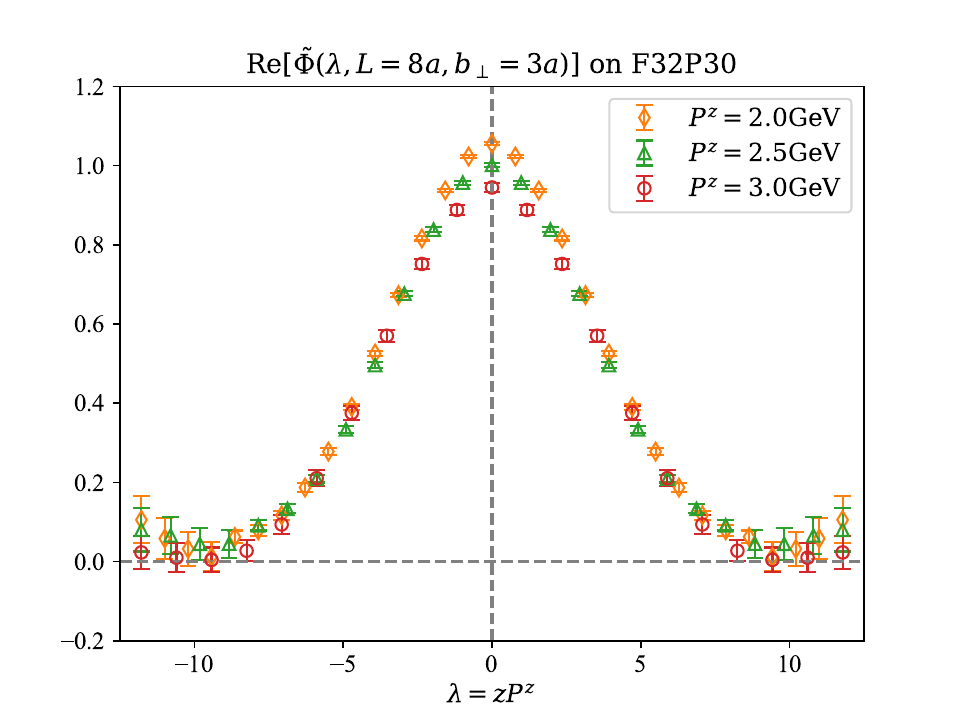}
\includegraphics[width=0.45\textwidth]{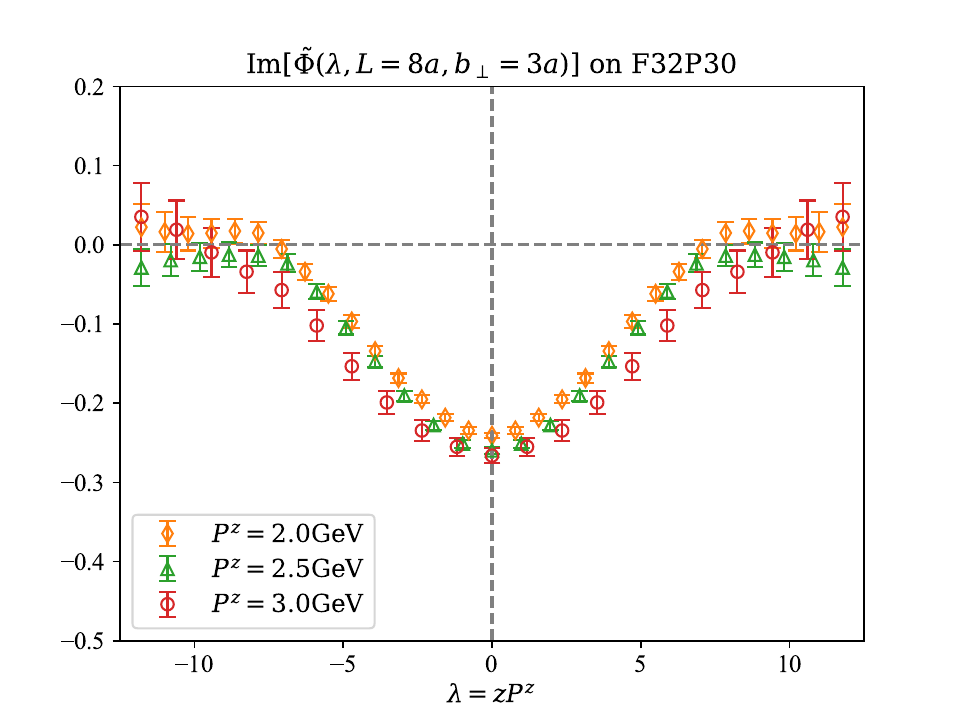}
\includegraphics[width=0.45\textwidth]{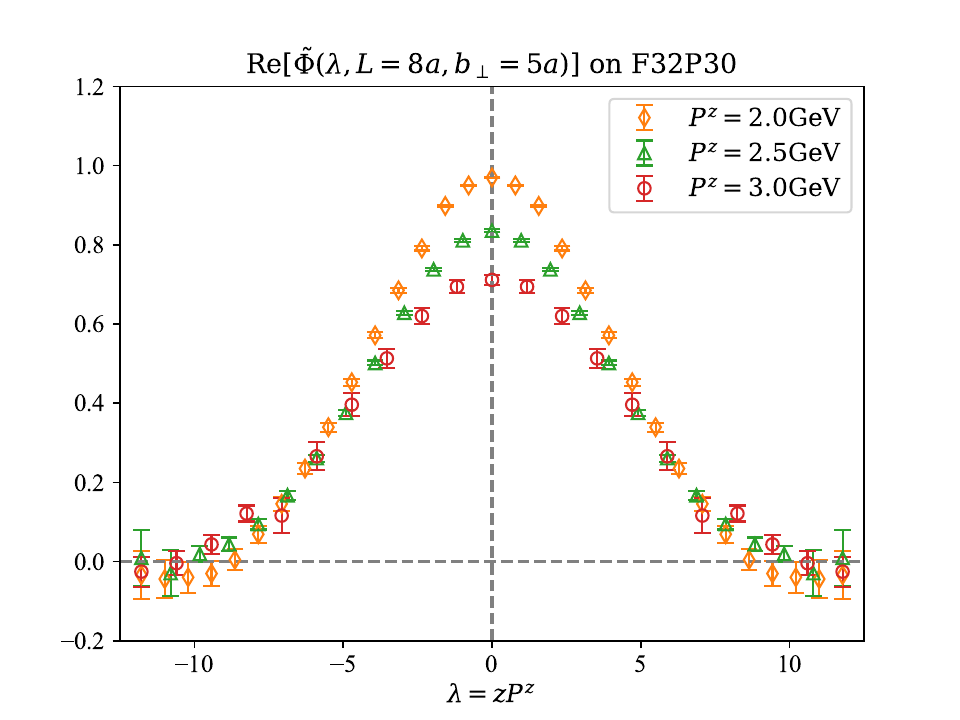}
\includegraphics[width=0.45\textwidth]{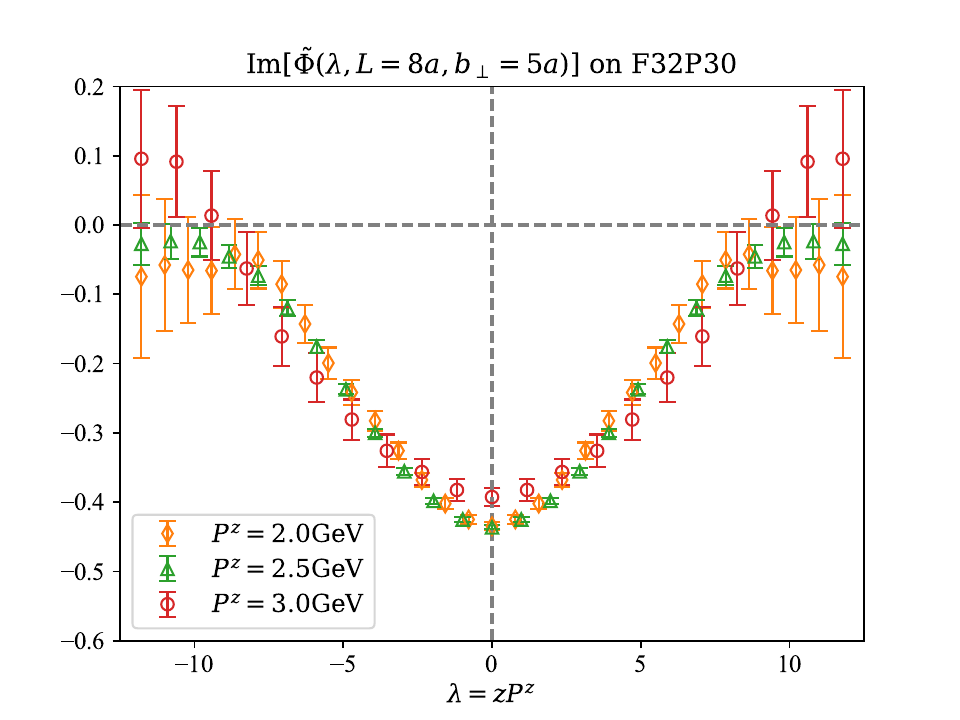}
\includegraphics[width=0.45\textwidth]{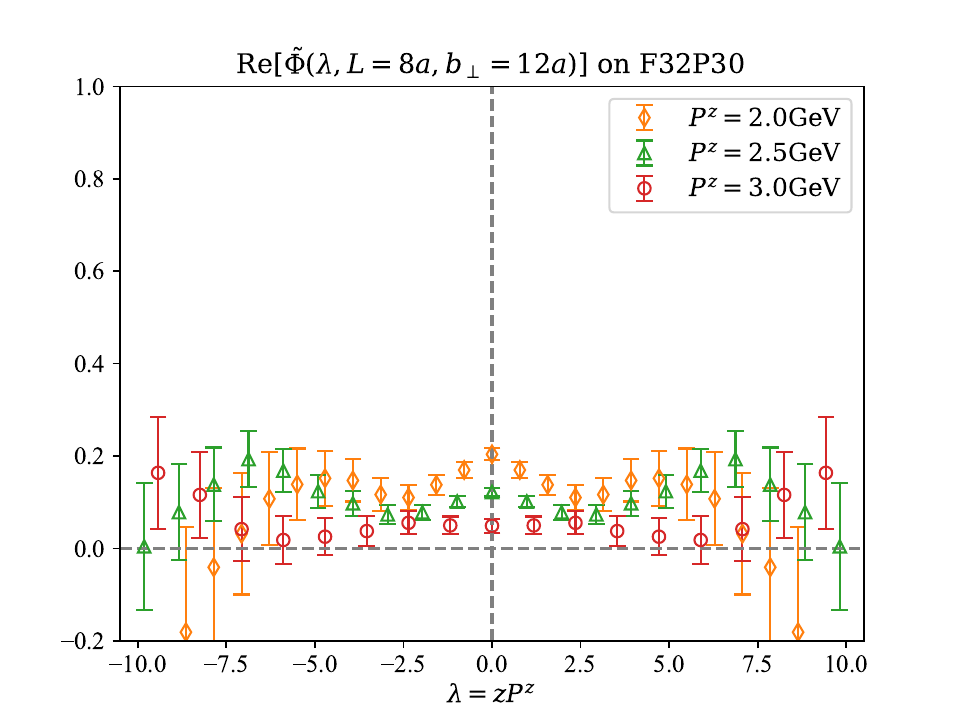}
\includegraphics[width=0.45\textwidth]{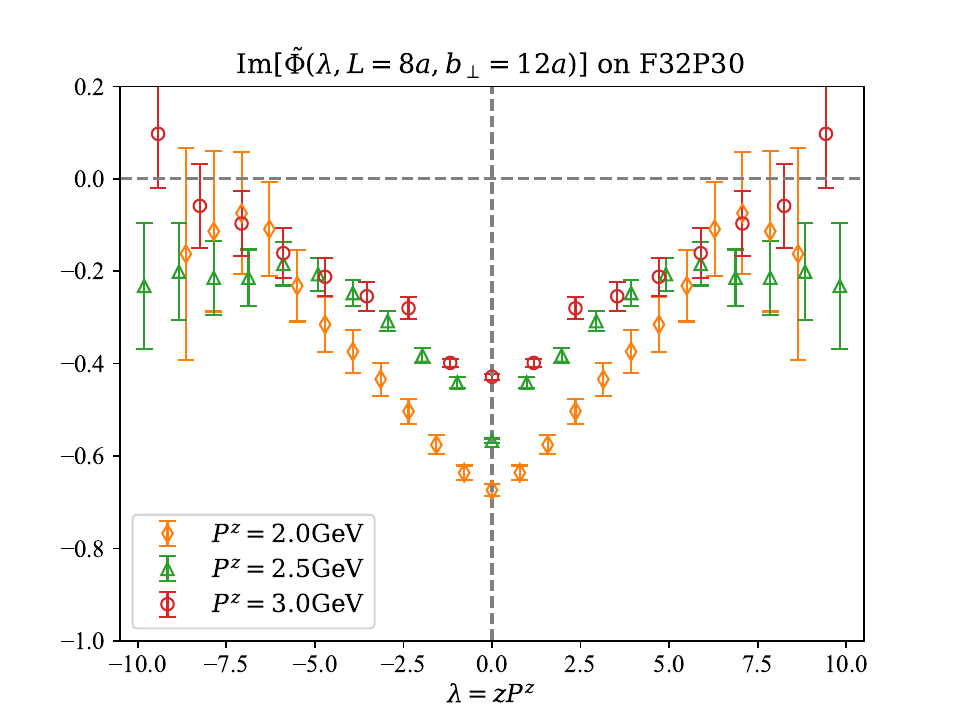}
\caption{Results from the F32P30 ensemble . A comparison of the results for $P^z = \lbrace 2.0, 2.5, 3.0 \rbrace \, \mathrm{GeV}$ at different $b_{\perp}$ values is presented in the three panels.} 
\label{fig:more_ME_4}
\end{figure}

\subsection{Results for quasi-TMDWFs}
More results for the quasi-TMDWFs on different ensembles are given in Fig.~\ref{fig:more_WFs_1}, ~\ref{fig:more_WFs_2}, ~\ref{fig:more_WFs_3} and ~\ref{fig:more_WFs_4}.

\begin{figure}
\centering
\includegraphics[width=0.45\textwidth]{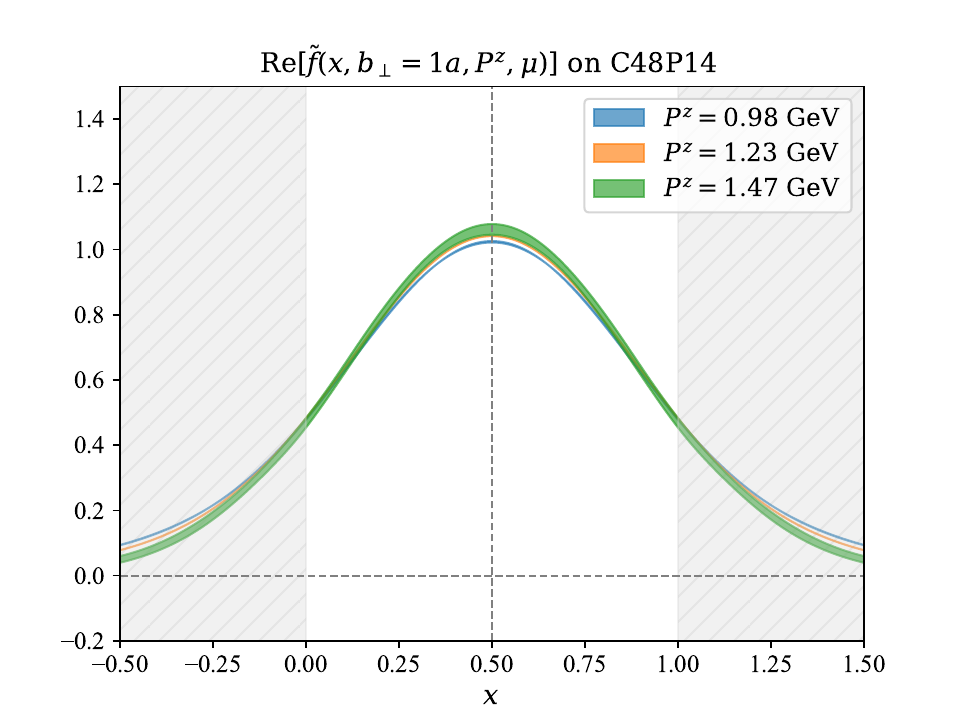}
\includegraphics[width=0.45\textwidth]{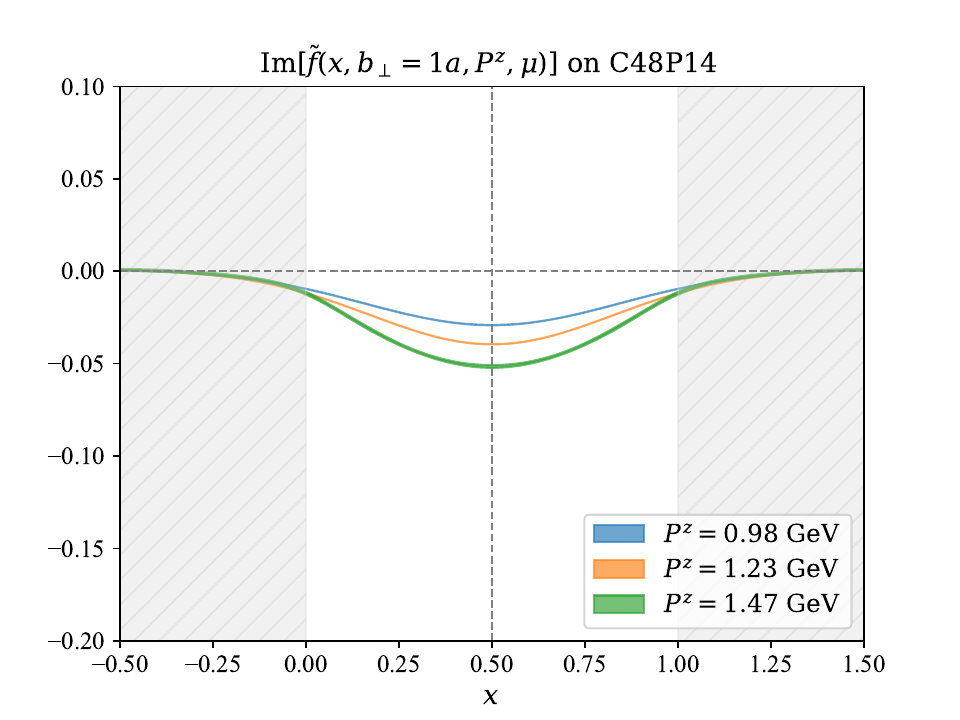}
\includegraphics[width=0.45\textwidth]{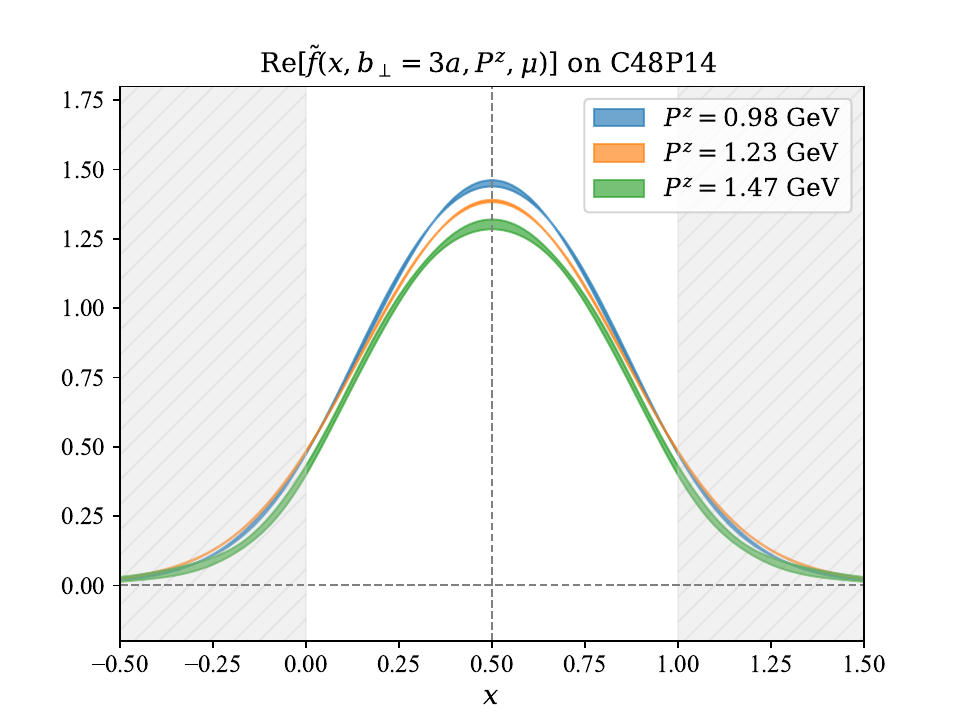}
\includegraphics[width=0.45\textwidth]{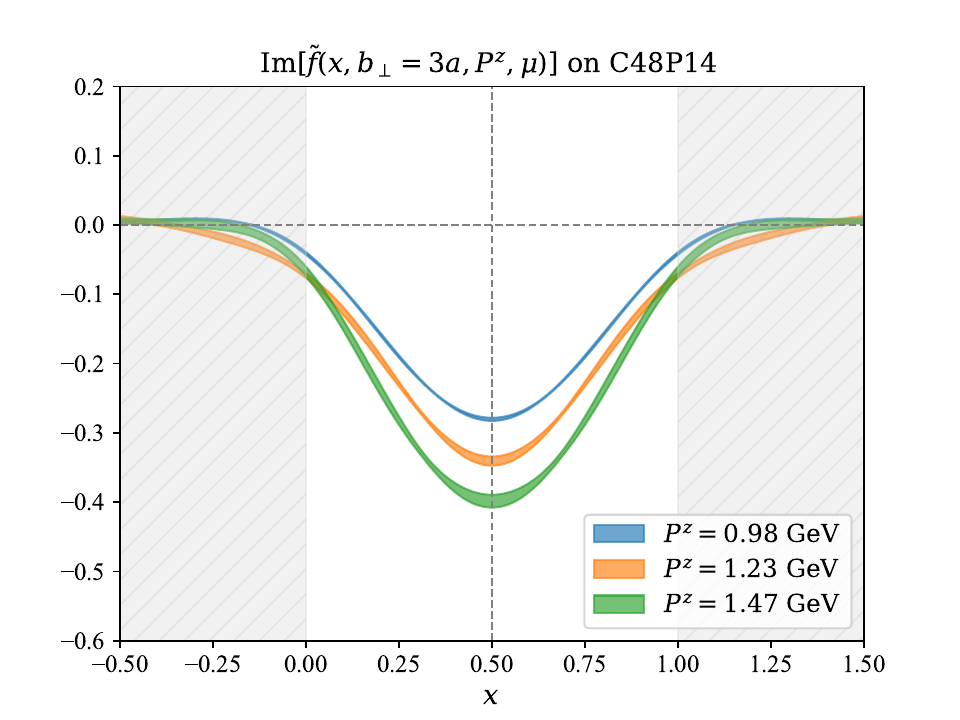}
\includegraphics[width=0.45\textwidth]{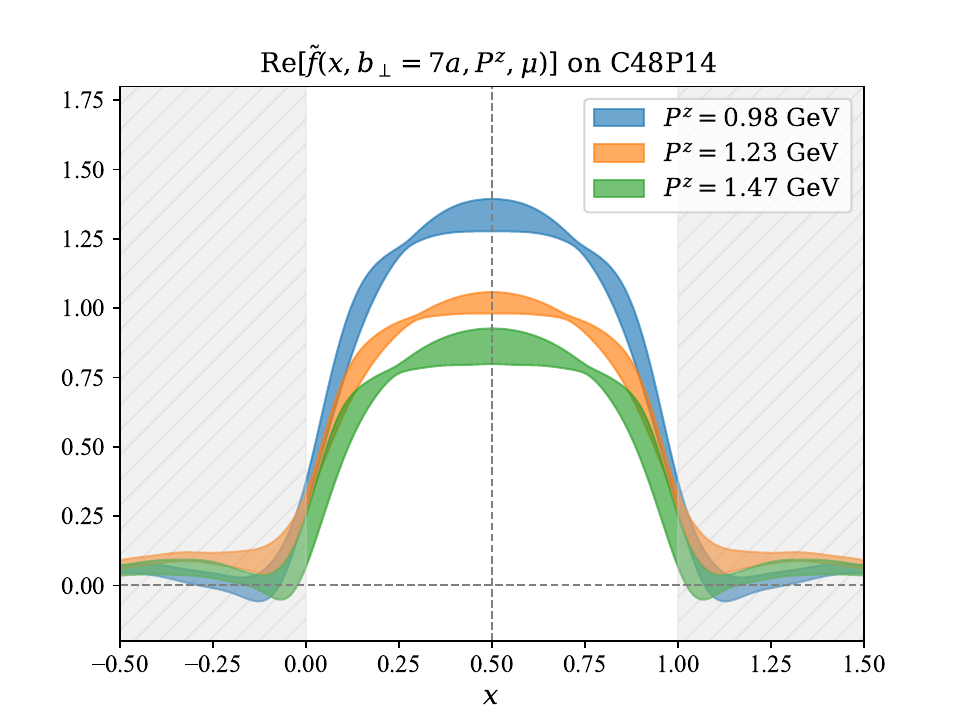}
\includegraphics[width=0.45\textwidth]{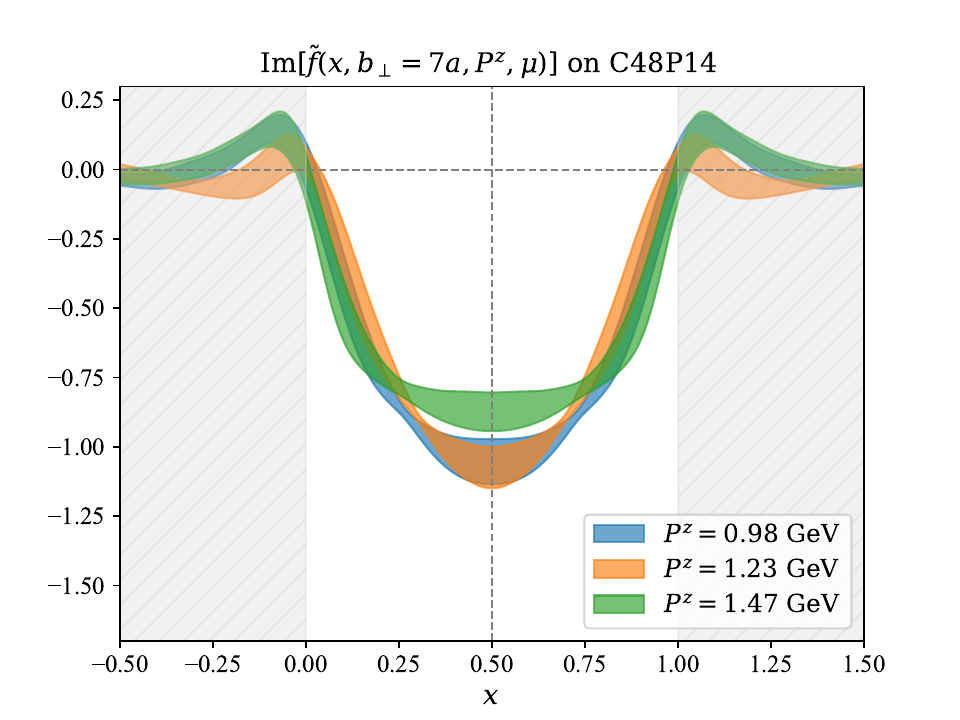}
\caption{Results from the C48P14 ensemble . A comparison of the results for $P^z = \lbrace 0.98, 1.23, 1.47 \rbrace \, \mathrm{GeV}$ at different $b_{\perp}$ values is presented in the three panels.} 
\label{fig:more_WFs_1}
\end{figure}

\begin{figure}
\centering
\includegraphics[width=0.45\textwidth]{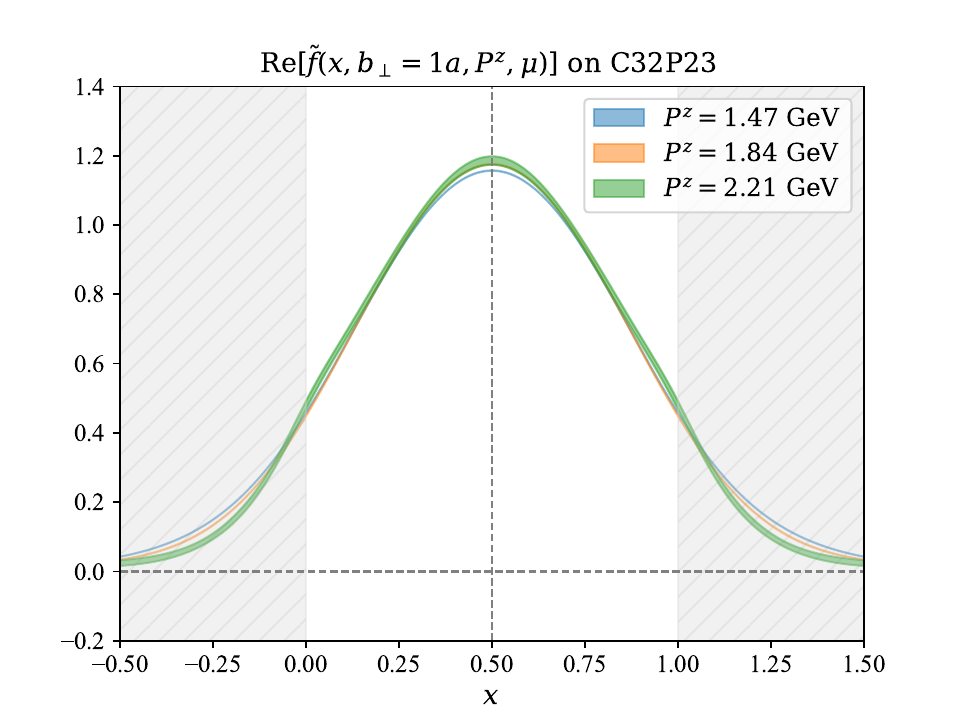}
\includegraphics[width=0.45\textwidth]{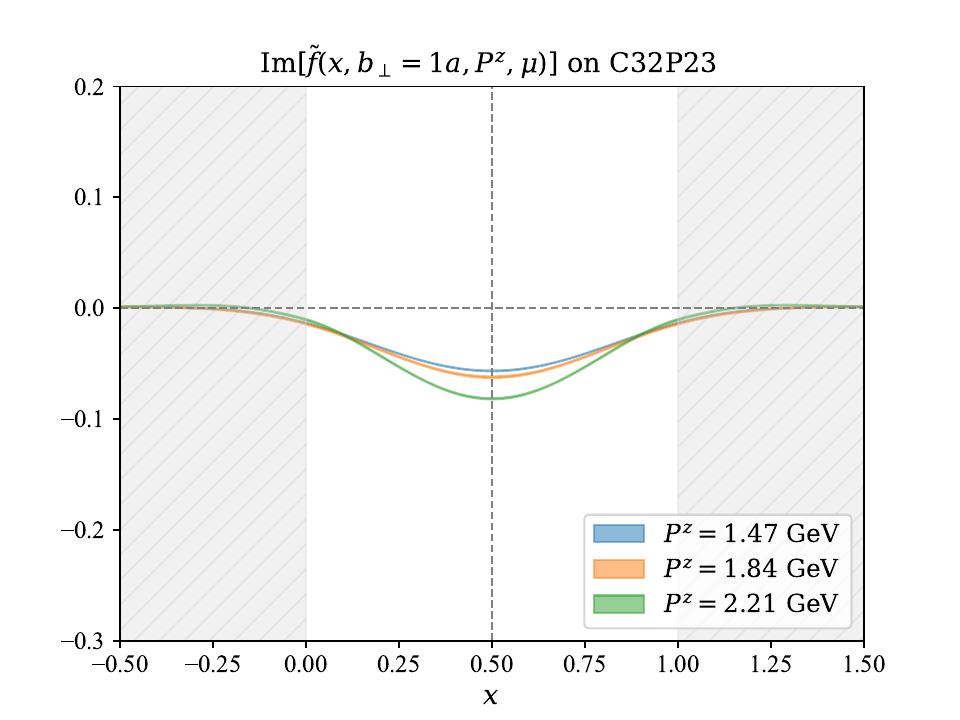}
\includegraphics[width=0.45\textwidth]{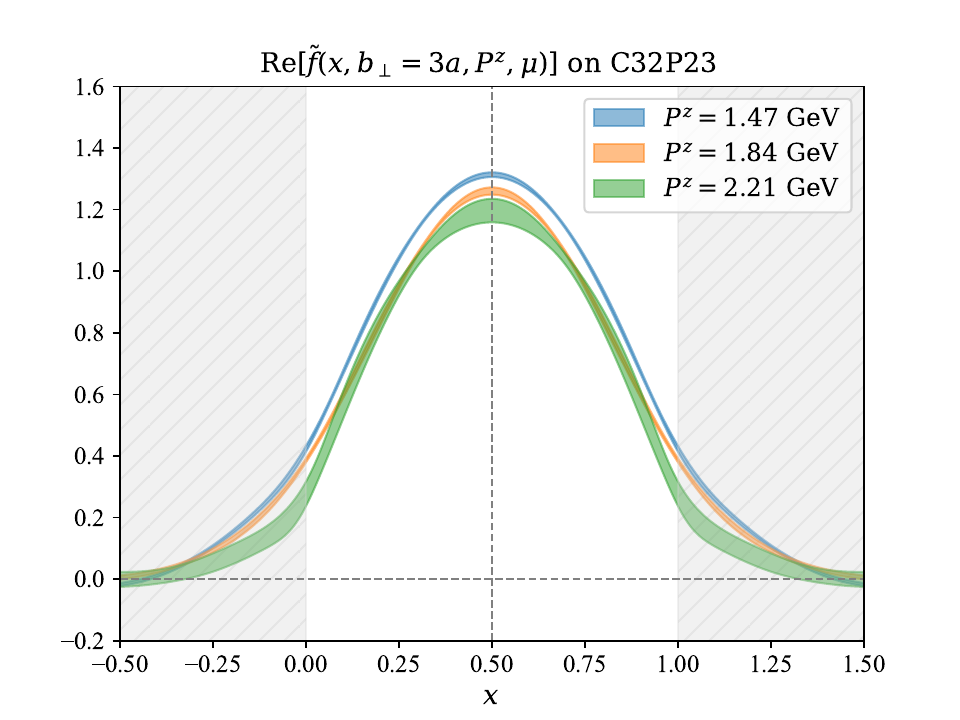}
\includegraphics[width=0.45\textwidth]{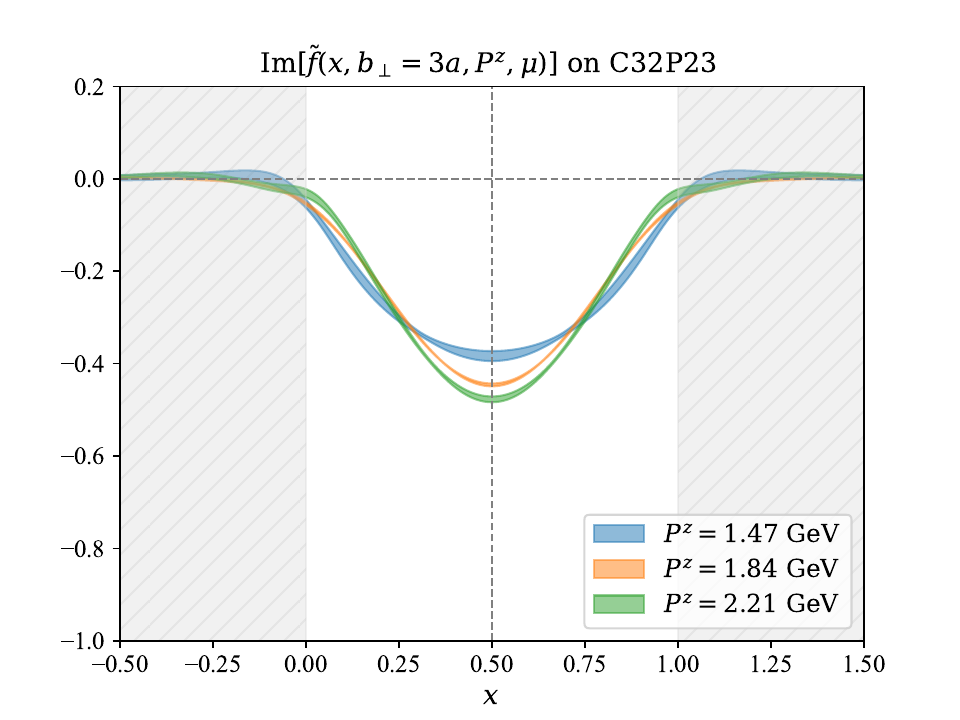}
\includegraphics[width=0.45\textwidth]{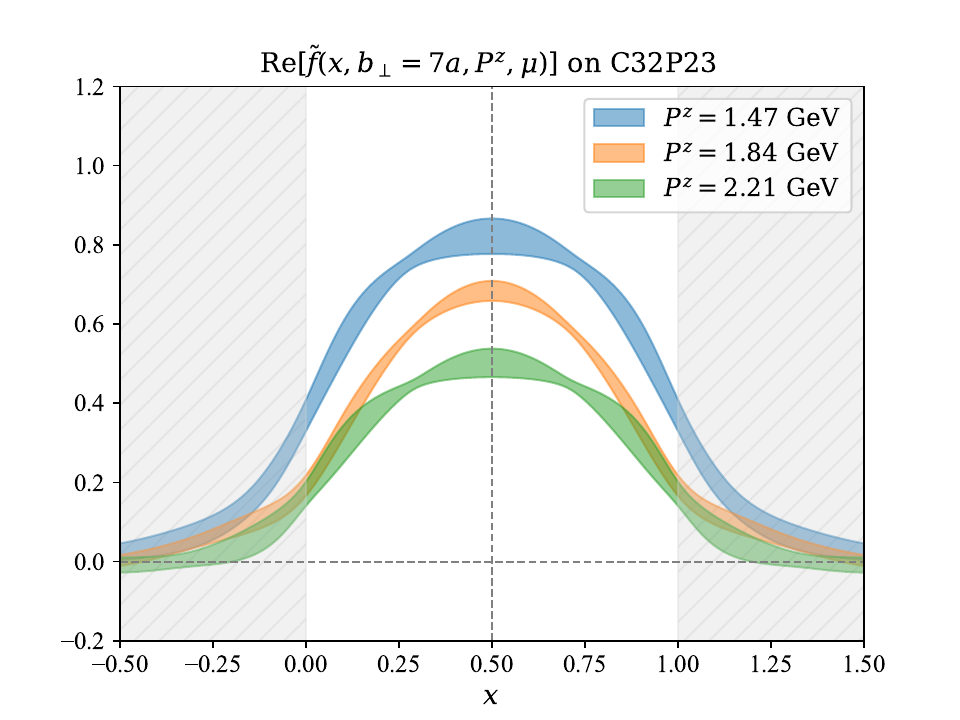}
\includegraphics[width=0.45\textwidth]{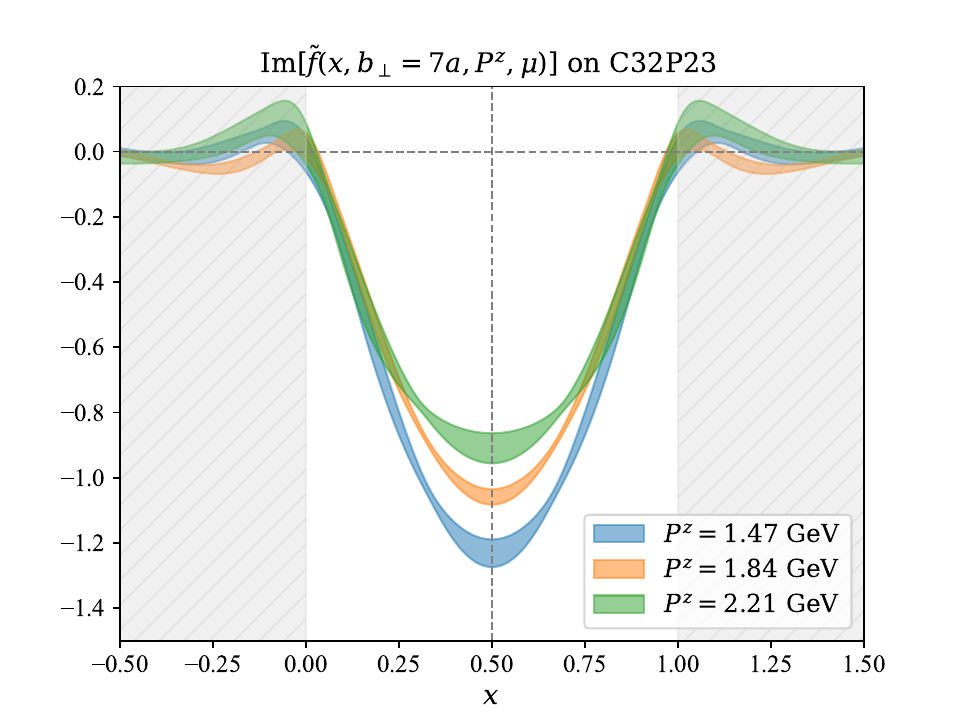}
\caption{Results from the C32P23 ensemble . A comparison of the results for $P^z = \lbrace 1.47, 1.84, 2.21 \rbrace \, \mathrm{GeV}$ at different $b_{\perp}$ values is presented in the three panels.} 
\label{fig:more_WFs_2}
\end{figure}

\begin{figure}
\centering
\includegraphics[width=0.45\textwidth]{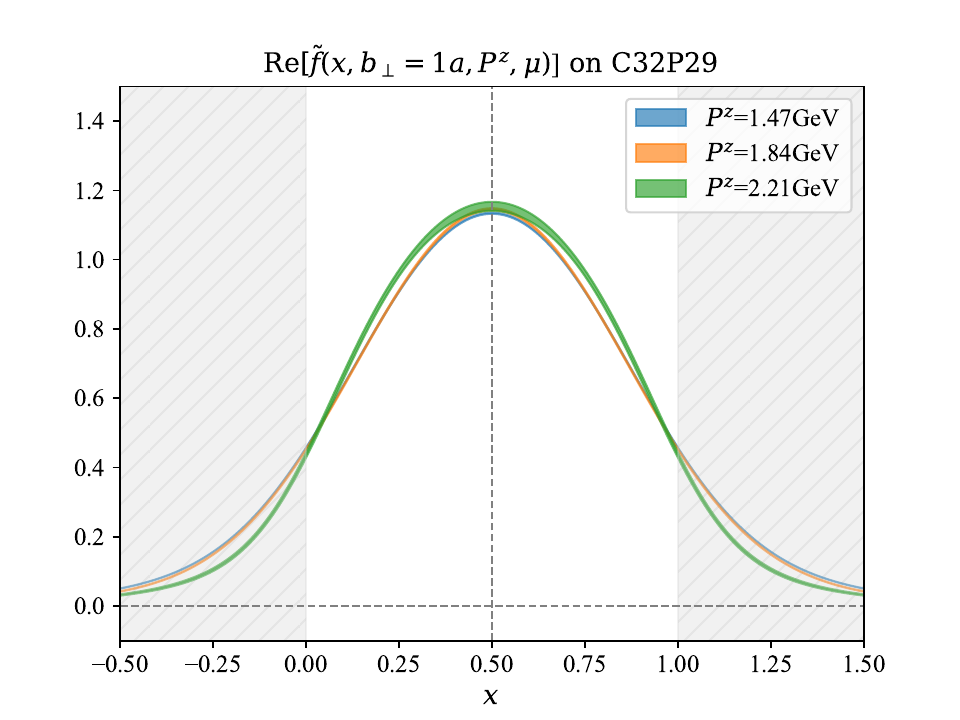}
\includegraphics[width=0.45\textwidth]{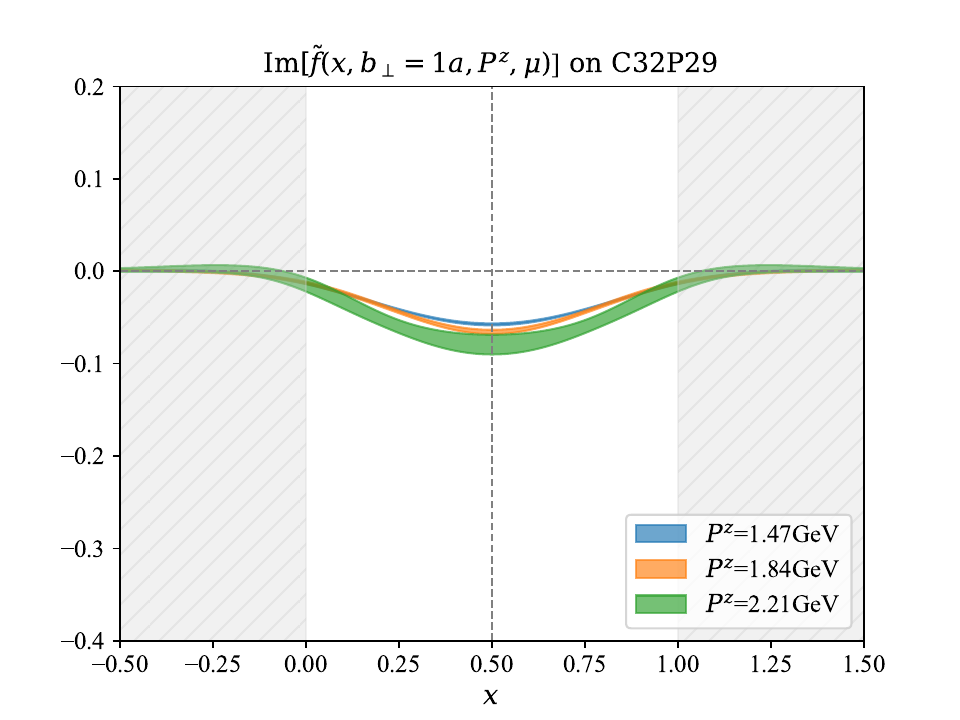}
\includegraphics[width=0.45\textwidth]{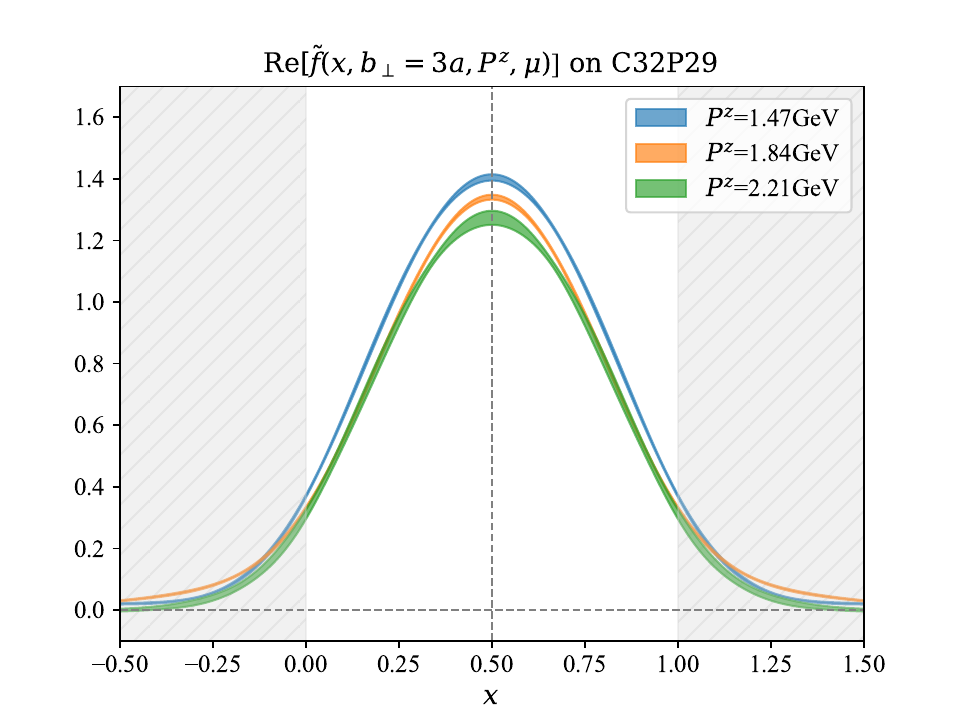}
\includegraphics[width=0.45\textwidth]{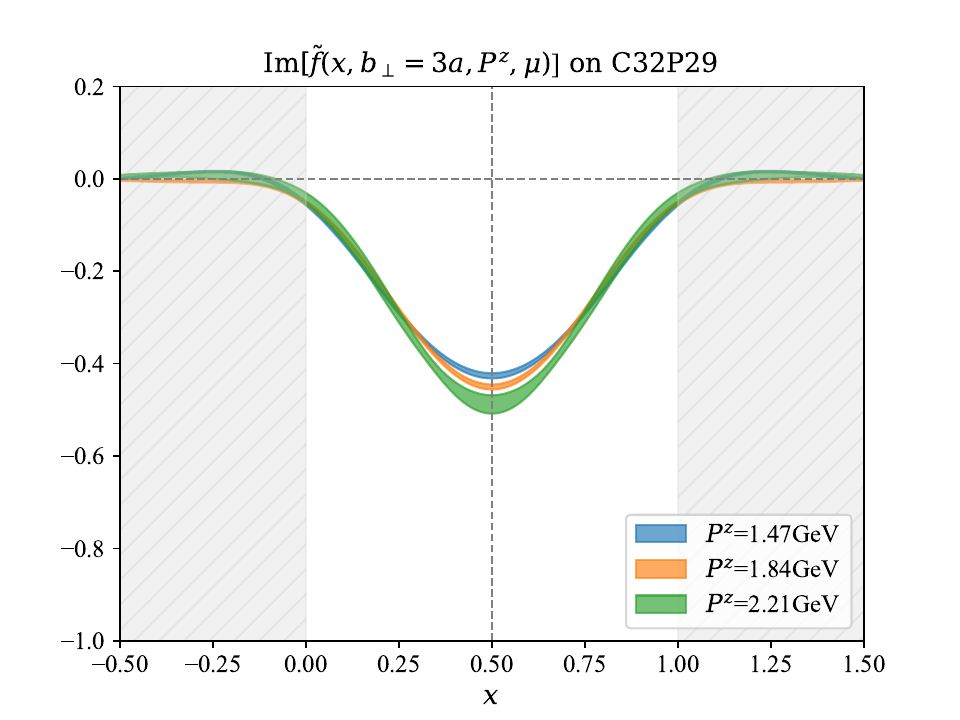}
\includegraphics[width=0.45\textwidth]{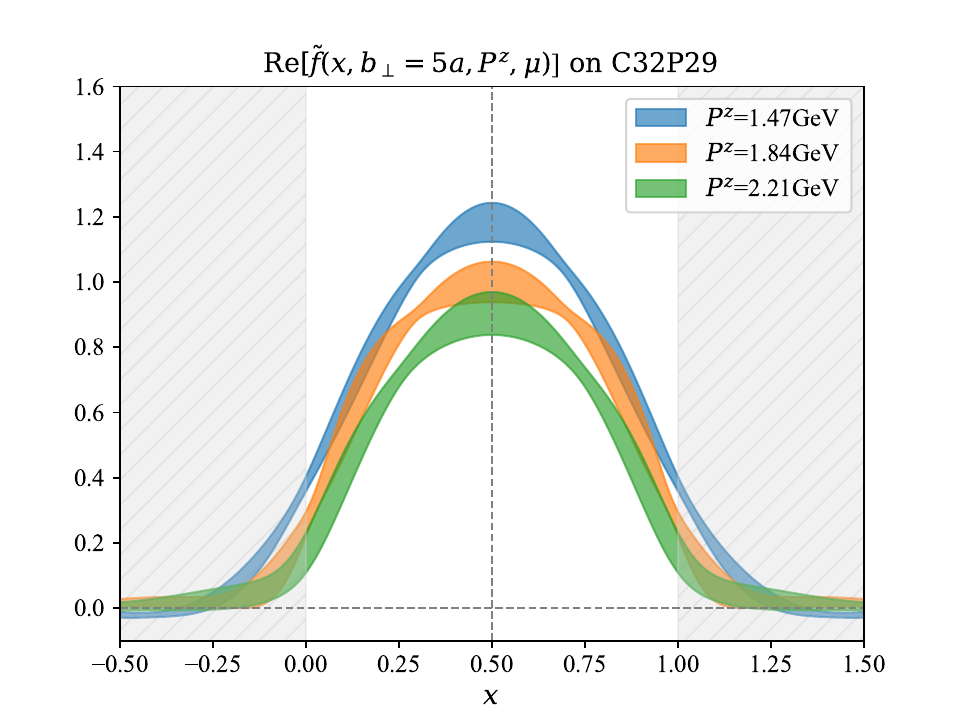}
\includegraphics[width=0.45\textwidth]{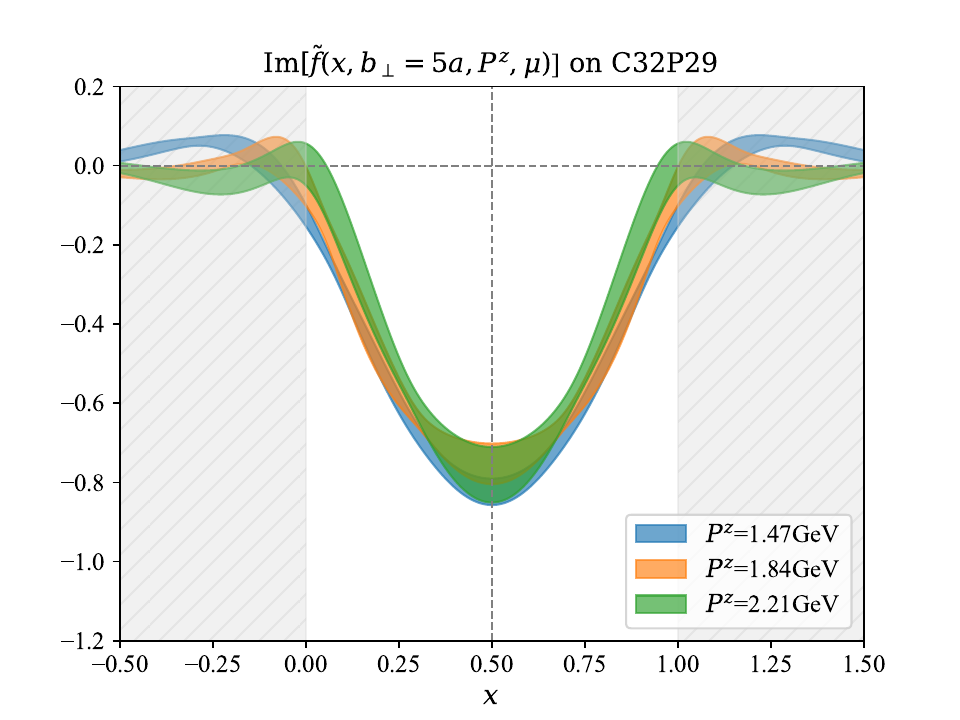}
\caption{Results from the C32P29 ensemble . A comparison of the results for $P^z = \lbrace 1.47, 1.84, 2.21 \rbrace \, \mathrm{GeV}$ at different $b_{\perp}$ values is presented in the three panels.} 
\label{fig:more_WFs_3}
\end{figure}

\begin{figure}
\centering
\includegraphics[width=0.45\textwidth]{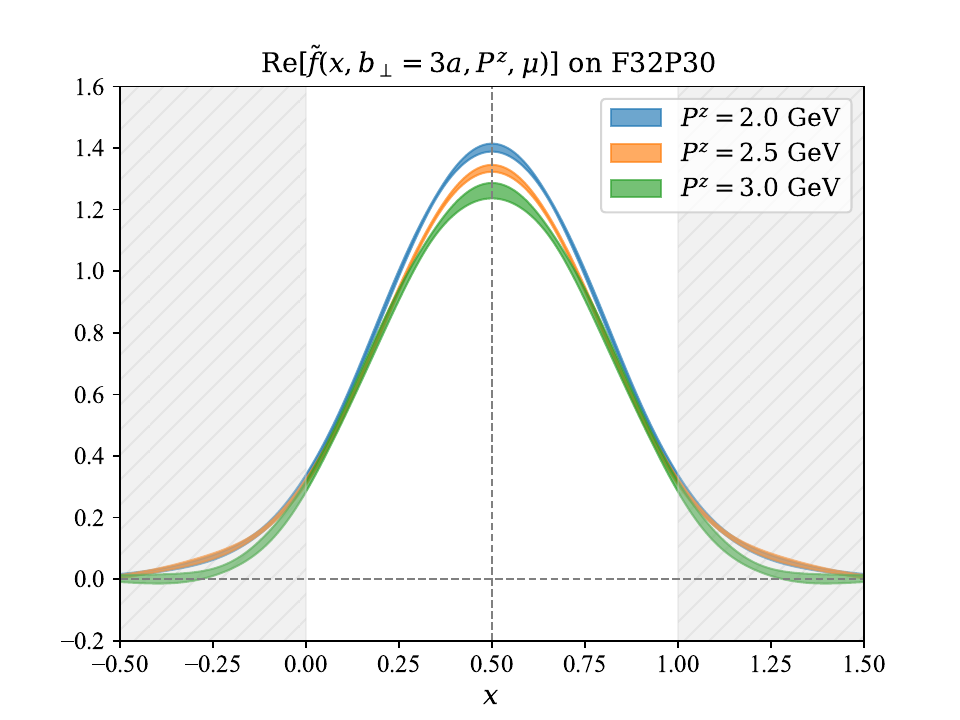}
\includegraphics[width=0.45\textwidth]{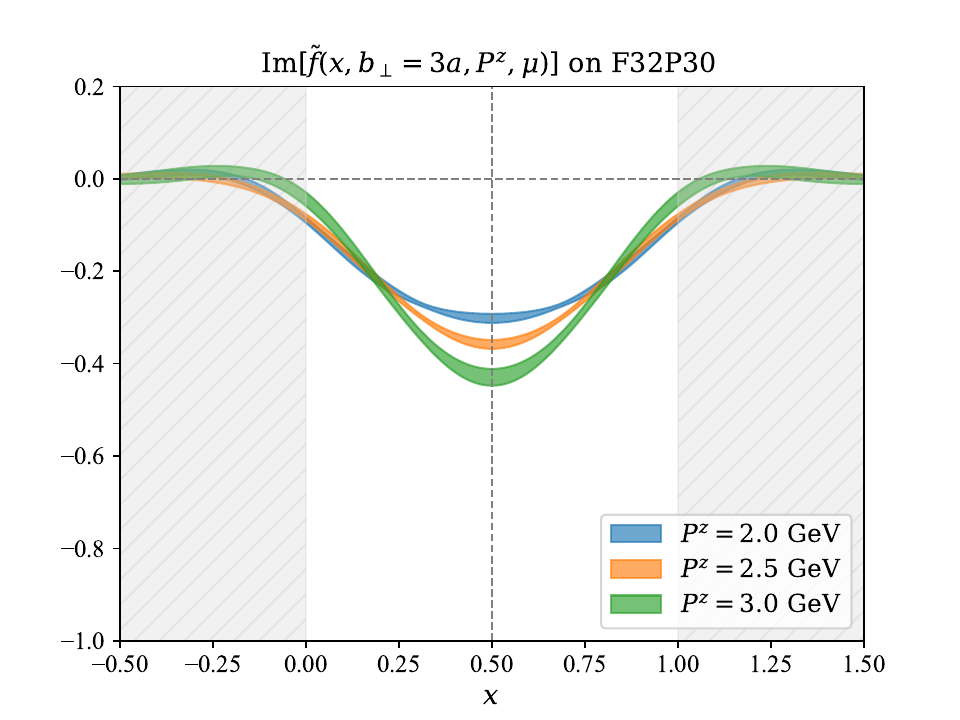}
\includegraphics[width=0.45\textwidth]{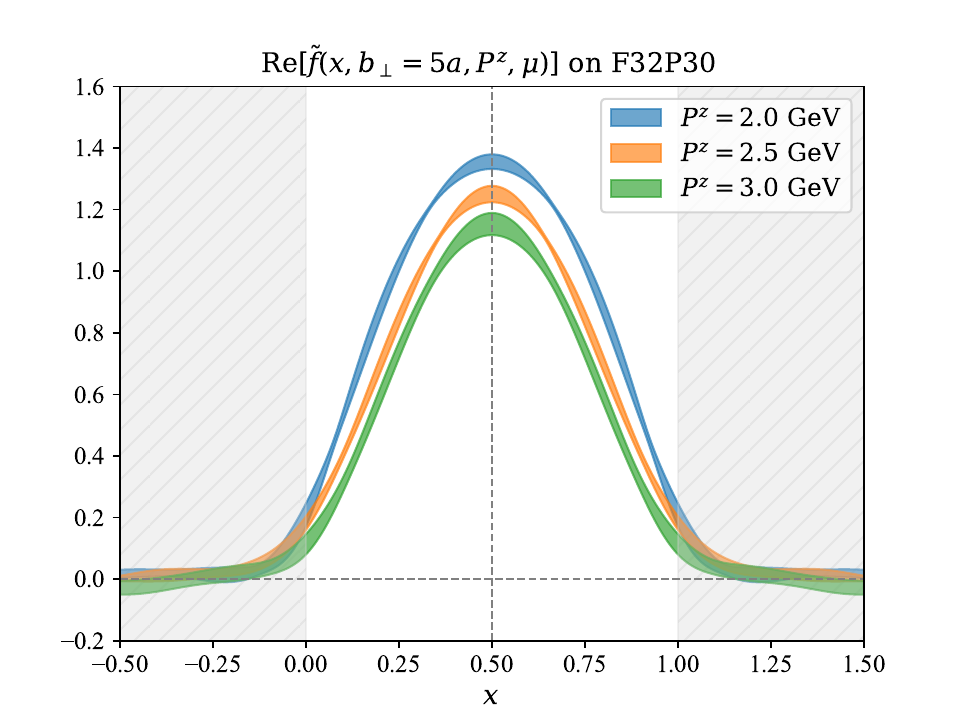}
\includegraphics[width=0.45\textwidth]{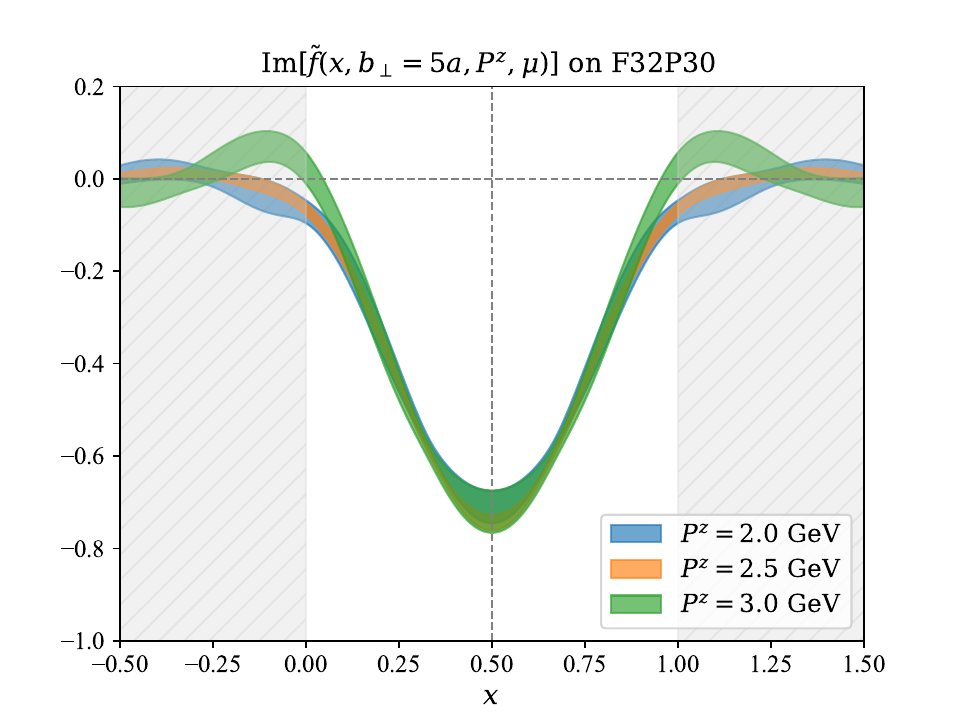}
\includegraphics[width=0.45\textwidth]{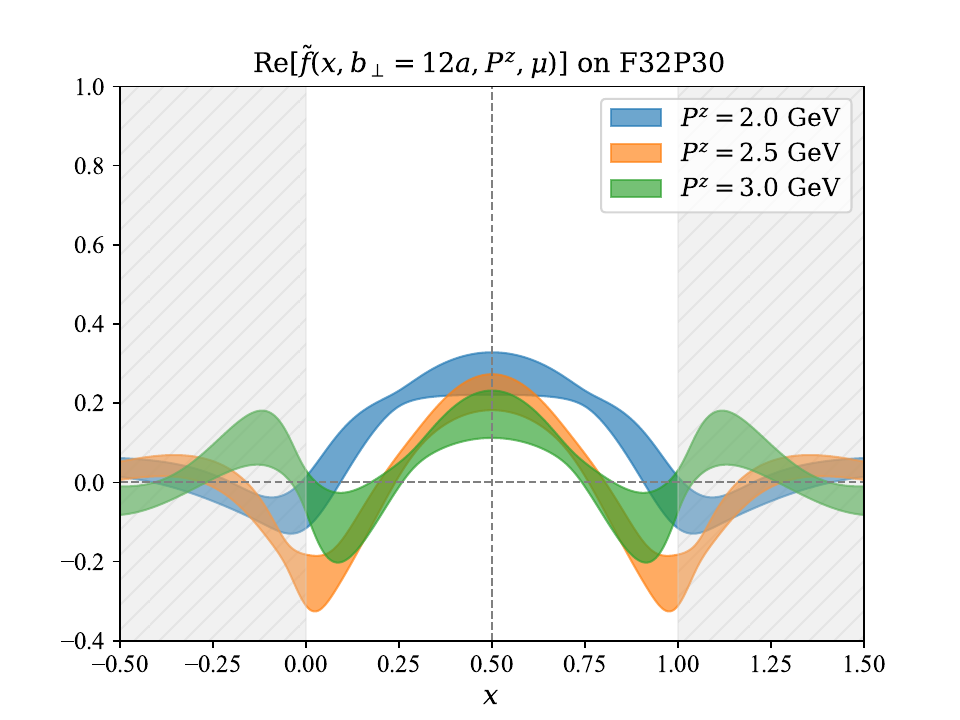}
\includegraphics[width=0.45\textwidth]{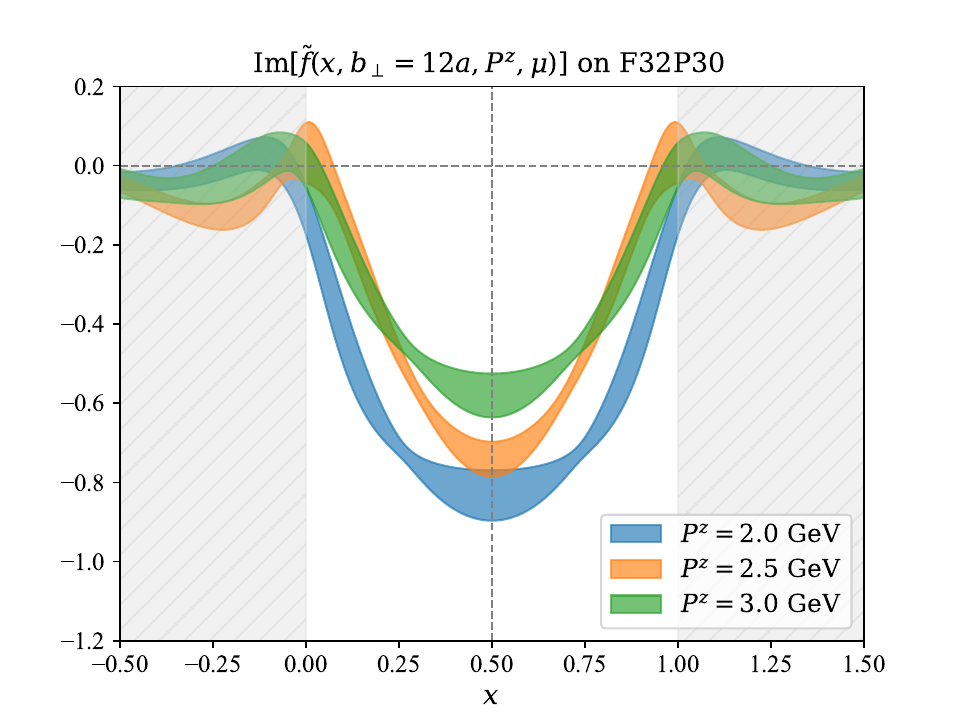}
\caption{Results from the F32P30 ensemble . A comparison of the results for $P^z = \lbrace 2.0, 2.5, 3.0 \rbrace \, \mathrm{GeV}$ at different $b_{\perp}$ values is presented in the three panels.} 
\label{fig:more_WFs_4}
\end{figure}

\subsection{Results for CS kernel}
More results for the CS kernel on different ensembles are given in Fig.~\ref{fig:more_CSK_1}, ~\ref{fig:more_CSK_2}, ~\ref{fig:more_CSK_3} and ~\ref{fig:more_CSK_4}.

\begin{figure}
\centering
\includegraphics[width=0.45\textwidth]{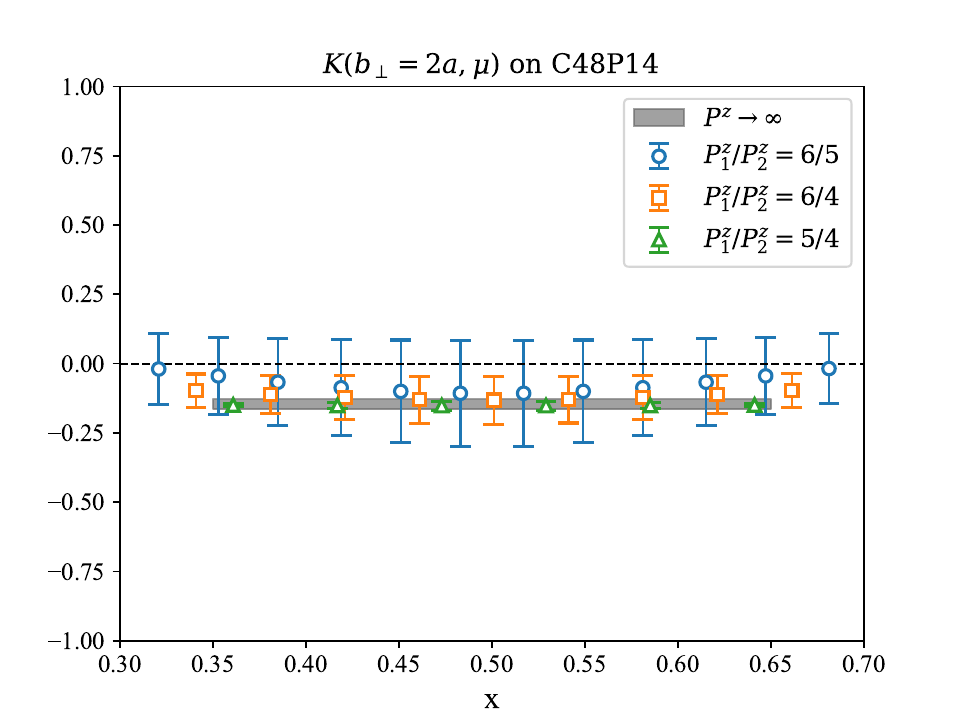}
\includegraphics[width=0.45\textwidth]{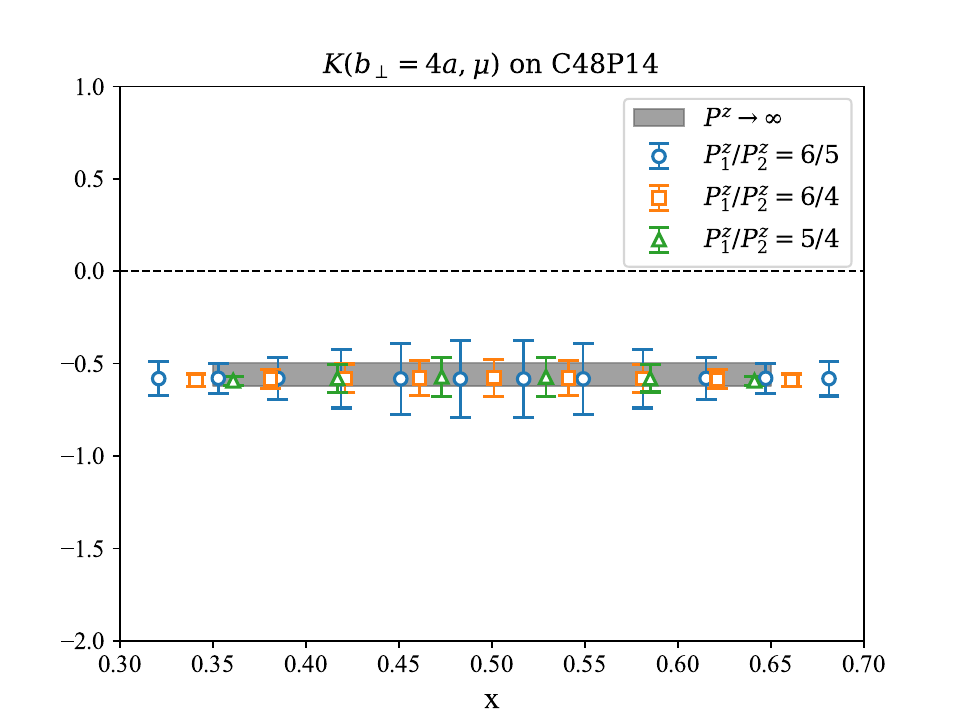}
\includegraphics[width=0.45\textwidth]{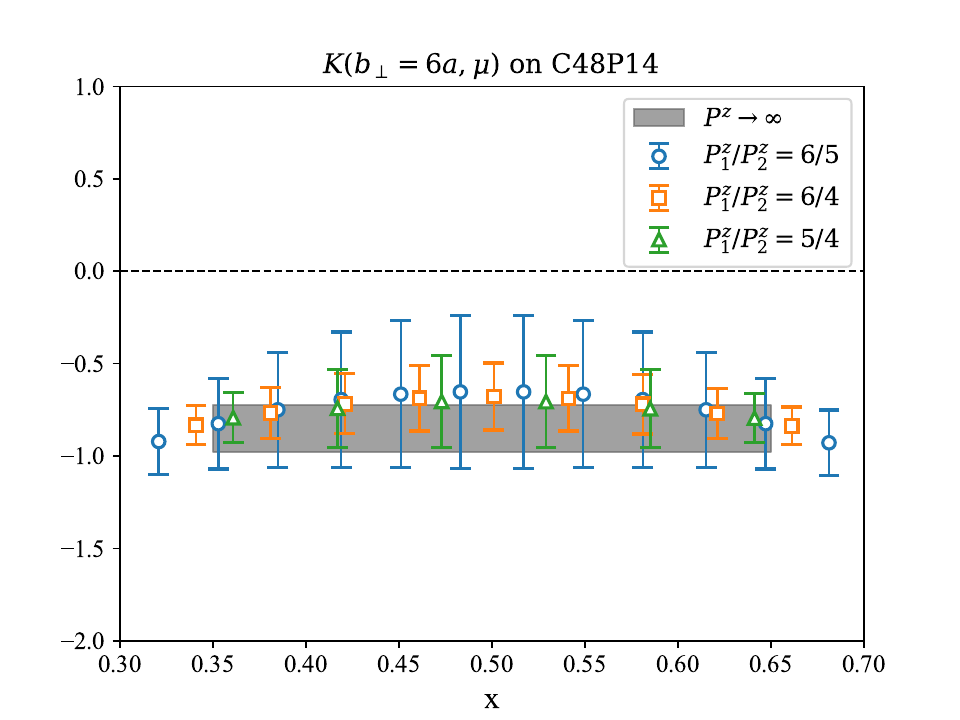}
\includegraphics[width=0.45\textwidth]{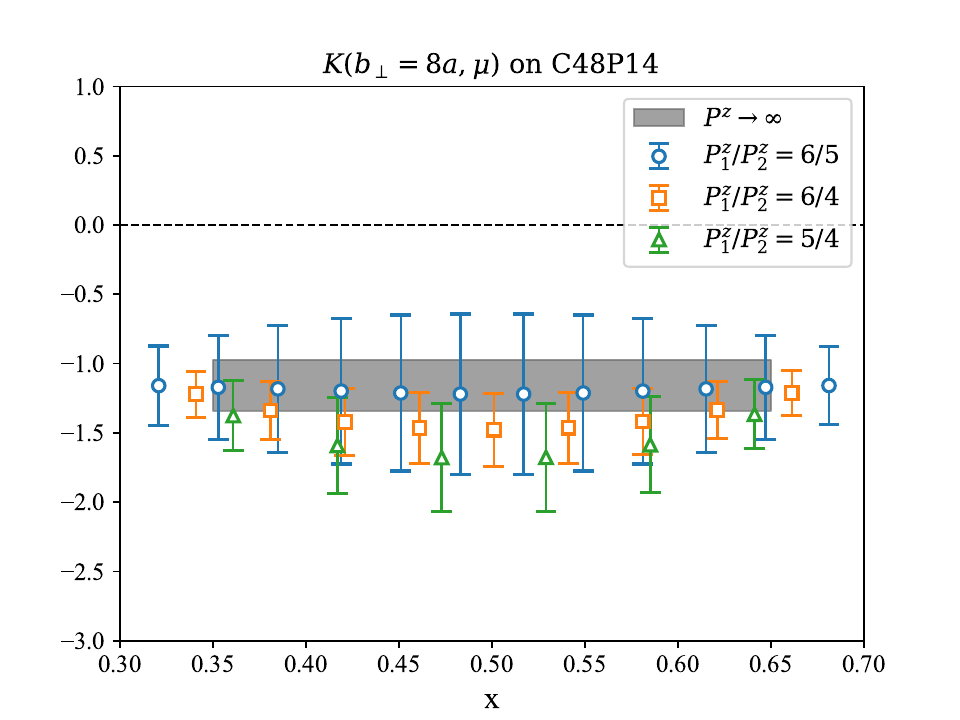}
\includegraphics[width=0.45\textwidth]{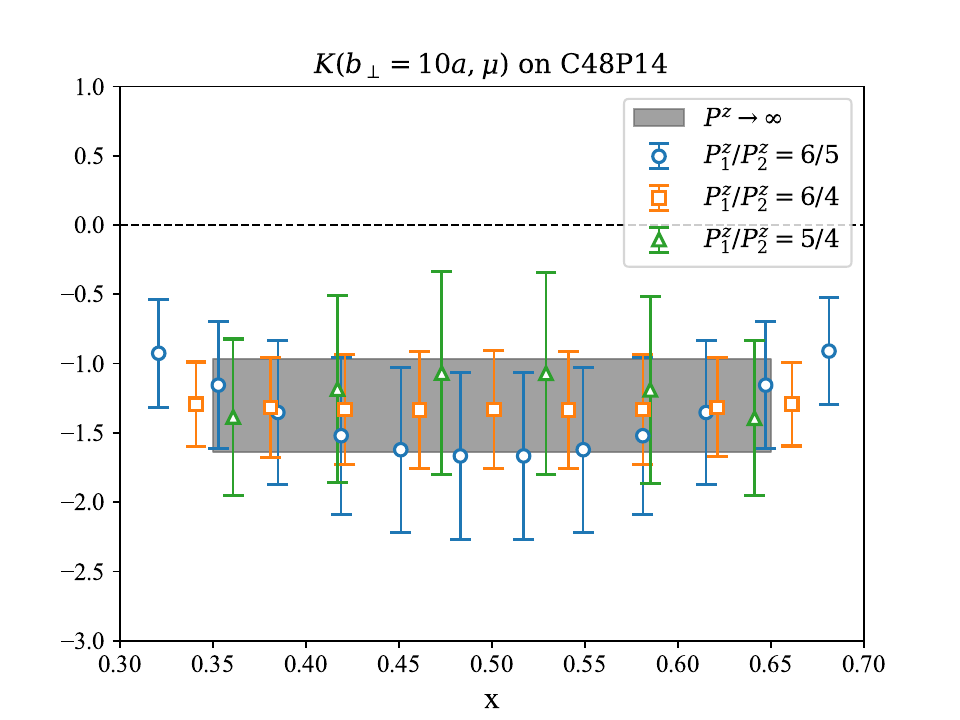}
\caption{The results of $b_\perp = \{2, 4, 6, 8, 10\} a$ on the C48P14 ensemble are shown for three different momentum combinations.} 
\label{fig:more_CSK_1}
\end{figure}

\begin{figure}
\centering
\includegraphics[width=0.45\textwidth]{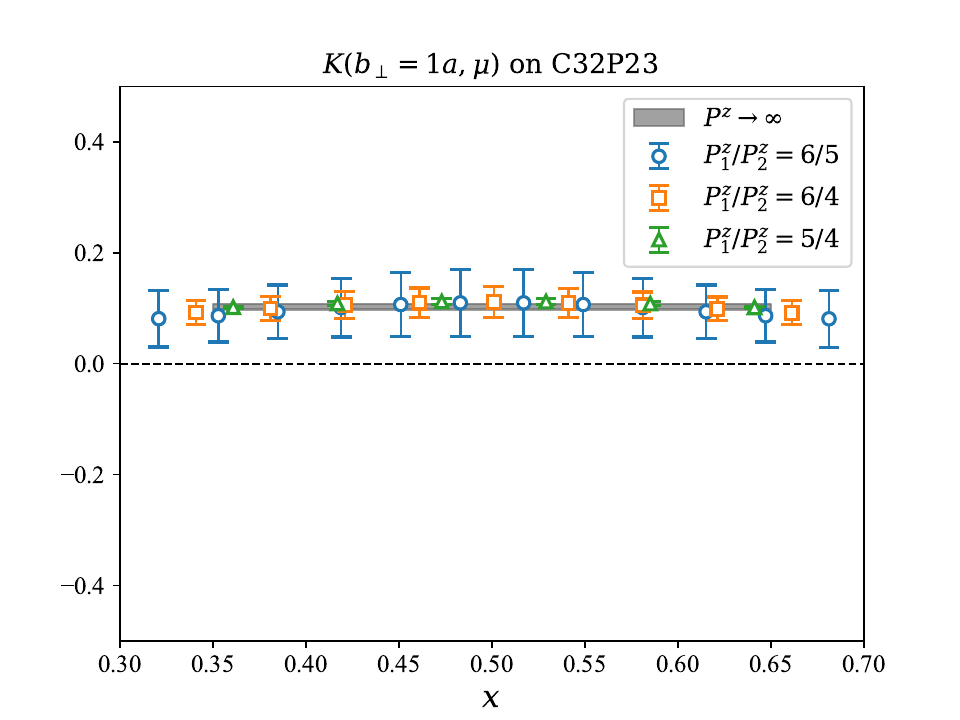}
\includegraphics[width=0.45\textwidth]{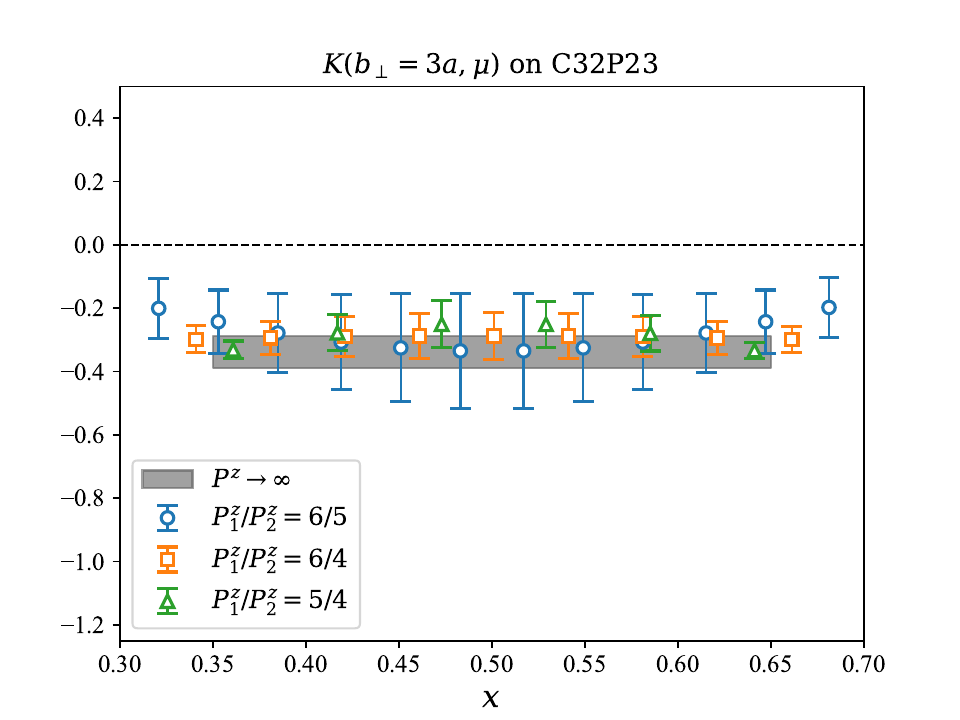}
\includegraphics[width=0.45\textwidth]{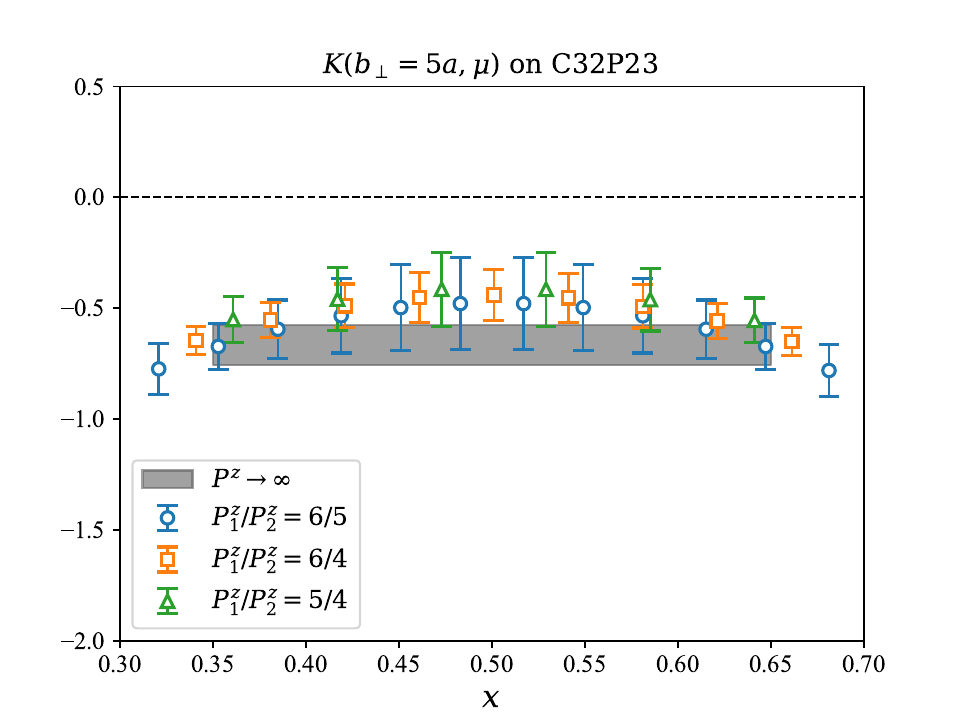}
\includegraphics[width=0.45\textwidth]{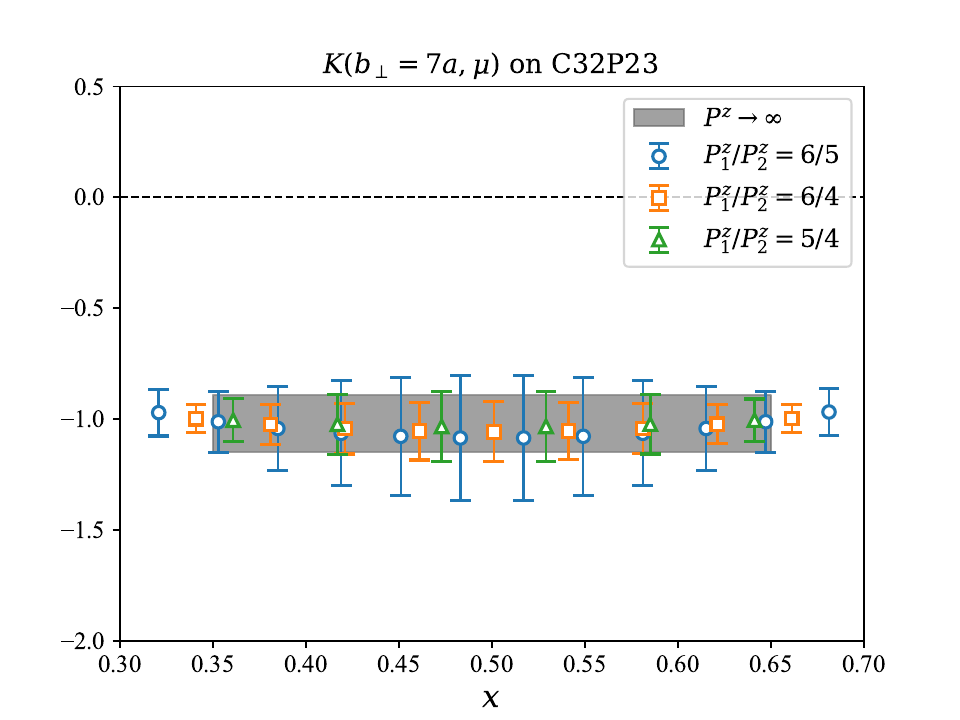}
\includegraphics[width=0.45\textwidth]{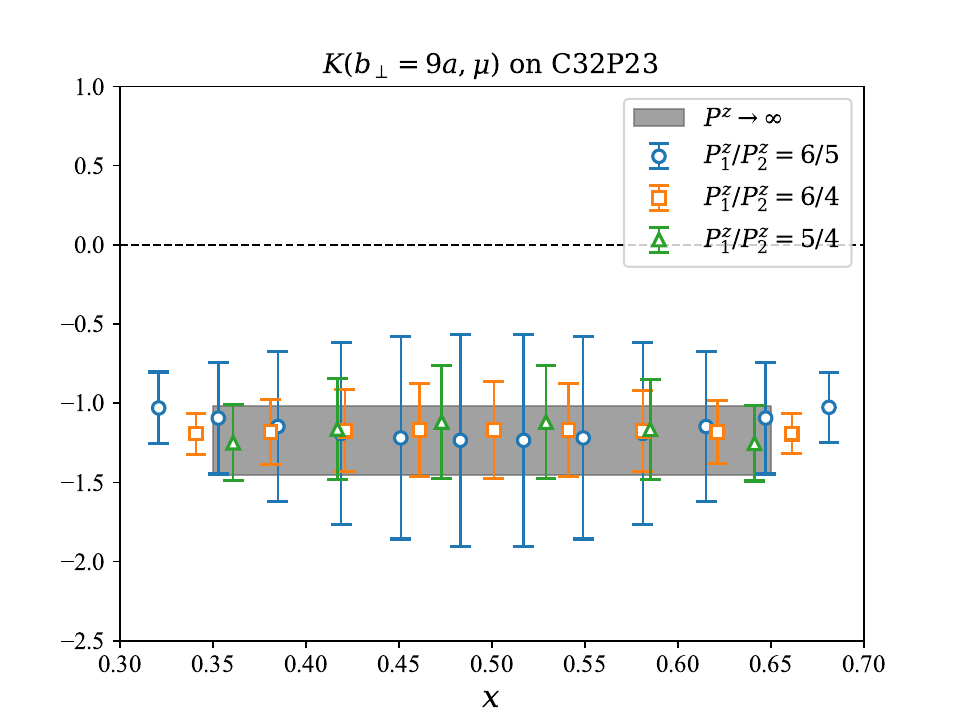}
\caption{The results of $b_\perp = \{1, 3, 5, 7, 9\}a$ on the C23P23 ensemble are shown for three different momentum combinations.} 
\label{fig:more_CSK_2}
\end{figure}

\begin{figure}
\centering
\includegraphics[width=0.45\textwidth]{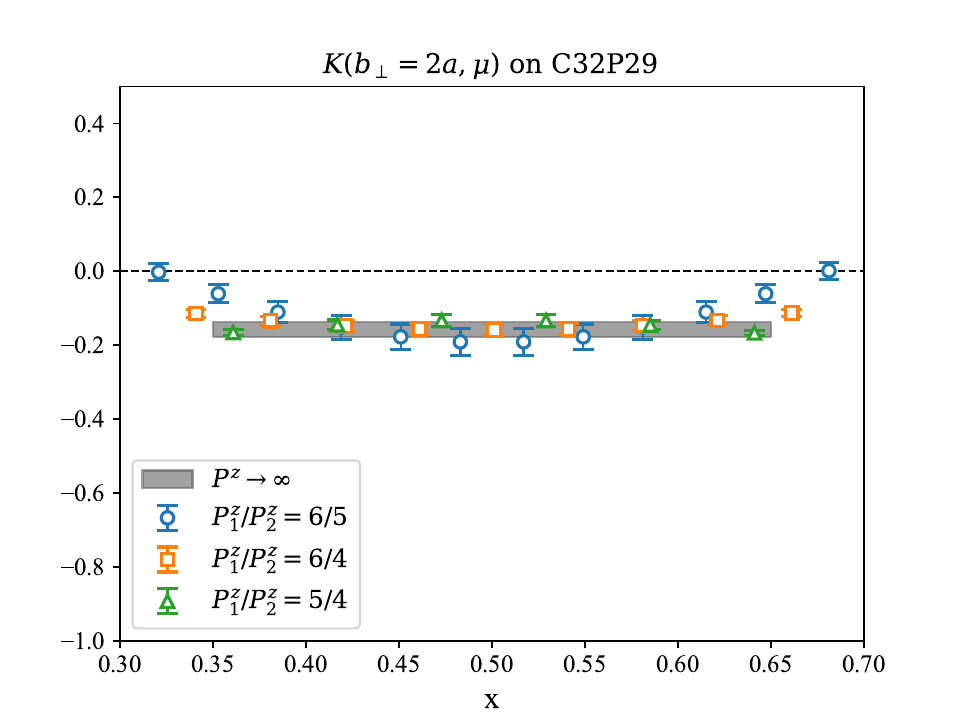}
\includegraphics[width=0.45\textwidth]{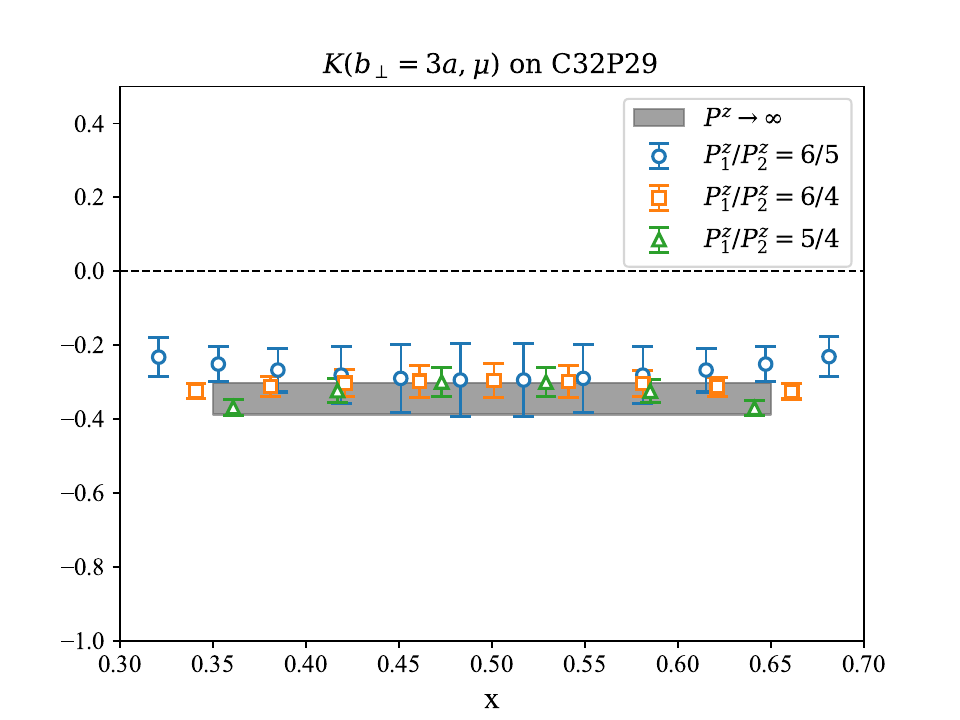}
\includegraphics[width=0.45\textwidth]{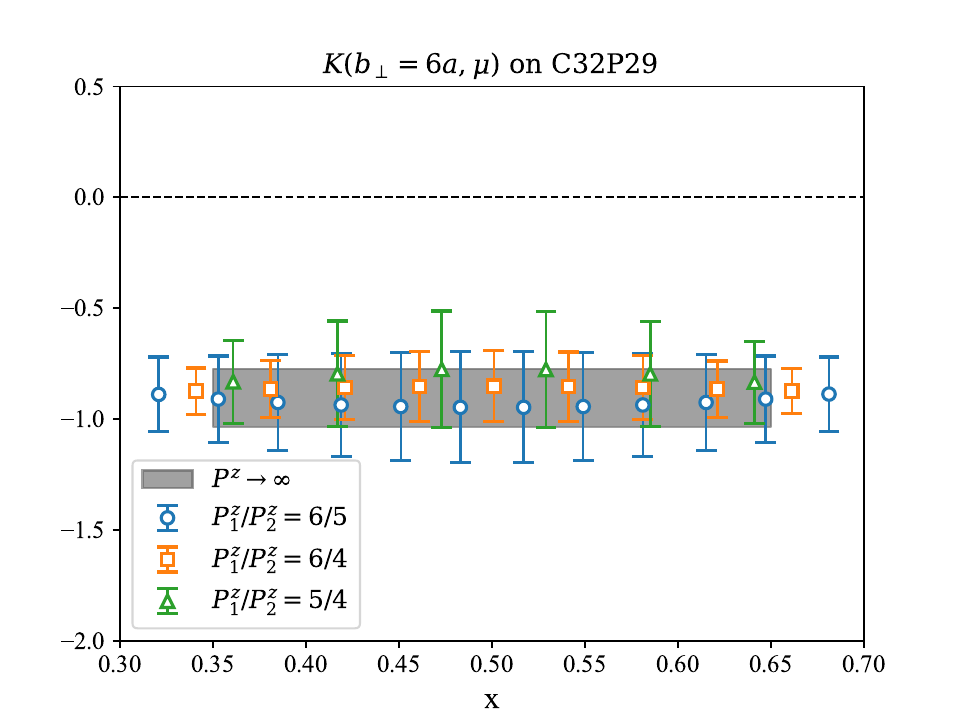}
\includegraphics[width=0.45\textwidth]{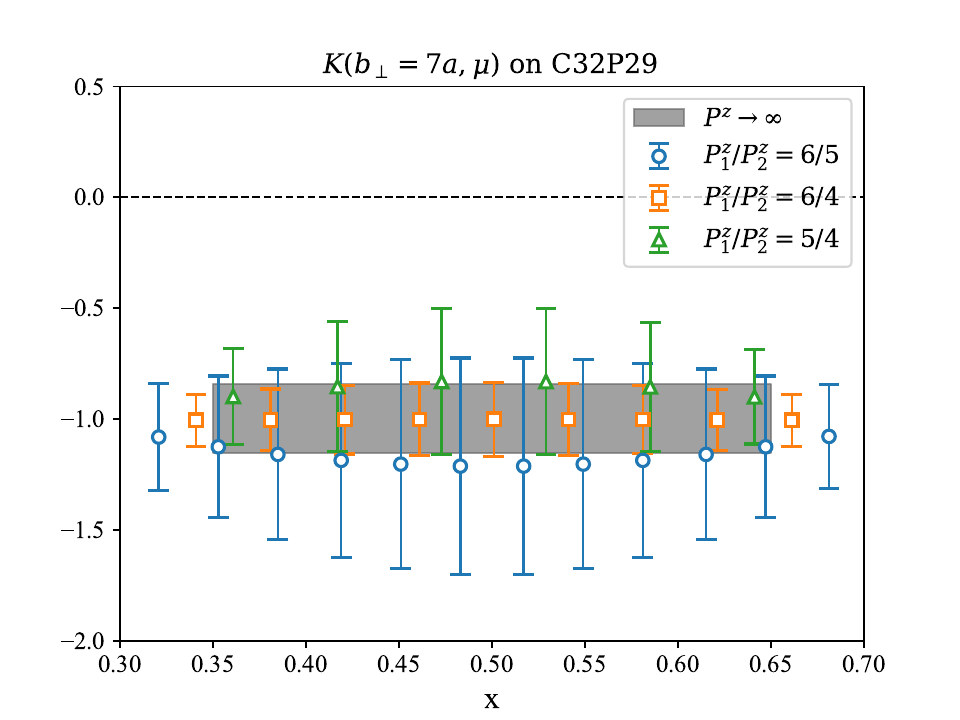}
\includegraphics[width=0.45\textwidth]{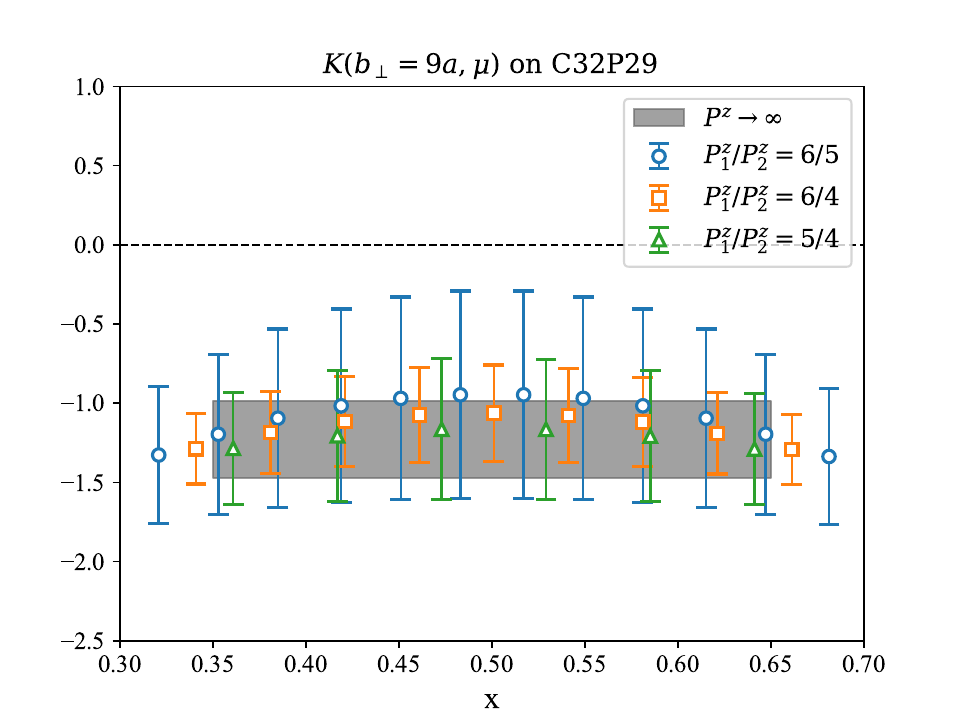}
\includegraphics[width=0.45\textwidth]{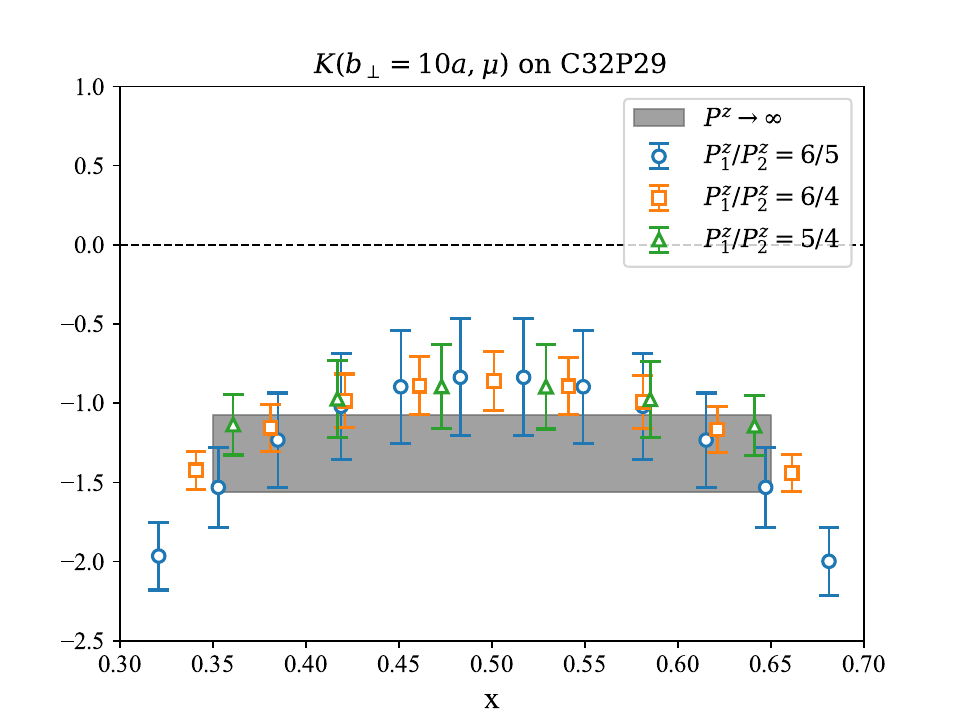}
\caption{The results of $b_\perp = \{2, 3, 6, 7, 9, 10\}a$ on the C32P29 ensemble are shown for three different momentum combinations.} 
\label{fig:more_CSK_3}
\end{figure}

\begin{figure}
\centering
\includegraphics[width=0.45\textwidth]{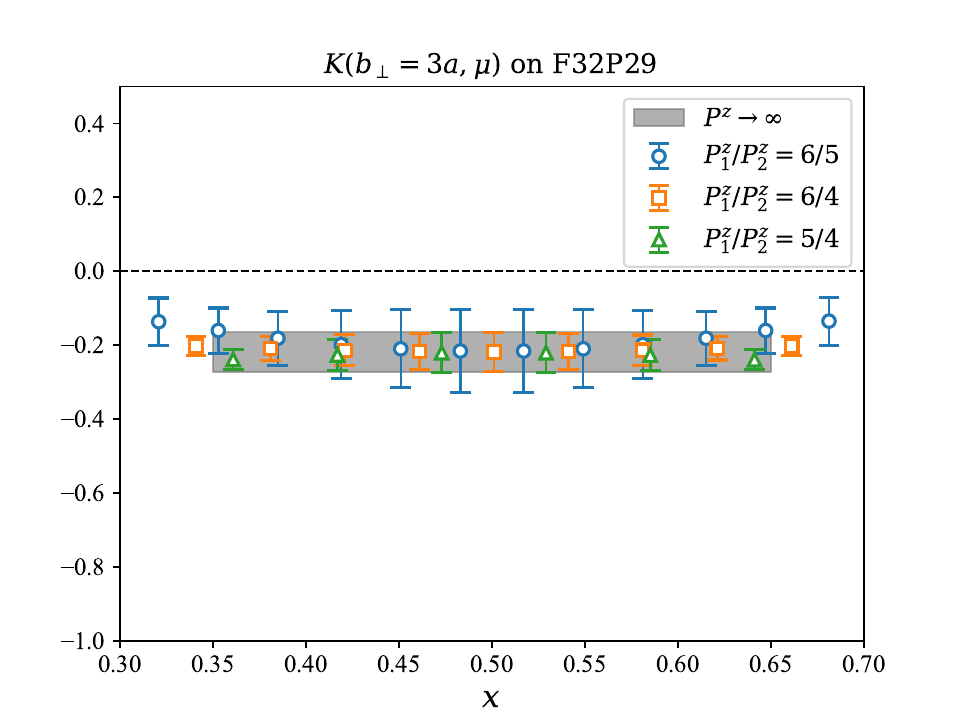}
\includegraphics[width=0.45\textwidth]{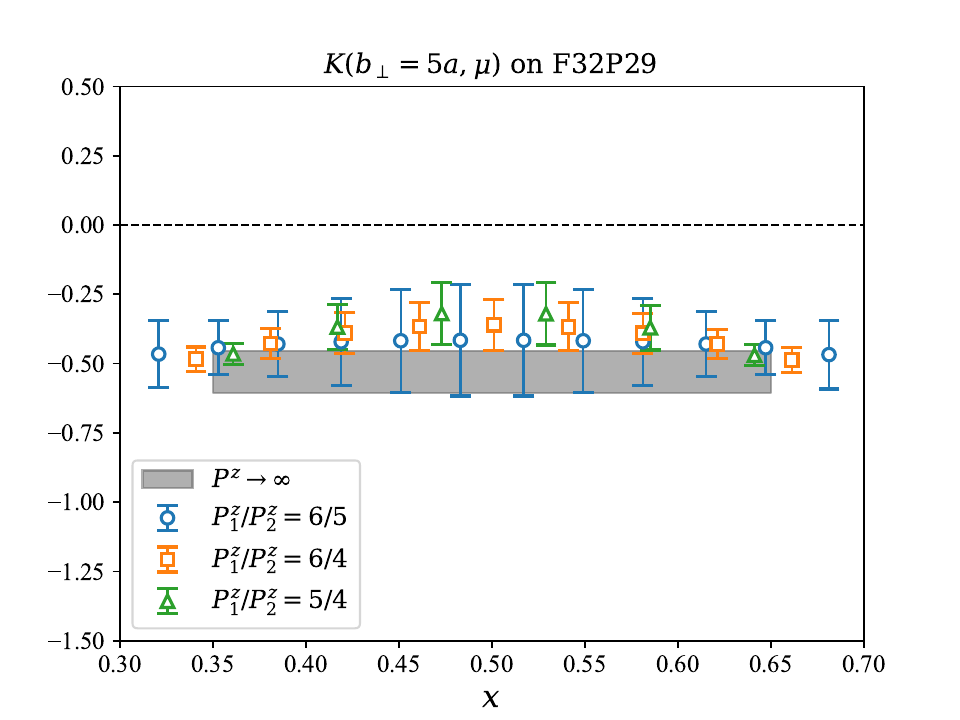}
\includegraphics[width=0.45\textwidth]{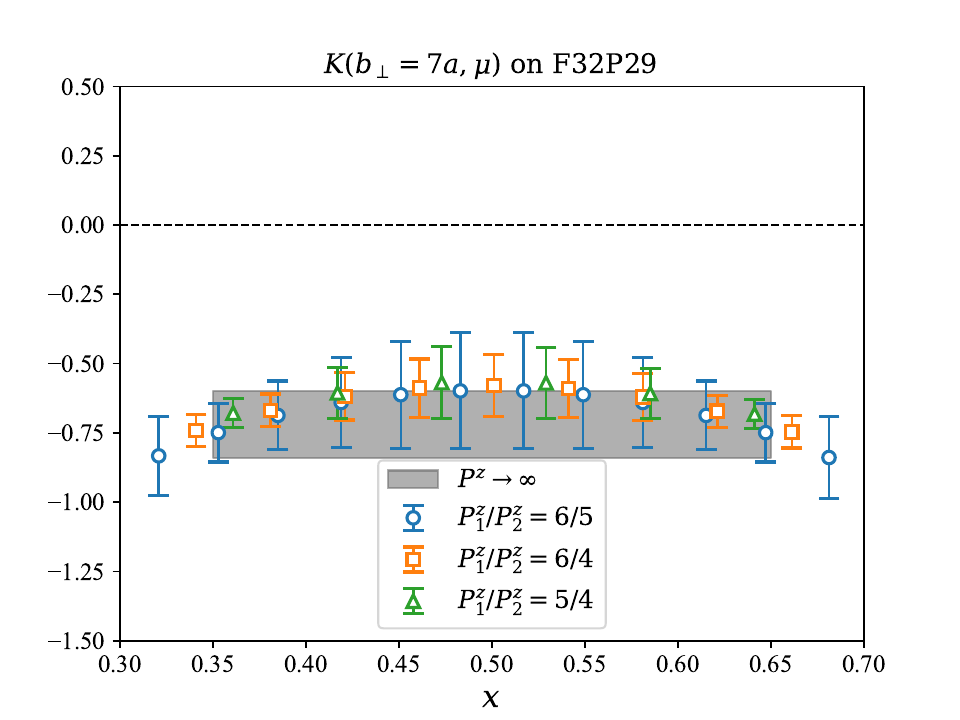}
\includegraphics[width=0.45\textwidth]{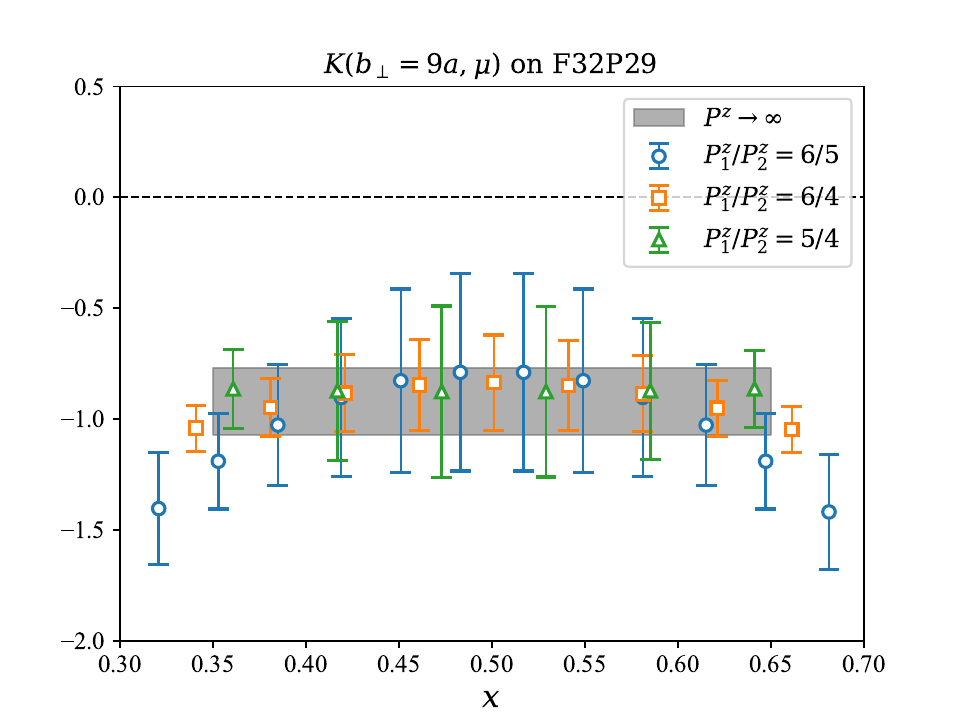}
\includegraphics[width=0.45\textwidth]{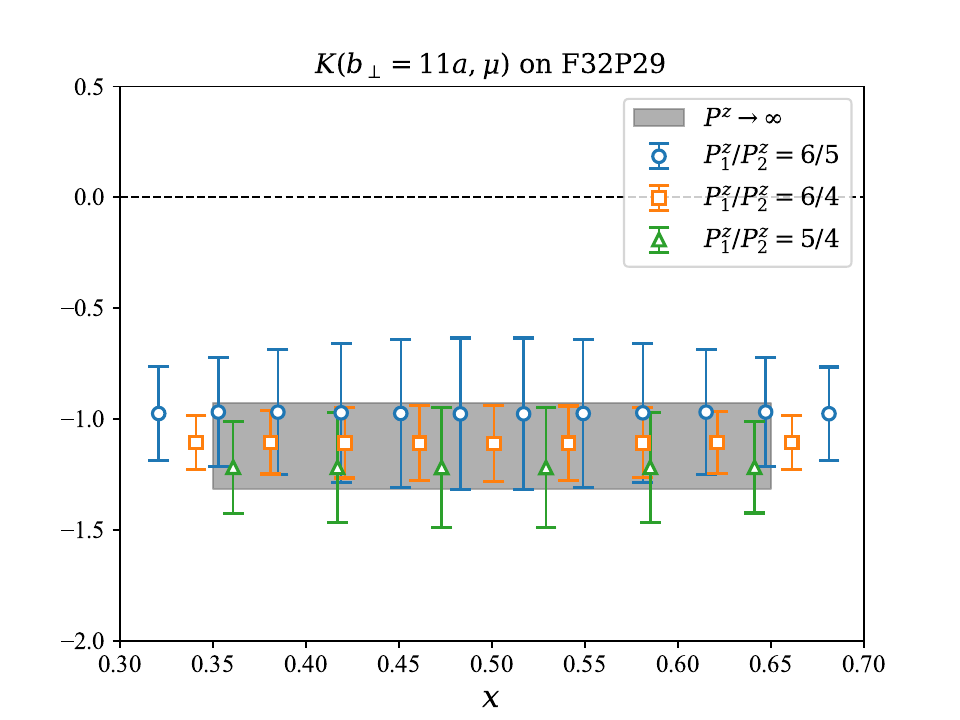}
\includegraphics[width=0.45\textwidth]{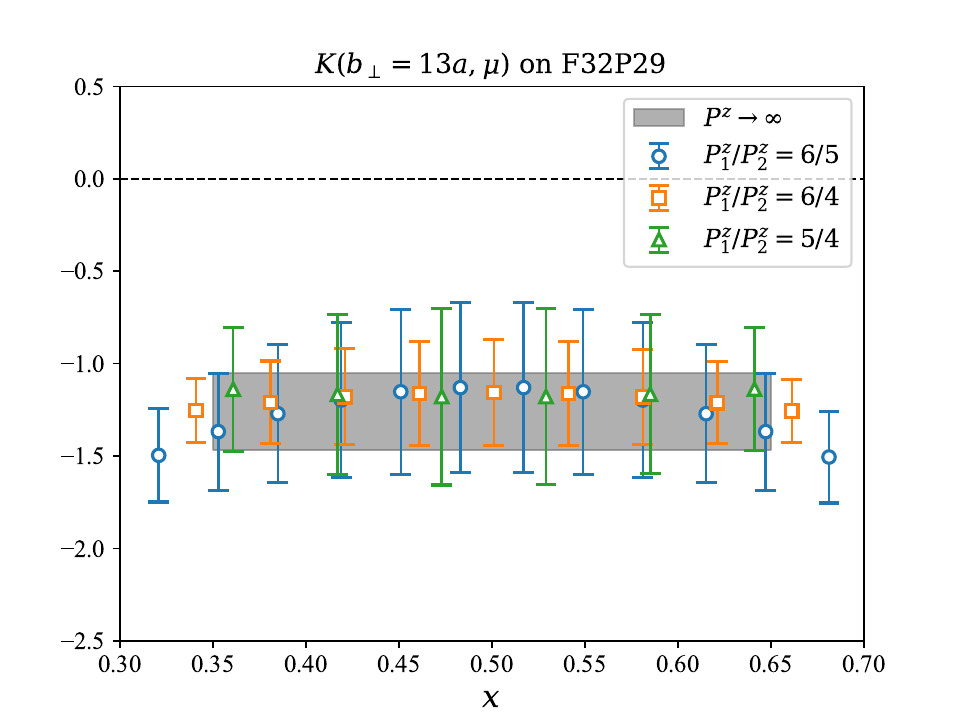}
\caption{The results of $b_\perp = \{3, 5, 7, 9, 11, 13\} a$ on the F32P30 ensemble are shown for three different momentum combinations.} 
\label{fig:more_CSK_4}
\end{figure}

\end{widetext}
\end{appendix}

\clearpage
\bibliographystyle{apsrev4-1} 
\bibliography{main_ref} 

\end{document}